\documentclass[traditabstract]{aa}

\usepackage{graphicx}
\usepackage{txfonts}
\usepackage{natbib}
\usepackage{enumerate}
\usepackage{bm}
\usepackage{hyperref}
\usepackage{fixltx2e}
\usepackage{afterpage}
\hypersetup{
  colorlinks=true,
  linkcolor=blue,
  citecolor=blue,
  urlcolor=blue,
}

\bibpunct{(}{)}{;}{a}{}{,} % to follow A\&A style

\newcommand{\sfr}{M$_\odot$\,yr$^{-1}$}

\newcommand{\mjysr}{MJy\,sr$^{-1}$}

\newcommand{\kms}{km\,s$^{-1}$}

\newcommand{\ropt}{$R_{\rm opt}$}
\newcommand{\rdisk}{$R_{\rm d}$}
\newcommand{\umin}{$U_{\rm min}$}
\newcommand{\umax}{$U_{\rm max}$}
\newcommand{\ubar}{$\langle U \rangle$}
\newcommand{\qpah}{$q_{\rm PAH}$}
\newcommand{\fpdr}{$f_{\rm PDR}$}
\newcommand{\redchi}{$\chi^2_{\nu}$}

\newcommand{\spit}{{\it Spitzer}}
\newcommand{\hers}{{\it Herschel}}

\newcommand{\logoh}{12$+$log(O/H)}

\newcommand{\mdust}{$M_{\rm dust}$}
\newcommand{\td}{$T_{\rm dust}$}
\newcommand{\tddl}{$T_{\rm dust(DL07)}$}
\newcommand{\betadl}{$\beta_{\rm DL07}$}
\newcommand{\hi}{\rm H{\sc i}}

\newcommand{\micron}{$\mu$m}

\newcommand{\sigmad}{$\Sigma_{\rm dust}$}
\newcommand{\taud}{$\tau_{\rm dust}$}

\def\tex {\ifmmode{{T}_{\rm ex}}\else{$T_{\rm ex}$}\fi}
\def\tmb {\ifmmode{{T}_{\rm mb}}\else{$T_{\rm mb}$}\fi}
\def\ci  {\ifmmode{{\rm C}{\rm \small I}}\else{C\ts {\scriptsize I}}\fi}
\def\hi  {\ifmmode{{\rm H}{\rm \small I}}\else{H\ts {\scriptsize I}}\fi}
\def\hh  {\ifmmode{{\rm H}_2}\else{H$_2$}\fi}
\def\kms    {\ifmmode{{\rm \ts km\ts s}^{-1}}\else{\ts km\ts s$^{-1}$}\fi}
\def\msun   {\ifmmode{{\rm M}_{\odot}}\else{M$_{\odot}$}\fi}
\def\msunpc   {\ifmmode{{\rm M}_{\odot}\,{\rm pc}^{-2}}\else{M$_{\odot}$\,pc$^{-2}$}\fi}
\def\msunyr   {\ifmmode{{\rm M}_{\odot}\,{\rm yr}^{-1}}\else{M$_{\odot}$\,yr$^{-1}$}\fi}
\def\lsun   {\ifmmode{{\rm L}_{\odot}}\else{L$_{\odot}$}\fi}
\def\zsun   {\ifmmode{{\rm Z}_{\odot}}\else{Z$_{\odot}$}\fi}

\begin{document}

\title{Cool dust heating and temperature mixing
in nearby star-forming galaxies\thanks{Based on {\it Herschel} observations.
{\it Herschel} is an ESA space observatory with science instruments
provided by European-led Principal Investigator consortia and with important
participation from NASA.}} 

\author{L.~K. Hunt \inst{\ref{inst:hunt}}
\and
B.~T. Draine\inst{\ref{inst:draine}}
\and
S.~Bianchi\inst{\ref{inst:hunt}}
\and
K.~D. Gordon\inst{\ref{inst:gordon},\ref{inst:gordonbelgium}}
\and
G.~Aniano\inst{\ref{inst:aniano},\ref{inst:draine}}
\and
D.~Calzetti\inst{\ref{inst:calzetti}}
\and
D.~A.~Dale\inst{\ref{inst:dale}}
\and
G.~Helou\inst{\ref{inst:helou}}
\and
J.~L. Hinz\inst{\ref{inst:hinz}}
\and
R.~C. Kennicutt\inst{\ref{inst:kennicutt}}
\and
H.~Roussel\inst{\ref{inst:roussel}}  
\and
C.~D. Wilson\inst{\ref{inst:wilson}}
\and
A.~Bolatto\inst{\ref{inst:bolatto}}
\and
M.~Boquien\inst{\ref{inst:boquien}}
\and
K.~V. Croxall\inst{\ref{inst:croxall}}
\and
M.~Galametz\inst{\ref{inst:galametz}}
\and
A.~Gil de Paz\inst{\ref{inst:gildepaz}}
\and
J.~Koda\inst{\ref{inst:koda}}
\and
J.~C. Mu\~noz-Mateos\inst{\ref{inst:munoz}}
\and
K.~M. Sandstrom\inst{\ref{inst:sandstrom},\ref{inst:sandstromarizona}}
\and
M. Sauvage\inst{\ref{inst:sauvage}}
\and
L.~Vigroux\inst{\ref{inst:vigroux}}
\and
S.~Zibetti\inst{\ref{inst:hunt}}
}

\offprints{L. K. Hunt}
\institute{INAF - Osservatorio Astrofisico di Arcetri, Largo E. Fermi, 5, 50125, Firenze, Italy
\label{inst:hunt}
\email{hunt@arcetri.astro.it} 
\and
Department of Astrophysical Sciences, Princeton University, Princeton, NJ 08544, USA \label{inst:draine}
\and
Space Telescope Science Institute, 3700 San Martin Drive, Baltimore, MD 21218, USA \label{inst:gordon}
\and
Sterrenkundig Observatorium, Universiteit Gent, Gent, Belgium \label{inst:gordonbelgium}
\and
Institut d'Astrophysique Spatiale, b\^atiment 121, Universit\'e Paris-Sud 11, CNRS UMR 8617, F-91405 Orsay, France \label{inst:aniano}
\and
Department of Astronomy, University of Massachusetts, Amherst, MA 01003, USA
\label{inst:calzetti}
\and
Department of Physics and Astronomy, University of Wyoming, Laramie, WY 82071, USA
\label{inst:dale}
\and
NASA Herschel Science Center, IPAC, California Institute of Technology, Pasadena, CA 91125, USA
\label{inst:helou}
\and
Steward Observatory, University of Arizona, 933 North Cherry Avenue, Tucson, AZ 85721, USA
\label{inst:hinz}
\and
Institute of Astronomy, University of Cambridge, Madingley Road, Cambridge CB3 0HA, UK
\label{inst:kennicutt}
\and
Institut d'Astrophysique de Paris, Sorbonne Universit\'es,
UPMC Univ. Paris 06, CNRS UMR 7095, 75014 Paris, France
\label{inst:roussel}
\and
Department of Physics and Astronomy, McMaster University, Hamilton, ON L8S 4M1, Canada
\label{inst:wilson}
\and
Department of Astronomy and Laboratory for Millimeter-wave Astronomy, University of Maryland, College Park, MD 20742, USA
\label{inst:bolatto}
\and
Institute of Astronomy, University of Cambridge, Madingley Road, Cambridge, CB3 0HA, UK
\label{inst:boquien}
\and
Department of Astronomy, The Ohio State University, 140 West 18th Avenue, Columbus, OH 43210, USA
\label{inst:croxall}
\and
European Southern Observatory, Karl-Schwarzchild-Str. 2, D-85748 Garching-bei-München, Germany
\label{inst:galametz}
\and
Departamento de Astrof\'isica, Universidad Complutense de Madrid, Avda. de la Complutense, s/n, E-28040 Madrid, Spain
\label{inst:gildepaz}
\and
Department of Physics and Astronomy, SUNY Stony Brook, Stony Brook, NY 11794-3800, USA
\label{inst:koda}
\and
European Southern Observatory, Casilla 19001, Santiago 19, Chile
\label{inst:munoz}
\and
Max-Planck-Institut fur Astronomie, Konigstuhl 17, D-69117 Heidelberg, Germany
\label{inst:sandstrom}
\and
Steward Observatory, University of Arizona, 933 N. Cherry Ave, Tucson, AZ 85721, USA
\label{inst:sandstromarizona}
\and
CEA, Laboratoire AIM, Irfu/SAp, Orme des Merisiers, 91191 Gif-sur-Yvette, France
\label{inst:sauvage}
\and
Institut d'Astrophysique de Paris, UMR 7095 CNRS, Universit\'e Pierre et Marie Curie, F-75014 Paris, France
\label{inst:vigroux}
}

\date{Received  2014/ Accepted  2014}

\titlerunning{Radial dust profiles in nearby galaxies}
\authorrunning{Hunt et al.}

\abstract{Physical conditions of the interstellar medium in galaxies are closely 
linked to the ambient radiation field and the heating of dust grains. 
In order to characterize dust properties in galaxies over a wide range of physical conditions,
we present here the radial surface brightness profiles of the entire sample of 61 galaxies 
from Key Insights into Nearby Galaxies: Far-Infrared Survey with \hers\ (KINGFISH).
The main goal of our work is the characterization of the grain emissivities,
dust temperatures, and interstellar radiation fields (ISRFs) responsible for heating the
dust.
We first fit the radial profiles with exponential functions in order to compare
stellar and cool-dust disk scalelengths, as measured by 3.6\,\micron\ and 250\,\micron\ 
surface brightnesses.
Our results show that the stellar and dust scalelengths are comparable, with a mean
ratio of 1.04, although several galaxies show dust-to-stellar scalelength ratios of 1.5 or more.
We then fit the far-infrared spectral energy distribution (SED) in each annular region
with single-temperature modified black bodies using both variable
(MBBV) and fixed (MBBF) emissivity indices $\beta$, as well as with physically motivated
dust models. 
%The physical dust models were also fit with MBBV and MBBF functions in the same way as the data.
The KINGFISH profiles are well suited to examine trends of dust temperature \td\ and $\beta$ because
they span a factor of $\sim$200 in the ISRF intensity heating the bulk of the dust mass, \umin.
%the local ultraviolet radiation field in the Solar Neighborhood; 
%values range from \umin\,=\,0.1 in dwarf galaxies and in the outskirts of virtually all galaxies
%to \umin$\sim$20 in the circumnuclear regions of starbursts.
Results from fitting the profile SEDs suggest that, on average, %in the KINGFISH galaxies,
\td, dust optical depth \taud, and \umin\ decrease with radius.
The emissivity index $\beta$ also decreases with radius in some galaxies, but in others
is increasing, or rising in the inner regions and falling in the outer ones.
%The fraction of Photon-Dominated Regions (PDRs) \fpdr\ tends to increase beyond the
%optical radius \ropt, implying that there may be extended ultraviolet disks responsible
%for heating the dust.
Despite the fixed grain emissivity (average $\beta\sim 2.1$) of the physically-motivated models,
they are well able to accommodate flat spectral slopes with $\beta\la 1$. 
An analysis of the wavelength variations of dust emissivities in both the data and the
models shows that
flatter slopes ($\beta\la 1.5$) are associated with cooler temperatures,
contrary to what would be expected from the usual \td-$\beta$ degeneracy.
This trend is related to variations in \umin\ since $\beta$ and
\umin\ are very closely linked over the entire range in \umin\ sampled by the KINGFISH
galaxies: low \umin\ is associated with flat $\beta\la 1$.
Both these results strongly suggest that the low apparent $\beta$ values (flat slopes) in MBBV fits
are caused by temperature mixing along the line-of-sight, rather than by intrinsic
variations in grain properties.
Finally, a comparison of dust models and the data show a slight $\sim$10\% excess
at 500\,\micron\ for low metallicity (\logoh$\la 8$) and low far-infrared surface
brightness ($\Sigma_{500}$).
}
\maketitle

%---------------------------------------------------------------

\section{Introduction}
\label{sec:intro}

The interstellar medium (ISM) is both the cradle and the
grave of star formation in galaxies. 
Gas is converted into stars in dense molecular clouds, 
and is expelled during a star's lifetime through
stellar winds and at the end of its evolution through supernovae.
Dust grains are created both during stellar evolution and in the ISM itself, and
act as a catalyst for molecule formation.
Dust also contributes significantly to the ISM energy budget through  
photoelectric heating.
For many years, the complex interplay among dust, gas, and star-formation processes 
was studied mainly through global properties, but the advent of infrared (IR) and
sub-millimeter (submm) satellites such as \spit\ and \hers,
together with ground-based facilities, has made
possible {\it resolved} studies of the dust and gas in large samples of
nearby galaxies beyond the Local Group.

Like the stars and the molecular component of the ISM
\citep[e.g.,][]{freeman70,bigiel08,leroy08}, dust is generally 
distributed in a disk, often with an exponential decline 
of surface density with radius
\citep{haas98,bianchi07,munoz09a,bianchi11,verstappen13,degeyter13}.
It is heated both by young stars from recent episodes of star formation, 
and by the more diffuse interstellar radiation field (ISRF) produced by
the quiescent underlying stellar population \citep[e.g.,][]{dl07}.
However, the intense ISRF in the
bulges of disk galaxies \citep{engelbracht10,groves12,draine14} can 
mimic, at least to some degree, the ISRF of star formation; 
consequently, warm dust emission may not necessarily be uniquely associated
with recent star formation \citep[e.g.,][]{sauvage92,bendo10,boquien11,bendo12}.
Whereas a global approach is usually unable to distinguish between these alternatives,
radial profiles of dust properties are a unique diagnostic for understanding
how dust is heated.
Such processes are important because of the relation of 
dust heating to star formation, the star-formation history (SFH) of a galaxy, and its structure.

In this paper, we assess the spatial variations of dust properties in galaxies with
radial surface brightness profiles of the sample from Key Insights into Nearby Galaxies: A Far-IR
Survey with \hers\ \citep[KINGFISH,][]{kennicutt11}.
Several previous papers have dealt with spatially-resolved dust heating in the KINGFISH sample
\citep[e.g.,][]{aniano12,galametz12,kirkpatrick14}, but they
were based on limited numbers of galaxies and pixel-by-pixel analyses.
% (2, 11, 20, respectively).
Here, we study the entire KINGFISH sample of 61 galaxies and assess radial
gradients through azimuthal averaging.
Thus, we can sample not only varying physical conditions within galaxies, 
comparing dense inner regions to tenuous outer ones, but also probe beyond typical
high surface-brightness boundaries because of the increased signal-to-noise (S/N) 
made possible by averaging over faint outer isophotes.
% along the outer lines-of-sight.

The main goal of this paper is the characterization of the grain emissivities,
dust temperatures, and the ISRFs responsible for heating the dust,
over a wider range of physical conditions than has been possible up to now. 
We also want to explore evidence for submm emission in excess 
of what would be expected from ``standard'' dust models
\citep[e.g.,][]{planck11a,galliano11,kirkpatrick13,galametz14}.
To achieve these goals, we adopt a multi-pronged approach
which comprises several data fits and fits of models themselves:
a modified blackbody (MBB) fitting of the far-infrared spectral energy distribution (SED)
at each radial data point within a galaxy;
fitting the SED at each radius with the physically-motivated models by \citet[][hereafter DL07]{dl07};
and, finally,
fitting the DL07 models with MBBs in the same way as the data were fit.
Both sets of MBB fits are two-fold: one with the dust power-law emissivity index $\beta$
left to vary, and another with $\beta$ fixed.
Our objective for fitting the DL07 models with MBBs is to connect
the physical parameters of the DL07 models 
with the approximation of a single dust temperature, as well as to
assess how well fixed-emissivity models such as DL07 can
accommodate apparent values of $\beta$ lower
than the intrinsic emissivity index assumed in the model.

The paper is structured as follows:
sample selection and ancillary data are described in Sect. \ref{sec:sample},
together with \hers\ image reduction and data preparation.
Sect. \ref{sec:profiles} explains the extraction of the radial profiles, and
the analysis of the disk scalelengths of the dust and stars;
Sect. \ref{sec:colors}  shows SPIRE and PACS colors of the radial profile data.
We outline the model fitting of the radial profile SEDs in Sect. \ref{sec:fitting},
and discuss the fitting results in two sections:
Sect. \ref{sec:radial} (radial trends) and
Sect. \ref{sec:discussion} (emissivity variations, temperature mixing, 
and assessment of the models).
Our conclusions are given in Sect. \ref{sec:conclusions}.

%---------------------------------------------------------------
\section{The sample and the data}
\label{sec:sample}

The KINGFISH sample includes 61 nearby galaxies
within 30\,Mpc, selected to cover the variety of observed galaxy
morphologies, and the range of masses and luminosities within each
morphology as well as a variation of dust opacities \citep{kennicutt11}.
57 of these galaxies derive from the earlier SIRTF Infrared
Nearby Galaxy Survey \citep[SINGS,][]{kennicutt03} which comprised 75
nearby galaxies.
Although both samples are heavily biased toward star-forming galaxies,
$\sim$16\% of the KINGFISH sample are early types, ellipticals and lenticulars (S0's).
Table \ref{tab:sample} lists the KINGFISH galaxies, together with some of
their observational parameters.

% -----------------------------------------------------------------
% --- Table 1: Sample
%
\begin{center}
\begin{table*}
      \caption[]{KINGFISH galaxy sample} 
\label{tab:sample}
%\resizebox{\linewidth}{!}{
%\addtolength{\tabcolsep}{7pt}
{%\small
\tiny
\begin{tabular}{llcrlccrrrlcl}
\hline
%\vspace{-2 mm} \\
\multicolumn{1}{c}{Name} & 
\multicolumn{2}{c}{Hubble Type} & 
\multicolumn{1}{c}{Distance$^{\mathrm a}$} &
\multicolumn{1}{c}{12$+$} &
\multicolumn{1}{c}{RA} & 
\multicolumn{1}{c}{Dec.} & 
\multicolumn{1}{c}{Major} &
\multicolumn{1}{c}{Minor} &
\multicolumn{1}{c}{PA} &
\multicolumn{1}{c}{PA} &
\multicolumn{1}{c}{Incl.} &
\multicolumn{1}{c}{Incl.} \\
& Morph. & T & \multicolumn{1}{c}{(Mpc)} 
& \multicolumn{1}{c}{log(O/H)} & \multicolumn{2}{c}{(J2000)} &
\multicolumn{2}{c}{Diameters$^{\mathrm b}$} & \multicolumn{1}{c}{(E from N)} & 
\multicolumn{1}{c}{Ref.$^{\mathrm c}$} &
\multicolumn{1}{c}{(degrees)} &
\multicolumn{1}{c}{Ref.$^{\mathrm c}$} \\
&&&&&&& \multicolumn{2}{c}{(arcmin)} & 
\multicolumn{1}{c}{(degrees)} \\
% see output from readsample.pl
\hline
\\
DDO053       & Im     &  10 &   3.61 &   7.60\ \ \ (0.11) & 08:34:07.2  & +66:10:54   &    1.5 &    1.3 &  132 & B06    &   31 & W08    \\             
DDO154       & IBm    &  10 &   4.30 &   7.54\ \ \ (0.09) & 12:54:05.2  & +27:08:59   &    3.0 &    2.2 &   50 & dB08   &   66 & dB08   \\             
DDO165       & Im     &  10 &   4.57 &   7.63\ \ \ (0.08) & 13:06:24.8  & +67:42:25   &    3.5 &    1.9 &   90 & NED    &   61 & LTW    \\             
HoI          & IABm   &  10 &   3.90 &   7.61\ \ \ (0.11) & 09:40:32.3  & +71:10:56   &    3.6 &    3.0 &    0 & NED    &   12 & W08    \\             
HoII         & Im     &  10 &   3.05 &   7.72\ \ \ (0.14) & 08:19:05.0  & +70:43:12   &    7.9 &    6.3 &   15 & NED    &   41 & W08    \\             
IC0342       & SABcd  &   6 &   3.28 &   8.80$^{\mathrm d}$ (0.10) & 03:46:48.5  & +68:05:46   &   21.4 &   20.9 &   37 & C00    &   31 & C00    \\    
IC2574       & SABm   &   9 &   3.79 &   7.85\ \ \ (0.14) & 10:28:23.5  & +68:24:44   &   13.2 &    5.4 &   56 & O08    &   53 & W08    \\             
M81DwB       & Im     &  10 &   3.60 &   7.84\ \ \ (0.13) & 10:05:30.6  & +70:21:52   &    0.9 &    0.6 &  140 & NED    &   48 & This paper \\         
NGC0337      & SBd    &   7 &  19.30 &   8.18\ \ \ (0.07) & 00:59:50.1  & -07:34:41   &    2.9 &    1.8 &  130 & NED    &   52 & This paper \\         
NGC0584      & E4     &  -4 &  20.80 &   8.43\ \ \ (0.30) & 01:31:20.7  & -06:52:05   &    4.2 &    2.3 &   55 & NED    &   58 & This paper \\         
NGC0628      & SAc    &   5 &   7.20 &   8.43$^{\mathrm d}$ (0.02) & 01:36:41.8  & +15:47:00   &   10.5 &    9.5 &   25 & NED    &   25 & This paper \\
NGC0855      & E      &  -5 &   9.73 &   8.29\ \ \ (0.10) & 02:14:03.6  & +27:52:38   &    2.6 &    1.0 &   67 & NED    &   70 & This paper \\         
NGC0925      & SABd   &   7 &   9.12 &   8.32$^{\mathrm d}$ (0.01) & 02:27:16.9  & +33:34:45   &   10.5 &    5.9 &  107 & dB08   &   66 & dB08   \\    
NGC1097      & SBb    &   3 &  14.20 &   8.58$^{\mathrm d}$ (0.01) & 02:46:19.0  & -30:16:30   &    9.3 &    6.3 &  131 & D09    &   32 & D09    \\    
NGC1266      & SB0    &  -2 &  30.60 &   8.29\ \ \ (0.30) & 03:16:00.7  & -02:25:38   &    1.5 &    1.0 &  108 & NED    &   49 & This paper \\         
NGC1291      & SBa    &   1 &  10.40 &   8.52\ \ \ (0.30) & 03:17:18.6  & -41:06:29   &    9.8 &    8.1 &  170 & NED    &   35 & This paper \\         
NGC1316      & SAB0   &  -2 &  21.00 &   8.77\ \ \ (0.30) & 03:22:41.7  & -37:12:30   &   12.0 &    8.5 &   50 & NED    &   46 & This paper \\         
NGC1377      & S0     &  -1 &  24.60 &   8.29\ \ \ (0.30) & 03:36:39.1  & -20:54:08   &    1.8 &    0.9 &   92 & NED    &   62 & This paper \\         
NGC1404      & E1     &  -4 &  20.20 &   8.54\ \ \ (0.30) & 03:38:51.9  & -35:35:40   &    3.3 &    3.0 &  162 & NED    &   25 & This paper \\         
NGC1482      & SA0    &  -2 &  22.60 &   8.44\ \ \ (0.10) & 03:54:38.9  & -20:30:09   &    2.5 &    1.4 &  103 & NED    &   57 & This paper \\         
NGC1512      & SBab   &   2 &  11.60 &   8.56\ \ \ (0.12) & 04:03:54.3  & -43:20:56   &    8.9 &    5.6 &   45 & NED    &   52 & This paper \\         
NGC2146      & SBab   &   2 &  17.20 &   8.68\ \ \ (0.10) & 06:18:37.7  & +78:21:25   &    6.0 &    3.4 &  123 & NED    &   57 & This paper \\         
NGC2798      & SBa    &   1 &  25.80 &   8.52\ \ \ (0.05) & 09:17:22.9  & +41:59:59   &    2.6 &    1.0 &  160 & NED    &   68 & D06    \\             
NGC2841      & SAb    &   3 &  14.10 &   8.72$^{\mathrm d}$ (0.12) & 09:22:02.6  & +50:58:35   &    8.1 &    3.5 &  153 & dB08   &   74 & dB08   \\    
NGC2915      & I0     &  10 &   3.78 &   7.94\ \ \ (0.13) & 09:26:11.5  & -76:37:35   &    1.9 &    1.0 &  129 & NED    &   59 & This paper \\         
NGC2976      & SAc    &   5 &   3.55 &   8.36\ \ \ (0.06) & 09:47:15.4  & +67:54:59   &    5.9 &    2.7 &  155 & dB08   &   65 & dB08   \\             
NGC3049      & SBab   &   2 &  19.20 &   8.53\ \ \ (0.01) & 09:54:49.5  & +09:16:16   &    2.1 &    1.1 &   25 & NED    &   61 & This paper \\         
NGC3077      & I0pec  &  10 &   3.83 &   8.64\ \ \ (8.64) & 10:03:19.1  & +68:44:02   &    5.4 &    4.5 &   45 & NED    &   33 & This paper \\         
NGC3184      & SABcd  &   6 &  11.70 &   8.65$^{\mathrm d}$ (0.02) & 10:18:17.0  & +41:25:28   &    7.4 &    6.9 &  135 & NED    &   16 & W08    \\    
NGC3190      & SAap   &   1 &  19.30 &   8.49\ \ \ (0.30) & 10:18:05.6  & +21:49:55   &    4.4 &    1.5 &  125 & NED    &   73 & This paper \\         
NGC3198      & SBc    &   5 &  14.10 &   8.49$^{\mathrm d}$ (0.04) & 10:19:54.9  & +45:32:59   &    8.5 &    3.3 &   35 & dB08   &   72 & dB08   \\    
NGC3265      & E      &  -5 &  19.60 &   8.39\ \ \ (0.06) & 10:31:06.8  & +28:47:48   &    1.0 &    0.7 &   73 & NED    &   46 & This paper \\         
NGC3351      & SBb    &   3 &   9.33 &   8.69$^{\mathrm d}$ (0.01) & 10:43:57.7  & +11:42:14   &    7.4 &    4.2 &   12 & T08    &   41 & W08    \\    
NGC3521      & SABbc  &   4 &  11.20 &   8.44$^{\mathrm d}$ (0.05) & 11:05:48.6  & -00:02:09   &   11.0 &    5.1 &  160 & dB08   &   73 & dB08   \\    
NGC3621      & SAd    &   7 &   6.55 &   8.33$^{\mathrm d}$ (0.02) & 11:18:16.5  & -32:48:51   &   12.3 &    7.1 &  165 & dB08   &   65 & dB08   \\    
NGC3627      & SABb   &   3 &   9.38 &   8.34\ \ \ (0.24) & 11:20:15.0  & +12:59:30   &    9.1 &    4.2 &  173 & dB08   &   62 & dB08   \\             
NGC3773      & SA0    &  -2 &  12.40 &   8.43\ \ \ (0.03) & 11:38:12.9  & +12:06:43   &    1.2 &    1.0 &  165 & NED    &   34 & This paper \\         
NGC3938      & SAc    &   5 &  17.90 &   8.42\ \ \ (0.30) & 11:52:49.4  & +44:07:15   &    5.4 &    4.9 &   28 & NED    &   25 & This paper \\         
NGC4236      & SBdm   &   8 &   4.45 &   8.17\ \ \ (0.30) & 12:16:42.1  & +69:27:45   &   21.9 &    7.2 &  162 & NED    &   72 & This paper \\         
NGC4254      & SAc    &   5 &  14.40 &   8.56$^{\mathrm d}$ (0.02) & 12:18:49.6  & +14:24:59   &    5.4 &    4.7 &   23 & NED    &   29 & This paper \\
NGC4321      & SABbc  &   4 &  14.30 &   8.61$^{\mathrm d}$ (0.07) & 12:22:54.9  & +15:49:21   &    7.4 &    6.3 &   30 & NED    &   32 & D06    \\
NGC4536      & SABbc  &   4 &  14.50 &   8.21\ \ \ (0.08) & 12:34:27.0  & +02:11:17   &    7.6 &    3.2 &  130 & NED    &   67 & This paper \\
NGC4559      & SABcd  &   6 &   6.98 &   8.32$^{\mathrm d}$ (0.02) & 12:35:57.7  & +27:57:35   &   10.7 &    4.4 &  150 & NED    &   66 & This paper \\
NGC4569      & SABab  &   2 &   9.86 &   8.58\ \ \ (0.30) & 12:36:49.8  & +13:09:46   &    9.5 &    4.4 &   23 & NED    &   64 & This paper \\
NGC4579      & SABb   &   3 &  16.40 &   8.54\ \ \ (0.30) & 12:37:43.5  & +11:49:05   &    5.9 &    4.7 &   95 & NED    &   38 & This paper \\
NGC4594      & SAa    &   1 &   9.08 &   8.54\ \ \ (0.30) & 12:39:59.4  & -11:37:23   &    8.7 &    3.5 &   90 & NED    &   69 & This paper \\
NGC4625      & SABmp  &   9 &   9.30 &   8.35\ \ \ (0.17) & 12:41:52.7  & +41:16:26   &    2.2 &    1.9 &   27 & NED    &   30 & This paper \\
NGC4631      & SBd    &   7 &   7.62 &   8.12\ \ \ (0.11) & 12:42:08.0  & +32:32:29   &   15.5 &    2.7 &   86 & NED    &   83 & This paper \\
NGC4725      & SABab  &   2 &  11.90 &   8.35\ \ \ (0.13) & 12:50:26.6  & +25:30:03   &   10.7 &    7.6 &   35 & NED    &   45 & This paper \\
NGC4736      & SAab   &   2 &   4.66 &   8.40$^{\mathrm d}$ (0.01) & 12:50:53.0  & +41:07:14   &   11.2 &    9.1 &  116 & dB08   &   41 & dB08   \\
NGC4826      & SAab   &   2 &   5.27 &   8.54\ \ \ (0.10) & 12:56:43.7  & +21:40:58   &   10.0 &    5.4 &  121 & dB08   &   65 & dB08   \\
NGC5055      & SAbc   &   4 &   7.94 &   8.59$^{\mathrm d}$ (0.07) & 13:15:49.3  & +42:01:45   &   12.6 &    7.2 &  102 & dB08   &   59 & dB08   \\
NGC5398      & SBdm   &   8 &   7.66 &   8.35\ \ \ (0.05) & 14:01:21.5  & -33:03:50   &    2.8 &    1.7 &  172 & NED    &   53 & This paper \\
NGC5408      & IBm    &  10 &   4.80 &   7.81\ \ \ (0.09) & 14:03:20.9  & -41:22:40   &    1.6 &    0.8 &   12 & NED    &   60 & This paper \\
NGC5457      & SABcd  &   6 &   6.70 &   8.73$^{\mathrm d}$ (0.03) & 14:03:12.6  & +54:20:57   &   28.8 &   26.9 &   39 & B81    &   18 & W08    \\
NGC5474      & SAcd   &   6 &   6.80 &   8.31\ \ \ (0.22) & 14:05:01.6  & +53:39:44   &    4.8 &    4.3 &   97 & NED    &   26 & This paper \\
NGC5713      & SABbcp &   4 &  21.40 &   8.48\ \ \ (0.10) & 14:40:11.5  & -00:17:20   &    2.8 &    2.5 &   11 & W08    &   33 & D06    \\
NGC5866      & S0     &  -2 &  15.30 &   8.47\ \ \ (0.30) & 15:06:29.5  & +55:45:48   &    4.7 &    1.9 &  128 & NED    &   68 & This paper \\
NGC6946      & SABcd  &   6 &   6.80 &   8.45$^{\mathrm d}$ (0.06) & 20:34:52.3  & +60:09:14   &   11.5 &    9.8 &   63 & dB08   &   33 & dB08   \\
NGC7331      & SAb    &   3 &  14.50 &   8.41$^{\mathrm d}$ (0.06) & 22:37:04.1  & +34:24:56   &   10.5 &    3.7 &  168 & dB08   &   76 & dB08   \\
NGC7793      & SAd    &   7 &   3.91 &   8.34$^{\mathrm d}$ (0.02) & 23:57:49.8  & -32:35:28   &    9.3 &    6.3 &  110 & dB08   &   50 & dB08   \\
\\
\hline
\end{tabular}
}
%}
\vspace{0.5\baselineskip}
\begin{description}
\item
[$^{\mathrm{a}}$]
The method of distance determination is given by \citet{kennicutt11}.
\item
[$^{\mathrm{b}}$]
Galaxy sizes are taken from 
NED\,=\,NASA/IPAC Extragalactic Database;
\item
[$^{\mathrm{c}}$]
References for PAs and inclination angles:
B06\,=\,\citet{begum06};
B81\,=\,\citet{bosma81};
C00\,=\,\citet{crosthwaite00};
D06\,=\,\citet{daigle06};
D09\,=\,\citet{davies09};
dB08\,=\,\citet{deblok08};
NED;
O08\,=\,\citet{oh08};
T08\,=\,\citet{tamburro08};
W08\,=\,\citet{walter08}.
\item
[$^{\mathrm{d}}$] 
Metallicity gradient from \citet{moustakas10}, except for
NGC\,5457 (M\,101) from \citet{li13}
and IC\,342 from \citet{pilyugin07}.
The latter has been adjusted to the mean slope ratio between 
\citet{moustakas10} and \citet{pt05}, namely $\Delta$O/H\,=\,$-0.49$.
\end{description}
\end{table*}
\end{center}
% -----------------------------------------------------------------

The stellar masses of KINGFISH galaxies span 5 orders of magnitude
(from $2\times10^6$\,\msun, DDO\,53, to $3\times10^{11}$\,\msun, NGC\,1316) with star-formation 
rates (SFRs) from $2\times10^{-3}$\,\sfr\ to $\sim9$\,\sfr\ \citep{skibba11,kennicutt11}.
Only NGC\,2146 and NGC\,1097 might be considered ``starbursts'' in terms
of their specific SFR (SFR divided by stellar mass, sSFR) of $\sim0.5$\,Gyr$^{-1}$; 
most of the remaining galaxies are star-forming galaxies along the ``main sequence''
of star formation \citep{salim07,noeske07,karim11}.
Some of the highest SFRs in the KINGFISH sample 
are found in the lenticulars (e.g., NGC\,1482, NGC\,1377, and NGC\,1266);
%however, these are not typical early-type galaxies.
Such intense star-formation activity is not particularly unusual in S0 galaxies
\citep[e.g.,][]{amblard14},
although these galaxies are not typical early-type systems.
NGC\,1377 is a nascent starburst with exceptionally warm dust and 
a deep silicate absorption feature at 9.7\,\micron\ 
\citep{vader93,laureijs00,roussel06}, and 
NGC\,1266 has shock-excited molecular gas entrained in a molecular outflow 
\citep[e.g.,][]{alatolo11,pellegrini13}.

%---------------------------------------------------------------
\subsection{Metallicity and metallicity gradients}
\label{sec:metallicity}

\citet{moustakas10} presented optical long-slit observations and  
measured oxygen abundances (O/H) and their radial gradients for galaxies in the SINGS sample.
We adopt these for the KINGFISH galaxies, using the \citet[][hereafter PT]{pt05} calibration
\citep[given in Table 1 of][]{kennicutt11}.
%listed in Table \ref{tab:sample}.
Four KINGFISH galaxies are missing from \citet{moustakas10};
for these we rely on abundances taken from the literature, which have been
reported to the same (PT) metallicity scale.
These include IC\,342 \citep{pilyugin07}, 
NGC\,2146 \citep{moustakas06}, NGC\,3077 \citep{storchi94,croxall09}, 
and NGC\,5457 \citep[M\,101,][]{li13}. 
Our metallicities for these galaxies are the same as those given by \citet{kennicutt11},
except for NGC\,3077 where we have adopted the determination by \citet{croxall09}.
We incorporate metallicity gradients in the profile analysis when available;
\citet{moustakas10} gives significant gradients for 17 KINGFISH galaxies.
M\,101 and IC\,342 also have well-determined metallicity gradients,
so that for 19 galaxies
we are able to incorporate abundance gradients in our radial analysis.
The adopted metallicities are also listed in Table \ref{tab:sample}.

%---------------------------------------------------------------
\subsection{Herschel PACS and SPIRE data}
\label{sec:herschel}

The KINGFISH Open-Time Key Project acquired far-infrared (FIR) images for 61 galaxies
with two instruments on board \hers\ \citep{pilbratt10}.
PACS \citep[Photodetector Array Camera 
\& Spectrometer\footnote{PACS has been developed by a consortium of institutes led by MPE
(Germany) and including UVIE (Austria); KU Leuven, CSL, IMEC (Belgium); CEA,
LAM (France); MPIA (Germany); INAF-IFSI/OAA/OAP/OAT, LENS, SISSA (Italy); IAC
(Spain). This development has been supported by the funding agencies BMVIT
(Austria), ESA-PRODEX (Belgium), CEA/CNES (France), DLR (Germany), ASI/INAF
(Italy), and CICYT/MCYT (Spain).},][]{poglitsch10} 
images at 70\,\micron, 100\,\micron\ and 160\,\micron\
were acquired in scan mode at medium speed (20\arcsec\,s$^{-1}$)
with two orthogonal scans in order
to better remove instrumental artefacts and transients. 
We tailored the PACS exposure times to account for 160\,\micron\
surface brightnesses at the optical radius, \ropt\ (the isophotal
$B$-band surface brightness), which we were
able to estimate thanks to \citet{munoz09b}; thus 
for each ``blue'' wavelength (70\,\micron, 100\,\micron) there were
three PACS scan repetitions for bright targets and six for faint ones.
Because of the simultaneous coverage at 160\,\micron,
PACS 160\,\micron\ has twice the number of scans as at the bluer wavelengths.
SPIRE 
\citep[Spectral and Photometric Imaging REceiver\footnote{SPIRE has 
been developed by a consortium of institutes led by Cardiff
University (UK) and including Univ. Lethbridge (Canada); NAOC (China); CEA, LAM
(France); IFSI, Univ. Padua (Italy); IAC (Spain); Stockholm Observatory
(Sweden); Imperial College London, RAL, UCL-MSSL, UKATC, Univ. Sussex (UK); and
Caltech, JPL, NHSC, Univ. Colorado (USA). This development has been supported
by national funding agencies: CSA (Canada); NAOC (China); CEA, CNES, CNRS
(France); ASI (Italy); MCINN (Spain); SNSB (Sweden); STFC, UKSA (UK); and NASA
(USA).},][]{griffin10}
mapping was performed with large-map mode at the nominal speed
of 30\arcsec\,s$^{-1}$ with, as for PACS, two orthogonal scans.
At the longest wavelengths, SPIRE maps are confusion limited \citep{nguyen10},
so for SPIRE we did not adopt two observing regimes.
Because of the large angular extent of IC\,342, one of the largest KINGFISH
galaxies, we acquired PACS$+$SPIRE maps in parallel mode with two
repetitions at slow speed (20\arcsec\,s$^{-1}$)\footnote{Despite its
size, the \hers\ images for NGC\,5457 (M\,101) were acquired in the same
way as the smaller galaxies, not in parallel mode.}.
We required both short PACS wavelengths, so used two repetitions in parallel mode.
All maps were at least 1.5 times the optical size of the galaxy,
with a minimum map size of 10\arcmin$\times$10\arcmin\ for the sake of efficiency.

The \hers\ Interactive Processing Environment \citep[HIPE, v.8;][]{hipe} was
used for the image processing up to Level 1.
Then for both PACS and SPIRE, images were converted from HIPE Level 1 to final maps
with the {\tt scanamorphos} algorithm (v. 16.9) \citep{roussel13}.
This method seemed to give the best estimate of faint diffuse low-surface
brightness flux with PACS, since before projecting the data onto a spatial grid,
it subtracts the brightness drifts caused by low-frequency flicker noise in the bolometers
and the thermal drifts of the detectors and the telescope.
%The adopted PACS pixel scale is 1\farcs4, 1\farcs7, and 2\farcs85 at 70, 100, and 160\,\micron,
%respectively.
After producing the final maps, both sets of images were corrected astrometrically
to be aligned with the MIPS 24\,\micron\ images (see below).

PACS calibrations correspond to the V7 responsivity calibration (global calibration
V65).
This has been accomplished by rescaling the original V6 fluxes obtained in the HIPE context
with multiplicative factors of 1.0, 1.0152, 1.0288 at 70\,\micron, 100\,\micron, and 160\,\micron,
respectively.
Finally, to take into account an internal inconsistency of mappers external to HIPE
(corrected in March, 2014), we multiplied the PACS 160\,\micron\ flux by 0.925\footnote{NGC\,584
was observed much later than the other KINGFISH galaxies, and thus only the factor of 0.925
was applied to the 160\,\micron\ flux for this galaxy.}.

SPIRE calibrations changed during the course of our analysis, so we multiplied 
the original (from HIPE v.8) fluxes by 0.9282 (250\,\micron), 0.9351 (350\,\micron), and 0.9195 (500\,\micron)
in order to correct the images to the revised flux scale
\citep[see][]{griffin13}\footnote{For more information, please
refer to the document posted at
http://www.astro.princeton.edu/$\sim$draine/ \\
\hbox{\quad\quad\quad\quad\quad\quad 
Notes\_re\_KINGFISH\_SPIRE\_Photometry.pdf.}}.
These factors also take into account an internal discrepancy 
that introduced an error in the surface brightness units of the SPIRE maps 
(see the KINGFISH Data Release 3 document,
\url{ftp://hsa.esac.esa.int/URD\_rep/KINGFISH-DR3/}).
Finally, to adjust to the latest HIPE 11 calibration,
SPIRE fluxes were further multiplied by factors of 1.0321, 1.0324, and 1.0181,
at 250\,\micron, 350\,\micron, and 500\,\micron, respectively.

SPIRE images of six of the 61 galaxies were not acquired in the context of the KINGFISH
program but rather by the \hers\ Reference Survey \citep[HRS,][]{boselli10}.
However, these images have been reduced with the KINGFISH pipeline so as to be consistent
with the remainder of the sample.
Further details on data acquisition and reduction are given by \citet{kennicutt11}.

%---------------------------------------------------------------
\subsection{Spitzer IRAC and MIPS data}
\label{sec:spitzer}

We included in the SED analysis data from \spit, namely
images acquired with IRAC \citep[InfraRed Array Camera,][]{fazio04},
and with MIPS \citep[Multiband Imaging Photometer,][]{rieke04}.
Most of these were taken in context of the SINGS survey \citep{kennicutt03}, but
many galaxies were also observed in the Local Volume Legacy \citep[LVL,][]{dale09},
and we adopted that reduction when available.
As mentioned above, four of the 61 KINGFISH galaxies are not in SINGS;
for these we retrieved IRAC and MIPS data from the \spit\ archive, and
reduced them using the LVL pipeline \citep{dale09}.
A correction for non-linearities in the MIPS 70\,\micron\ images was applied as
described by \citet{dale09} and \citet{gordon11}, and extended-source
corrections were applied to the IRAC fluxes as described in \citet{aniano12}.

%---------------------------------------------------------------
\section{The radial brightness profiles}
\label{sec:profiles}

Before extracting the elliptically averaged profiles,
the images were processed to ensure that the SEDs would
be reliable representations of the galaxies.
Following \citet{aniano12},
first the background light was estimated and subtracted by
fitting tilted planes to the empty sky regions of each
image. 
Bright sources and faint background galaxies were masked out
a priori.
The procedure was iterated in a clipping algorithm by successively 
minimizing the dispersion of the candidate background pixels around 
the best-fit plane, and upon convergence is subtracted from the image 
using the native pixel grid.
Then, all images are convolved to a common point-spread function (PSF),
in our case the lowest resolution, MIPS160, with a FWHM of 38\farcs8.
This choice enables inclusion in the SED of all possible instruments (IRAC, MIPS,
PACS, SPIRE) and is considered the ``gold standard'' by \citet{aniano12}.
Finally, the images were rebinned to a common pixel size, 18\arcsec\
(roughly half the FWHM of the MIPS160 beam), and astrometrically aligned. 
Care was taken to ensure that the final maps including all instruments
covered at least 1.5 times the optical size of the galaxy.
Further details of the comprehensive image analysis are given by \citet{aniano12}.
The width of the PSF (38\farcs8) corresponds to 1.86\,kpc at the
median KINGFISH sample distance of 9.9\,Mpc.  

\begin{figure*}[!ht]
\centerline{
\includegraphics[angle=0,width=0.5\linewidth,bb=18 280 592 718]{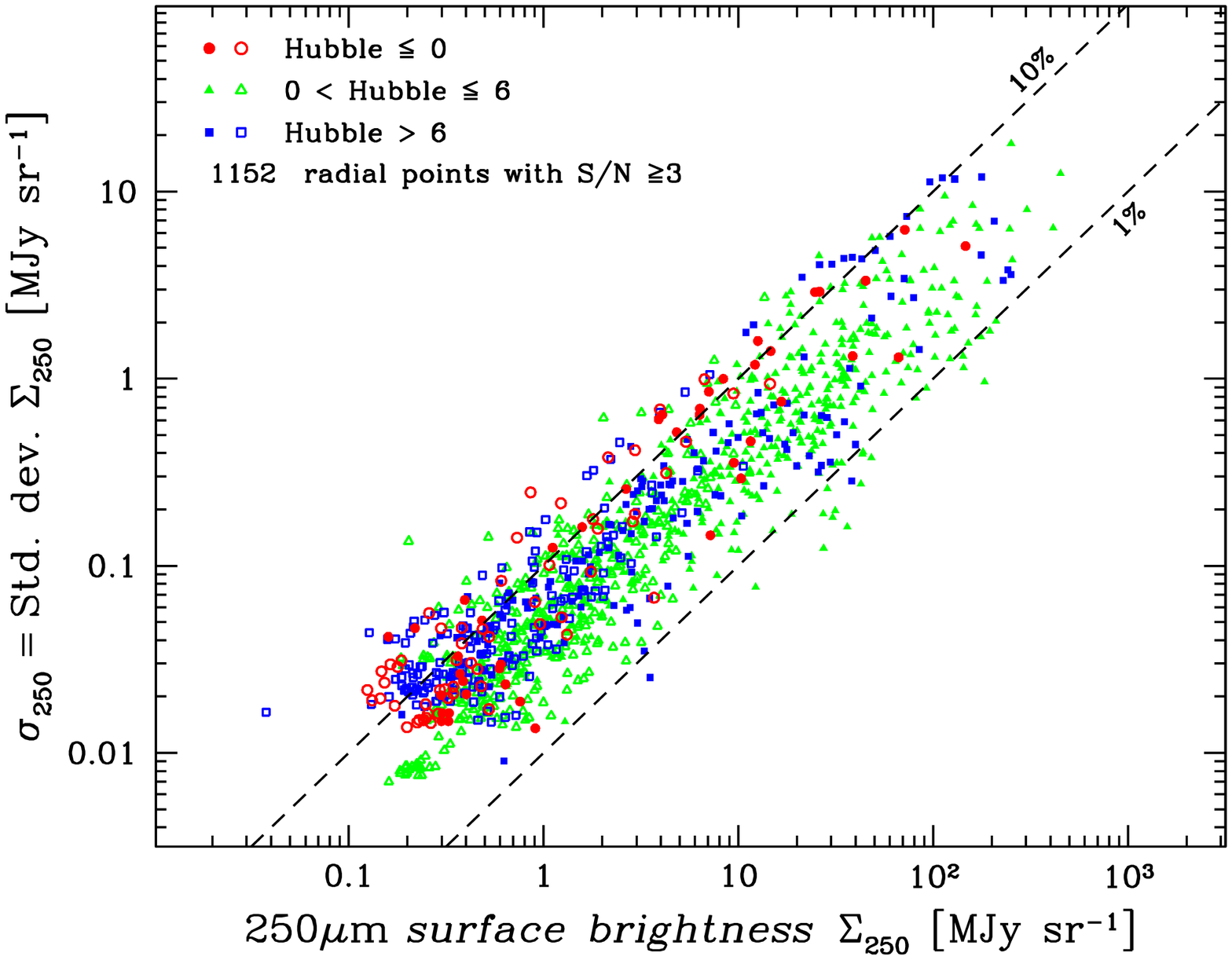}
\hspace{0.1cm}
\includegraphics[angle=0,width=0.5\linewidth,bb=18 280 592 718]{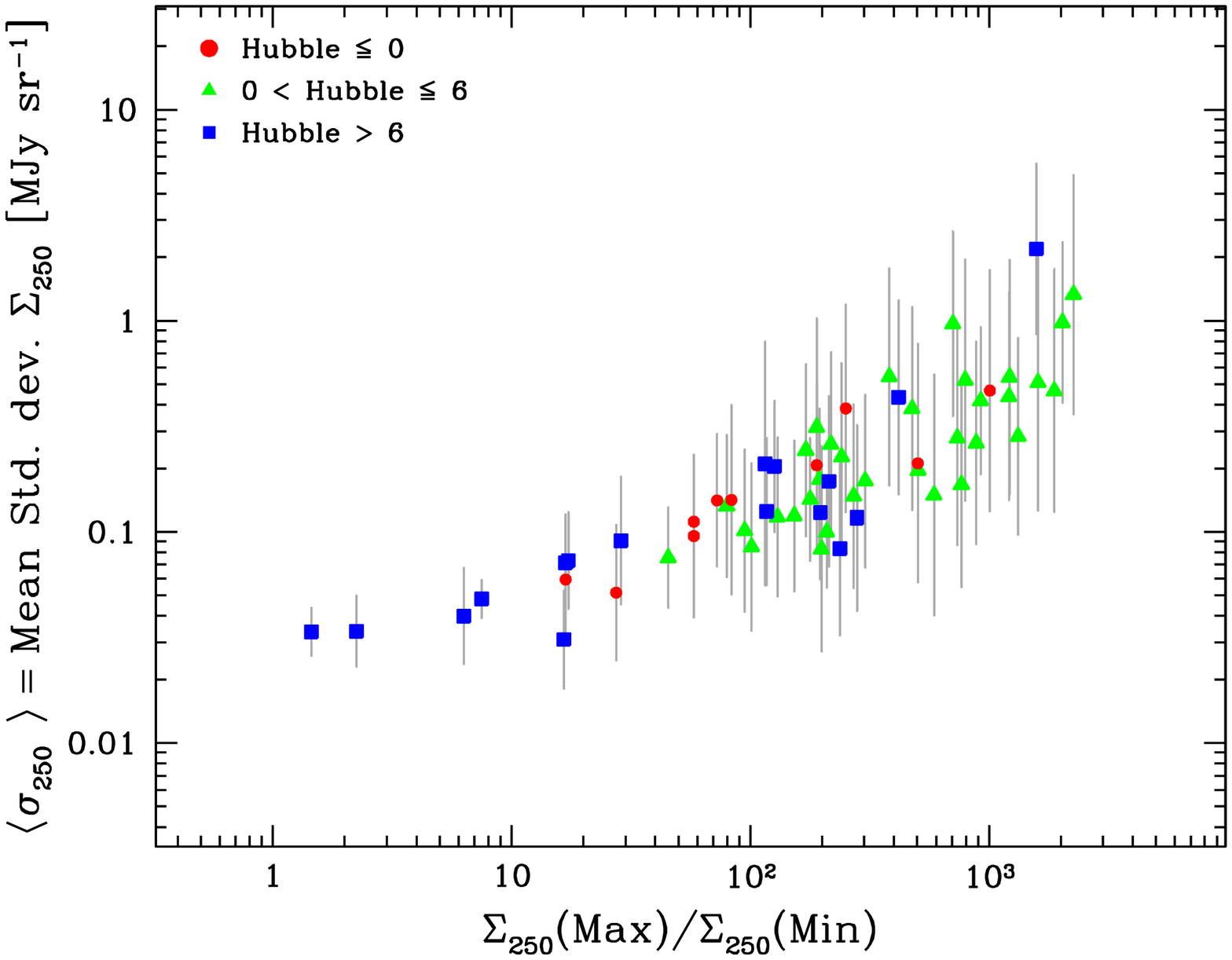}
}
\caption{Standard deviation $\sigma_{250}$ of the 250\,\micron\ surface brightness
within an annulus  vs.  250\,\micron\ surface brightness $\Sigma_{250}$ in the annulus (left panel),
and the mean  $\sigma_{250}$ over all the profiles in a galaxy vs. the ratio of maximum
and minimum $\Sigma_{250}$ in the galaxy (right panel).
Only radial points with S/N$\geq$3 are shown.
In the left panel,
the dashed lines correspond to 10\% variation (upper) and 1\% variation (lower)
relative to $\Sigma_{250}$.
The error bars in the right panel reflect the standard deviation within
a given profile. 
In both panels, points are plotted by Hubble type with (red) circles corresponding
to early types (T$\leq$0), (green) triangles to spirals (0$<$T$\leq$6), and
(blue) squares to late types (T$>$6); in the left panel
filled symbols correspond to $R$/\ropt$\leq$0.8 and open ones to larger radii.
\label{fig:sb}
}
\end{figure*}
% ---------------------

\subsection{Extracting the profiles}
\label{sec:extracting}

From these images, elliptically-averaged radial surface brightness profiles were 
extracted using the IRAF task {\tt ellipse}. 
Centers, position angles (PAs), and ellipticities ($1-b/a$, where $a$
and $b$ are the major and minor axes of the galaxy) were kept fixed
to the values reported in Table \ref{tab:sample}.
To circumvent possible slight (fraction of a pixel) misalignment despite the
astrometric corrections described above,
the profiles were extracted centered on the RA and Dec. coordinates as given
in Table \ref{tab:sample}.
The width (along the major axis) of each annulus is the same as the pixel size,
18\arcsec, and the
radial profile extraction extends in linear increments to at least 1.5 times the 
optical radius, \ropt.
%\footnote{We did not correct the pixels for inclination
%so they are square in the plane of the sky, but rectangular in the plane of the galaxy.}.

As shown in Table \ref{tab:sample}, when possible,
we adopted PAs and inclination angles $i$ determined kinematically. 
When these were not available,
inclinations $i$ were calculated photometrically as a function of axial ratio
according to \citet{hubble26}:
\begin{equation}
cos^2\ i\,=\,\frac{(b/a)^2 - q_0^2}{1-q_0^2}
\end{equation}
For Hubble types $T<4$, we take $q_0\,=\,0.2$ 
for the intrinsic flattening of the galactic disk \citep{holmberg58},
while Hubble types Sbc and later ($T\geq4$) were assumed
to be intrinsically more flattened with $q_0\,=\,0.13$
\citep[e.g.,][]{giovanelli94,dale97,murphy08}.
To avoid misrepresenting the elliptically averaged brightness profiles
of highly inclined galaxies by including points too close to the projected
minor axis, we applied a masking technique during ellipse fitting.
For galaxies with inclination $i\geq60^\circ$, 
we generated masks which removed from consideration in the elliptical
averages a wedge-shaped subset of points around the minor axes.
After several tests, the opening half-angle of the wedge was defined to be 
30$^\circ$ along the
minor axis for $i\,=\,60^\circ$ and to vary linearly to a maximum of 80$^\circ$
at the highest inclinations.
Such a technique is particularly important for galaxies such as NGC\,4631 and NGC\,7331, 
both of which have $i>75^\circ$.
Table \ref{tab:sample} gives the parameters used for the ellipse extraction,
together with the references for PAs and inclinations.

Uncertainties in the surface brightnesses as a function of radius were calculated as the 
quadrature sum of the variation along the elliptical isophotes 
and the calibration fractional uncertainties as given by \citet{aniano12}.
The latter are 0.083, 0.071, 0.221, 0.167 (i.e., 8.3\%, 7.1\%,
22.1\%, 16.7\%) for the four IRAC channels, respectively,
and 0.10 for MIPS, PACS, and SPIRE.

The MIPS~70 and MIPS~160 filters are not identical to the PACS~70 and PACS~160
filters, so that the two instruments will report different flux densities for 
the same nominal wavelength.
For MBB spectra with $\beta$\,=\,2, and $T$ varying from 15 to 30\,K,
the PACS/MIPS(70) flux ratio should vary from 1.059 to 0.965,
while the PACS/MIPS(160) ratio varies from 0.919 to 0.944.
For our sample
the PACS/MIPS(70) mean flux ratio\,=\,1.11$\pm$0.23, and the mean
flux ratio PACS/MIPS(160)\,=\,0.90$\pm$0.19.
These mean values are roughly consistent with the estimated 10\% calibration
uncertainties for MIPS and PACS, but in individual cases the 
MIPS and PACS flux density ratios can differ from the expected
ratio by more than this.
Thus in order not to unduly confound the fitting procedure,
the uncertainties for the MIPS and PACS data were taken to be
0.5 of the absolute value of the difference in fluxes between the two instruments;
the uncertainties for PACS 100\,\micron\ data were adjusted to be the mean of
those at 70\,\micron\ and 160\,\micron.

The uncertainties of the annular flux extractions were calculated as the
error in the mean, namely the standard deviation $\sigma_{\lambda}$ of the distribution along
the elliptical isophote circumference 
divided by the square root of the number of pixels in the circumference, $N_{\rm isophote}$;
this is then added in quadrature to the calibration uncertainties to obtain
the total error.
In the outer regions, the uncertainties are generally dominated
by calibration errors because of the relatively large numbers of pixels.
We have not considered the errors in the sky subtraction because of the
complex method used to subtract background emission combined
with the image degrading produced by convolution to the common MIPS\,160\,\micron\
resolution \citep{aniano12}.
However, we have measured the noise of the sky subtraction of the original
images by measuring the sky in 20 empty regions around the galaxies. 
The variation is in all cases much smaller than either the calibration uncertainties or
the standard deviation of the elliptical isophotes.

\subsection{Variation within the radial profiles and dynamic range}
\label{sec:variation}

We know galaxies have structure (e.g., spiral arms, bars) not described by their radial profiles.
Here we investigate the internal variation of the profiles for each galaxy
and the dynamic range of their radial gradients by comparing the surface-brightness dispersion 
{\it within} each annulus to the ratio of the profile maximum and minimum brightness. 
For this analysis, we have chosen the brightness at 250\,\micron, in order
to maximize the number of data points (see Sect. \ref{sec:mbbdata}).
Fig. \ref{fig:sb} shows the results of these comparisons on the convolved images.
In the left panel, all radial points are shown with the standard deviation
$\sigma_{250}$ of each annulus plotted against the surface brightness $\Sigma_{250}$
of that annulus\footnote{These standard deviations $\sigma$ are not the same as the
errors in the mean described in the previous section because they have not
been divided by $\sqrt{N_{\rm isophote}}$.}.
Both $\Sigma_{250}$ and  $\sigma_{250}$ vary by roughly 3 orders of magnitude 
over the galaxies in the KINGFISH sample,
with the standard deviation roughly proportional to the surface brightness. 
Nevertheless, at a given surface brightness, $\sigma_{250}$ can vary by almost a factor
of 10, but at an amplitude between 10 and 100 times smaller than the surface brightness
itself.
Thus, the dynamic range of the profiles exceeds the internal variation by a factor
of 10 or more.

This is perhaps better illustrated in the
right panel of Fig. \ref{fig:sb} where one point is given for each galaxy:
the average variation $\langle\sigma_{250}\rangle$ is plotted against the dynamic range of the 
profile calculated as the ratio of maximum and minimum surface brightnesses within it.
The error bars correspond to the %{\it maximum} and {\it minimum} standard deviation
mean standard deviation of each radial annulus in the galaxy.
The internal dispersions of the annuli are much smaller than their dynamic range,
except for four dwarf-irregular galaxies with rather flat profiles
($\Sigma_{250}$(Max)$\Sigma_{250}$/(Min)\,$\la$10, e.g., DDO\,154, DDO\,165 with only one radial point,
DDO\,53, Holmberg\,II).
In these cases, the average $\sigma_{250}$ is comparable
to the lowest surface brightnesses, but 3 to 5 times fainter than
the maximum $\Sigma_{250}$. 
Except for these extreme cases, the internal dispersion within each
annulus is much smaller than the radial gradient, so that
our analysis should reflect real radial trends, 
rather than just random noise introduced by internal variation within the rings.

%---------------------------------------------------------------
\subsection{Radial distributions of dust as exponential disks}
\label{sec:radial_disks}
\begin{figure*}[!ht]
\centerline{
\includegraphics[angle=0,width=0.33\linewidth,bb=18 300 592 718]{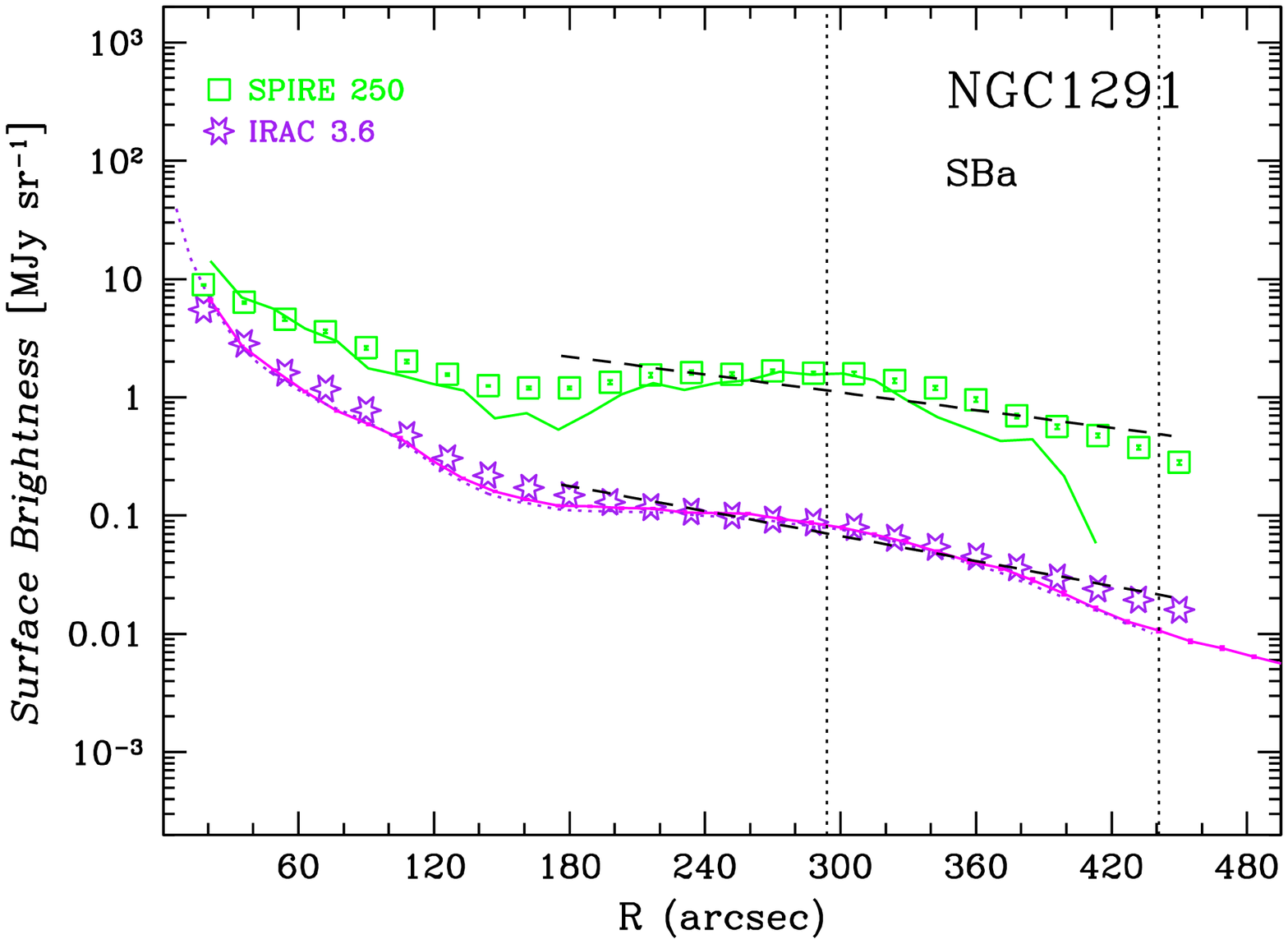}
\hspace{-0.4cm}
\includegraphics[angle=0,width=0.33\linewidth,bb=18 300 592 718]{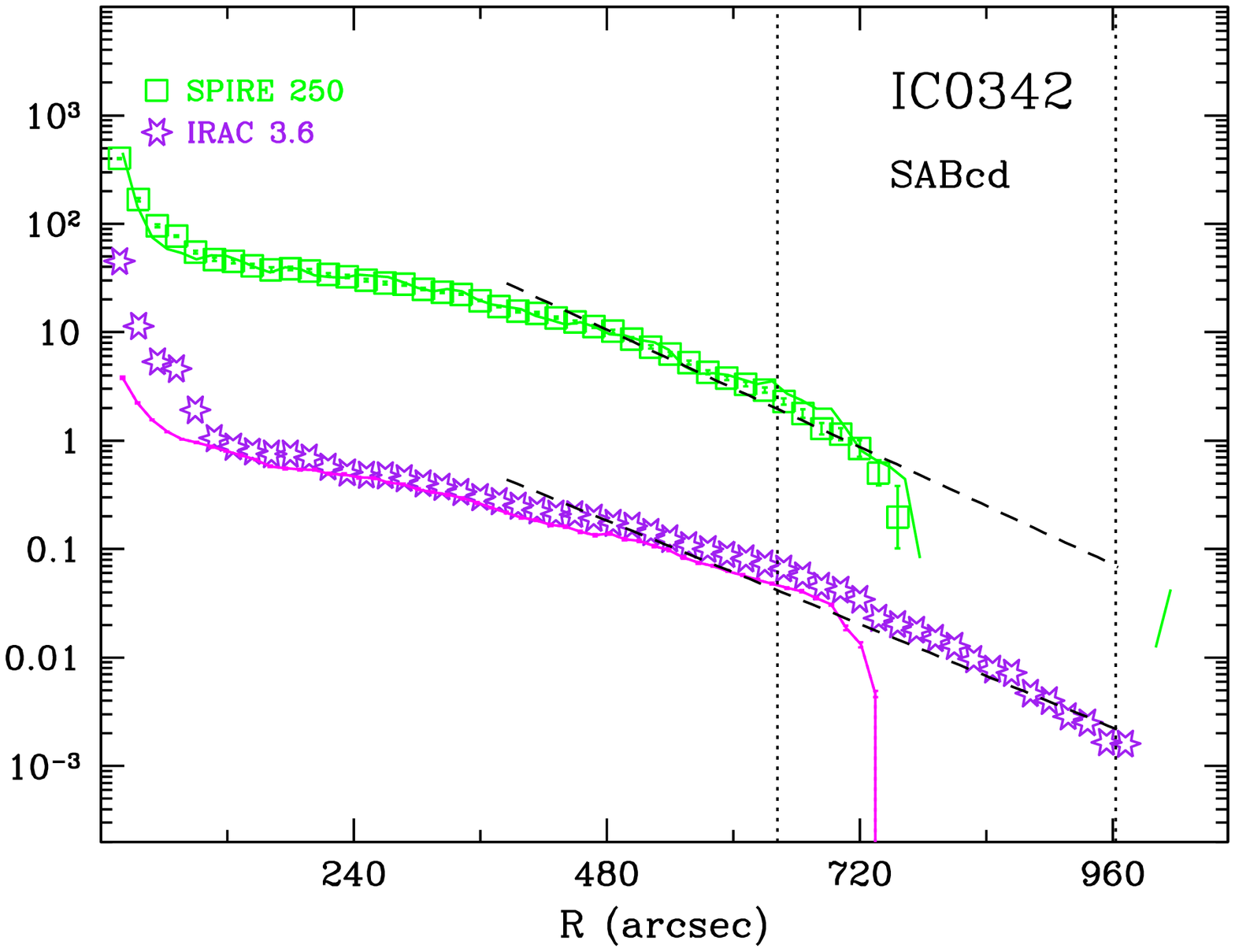}
\hspace{-0.4cm}
\includegraphics[angle=0,width=0.33\linewidth,bb=18 300 592 718]{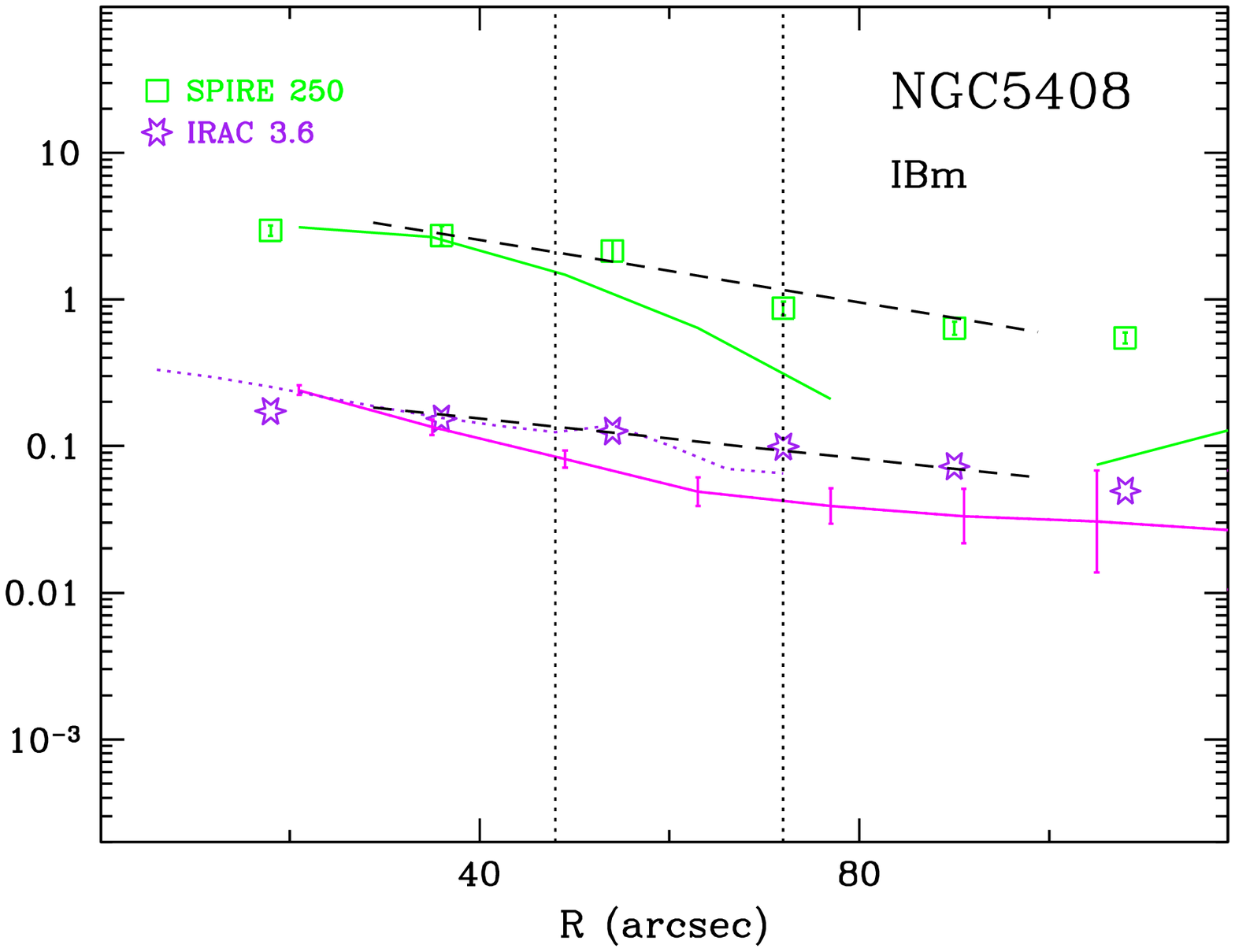}
}
\caption{Radial surface brightness profiles at IRAC 3.6\,\micron\ and SPIRE 250\,\micron\
shown in log-linear scale (c.f., Fig. \ref{fig:radial} with log-log)
for NGC\,1291 (left panel), IC\,342 (middle), and NGC\,5408 (right).
Open (green) squares show the 250\,\micron\ profiles and open (purple) stars 3.6\,\micron.
The IRAC 3.6\,\micron\ profiles from from Mu{\~n}oz-Mateos (priv. communication) correspond to solid
(purple) curves, and the 250\,\micron\ profiles extracted in a similar way
\citep[i.e., without the convolution by][]{aniano12} 
as solid (green) curves.
The dashed regression lines indicate the best-fit disk model fitted from 0.6*\ropt;
The vertical dotted lines indicate the optical radius, \ropt, and 
1.5\,\ropt. 
\label{fig:exponential}
}
\end{figure*}

Because dust is generally distributed in an exponential disk 
\citep[e.g.,][]{haas98,bianchi07,munoz09a},
we have fit the radial profiles at 3.6\,\micron\ and 250\,\micron\
with a simple exponential
%$\Sigma = \Sigma_0 \exp^{-R/\alpha}$
%$\Sigma = \Sigma_0 \exp^{-R/R_{\rm d}}$
$\Sigma = \Sigma_0 \exp({-R/R_{\rm d}})$
starting from normalized radius $R$/\ropt$\geq$0.6.
These exponential fits are not intended as true bulge-disk decompositions,
but rather as a characterization of the cool-dust distribution (at 250\,\micron) relative
to the stars (traced by 3.6\,\micron\ emission).
Because we want to avoid contamination by the bulge, the fits consider only the
external regions of the galaxy, 
with $R$/\ropt$\geq$0.6,
an arbitrary value chosen because the contribution from the bulge is usually
negligible at these radii, even in early-type spirals \citep[e.g.,][]{moriondo98}.
PAH emission at 3.3\,\micron\ in very dusty and active star-forming galaxies 
could contribute significantly to the 3.6\,\micron\ flux, which may produce
a slightly tighter correlation with 250\,\micron\ emission than would
be expected from a comparison of pure stellar emission and dust
\citep[e.g.,][]{zibetti11};
however, this is not expected to affect our conclusions.
Representative fits\footnote{The entire suite of exponential-fit plots is available as part of the on-line data set.}
are illustrated graphically in Fig. \ref{fig:exponential}.

\begin{figure}[!ht]
\centerline{
\includegraphics[angle=0,width=\linewidth,bb=18 200 592 718]{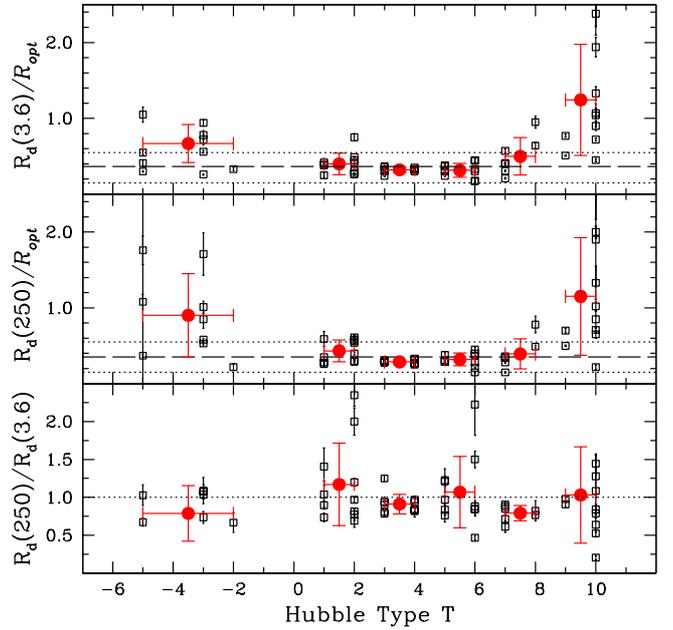}
}
\caption{Disk scalelengths normalized to \ropt\ as a function of Hubble type T.
The top panel shows the individual 3.6\,\micron\ scalelengths \rdisk$_{3.6}$,
the middle the individual 250\,\micron\ scalelengths \rdisk$_{250}$,
and the bottom their ratio.
In each panel,
the large circles with error bars report the means over a small
range of Hubble types as indicated by the horizontal error bars;
the vertical error bars are the standard deviations.
The horizontal dashed lines in the top two panels show the mean scalelength
obtained by averaging over spirals only (1\,$\leq$\,T\,$\leq$5); the 
dotted lines show the range from 0.25 to 0.45.
In the bottom panel, the dotted line shows a scalelength ratio of unity.
\label{fig:disks}
}
\end{figure}

The resulting normalized disk scalelengths\footnote{Because 
some dwarf galaxies (DDO\,165, Holmberg\,I, M\,81dwB) and one elliptical (NGC\,1404) 
have only one point in the multi-wavelength SED radial profiles, 
these galaxies are not considered in the exponential fit analysis.}
\rdisk/\ropt\ are shown in Fig. \ref{fig:disks}
where the top and middle panels show the scalelengths of the 3.6\,\micron\ 
and 250\,\micron\ profiles, and the bottom panel their ratio.
The mean values in the two top panels, shown as horizontal dotted lines,
correspond to the means over only spirals (1\,$\leq$\,T\,$\leq$5),
and are similar for both wavelengths: 
\rdisk/\ropt\,=\,0.37$\pm$0.11 for 3.6\,\micron\
and
\rdisk/\ropt\,=\,0.35$\pm$0.12 for 250\,\micron.
These normalized disk scalelength are larger than, although comparable to, those found for 
stars in spiral galaxies from detailed two-dimensional bulge-disk decompositions
\citep[$0.25-0.30$, e.g.,][]{giovanelli94,moriondo98,hunt04};
they are also slightly larger than, although within the uncertainties, the
median value, 0.29, of the dust scalelengths in SINGS galaxies found by
\citet{munoz09a}.
The slight increase in the disk scalelengths with respect to previous
work could be due to the method of sky subtraction in the convolved images
we are using here; if less sky were subtracted, disks would be shallower 
with larger scalelengths.
Nevertheless, the similar processing of both the 3.6\,\micron\ and 250\,\micron\ images
should obviate potential biases that could affect our conclusions.

The salient point is that the exponential distributions
of the cool dust and the stars are similar, as can be seen
in the bottom panel of Fig. \ref{fig:disks}.
The mean of the ratio of the 250\,\micron\ and 3.6\,\micron\ scalelengths is 0.96$\pm$0.36;
the median of the ratio is even smaller, 0.88, with quartile spreads of $\sim$0.2.
There is a tendency for both early-type galaxies (T$<$0, 10 galaxies) and very late-type
galaxies (T$\ga$9, 10 galaxies) to have larger exponential scalelengths, both in the dust
and in the stars.
Three spiral galaxies, NGC\,5457 (1.5), NGC\,5474 (2.2), NGC\,4826 (2.3), have
scalelength ratios $\ga$1.5, so in some cases, spiral dust disks are more extended
than their stellar ones.

In general, however,
our results are roughly consistent with the finding of \citet{munoz09a} that 
the dust distribution can be more extended than the stars by $\sim$10\% at most.
On the other hand, radiative transfer models of dust and stars in edge-on spirals
suggest that dust can be significantly more extended \citep{xilouris99,bianchi07,holwerda12a},
although this is not a general rule \citep[e.g.,][]{bianchi11}.
Analyses of the radial trends of dust extinction also imply that the dust 
distribution extends beyond the stellar one \citep[e.g.,][]{holwerda05}.
With our simple analysis,
we find that both dust and stars are distributed in an exponential decline
with similar scalelengths; the cool dust in most KINGFISH galaxies 
traced by 250\,\micron\ emission has a similar distribution to the
stellar one.
However, the stellar populations traced by IRAC 3.6\,\micron\ emission include
both main sequence and red giant stars,
but some studies have shown that dust emission is more closely related to
somewhat younger populations as traced by the optical $B$ band
\citep[e.g.,][]{alton98}.
Comparison of radial gradients of SFR, stellar age, or SFH would be necessary
to link our result with scenarios for inside-out disk growth \citep[e.g.,][]{williams09}. 
%More statistics with detailed modelling and deep observations of dust emission
%combined with SFR and SFH measurements
%are needed to better resolve this question.

%---------------------------------------------------------------
\section{Radial far-infrared colors}
\label{sec:colors}

\begin{figure*}[!ht]
\centerline{
\includegraphics[angle=0,height=0.33\linewidth,bb=18 200 592 718]{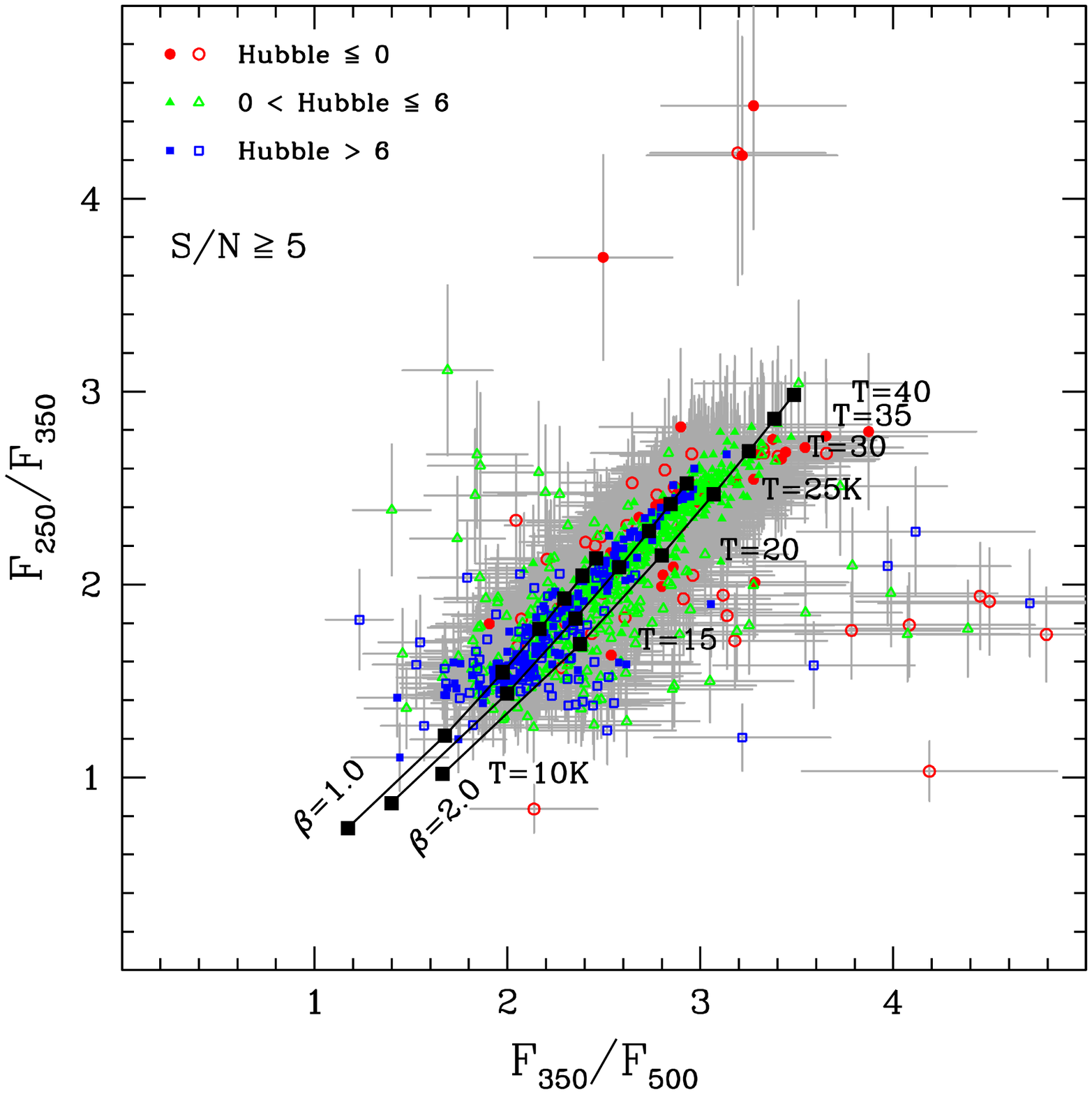}
%\hspace{-2.0cm}
\hspace{-1.2cm}
\includegraphics[angle=0,height=0.33\linewidth,bb=18 200 592 718]{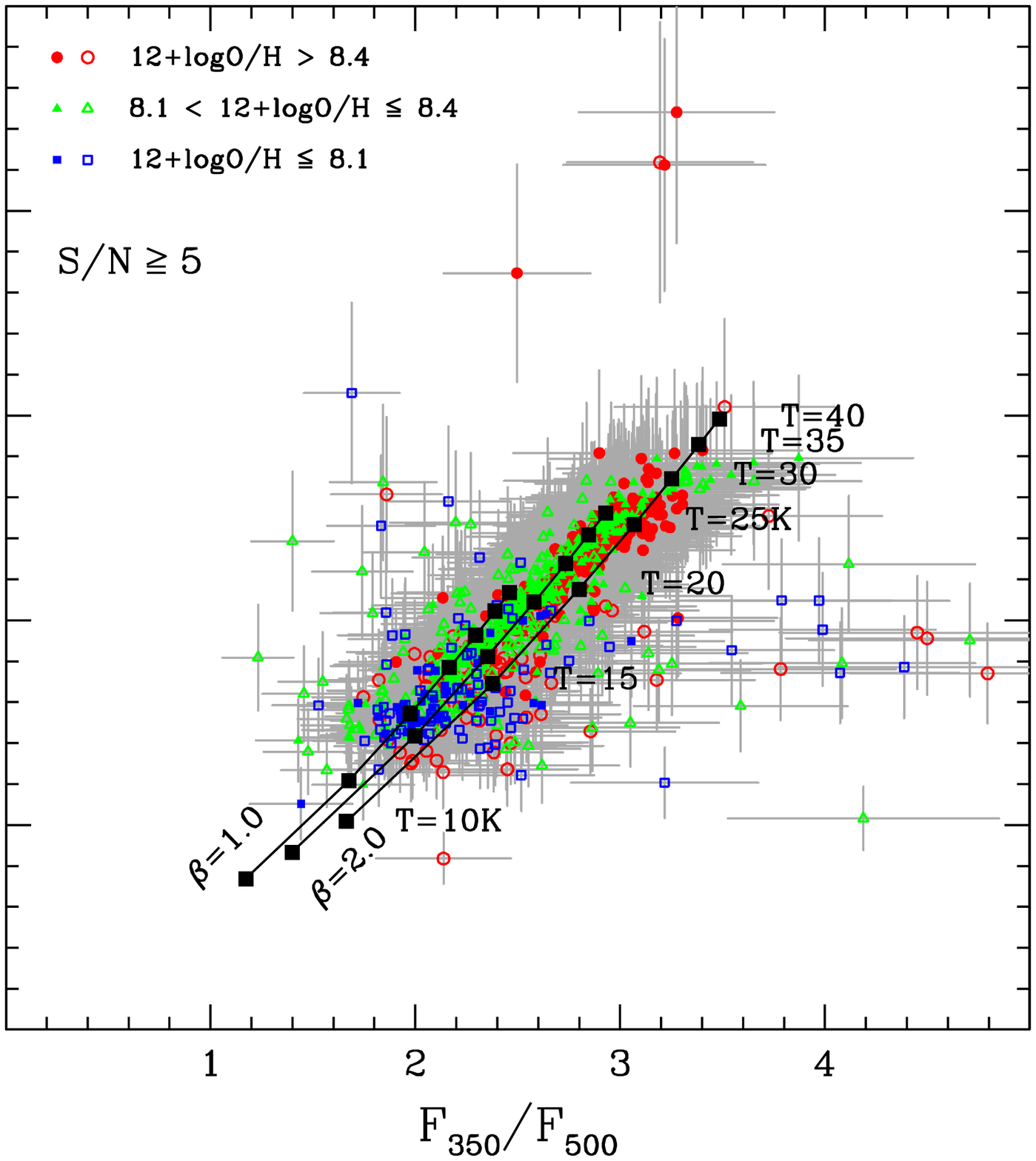}
%\hspace{-2.0cm}
\hspace{-1.2cm}
\includegraphics[angle=0,height=0.33\linewidth,bb=18 200 592 718]{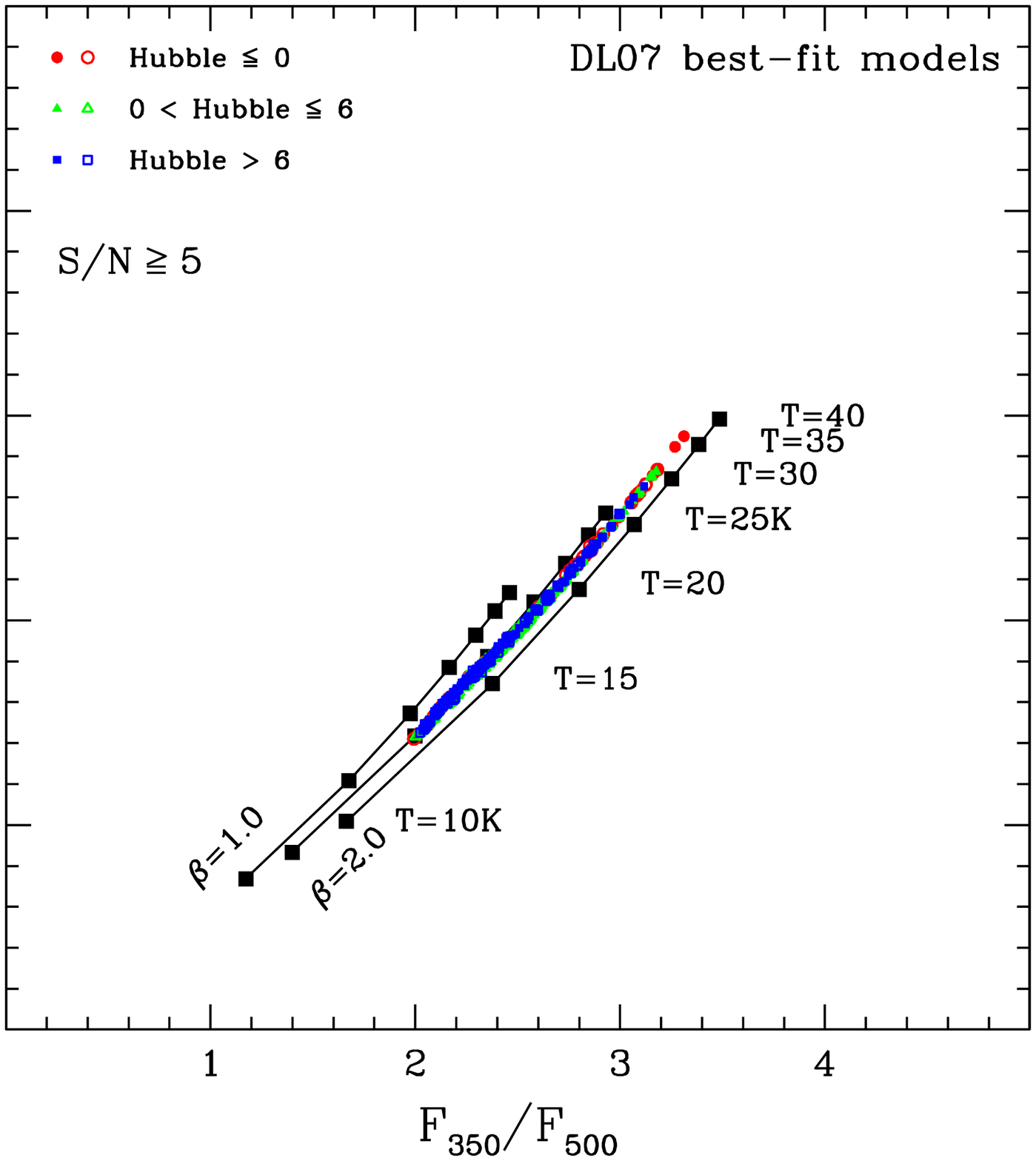}
}
\caption{SPIRE flux ratios of the radial profiles with S/N\,$\geq$\,5.
In the left and right panels, points are plotted by Hubble type with (red) circles corresponding
to early types (T$\leq$0), (green) triangles to spirals (0$<$T$\leq$6), and
(blue) squares to late types (T$>$6).
In the middle panel, points are distinguished by O/H with (red) circles showing
\logoh$>$8.4, (green) triangles 8.0$<$\logoh$\leq$8.4, and (blue) squares \logoh$\leq$8.1.
Filled symbols correspond to positions with normalized (to optical radius \ropt) radii 
within $R$/\ropt$\leq$0.8, and open symbols to larger radii.
The right panel shows DL07 best-fit models only.
In all panels, MBBF fits are shown with three values of emissivity index $\beta$: from left
to right $\beta$\,=\,1.0, $\beta$\,=\,1.5, $\beta$\,=\,2.0, respectively.
Temperatures are also labeled from \td\,=\,10\,K to \td\,=\,40\,K. 
\label{fig:spirecolors}
}
\end{figure*}

Before pursuing the relation between the physical parameters driving the DL07
models, and their connection with MBBV fits, we present here color-color diagrams
for the $\sim$900 SEDs from the radial profile data independently of their position.
Figure \ref{fig:spirecolors} shows the SPIRE flux ratios,
F$_{250}$/F$_{350}$ vs. F$_{350}$/F$_{500}$.
The left and middle panels show the data, coded by either Hubble type, T, or 
oxygen abundance, \logoh, and the right panel shows the best-fit DL07 models
described in Sect. \ref{sec:dl07}.
All panels show only data with S/N\,$\geq$\,5;
we have used this relatively high S/N threshold because we are examining colors
and want to ensure their reliability.
Single-temperature MBBF models are also illustrated with three values of
emissivity index $\beta$; 
%the MBB values have been reported (with an inverse correction) 
%to the scale of the SPIRE observed flux calibration.
the MBB models have been % corrected %(with an inverse correction) 
%to the SPIRE observed flux calibration (rather than correcting the fluxes themselves).
integrated over the SPIRE response function for comparison with the observed
SPIRE fluxes.
Dust emitting at a single temperature with a single $\beta$ would occupy
a single point on such a plot.

\begin{figure*}[!ht]
\centerline{
\includegraphics[angle=0,height=0.33\linewidth,bb=18 200 592 718]{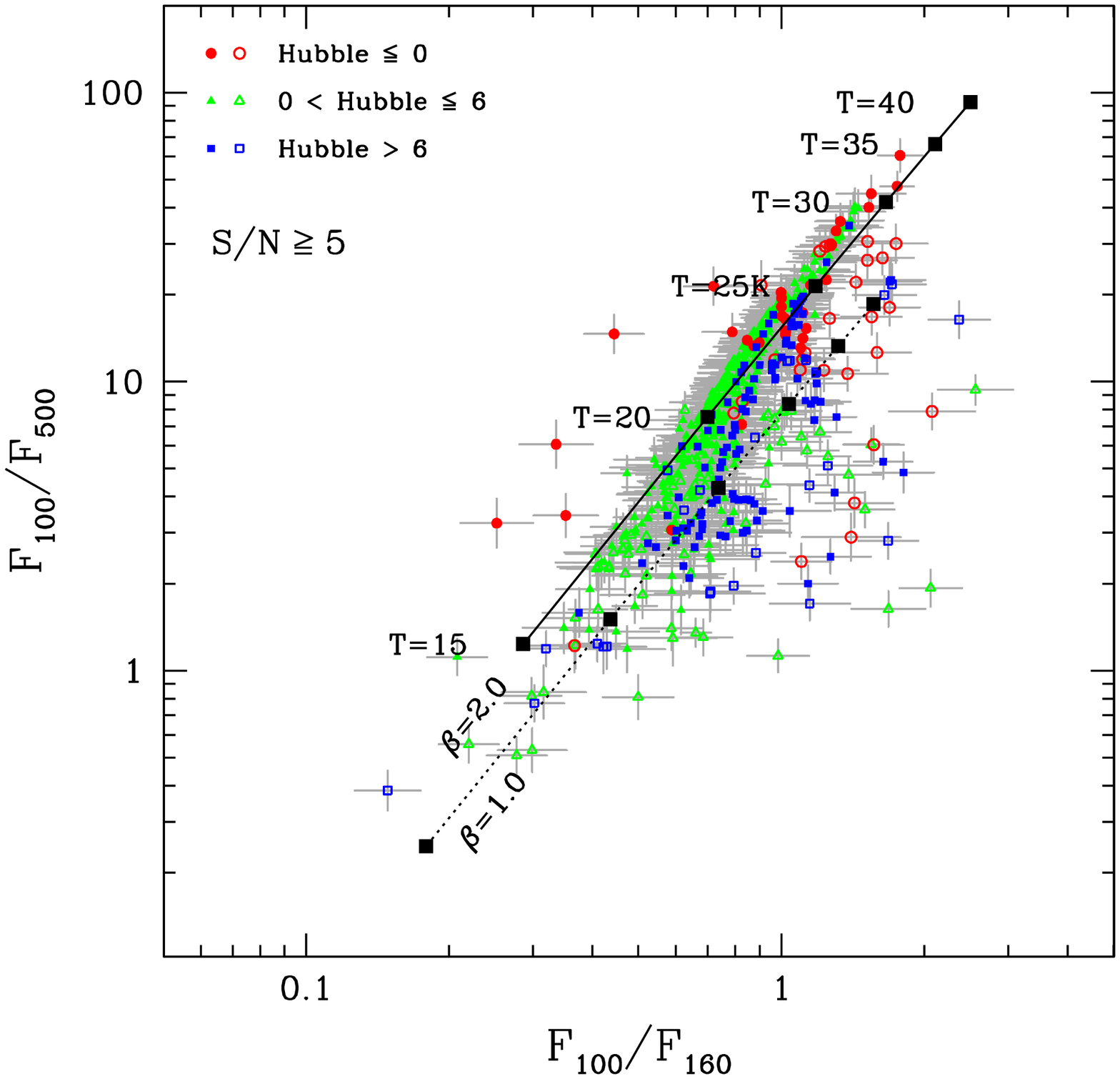}
%\hspace{-2.0cm}
\hspace{-1.2cm}
\includegraphics[angle=0,height=0.33\linewidth,bb=18 200 592 718]{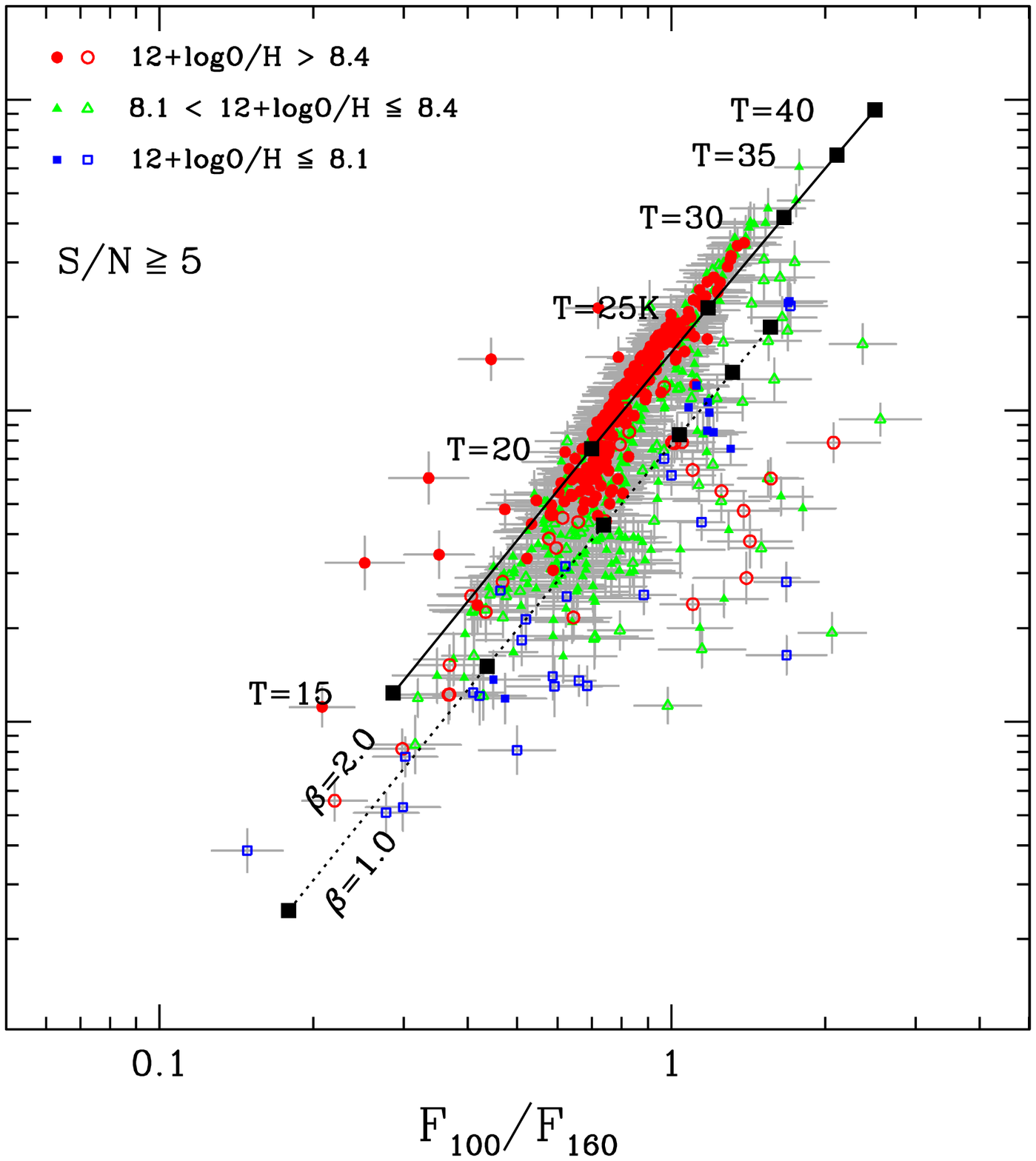}
%\hspace{-2.0cm}
\hspace{-1.2cm}
\includegraphics[angle=0,height=0.33\linewidth,bb=18 200 592 718]{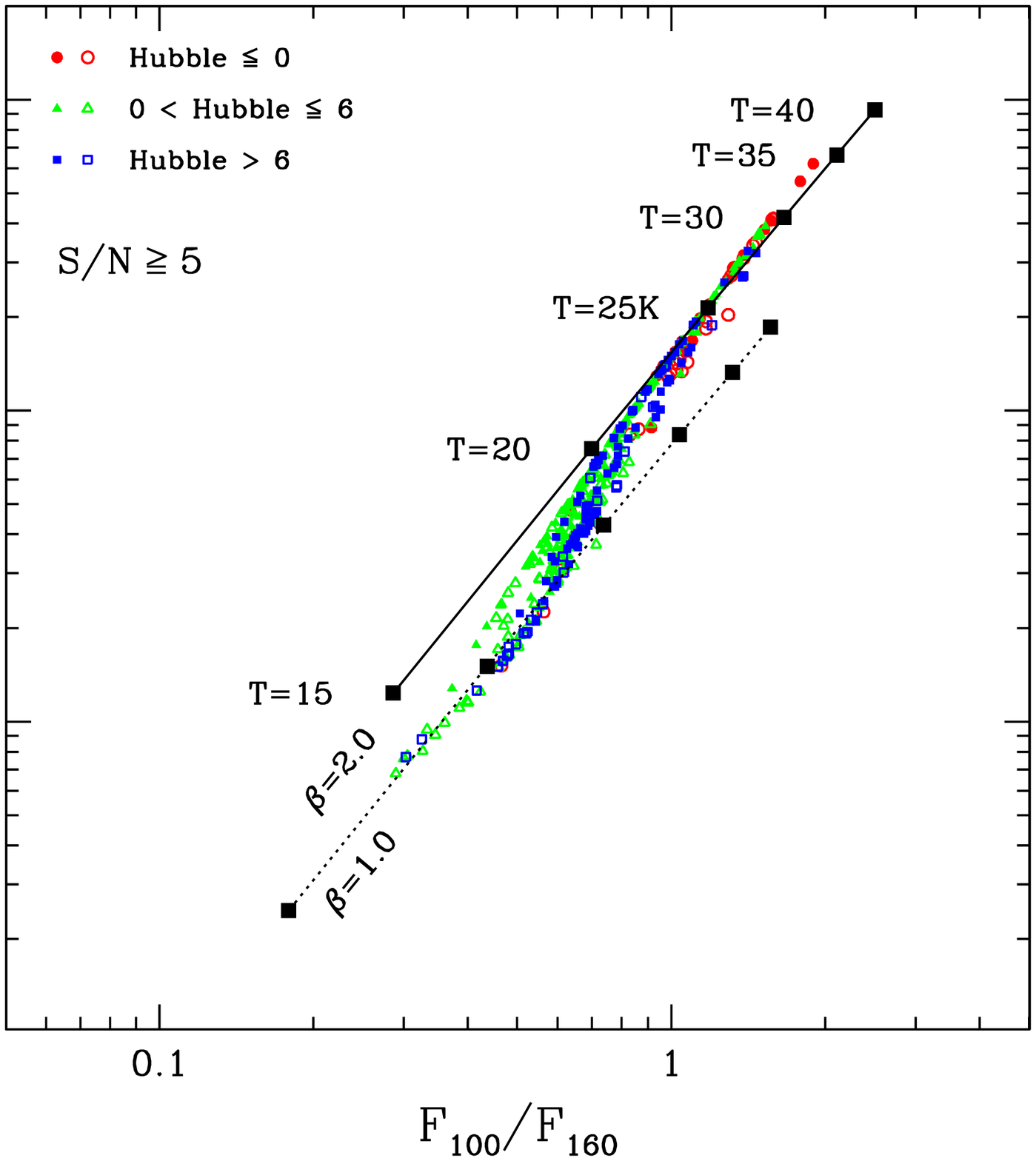}
}
\caption{PACS (100\,\micron, 160\,\micron) and SPIRE 500\,\micron\ flux ratios of the radial profiles with S/N\,$\geq$\,5.
As in Fig. \ref{fig:spirecolors},
in the left and right panels, points are plotted by Hubble type with (red) circles corresponding
to early types (T$\leq$0), (green) triangles to spirals (0$<$T$\leq$6), and
(blue) squares to late types (T$>$6).
In the middle panel, points are distinguished by O/H with (red) circles showing
\logoh$>$8.4, (green) triangles 8.0$<$\logoh$\leq$8.4, and (blue) squares \logoh$\leq$8.1.
Filled symbols correspond to positions with normalized (to optical radius \ropt) radii 
within $R$/\ropt$\leq$0.8, and open symbols to larger radii.
The right panel shows DL07 best-fit models only.
In all panels, MBBF fits are shown with two values of emissivity index $\beta$: 
$\beta$\,=\,1.0 (dotted line) and  $\beta$\,=\,2.0 (solid line).
Temperatures are also labeled from \td\,=\,15\,K to \td\,=\,40\,K. 
\label{fig:pacsspirecolors}
}
\end{figure*}

\begin{figure*}[!t]
\centerline{
\hbox{
\includegraphics[angle=0,height=0.33\linewidth,bb=18 200 592 718]{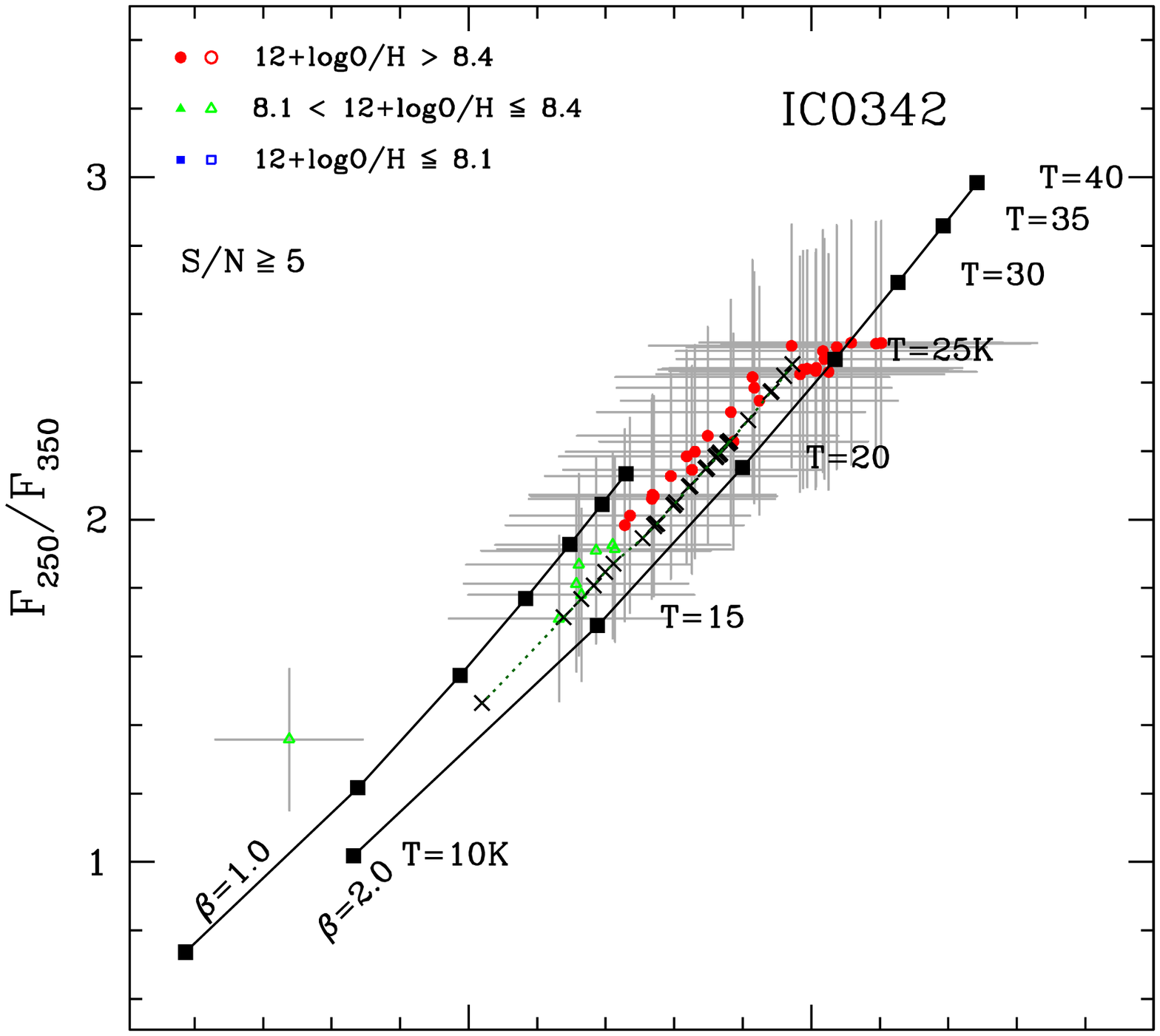}
%\hspace{-2.0cm}
\hspace{-1.3cm}
\includegraphics[angle=0,height=0.33\linewidth,bb=18 200 592 718]{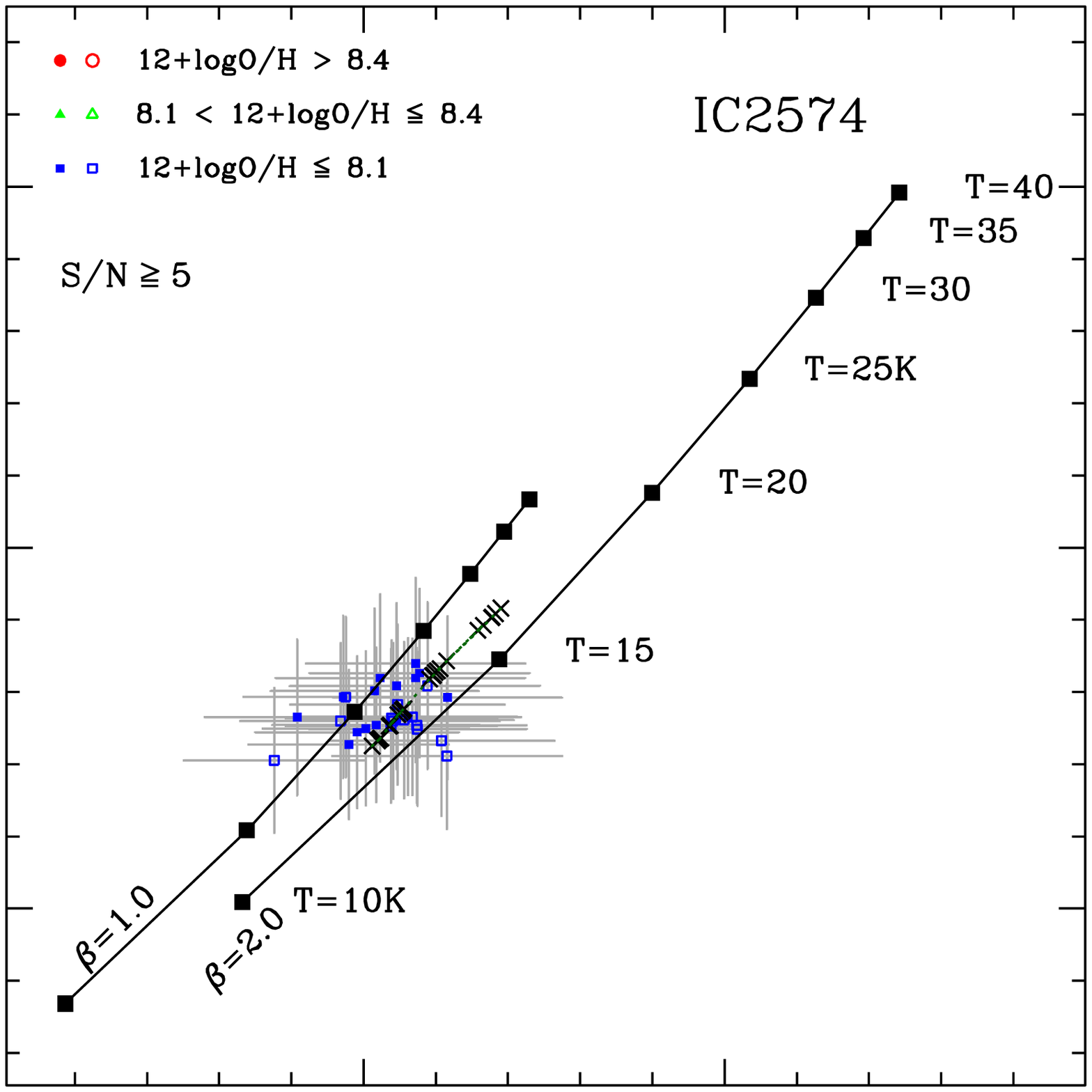}
%\hspace{-2.0cm}
\hspace{-1.3cm}
\includegraphics[angle=0,height=0.33\linewidth,bb=18 200 592 718]{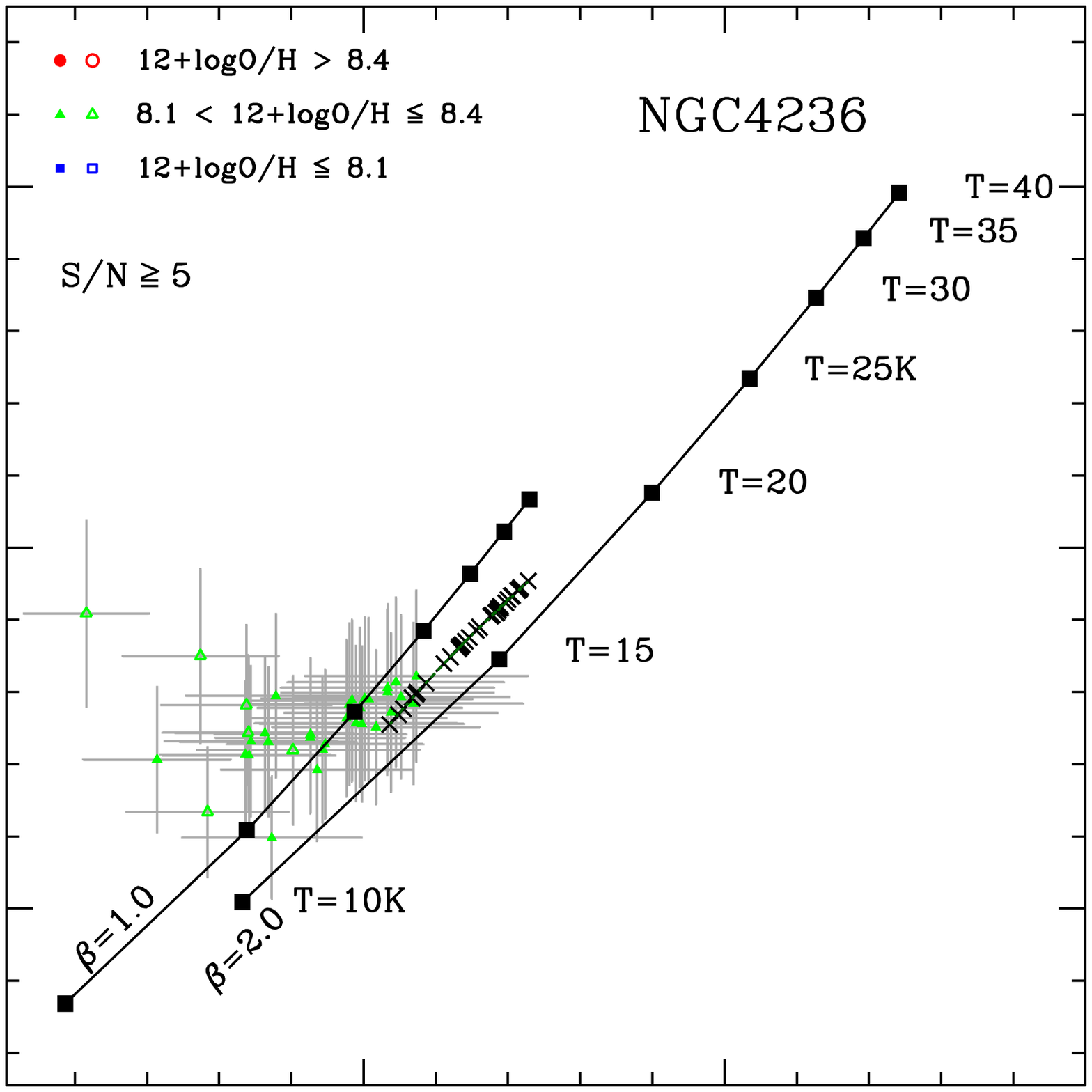}
}
}
\vspace{-0.52cm}
\centerline{
\hbox{
\includegraphics[angle=0,height=0.33\linewidth,bb=18 200 592 718]{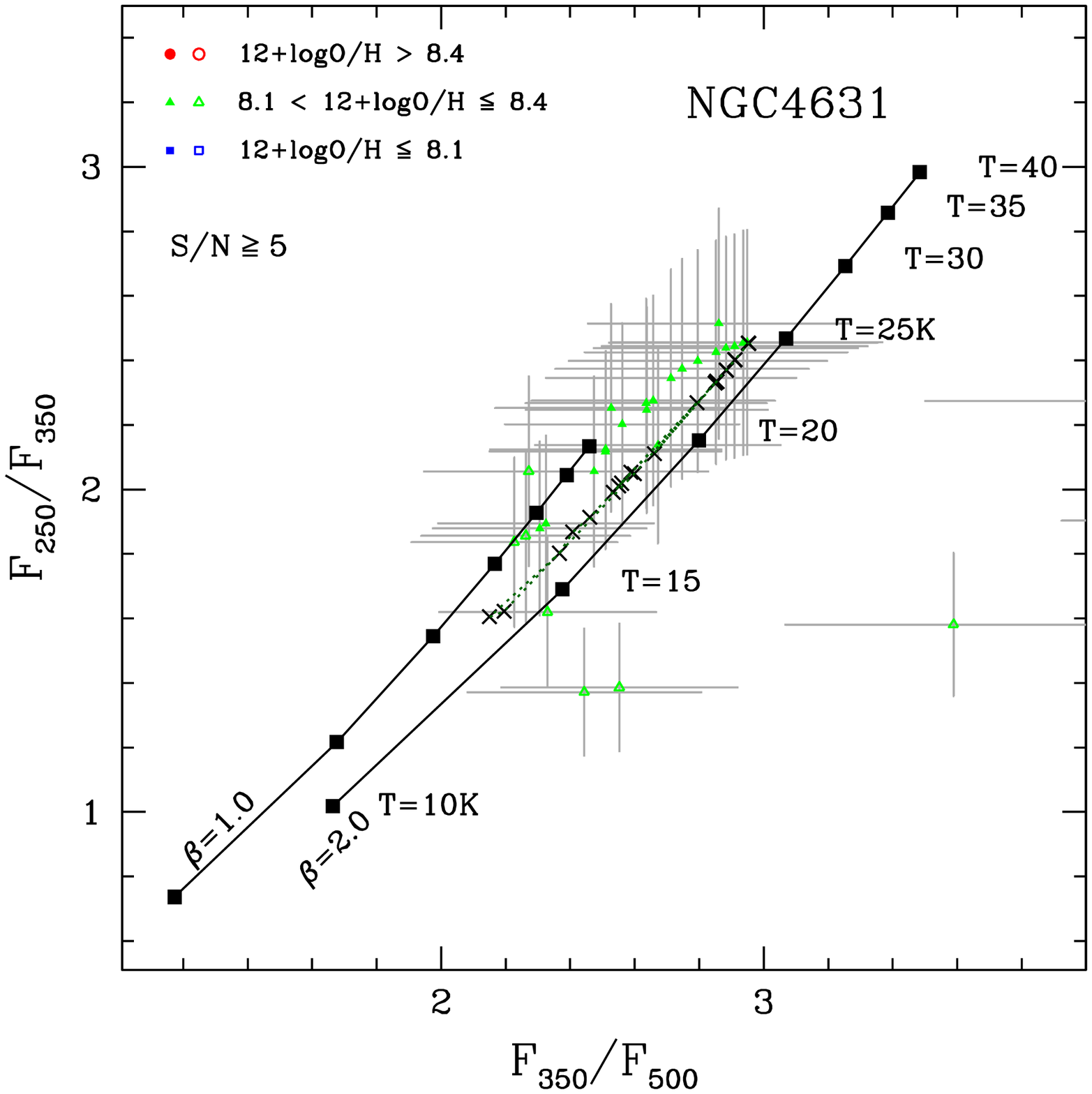}
%\hspace{-2.0cm}
\hspace{-1.3cm}
\includegraphics[angle=0,height=0.33\linewidth,bb=18 200 592 718]{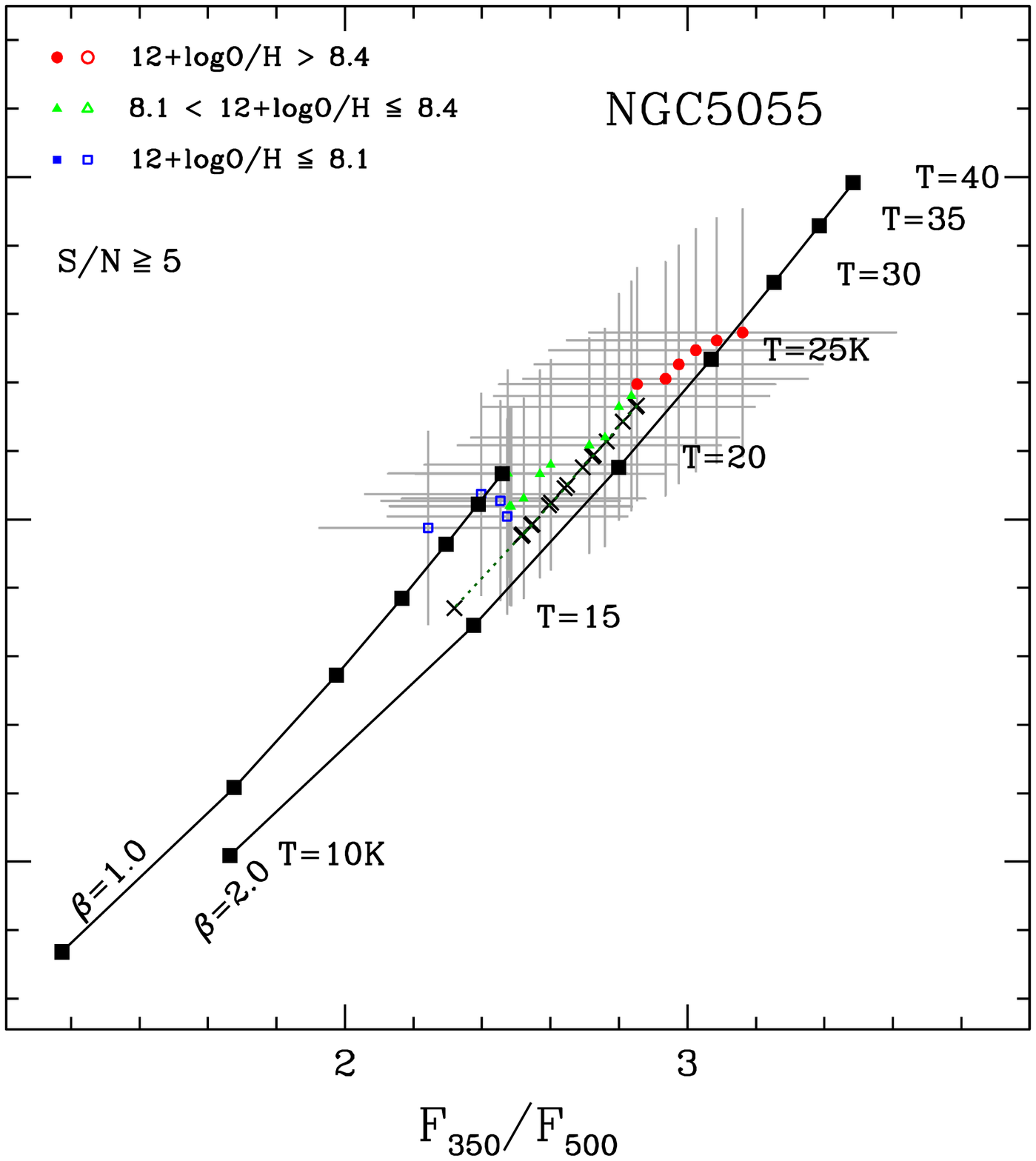}
%\hspace{-2.0cm}
\hspace{-1.3cm}
\includegraphics[angle=0,height=0.33\linewidth,bb=18 200 592 718]{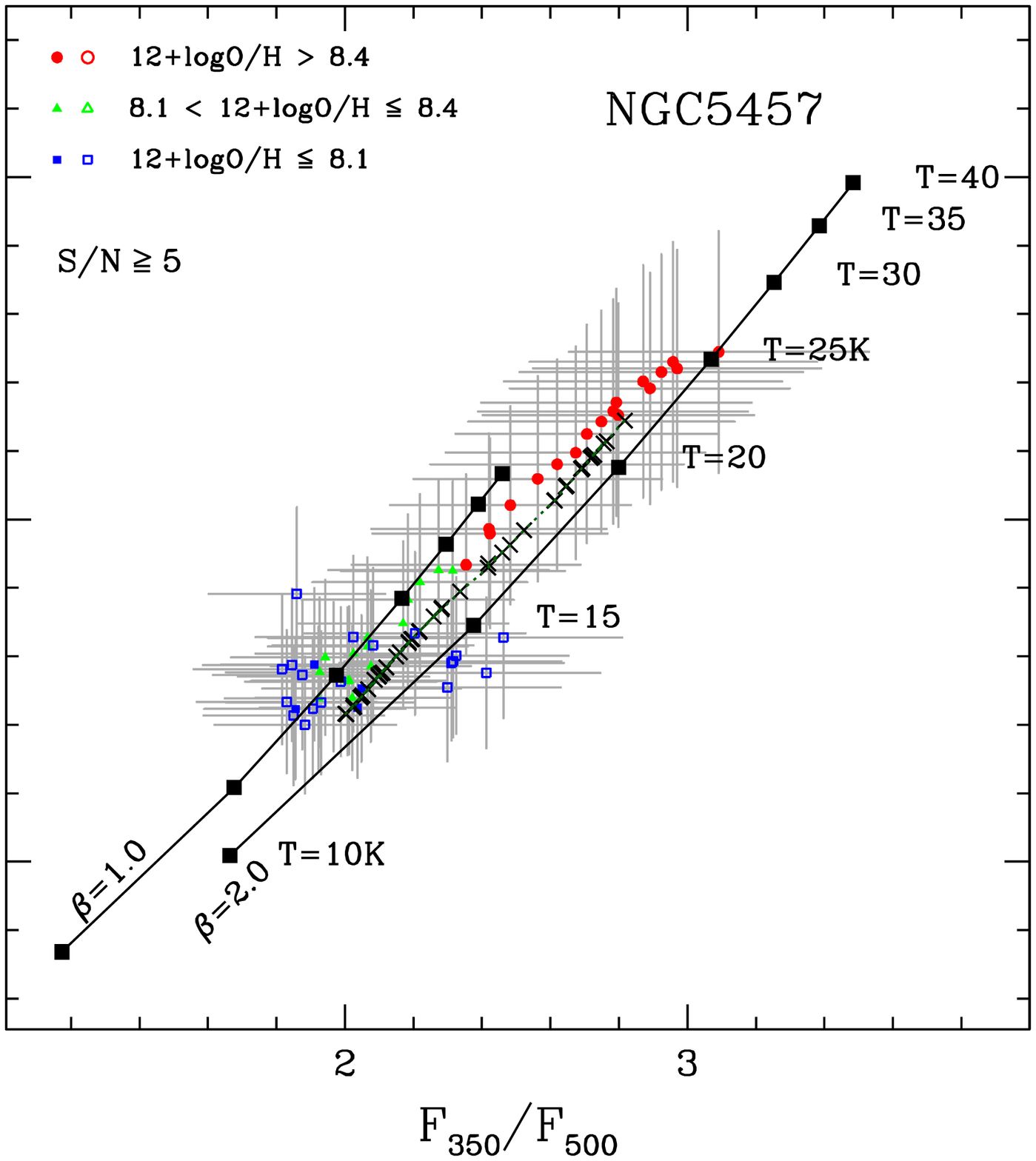}
}
}
\caption{SPIRE colors of individual galaxies with more than 30 radial data points.
Data are distinguished by O/H with (red) circles showing
\logoh$>$8.4, (green) triangles 8.0$<$\logoh$\leq$8.4, and (blue) squares \logoh$\leq$8.1.
Filled symbols correspond to positions with normalized (to optical radius \ropt) radii 
within $R$/\ropt$\leq$0.8, and open symbols to larger radii.
DL07 best-fit models for the data are shown with $\times$, connected by a dotted line. 
In all panels, MBBF fits are given for two values of emissivity index $\beta$: 
$\beta$\,=\,1.0 (dotted line) and  $\beta$\,=\,2.0 (solid line).
Temperatures are also labeled from \td\,=\,15\,K to \td\,=\,40\,K. 
\label{fig:spirecolors_indiv}
}
\end{figure*}

Figure \ref{fig:spirecolors} illustrates that the data have a wide range
of apparent dust temperatures and generally fall 
between $\beta$\,=\,1 and $\beta$\,=\,2, as expected from previous work
\citep[e.g.,][]{boselli12}.
Nevertheless, even at this significant S/N\,$\geq$\,5, there are clear
variations.
Fig. \ref{fig:radial_colors} plots SPIRE colors against galactocentric distance
and shows that most of the data that might indicate a potential 500\,\micron\ excess
(e.g., F$_{350}$/F$_{500}\la1.5$) are in the external regions of the galaxies ($R$/\ropt\,$>$0.8).
However, not all galaxies with a possible excess
are either low metallicity (\logoh$\leq$8.1) or late type (T$>$6).
The search for long-wavelength excesses will be discussed in Sect. \ref{sec:models}.

Figure \ref{fig:spirecolors} also demonstrates the unexpected result that
SPIRE colors are not necessarily indicative of the average emissivity
index of the dust.
The range in colors spanned by the best-fit DL07 models (see Sect. \ref{sec:dl07}) is narrower than
observed, and 
the best-fit DL07 colors in the right panel do not fall on $\beta$\,=\,2,
as would be expected given the mean DL07 value of $\beta$ of 2.08 \citep{dl07,bianchi13}.
The apparent emissivity index of the DL07 models in this wavelength range is
flatter, $\beta\sim$1.5 with a maximum value toward warmer temperatures of $\beta\sim$1.7 at SPIRE wavelengths.
%This behavior does not depend on the mixing due to the different
%temperatures attained by grains of different radii and composition,
%even under the same intensity of the ISRF. 
This behavior depends on the
intrinsic grain properties of dust in the DL07 models at SPIRE wavelengths.
Although the mean emissivity of the DL07 models over the whole PACS and
SPIRE range can be globally described by a power law with $\beta\sim2.1$,
the model emissivity at $\lambda\geq$250\,\micron\ is better described by a broken
power law, with a slightly steeper $\beta$ for the 250/350
ratio than for the 350/500 ratio ($\beta$ decreasing from 2.2 to 2.0).
These intrinsic properties of the dust model, leading to a 350\,\micron\
emissivity 4\% smaller than predicted by the $\beta$\,=\,2.1 power-law
fit, are responsible for the colors of the DL07 best-fit models.
The reason for the change in slope at the SPIRE wavelengths is
due mostly to modifications in the optical properties of astronomical
silicates, which were made by \citet{li01} to provide a better
match between the model and the FIRAS high-latitude Galaxy spectrum.
These results imply that IR colors over a small wavelength range
should not be used to infer intrinsic dust properties.

PACS$+$SPIRE colors are given in Fig. \ref{fig:pacsspirecolors}
where we plot PACS/SPIRE F$_{100}$/F$_{500}$ vs. PACS F$_{100}$/F$_{160}$
\citep[see also][]{cortese14}. 
As in Fig. \ref{fig:spirecolors}, the left panel shows the data coded
by Hubble type, the middle by \logoh, and the right panel shows the best-fit DL07 models
(see Sect. \ref{sec:dl07}).
Single-temperature MBBF models are illustrated here with two values of
emissivity index $\beta$; 
as before, 
%the MBB models have been corrected %(with an inverse correction) 
%to the SPIRE observed flux calibration (rather than correcting the fluxes themselves).
the MBB values have been integrated over the PACS and
SPIRE response functions to be consistent with %reported (with an inverse correction) 
the scale of the PACS$+$SPIRE observed flux calibrations.
Again, dust emitting at a single temperature with a given emissivity index $\beta$
would appear as a single point.

Figure \ref{fig:pacsspirecolors} shows that for a given PACS color (horizontal axis)
%the PACS$+$SPIRE color 
F$_{100}$/F$_{500}$ (vertical) can 
slightly exceed the ratios expected from a single-temperature MBB.
In addition, the same 
%PACS$+$SPIRE 
F$_{100}$/F$_{500}$ color can also fall well below expectations,
implying perhaps a 500\,\micron\ excess.
In any case, there is significant variation, and 
these trends will be discussed further in Sect. \ref{sec:models}.

The best-fit DL07 models given in the right panel of Fig. \ref{fig:pacsspirecolors}
and described in Sect. \ref{sec:dl07}
change apparent emissivity index as a function of \td.
For low temperatures (\td$\la$25\,K), $\beta$ is lower than the nominal
value of $\sim$2, reaching an extreme of $\beta\sim$1 for \td$\la$20\,K.
At higher temperatures, the best-fit DL07 models fall along the $\beta$\,=\,2
MBB fits, in accordance with the mean emissivity value.
These simple color-color plots imply that fixed-emissivity models can
well approximate dust emission that is well fit by low values of $\beta$.
We discuss possible reasons why in Sect. \ref{sec:discussion}.

The FIR colors of galaxies vary with position, typically as a signature
of radial temperature gradients.
Such trends are illustrated in Fig. \ref{fig:spirecolors_indiv} where we
show the SPIRE flux ratios of individual galaxies with more than 30 radial data points:
IC\,342, IC\,2574, NGC\,4236, NGC\,4631, NGC\,5055, NGC\,5457.
%Similarly to the trends shown in Fig. \ref{fig:spirecolors}, 
Observed SPIRE flux ratios correspond to cooler temperatures (occurring
in the outer regions of the galaxies
and apparently flatter emissivity $\beta\sim$1). 
The DL07 best-fit models in Fig. \ref{fig:spirecolors_indiv}
reproduce the SPIRE colors to within the uncertainties, although there
may be systematic variations.
We will examine these in more detail in Sect. \ref{sec:models}.

%---------------------------------------------------------------
\section{Modeling the radial dust SEDs}
\label{sec:fitting}

\hers\ has enabled the analysis of the cool dust in galaxies
in a detail that was not previously possible.
Although much work has focused on simple MBB fits of
the dust SED, the application of more physically motivated dust models 
such as those of \citet{dl07} can give greater physical insight.
Here we combine both approaches and fit the SED of each annular region
in each galaxy with three sets of models:
the DL07 templates and 
MBB fits with variable (MBBV) and fixed (MBBF) emissivity index $\beta$. 
In addition, to better comprehend the relation between the ISRF and the
apparent properties of dust grains that emerge from MBB fits, we fitted the 
best-fit DL07 models themselves with the two kinds of MBB fits.
Thus, in total, we will analyze five sets of fits:
three sets of model fits to data (MBBV, MBBF, DL07)
and two sets of fits (MBBV, MBBF) to the best-fit DL07 models. 
The profile extraction gives 1166 radial data points, although we apply
S/N thresholds (see below) so in the analysis only a subset of these is considered 
($\sim$800 data points). 

\begin{figure*}[!ht]
\centerline{
\hbox{
\includegraphics[width=0.45\linewidth,bb=18 308 588 716]{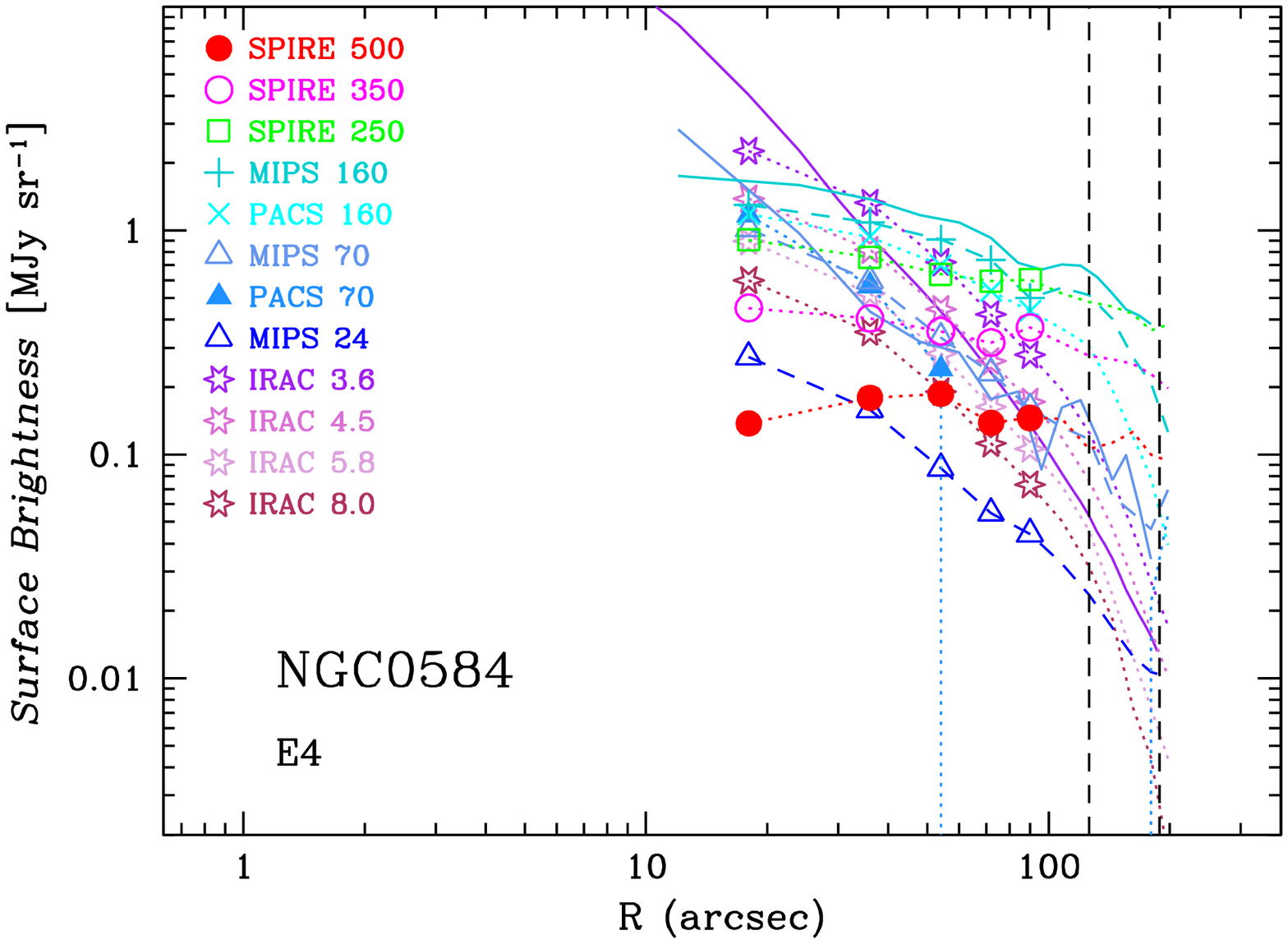}
%\hspace{0.1\linewidth}
\hspace{0.05\linewidth}
\includegraphics[width=0.45\linewidth,bb=18 167 592 718]{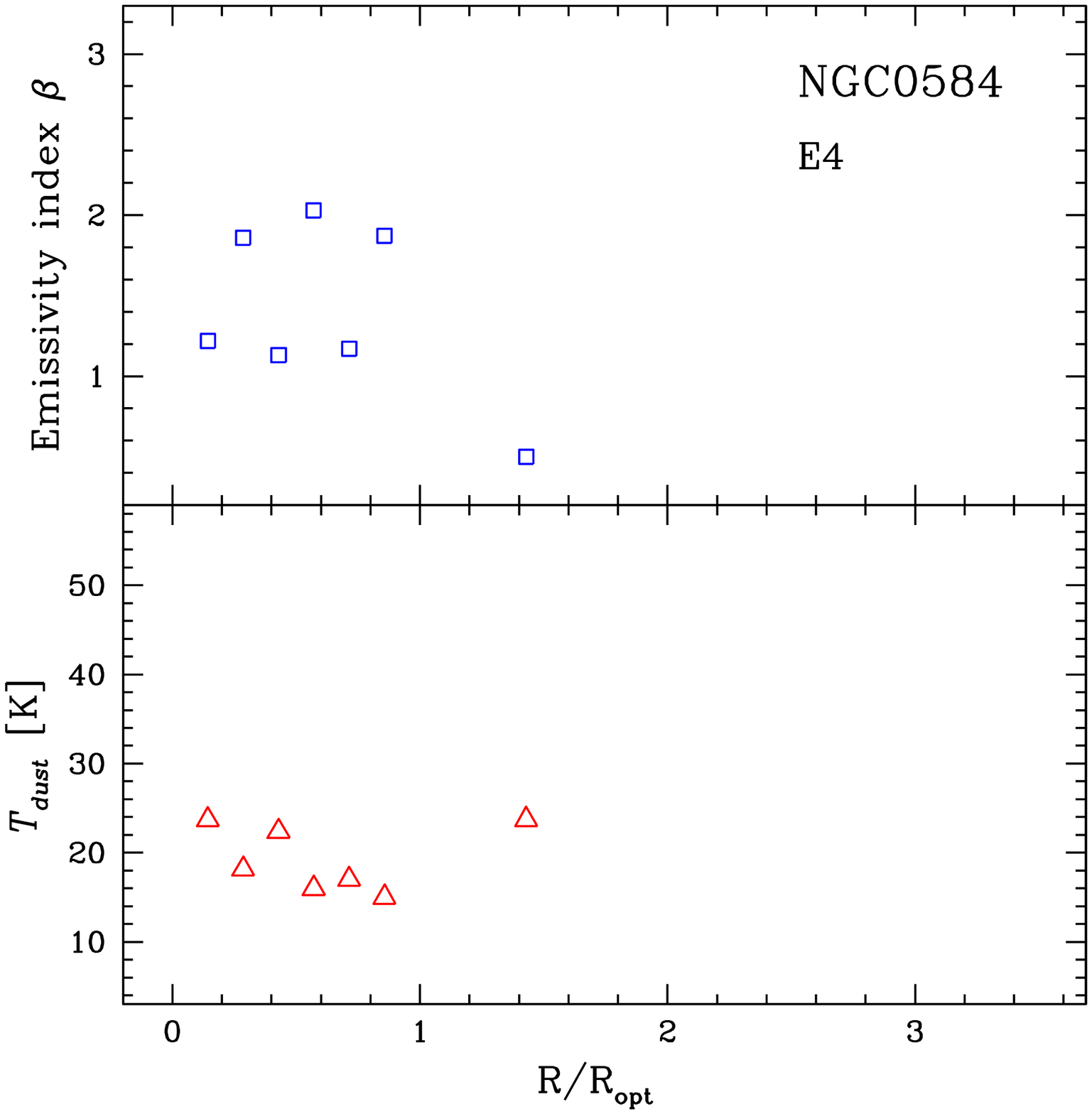}
}
}
\vspace{2.15\baselineskip}
\vspace{-1.15\baselineskip}
\centerline{
\hbox{
\includegraphics[width=0.22\linewidth,bb=18 308 588 716]{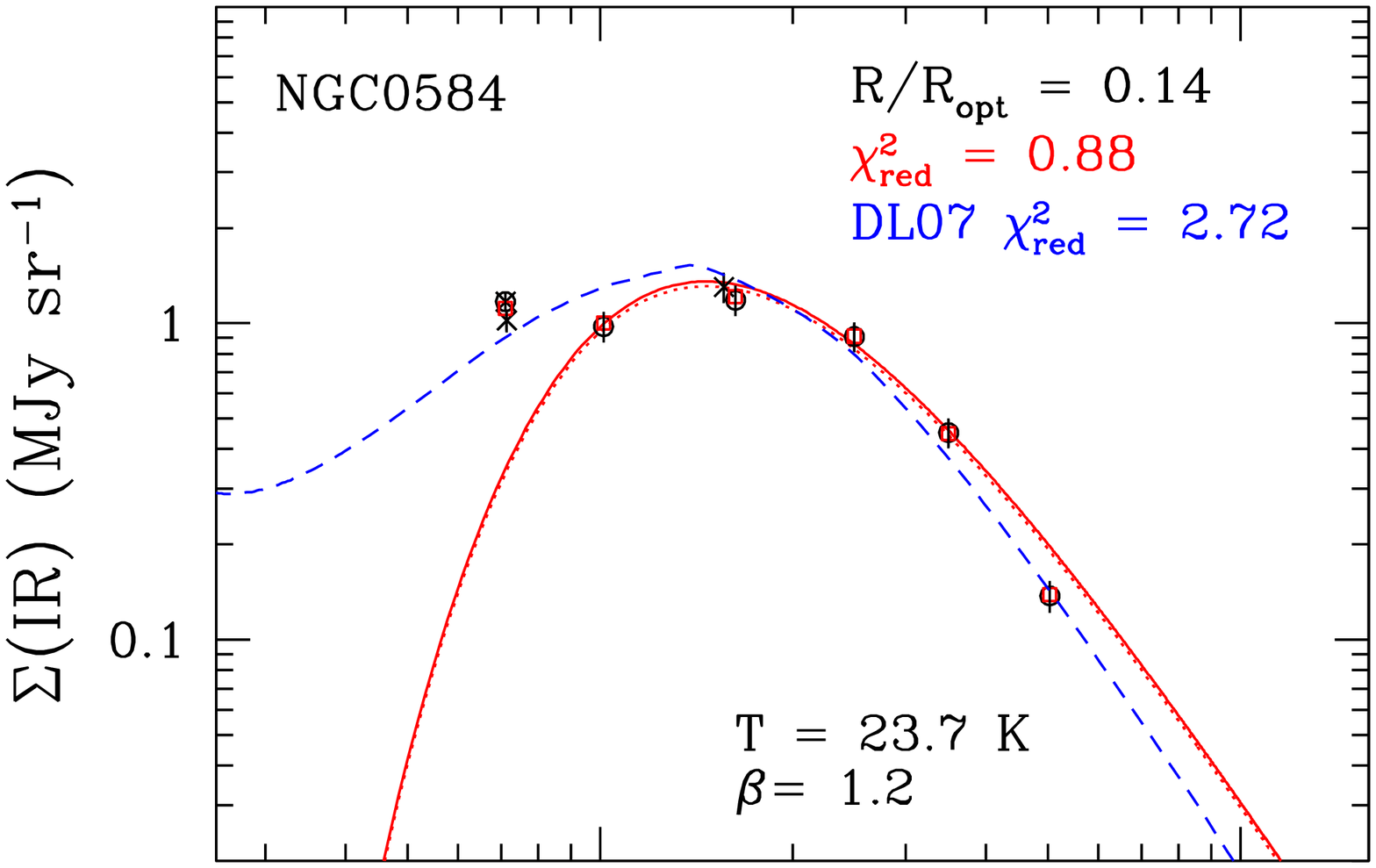}
\hspace{-0.045\linewidth}
\includegraphics[width=0.22\linewidth,bb=18 308 588 716]{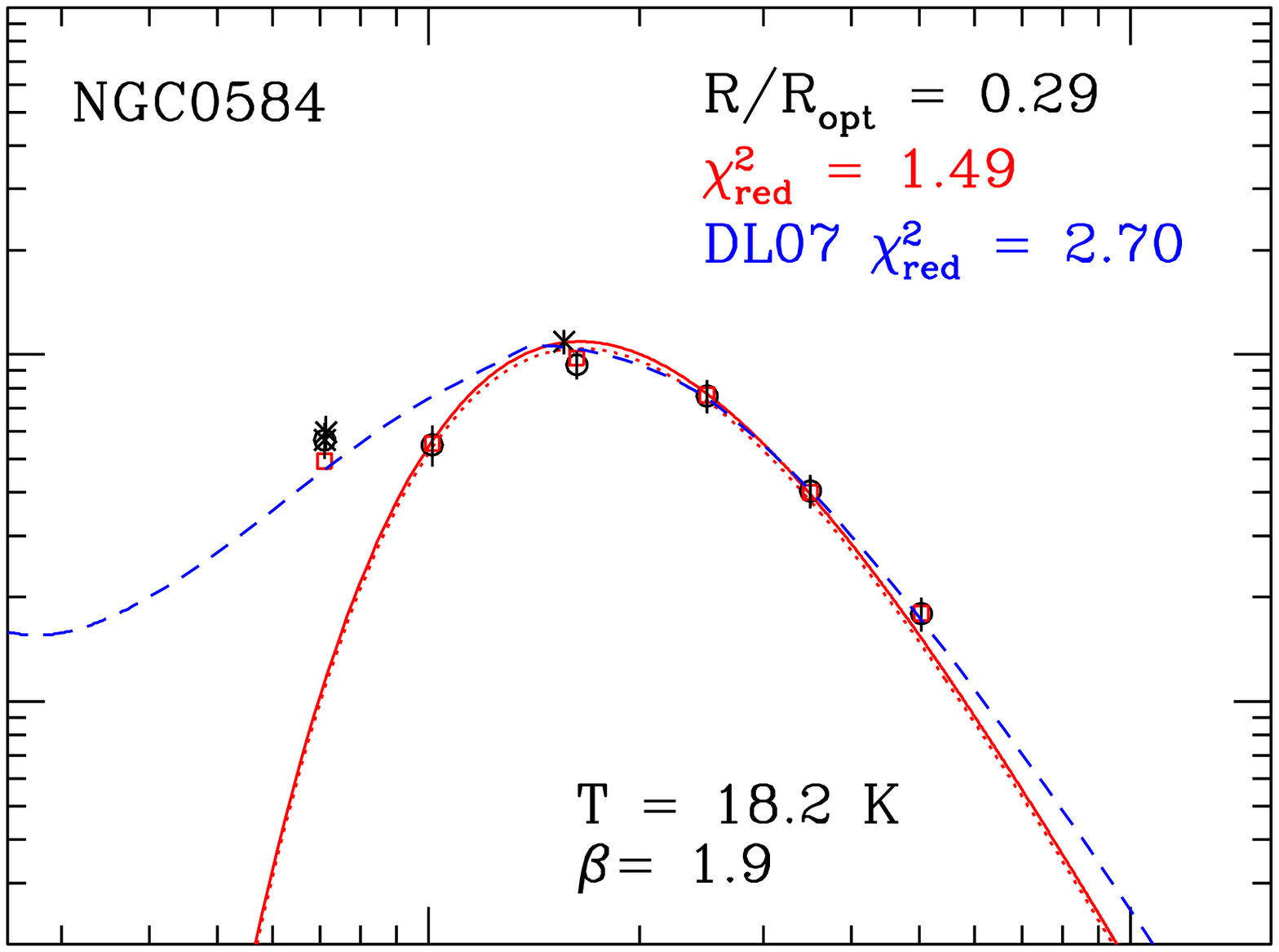}
\hspace{-0.045\linewidth}
\includegraphics[width=0.22\linewidth,bb=18 308 588 716]{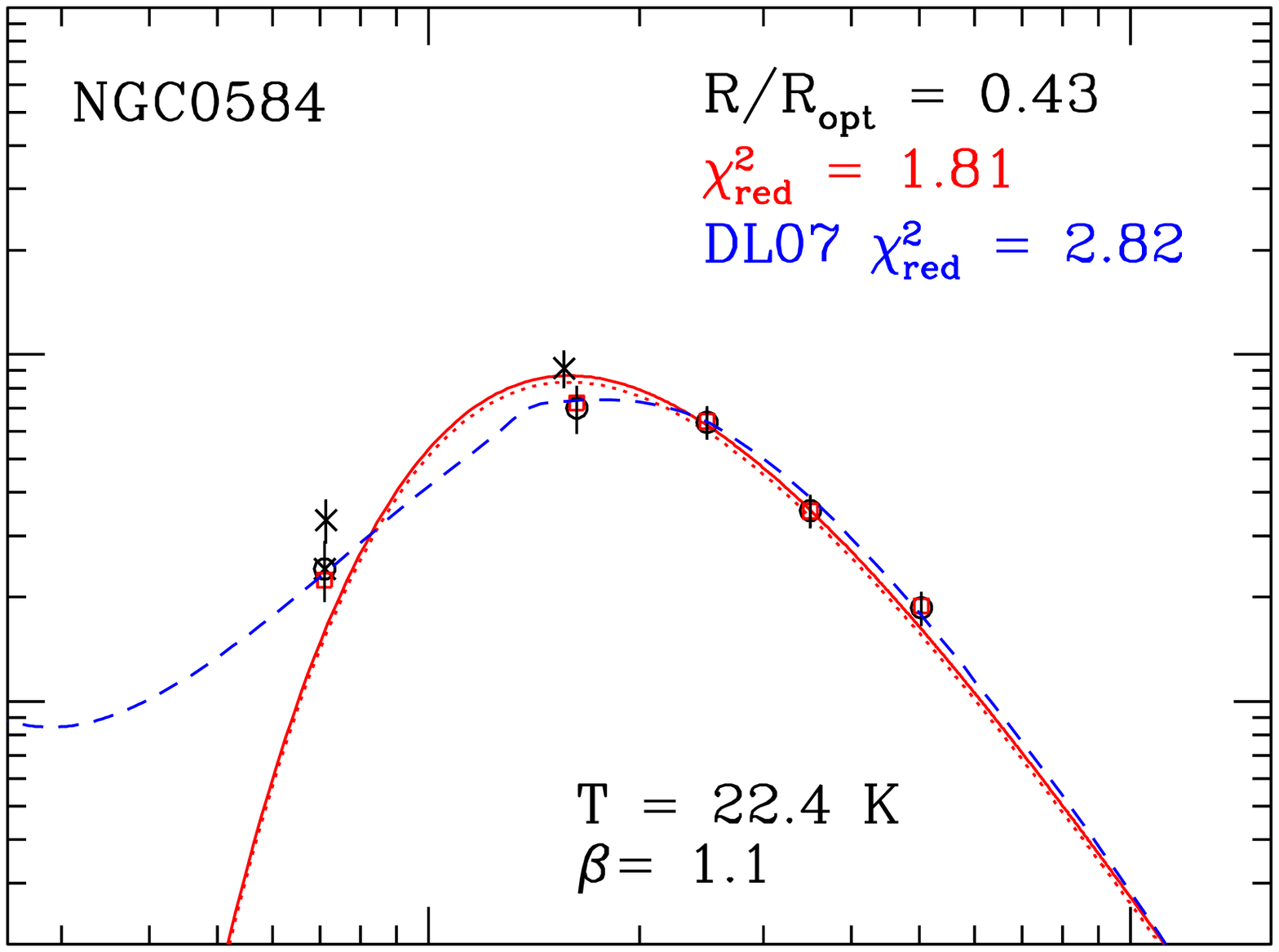}
\hspace{-0.045\linewidth}
\includegraphics[width=0.22\linewidth,bb=18 308 588 716]{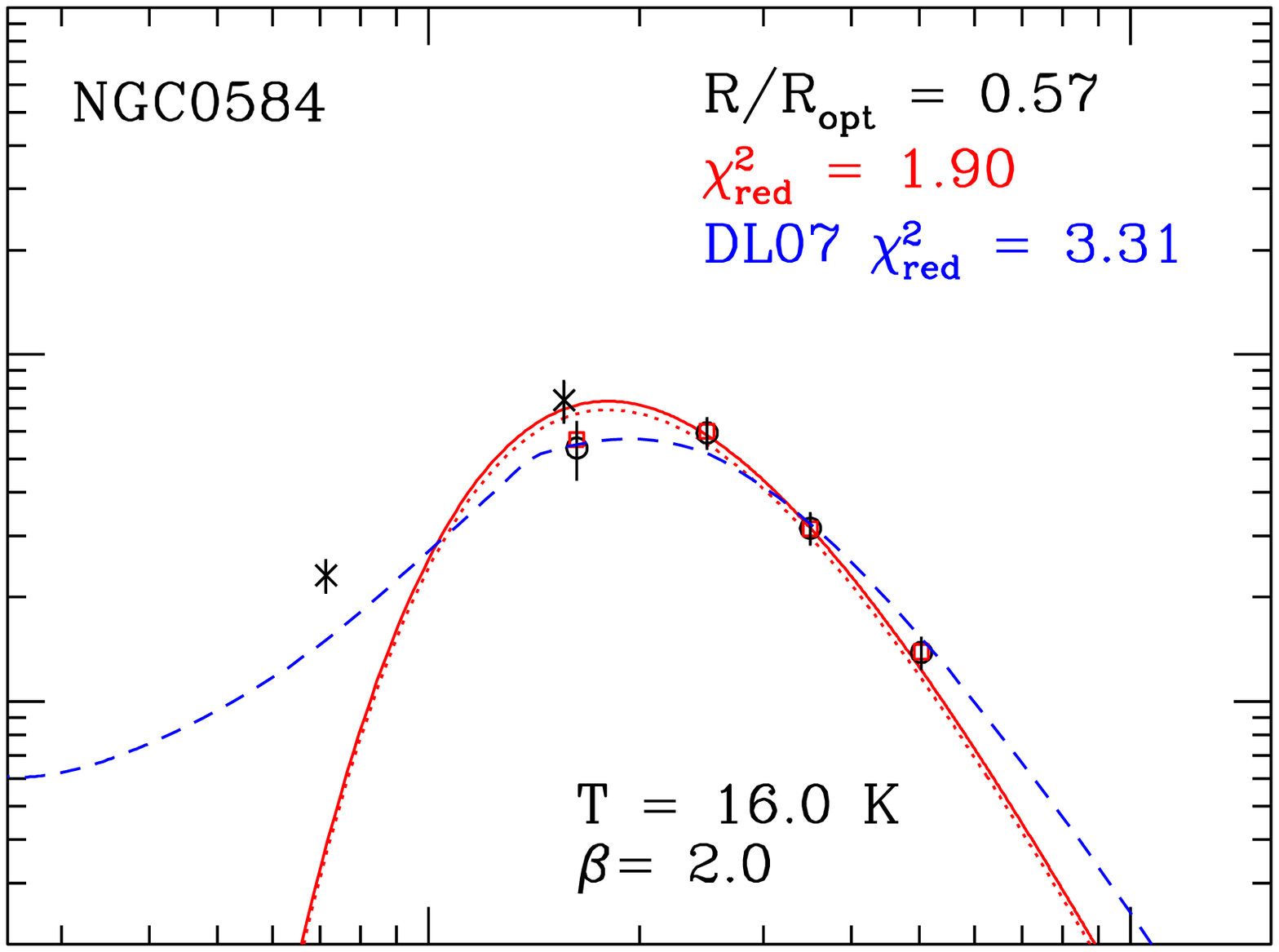}
\hspace{-0.045\linewidth}
\includegraphics[width=0.22\linewidth,bb=18 308 588 716]{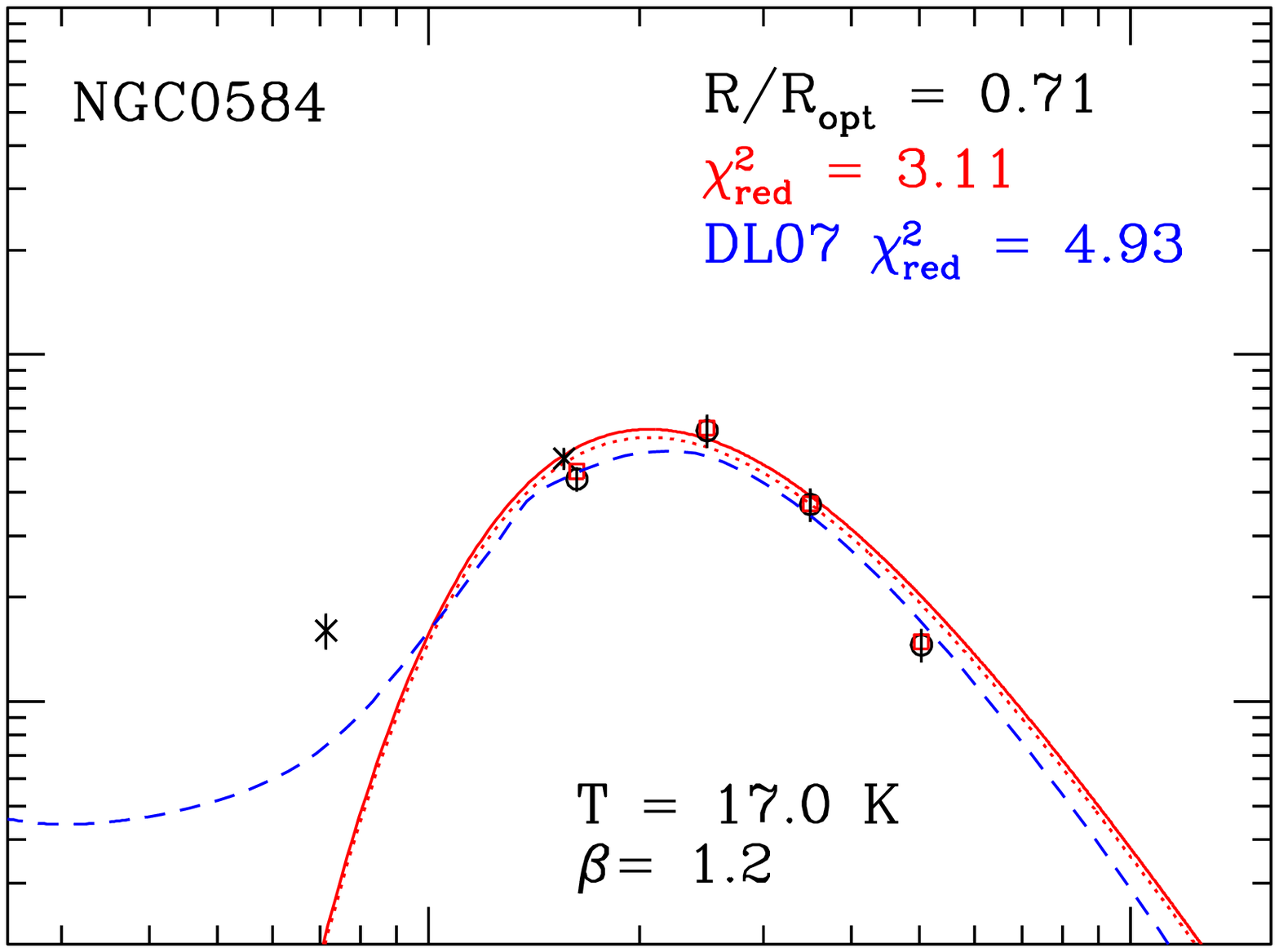}
}
}
\vspace{-1.15\baselineskip}
\centerline{
\hbox{
\includegraphics[width=0.22\linewidth,bb=18 308 588 716]{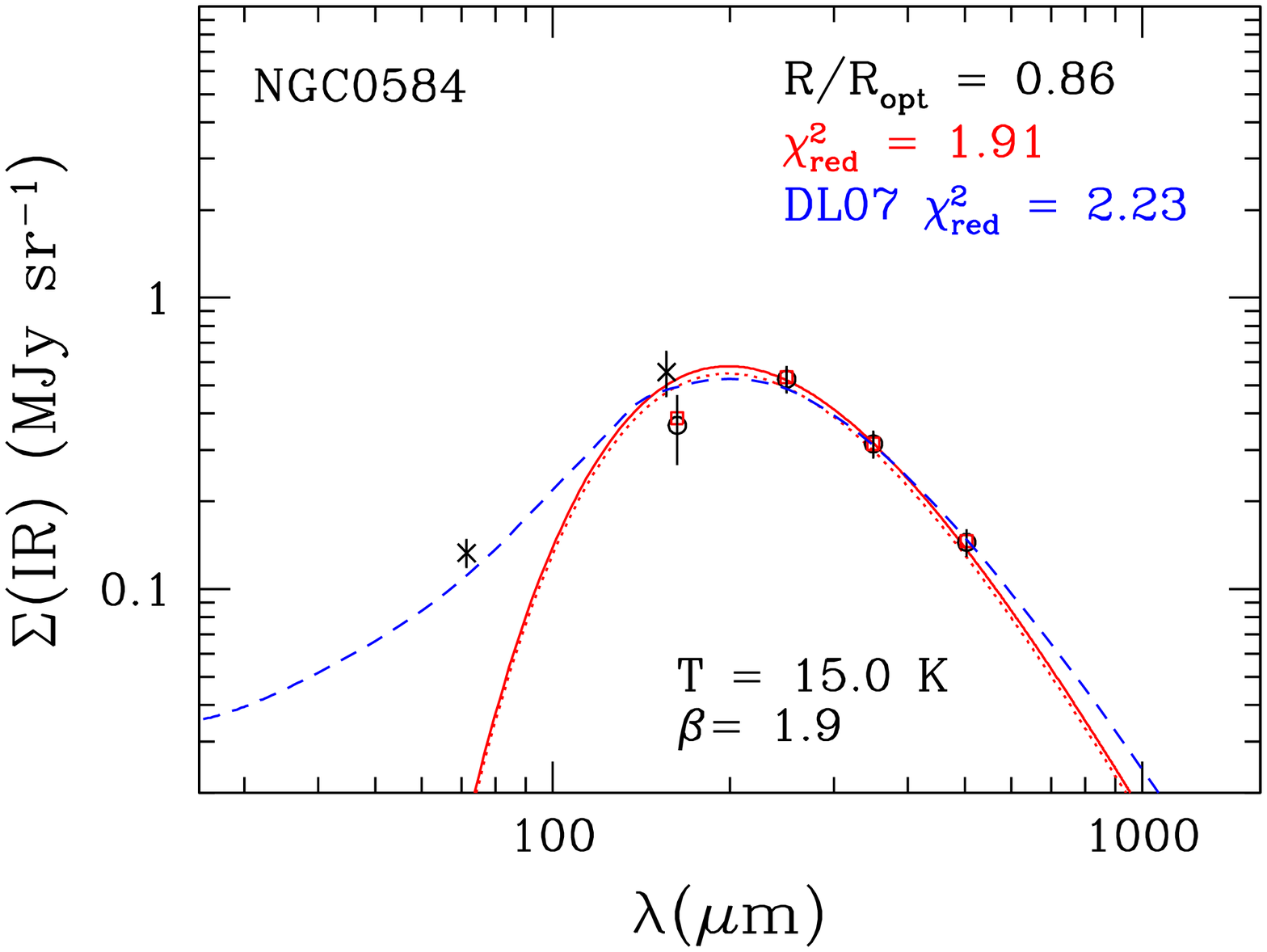}
\hspace{-0.045\linewidth}
\includegraphics[width=0.22\linewidth,bb=18 308 588 716]{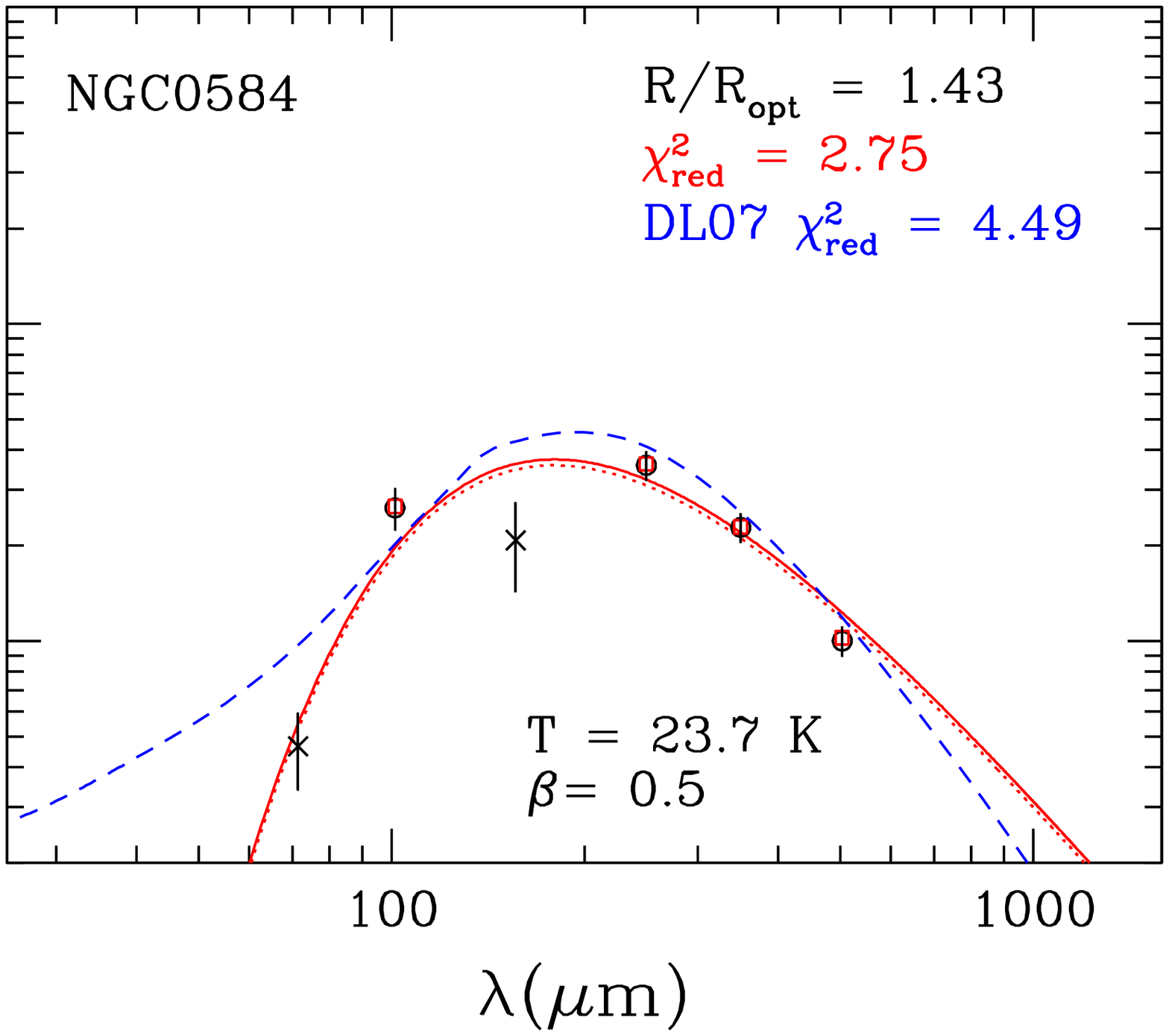}
}
}
\caption{{\bf (a)} NGC\,584 (E4) radial profiles. 
Top left panel: Observed surface brightness profiles, $\Sigma_{\rm IR}$, for PACS, MIPS, SPIRE, and IRAC 3.6\,\micron\
shown in log-log space.
Symbols distinguish the different wavelengths for radial bins $\leq$100\arcsec.
The vertical dashed lines illustrate \ropt\ and 1.5\,\ropt.
The solid curves give the radial profiles at 3.6\,\micron,
24\,\micron, 70\,\micron, and 160\,\micron\ from Mu{\~n}oz-Mateos (priv. communication)
%\citet{munoz09a} 
which were not convolved
to the MIPS 160\,\micron\ resolution as are the profiles presented here.
Top right panel: MBBV radial temperature trends \td\ and
emissivity index trends $\beta$ vs. normalized radii, R/R$_{\rm opt}$.
Sets of fluxes to fit are defined over $\geq$ { 4 wavelengths.}
The bottom panels show each MBBV SED fit as a (red) solid line, $\Sigma_{\rm IR}$ vs. wavelength,
with best-fit temperature and emissivity index $\beta$.
\hers\ data are shown as open circles and (red) squares with squares corresponding to
the calibration corrections inferred from the MBB fits;
MIPS data are shown as $\times$.
As discussed in the text, neither the 70\,\micron\ nor the 500\,\micron\ data
points are included in the MBB fits.
Also shown are the best-fit DL07 models as a (blue) dashed line.
\redchi\ values are given for both fits in the upper right corner of each plot.
}
\label{fig:radial}
\end{figure*}

%---------------------------------------------------------------
\subsection{Modified blackbody fitting}
\label{sec:mbb}

The SED of dust grains in local thermal equilibrium 
%emitting in an optically thin regime 
can be represented by a MBB:
%$$
\begin{equation}
F_\nu\,=\, B(T_{\rm dust},\nu)\ \Omega\ (1-e^{-\tau_\nu})\approx B(T_{\rm dust},\nu)\,\Omega\,\tau_\nu
\end{equation}
%$$
\noindent
where $F_\nu$ is the observed monochromatic flux, 
$B(T_{\rm dust},\nu)$ is the Planck function,
$\Omega$ is the solid angle of the observing beam, 
and $\tau_\nu$ is the dust opacity.
The approximation of $F_\nu \propto B(T_{\rm dust},\nu)\,\tau_\nu$ holds in an
optically thin regime which is what we will assume for the cool
dust component of the KINGFISH galaxies.
The opacity $\tau_\nu$ is directly proportional to the mass attenuation
coefficient (alternatively emissivity or the grain absorption
cross-section per unit mass): 
%$$
\begin{equation}
\tau_\nu\,=\,\Sigma_{\rm dust}\,\kappa_\nu\,=\,\Sigma_{\rm dust}\,\kappa_0\,\left( \frac{\nu}{\nu^0} \right)^\beta
\end{equation}
%$$
where \sigmad\ is the mass surface density of the dust and
$\kappa_\nu$ % (in units of m$^2$\,kg$^{-1}$)
is assumed to have a power-law dependence $\propto\nu^\beta$
\citep{hildebrand83}, 
normalized to the frequency $\nu_0$ corresponding to 250\,\micron.
For the DL07 Milky Way dust models \citep{dl07}\footnote{We will not
be using $\kappa_0$ in this paper because we are not calculating dust masses,
but give them here for completeness.}, 
$\kappa_0(250$\micron$)$\,=\,0.40\,m$^2$\,kg$^{-1}$ and
$\kappa_0(160$\micron$)$\,=\,1.00\,m$^2$\,kg$^{-1}$.
Hence, for optically thin dust, we can write:
%$$
\begin{equation}
F_\nu\,=\, \Sigma_{d}\ B(T_{\rm dust},\nu)\ \Omega\ \kappa_0\,\left( \frac{\nu}{\nu^0} \right)^\beta
\end{equation}
%$$
%where \nd\ is the total dust mass surface density.
%$\kappa_0$ is the value of the emissivity at $\nu_0$. 
Such a description is not generally realistic because of temperature
mixing along the line-of-sight (LOS) and because of the distribution
of dust sizes, densities, and compositions in the general grain population.
Nevertheless, despite its limitations, the MBB has been shown to provide a 
relatively good approximation of observed cool dust SEDs,
at least for dust-mass determinations \citep[e.g.,][]{bianchi13}.

The dust emissivity index, $\beta$, 
depends on the physical properties and chemical composition of the grains,
and may also depend on environment and temperature
\citep{mennella95,mennella98,stepnik03,paradis09,coupeaud11}.
For wavelengths $\lambda\ga100$\,\micron, typical dust compositions with
carbonaceous and silicate grains in diffuse high-latitude clouds 
have $\beta\approx2$ \citep{li01,dl07}.
Detailed observations of the Milky Way show that $\beta\sim1.5-1.8$ in
the Galactic plane and the diffuse halo \citep{planck11d,planck14}, but 
can be as high as $\beta\sim$2.8 in Galactic cold clumps \citep{planck11b}.
Global $\beta$ values from 1 to 2 are found in external galaxies
\citep{boselli12,dale12,auld13}, and
spatially resolved studies show a similar spread of $\beta$ even within a galaxy
\citep{smith12,galametz12,kirkpatrick14}.
Some galaxies tend to have systematically lower values of $\beta\sim1.5-1.7$
\citep[e.g., the SMC, LMC, M\,33:][respectively]{galliano11,planck11a,tabata14}.
These variations have been attributed to different metallicity or dust heating,
but such causes are currently difficult to prove.

Studies of the Milky Way also show that the spectrum of the dust emission 
%in dense environments is better approximated
%by a steeper $\beta$ ($\sim2.4-2.5$) for $100\la\lambda\la600$\,\micron\
%and a flatter value ($\sim1.5$) for $\lambda\ga600$\,\micron\
%\citep{pollack94,paradis09}. 
tends to flatten toward longer wavelengths so that 
$\beta$ in the FIR ($\lambda\la850$\,\micron) is $\sim$0.2 larger than in the submm 
($\lambda\ga850$\,\micron) regime 
\citep{planck14}.
Observationally, there is apparently also a temperature dependence;
higher (steeper) $\beta$ values tend to be associated with cooler dust temperatures,
and lower (flatter) ones with warmer dust
\citep{pollack94,paradis09}. 
Some laboratory experiments show similar results for the flattening of $\beta$
at long wavelengths and at high temperatures
\citep[e.g.,][]{mennella95,agladze96,mennella98,coupeaud11}, 

Despite these seemingly consistent results for the behavior of 
dust temperature \td\ and emissivity index $\beta$, they have been 
challenged by various groups. 
LOS temperature mixing and measurement noise 
have been proposed as the causes of correlated 
variations of temperature and $\beta$ in sightlines in the Galaxy
\citep{shetty09a,shetty09b,juvela12,juvela13}. 
Such correlations, namely that large $\beta$ (steep slopes) tends to 
be associated with cool \td,
emerge because equivalently good MBB fits (as measured by \redchi)
can give different, but related, values of $\beta$ and \td.
Such degeneracy between $\beta$ and \td\ in $\chi^2$ space
is not unexpected when curve-fitting algorithms
are used to simultaneously fit \td\ and $\beta$, because the derivatives
of the MBB function with respect to these parameters are correlated.
Thus parameter estimation has become another topic of debate
\citep[e.g.,][]{kelly12}.

In any case, results all indicate that the fitted dust emissivities
inferred from a MBB fit are not straightforward to interpret.
This is because both \td\ and $\beta$ are luminosity-weighted 
{\it apparent} values that may not reflect intrinsic grain properties.
Moreover, wavelength coverage, fitting technique, and data quality all influence
the outcome of MBB fitting, meaning that results must be carefully
assessed a posteriori.

\subsection{MBB fits of the radial-profile SED data}
\label{sec:mbbdata}

Other papers have relied on more sophisticated methods for MBB fitting, 
including multiple temperature components \citep[e.g.,][]{kirkpatrick14,tabata14},
additional long-wavelength coverage \citep{galametz14},
and broken power-law emissivities \citep{gordon14}. 
Here we adopt the simplest approach of a single
temperature \td\ and a single emissivity $\beta$. 
As mentioned above, we pursue MBBV fits with variable $\beta$ 
and MBBF fits where we fix $\beta\,=\,2$. 

To be considered in the SED to be fit, each radial data point 
was required to have signal-to-noise (S/N) ratio $\geq3$.
Because 70\,\micron\ emission may be contaminated by non-equilibrium
emission from stochastically-heated grains
\citep[e.g.,][]{draine01,li01,compiegne11},
the fit did not include the MIPS or PACS 70\,\micron\ 
(or any shorter wavelength) points.
We however required the fit to
not exceed the average 70\,\micron\ flux plus its $1\sigma$ uncertainty.
The SPIRE 500\,\micron\ data point was excluded from the fit in order
to examine long-wavelength model predictions.
Thus, the best-case SED would have 5 data points (2 PACS, 1 MIPS, 2 SPIRE).
By not constraining the MBB at long wavelengths, we 
enable a potential search for a submm excess 
\citep[e.g.,][]{planck11a,galliano11,kirkpatrick13,galametz14},
as discussed in Sect. \ref{sec:excess}.

Because three parameters (normalization, \td, $\beta$) are needed for
the MBBV fits, as recommended by \citet{shetty09b},
we did not fit any profiles with fewer than 4 (S/N$\geq3$) data points. 
Photometric color corrections were applied to the MBB models before fitting, 
rather than applied directly to the data\footnote{See 
%http://herschel.esac.esa.int/twiki/bin/view/Public/\\
%SpireCalibrationWeb?template=viewprint
%and
%http://herschel.esac.esa.int/twiki/bin/view/Public/ \\
%PacsCalibrationWeb?template=viewprint.
the online documentation at \\
http://herschel.esac.esa.int/twiki/bin/view/Public. 
}. 
These take into account the \hers\ color-dependent beam sizes (PACS, SPIRE) and
also the observed spectral distribution of the source
across the instrumental band-passes (IRAC, MIPS, PACS, SPIRE).
Except for very low or very high temperatures, these corrections are usually 
on the order of a few percent.

The best-fit parameters were determined by minimizing reduced \redchi.
Traditional curve fitting uses $\chi^2$ gradients in parameter space to
find the minimum, but 
the partial derivatives of temperature and emissivity index are correlated
because of the mathematical form of the MBB function. 
Thus, such a procedure could induce potential correlations even when none are
present in the data.
Therefore,
to ameliorate as far as possible any spurious correlation
between \td\ and $\beta$, 
%rather than a traditional $\chi^2$-minimizing curve fitting algorithm,
we adopted a two-pass grid method 
of stepping through possible values of $\beta$ ($0.5\leq\beta\leq3$)
and \td\ (5\,K$\leq$\td $\leq$45\,K). 
Parameter uncertainties were calculated by assuming that
the Hessian matrix is diagonal, and performing analytically
second-order differentiation of the expression for $\chi^2$
with respect to \td\ and $\beta$.
Since the two parameters \td\ and $\beta$ are correlated (making the Hessian
not diagonal), the uncertainties will be underestimated.

We adopted the same algorithm for MBBF fits, except for fixing
$\beta\,=\,2$.
%To facilitate comparison with the MBBV fits, 
Again,
a S/N$\geq$3 was required for each radial data point. 
%Despite the reduction in the number of fitting parameters,
In the MBBF case, there is one fewer parameter to fit, and 
reduced \redchi\ was calculated accordingly.
The fits are in any case rather under-constrained. 

The S/N requirements result in 920 radial data points with sufficient SPIRE S/N
at all three wavelengths, and 766 with sufficient PACS S/N at both 100\,\micron\
and 160\,\micron\ (817 profile points have PACS 160\,\micron\ $\geq$3).
In the following analysis, only MBBV and MBBF fits with reduced \redchi\
$\leq$2 are considered.
This, together with the S/N requirement for individual surface brightness points
(S/N$\geq$3), results in
818 SED radial profile MBBV fits 
and 827 MBBF fits (of a total of 1166 radial data SEDs for 61 galaxies).
%There are 920 profiles having at least 4 data points with S/N$\geq$3 and \redchi$\leq$12.
Given the similar number of good (low \redchi) MBBF fits,
we conclude that the MBBF (fixed-emissivity) fits are as able 
as the MBBV ones to approximate the true data SEDs.

%---------------------------------------------------------------
\subsection{Physically realistic dust models}
\label{sec:dl07}

To better approximate the true physical conditions of the dust
grains, we also fitted the radial profile data SEDs with DL07 models. 
These models assume that dust is composed of a mixture of silicates,
graphites, and polycyclic aromatic hydrocarbons (PAHs), with
size distributions \citep{weingartner01} that reproduce the wavelength-dependent
extinction observed in the Milky Way.
The dust is heated by an ISRF with a Milky-Way like spectrum
\citep{mathis83}, 
with a distribution of starlight intensities.

The starlight intensity distribution is described as a truncated power law 
with four parameters:
\umin, the minimum ISRF level;
\umax, the upper limit of ISRF level;
$\alpha$, the power-law index of the ISRF distribution;
and $\gamma$, the fraction of the dust mass exposed to starlight
with intensities greater than \umin. 
The fraction of the dust mass exposed to a distribution of starlight
intensities $U$ is defined as:
%$$
\begin{equation}
\frac{dM_{\rm dust}}{dU}\,=\,(1-\gamma) M_{\rm dust}\,\delta(U-U_{\rm min}) 
+ \gamma M_{\rm dust}\,\frac{\alpha-1}{(U^{1-\alpha}_{\rm max} - U^{1-\alpha}_{\rm min})} U^{-\alpha}
\label{eqn:dl07}
\end{equation}
%$$
\noindent
where the power-law index, $\alpha\neq$1, and \mdust\ is the total dust mass.
Remaining parameters in the DL07 fits include
\qpah, the fraction of dust mass composed of PAHs, and
\fpdr, the fraction of dust emission radiated by grains
exposed to intense levels of ISRF ($U>10^2$). More details
and recipes for calculating \qpah\ and \fpdr\ are given by \citet{dl07}.

The functional form of the DL07 models is similar to the power-law distribution
formulated by \citet{dale01} and \citet{dale02}, but with an added
delta-function component of dust heated by \umin.
This is the component that dominates the dust mass,
representing the dust mass in the general diffuse ISM.
The shape of the dust emission spectrum is governed by
\qpah, \umin, and $\gamma$; \qpah\ and $\gamma$ generally determine the
short-wavelength mid-infrared (MIR) region 
($\la$70\,\micron) while \umin\ the long-wavelength FIR regime ($\ga$70\,\micron).
A useful parameter to link total dust mass \mdust\ with the luminosity
of the dust emission is \ubar, or the mean starlight intensity; 
it is uniquely defined by \umin, \umax, and $\gamma$.
As shown by \citet{dl07}, \ubar\ is inversely proportional to the mass-to-light
ratio of the dust.
Like the MBB fits to the data, the DL07 
models were integrated over the instrumental response functions for comparison
with observed fluxes.

There are 13 data points in the DL07 fits (4 IRAC, 3 each MIPS, PACS, SPIRE);
hence the fits are somewhat over-constrained,
although there are several free parameters including 
the ISRF starlight intensity determined by the IRAC fluxes.
%\footnote{The starlight is assumed to have the spectrum 
%estimated by \citet{mathis83}.}. 
%However, the DL07 models 
%place no explicit constraints on the long-wavelength 
%regime of the dust SEDs. 
However, the long-wavelength regime of the dust SEDs is only part 
of the entire spectrum fit by the DL07 models.
Consequently, the quality of the fit depends on the dust optical properties 
and albedo assumed for the DL07 grain populations.
%The MBB fits have fewer degrees of freedom than the DL07 models and will,
%almost by definition, provide a closer approximation of the observed SED in the FIR.
The MBB fits are required only to fit the 100\,\micron$\leq \lambda \leq$350\,\micron
SED, and will
almost by definition, provide a closer approximation of the observed SED over this
limited wavelength range.
In the following, as for the MBBV and MBBF data fits, we will consider only the
DL07 fits with \redchi$\leq$2; there are 548 of these.

\subsection{MBB fits of the DL07 radial-profile SEDs}
\label{sec:mbbdl07}

%In order to better understand how
%well the data are constrained by the DL07 and MBB models,
We have also fit the best-fit DL07 models with MBBs.
Such a procedure will not only link the physical parameters 
of the DL07 models (e.g., \umin, \ubar, \qpah) 
to the approximation of a single dust temperature, but will also explicitly
demonstrate that even fixed-emissivity models such as DL07 can
well approximate values of $\beta$ much {\it lower} 
than the intrinsic emissivity index assumed in the model.

Accordingly, for each radial SED,
we have fit the best-fit DL07 models with both the %two ``flavors'' of MBB fits, namely the 
variable-emissivity MBBV and the fixed-emissivity MBBF models.
To each DL07 model point, we have assigned the same error as the analogous
data point, and the same wavelengths have been included, 
namely 100\,\micron$\leq\lambda<$500\,\micron.
\redchi\ is also calculated in the same way as for the MBBV and MBBF fits
of the data themselves.
Because neither model is a true observable, for these fits
no color corrections have been applied to either the MBB or DL07 models. 
Hereafter, in order to distinguish the MBBV best-fit parameters for the
data, $\beta$ and \td, the best-fit DL07 model MBBV values for emissivity index and dust 
temperature will be referred to as \betadl\ and \tddl, respectively.

Thus, as mentioned above, for each of 1166 radial SEDs in 61 galaxies, we have 
five fits: MBBV$+$MBBF fits of the data, DL07 fits of the data, and MBBV$+$MBBF 
fits of the best-fit DL07 models.
In truth, however, we analyze the full set of five fits for 548 SEDs, because
of the constraint on \redchi\ (see above).

%---------------------------------------------------------------
\section{Results of the SED modeling}
\label{sec:radial}

The resulting SEDs for the annular regions of selected galaxies are given in Fig. \ref{fig:radial}
where each observed SED is plotted together with the best-fit MBBV and
DL07 models\footnote{As for Fig. \ref{fig:exponential},
the entire suite of plots is available as part of the on-line data set,
together with the SED data themselves. 
Figures \ref{fig:radial}(b)-(d) are found in Appendix \ref{app:profiles},
and are thus renumbered, e.g., Fig. \ref{fig:appradial}.}.
We show an example of each morphological type: 
an elliptical, NGC\,584; a lenticular NGC\,1266;
a spiral NGC\,6946; and a dwarf irregular Holmberg\,II
(the latter three are found in Fig. \ref{fig:appradial}).
The upper panels show the brightness panels at various wavelengths (left)
and the radial run of \td\ and emissivity index $\beta$ (right).
In the lower panels, individual data points are required 
to have S/N\,$\geq$\,3.0, and
the 70\,\micron\ (PACS$+$MIPS) and 500\,\micron\ points are excluded from the fits
(see Sect. \ref{sec:fitting}).

Figs. \ref{fig:radial}a and 
%\ref{fig:radial}b 
%(in truth Fig. ...)
\ref{fig:appradial}b
clearly show that \hers\ is able
to trace dust well beyond \ropt\ even in early-type galaxies.
We are able to measure dust SEDs in NGC\,1266, a lenticular SB0, out to $\ga$2\,\ropt,
and in NGC\,584, an E4, to 1.6\,\ropt.
In general, the dust emission is well fit by both the MBBV and DL07 models;
the DL07 models are able to accommodate fitted emissivity $\beta\la 2.5$ 
(see e.g., the inner regions of NGC\,6946 in Fig. \ref{fig:appradial}c).
Moreover,
the DL07 models have an average dust emissivity index $\beta\sim$2.1 \citep{bianchi13} but are able
to reproduce low values of $\beta\sim0.5-1$ because of
the range of ISRF intensities intrinsic to the models.
This behavior is particularly evident in the outer regions of NGC\,6946
(Fig. \ref{fig:appradial}c) and for almost the entirety of Holmberg\,II (Fig. \ref{fig:appradial}d).
Thus, low values of apparent best-fit emissivity index $\beta\la 1.5$ cannot be interpreted as
necessarily due to intrinsic properties of grain populations.
However,
the steep FIR SED slopes implied by high values $\beta\ga2.5$ are not well reproduced
by the DL07 models (e.g., NGC\,5457, see Fig. \ref{fig:radial_beta}) and may be intrinsic.
Possible reasons for this behavior will be discussed in Sect. \ref{sec:discussion}.

%---------------------------------------------------------------
\subsection{Dust optical depth and temperature}
\label{sec:radial_opticaldepth}

\begin{figure}[!ht]
\centerline{
\includegraphics[angle=0,width=\linewidth,bb=18 280 592 718]{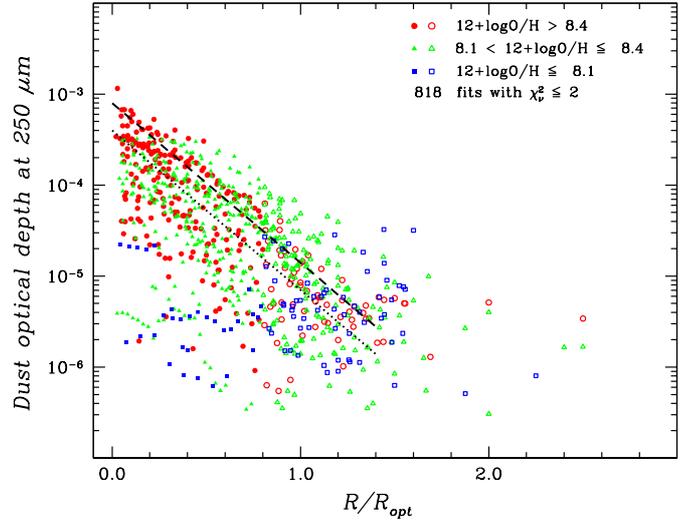}
}
\caption{Face-on 250\,\micron\ optical depth \taud\ obtained from MBBV fits plotted against normalized
radius, $R$/\ropt; \taud\ has been corrected for inclination as described in the text.
Points are coded by their oxygen abundance, with filled symbols corresponding
to $R$/\ropt$\leq$0.8 and open ones to larger radii.
Only fits with \redchi$\leq$2 are shown.
Both dashed and dotted lines show the mean 250\,\micron\ exponential scalelength
(\rdisk/\ropt\,=\,0.57$\pm$0.5) derived as the average over all galaxies;
the offset between the two lines corresponds to the mean difference between face-on
orientation (dotted line) and the observed one (dashed line, with \taud\ not corrected for inclination).
\label{fig:radial_tauropt}
}
\end{figure}

\begin{figure*}[!ht]
\centerline{
%\includegraphics[angle=0,width=0.5\linewidth,bb=18 280 592 718]{Temp_Ropt.eps}
%\hspace{-1.4cm}
\includegraphics[angle=0,width=0.5\linewidth,bb=0 100 503 503]{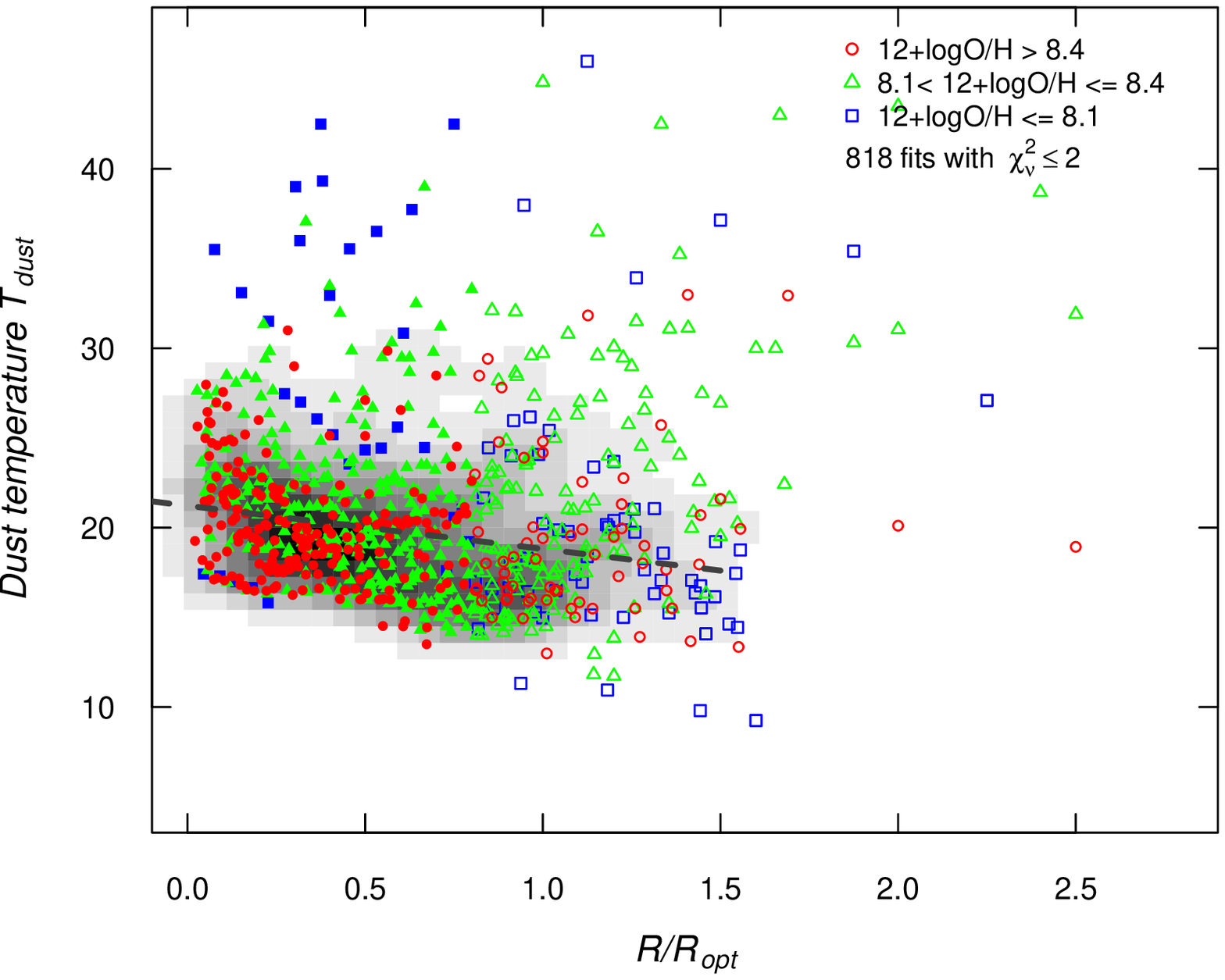}
\hspace{-1.7cm}
\includegraphics[angle=0,width=0.5\linewidth,bb=0 100 503 503]{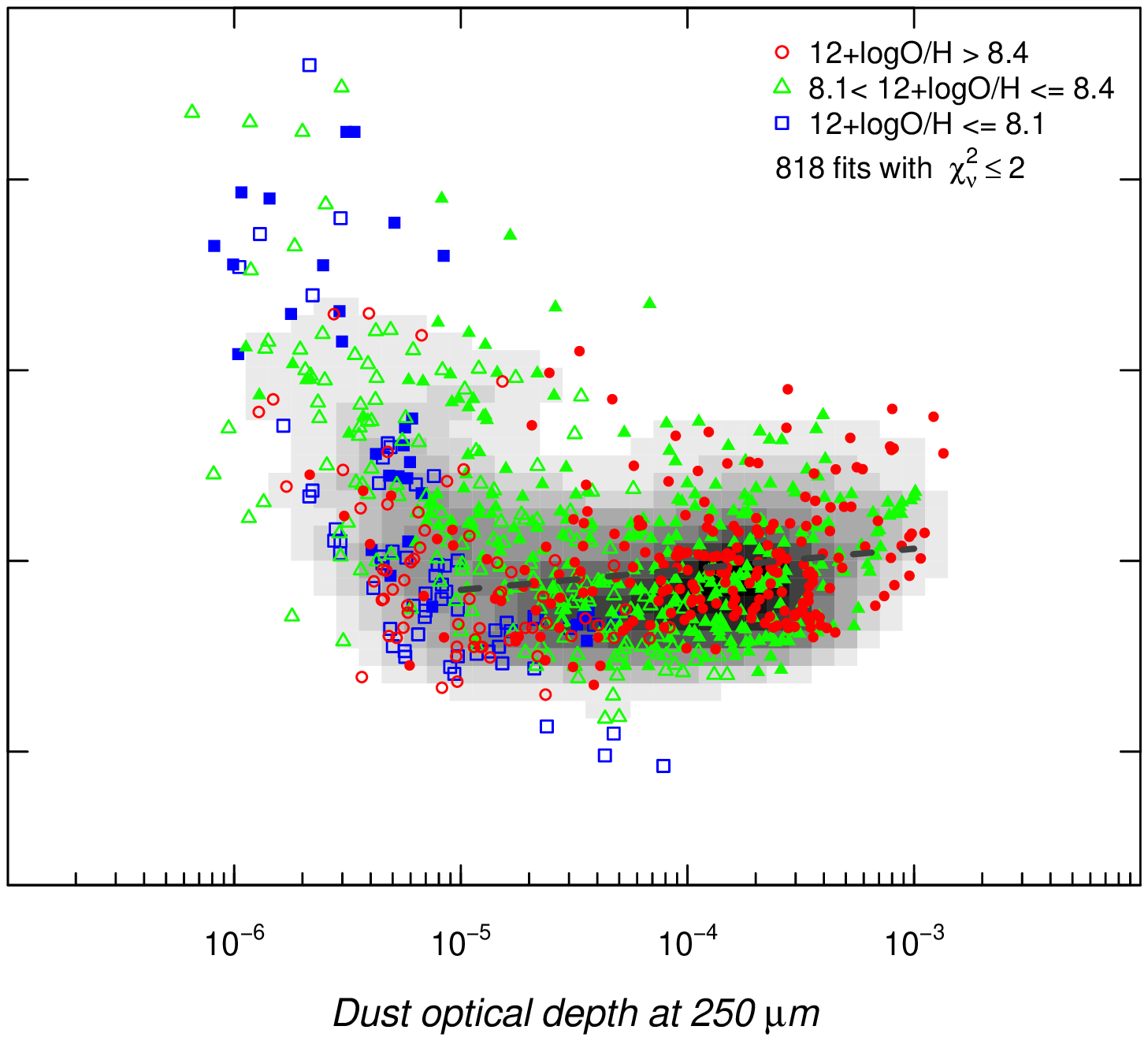}
}
\caption{Left panel: dust temperature \td\ versus normalized
galactocentric distance $R$/\ropt; right panel: \td\ versus \taud\
evaluated at 250\,\micron.
%In the left panel, 
As in Fig. \ref{fig:radial_tauropt},
points are coded by their oxygen abundance, 
with filled symbols corresponding
to $R$/\ropt$\leq$0.8 and open ones to larger radii.
The underlying greyscale shows the two-dimensional density distribution of the data.
%In the right panel, points are instead coded by Hubble type,
%again with filled symbols corresponding
%to $R$/\ropt$\leq$0.8 and open ones to larger radii.
The dashed lines are unweighted linear regressions to the data points, excluding
outliers with \td$\geq$30\,K (in the left panel,
\td\,=\,$-2.41\ R/R_{\rm opt} + 21.2$).
\label{fig:radial_tempropt_tau}
}
%\end{figure*}
%\begin{figure*}[!h]
\vspace{0.092\baselineskip}
\setcounter{figure}{9}
\centerline{
%\includegraphics[angle=0,height=0.40\linewidth,bb=18 260 592 718]{Beta_Ropt.eps}
%\hspace{-1.75cm}
%\includegraphics[angle=0,height=0.40\linewidth,bb=18 260 592 718]{Beta_Ropt_NegativeRadial_nolab.eps}
\includegraphics[angle=0,height=0.35\linewidth,bb=18 260 592 718]{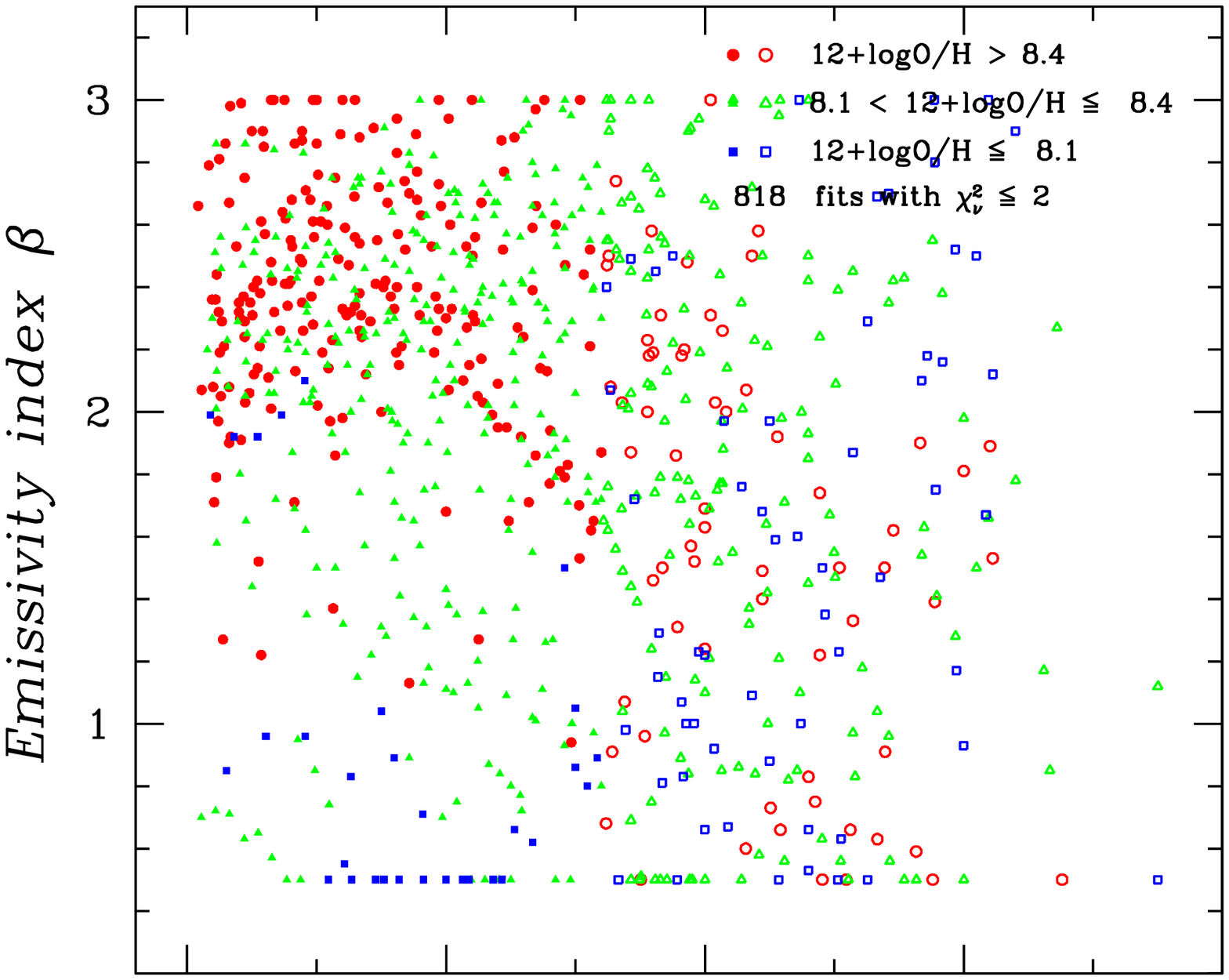}
\hspace{-1.55cm}
\includegraphics[angle=0,height=0.35\linewidth,bb=18 260 592 718]{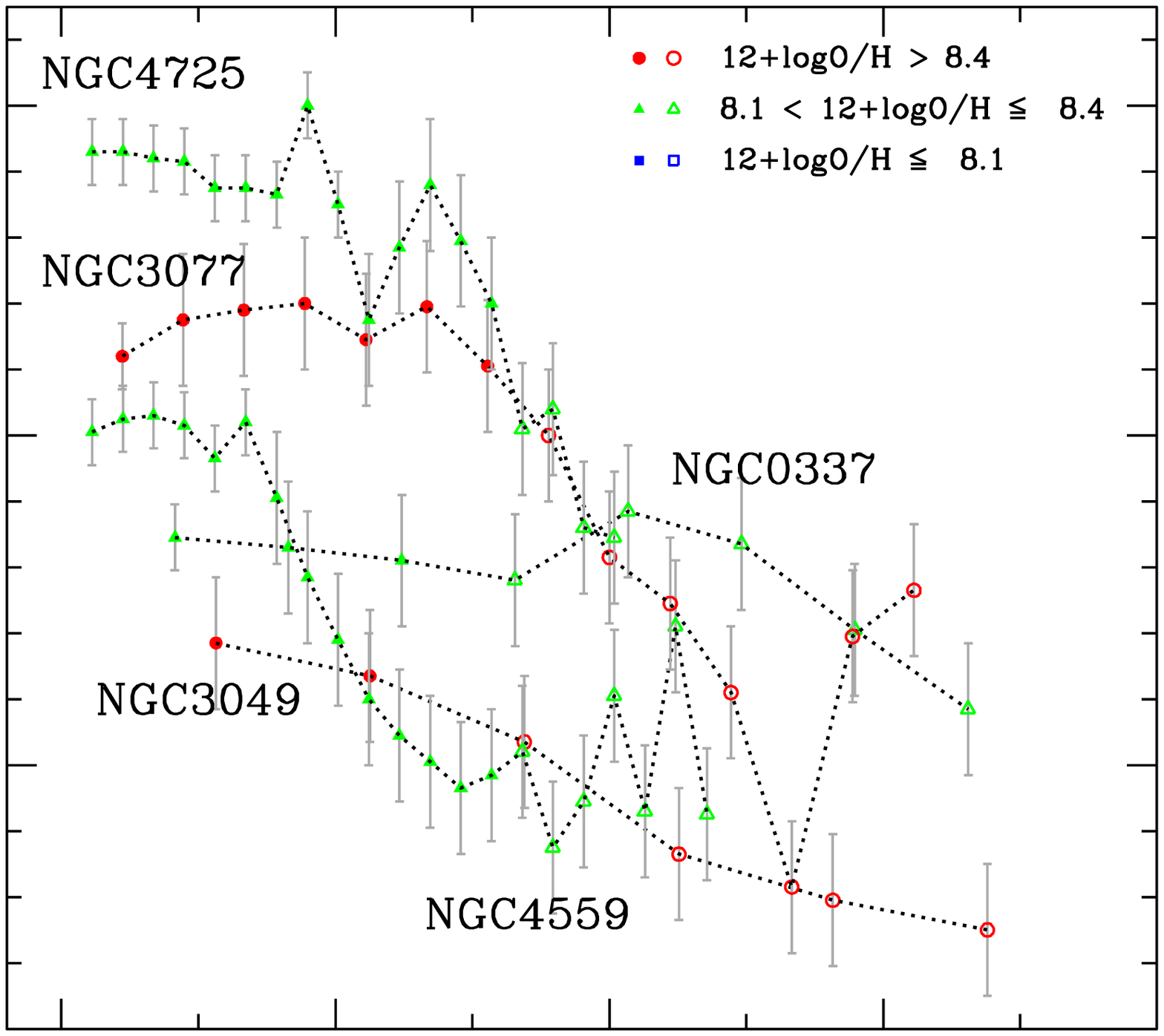}
}
%\vspace{-0.62cm}
\vspace{-0.53cm}
\centerline{
%\includegraphics[angle=0,height=0.40\linewidth,bb=18 260 592 718]{Beta_Ropt_PosNegRadial.eps}
%\hspace{-1.75cm}
%\includegraphics[angle=0,height=0.40\linewidth,bb=18 260 592 718]{Beta_Ropt_PositiveRadial.eps}
\includegraphics[angle=0,height=0.35\linewidth,bb=18 260 592 718]{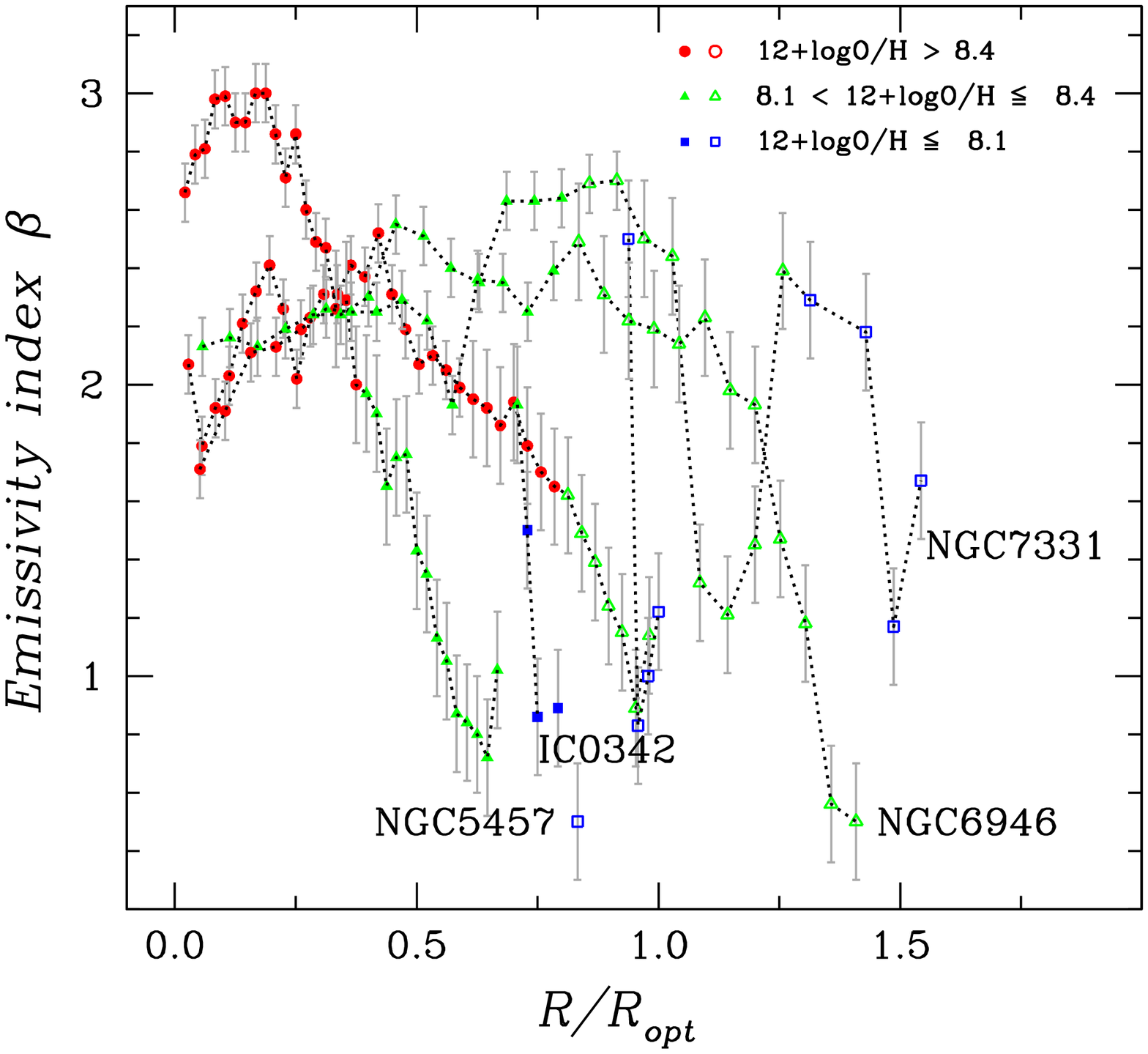}
\hspace{-1.55cm}
\includegraphics[angle=0,height=0.35\linewidth,bb=18 260 592 718]{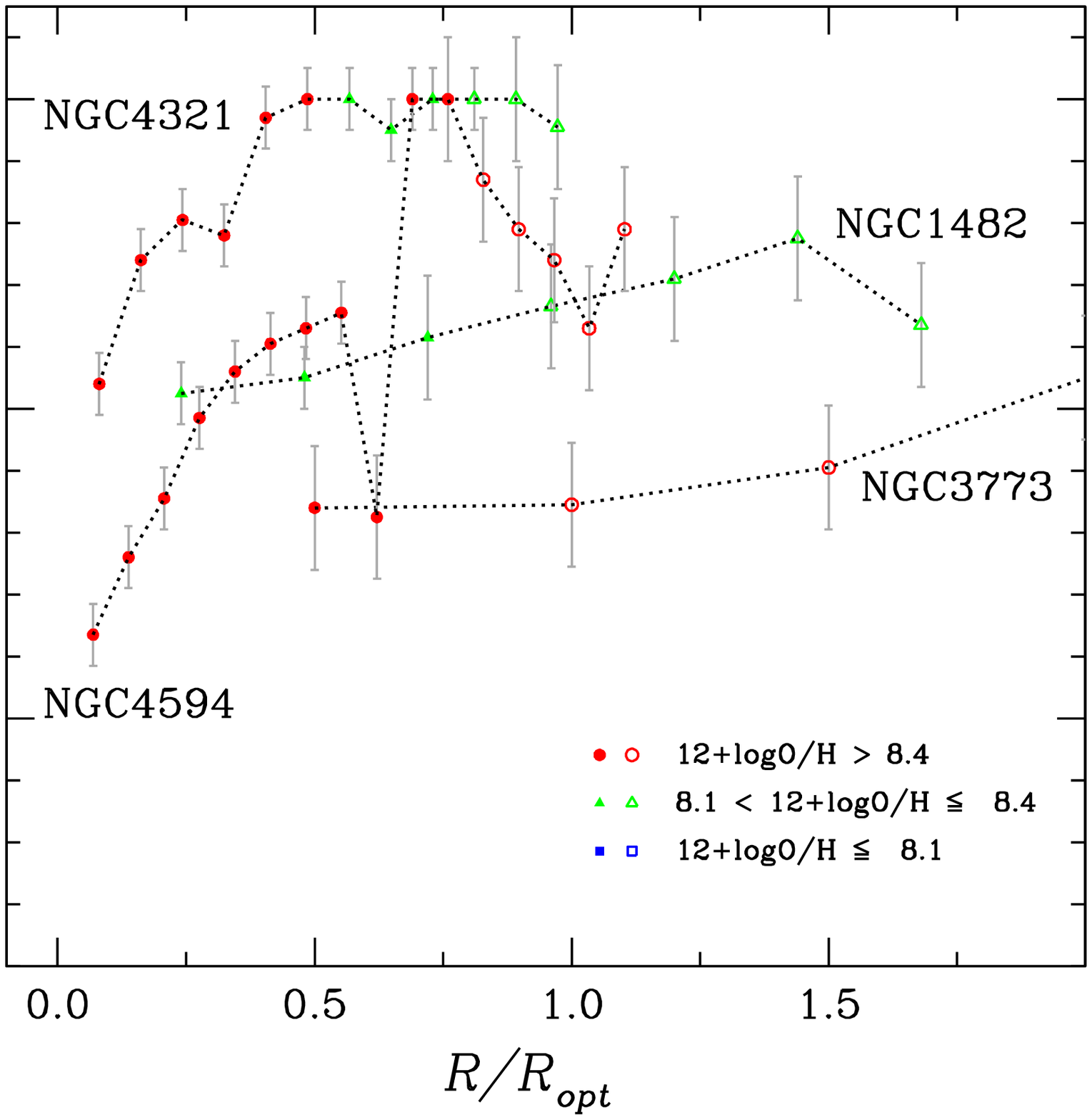}
}
\caption{Emissivity index $\beta$ gradients with galactocentric distance $R$/\ropt.
The upper-left panel shows all galaxies together, while individual galaxies are shown
in the remaining panels.
The upper-right panel illustrates examples of negative radial gradients,
the lower-left rising then falling, and the lower-right panel positive radial gradients.
In all panels, as in Fig. \ref{fig:radial_tauropt},
points are coded by their oxygen abundance, 
with filled symbols corresponding
to $R$/\ropt$\leq$0.8 and open ones to larger radii.
\label{fig:radial_beta}
}
\end{figure*}

Another way to illustrate the exponential trend of the dust in KINGFISH
galaxies is with dust optical depth \taud\ and its dependence on 
galactocentric distance.
Figure \ref{fig:radial_tauropt} shows \taud\ evaluated at 250\,\micron\
(with MBBV temperatures) plotted against normalized radius, $R$/\ropt.
Only those MBBV fits with \redchi$\leq$2 are shown, because of the need
for reliable temperatures to calculate \taud.
The surface brightnesses in \taud\ have been corrected for inclination to
correspond to face-on orientation.
The dotted and dashed lines show a negative slope with (the inverse of) \rdisk/\ropt\,=\,0.57, 
the mean over all the galaxies (this is shallower than the mean of 0.35 
for only spirals discussed in Sect. \ref{sec:radial_disks}).
If the surface brightnesses in the \taud\ calculation are not corrected for
inclination, the noise in the radial trend remains roughly the same but 
the mean dust optical depth is raised by a factor of 2, $\sim$\,0.3\,dex,
as shown by the dashed regression line. 
Although there is much variation at small radii, the dust
optical depth falls off as expected, in a roughly exponential distribution.

Like \taud, dust temperature \td\ also decreases with radius, but
much more gradually.
Such a trend can be appreciated in the individual upper right panels in Fig. \ref{fig:radial},
but is shown explicitly in Fig. \ref{fig:radial_tempropt_tau} where we have
plotted \td\ versus $R$/\ropt\ in the left panel and against \taud(250) in
the right.
As in Fig. \ref{fig:radial_tauropt}, 
only those radial SEDs with MBBV \redchi$\leq$2 are shown.
The dashed regression line is derived from a linear fit of
\td\ with radius: \td\,=\,$-2.41\ R/R_{\rm opt} + 21.2$. 
Although there is much variation, there is a significant trend
for dust to be cooler in the outer regions of the KINGFISH galaxies,
as also found by \citet{galametz12}.
The right panel of Fig. \ref{fig:radial_tempropt_tau} shows that
dust also tends to be cooler at low dust optical depth \taud, 
but again with much scatter.

In some cases, very low \taud\ is associated with warm temperatures, perhaps
associated with star-forming regions in late-type galaxies
or in extended tenuous disks.
We have confirmed that this effect is not merely due to sensitivity limits.
The lowest $\Sigma_{250}$ surface brightnesses (S/N$\geq$3) measured in our sample, 
$\Sigma_{250}\sim$\ 0.13-0.15\,\mjysr,
are associated with PACS or MIPS 160\,\micron\ surface brightnesses (S/N$\geq$3)
$\ga$0.05\,\mjysr, slightly lower than what would be expected in the worst-case
scenario of variable-emissivity MBB dust emission with $\beta$\,=\,0.5 and \td\,=\,15\,K, namely
$\Sigma_{160}\sim$\ 0.07\,\mjysr. 
However, the lowest values of $\Sigma_{250}$
are not necessarily those with the 
lowest values of \taud\ (at 250\,\micron), $\la 1\times10^{-6}$. 
For \taud\ in this range (31 radial data points), the average $\Sigma_{250}$\,=\,0.5\,\mjysr\
(with the lowest value of $\Sigma_{250}\sim$\,=\,0.13\,\mjysr\ as above).
The mean PACS or MIPS 160\,\micron\ surface brightness in this low \taud\ regime is 
$\sim$0.7\,\mjysr, with the lowest value being $\sim$0.07\,\mjysr, associated
with $\Sigma_{250}$\,=\,0.23\,\mjysr.
With dust having $\beta$\,=\,0.5 and \td\,=\,15\,K, and this $\Sigma_{250}$, we would
expect $\Sigma_{160}\sim$\ 0.15\,\mjysr, well above the lowest observed value.
Thus, we conclude that the KINGFISH profile sensitivity limits are sufficient
to detect, if it were present, dust with low opacity (\taud$\la1\times10^{-6}$), 
cold temperatures (\td$\sim$15\,K) and very flat ($\beta$\,=\,0.5) spectral distributions.
As a result, the high-temperature, low-opacity dust shown in Fig. \ref{fig:radial_tempropt_tau}
seems real, perhaps due to the low metallicities and high values of
\fpdr\ associated with this high-\td\, low-\taud\ regime
(see also Sect. \ref{sec:radial_dl07_pah} and Fig. \ref{fig:radial_dl07}).

%---------------------------------------------------------------
\subsection{Radial emissivity variations}
\label{sec:radial_emissivity}

Unlike \td\ and \taud, emissivity index $\beta$ shows no clear trend when considered
globally as shown in the upper left panel of Fig. \ref{fig:radial_beta}.
The radial behavior $\beta$ can vary significantly from one object to another. 
Many KINGFISH galaxies have negative radial gradients
(see upper-right panel of Fig. \ref{fig:radial_beta}).
Such a trend was already observed in the late-type spiral (Scd),
M\,33 \citep{tabata14}, at least over its inner disk, and is also
observed in many of the late-type galaxies of our sample but not only in
late-type objects:
NGC\,4725 and NGC\,3049 are earlier spiral types (SABab, SBab, respectively).
However, there are also other kinds of radial variations, for instance a $\beta$ that first rises
toward $\sim R$/\ropt\ then falls at $R\ga$\ropt.
Some of the late types (e.g., IC\,342, NGC\,6946, also Scd) show such ``rising/falling'' $\beta$ 
behavior (see lower-left panel of Fig. \ref{fig:radial_beta}).
Finally, some galaxies show a $\beta$ that rises almost monotonically 
toward large radii (see lower right panel of Fig. \ref{fig:radial_beta}).
Thus, the dependence of $\beta$ behavior with Hubble type is not clear.
Nor do the trends seem to depend on metallicity, since there is no clear correlation with
either emissivity index or type of radial decline.
The physical meaning of $\beta$ variations will be discussed in Sect. \ref{sec:discussion}.

%---------------------------------------------------------------
\subsection{Radiation field and dust heating }
\label{sec:radial_dl07}

\begin{figure*}[!ht]
\centerline{
\hbox{
\includegraphics[angle=0,width=0.40\linewidth,bb=0 0 503 503]{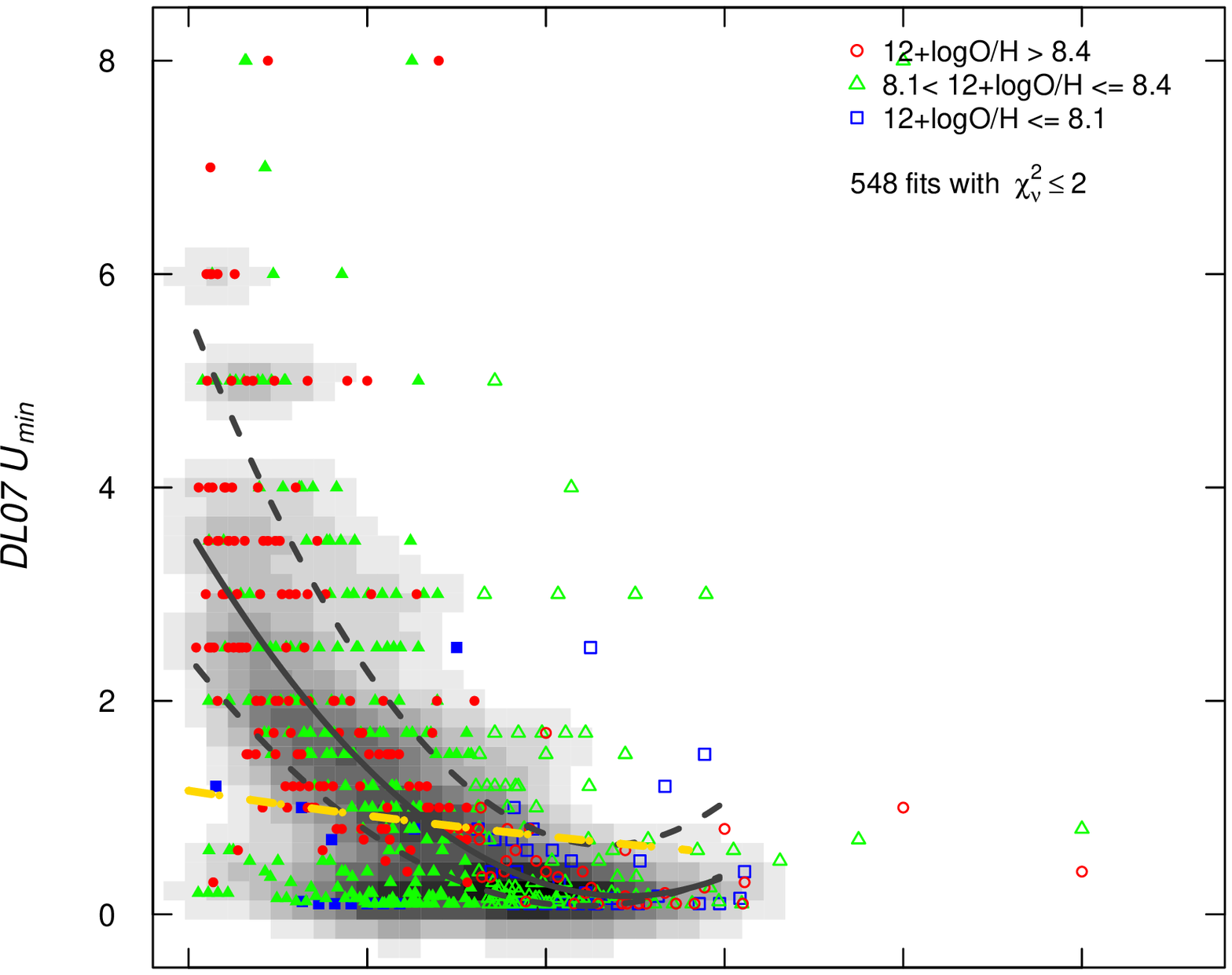}
\hspace{0.1cm}
\includegraphics[angle=0,width=0.40\linewidth,bb=0 0 503 503]{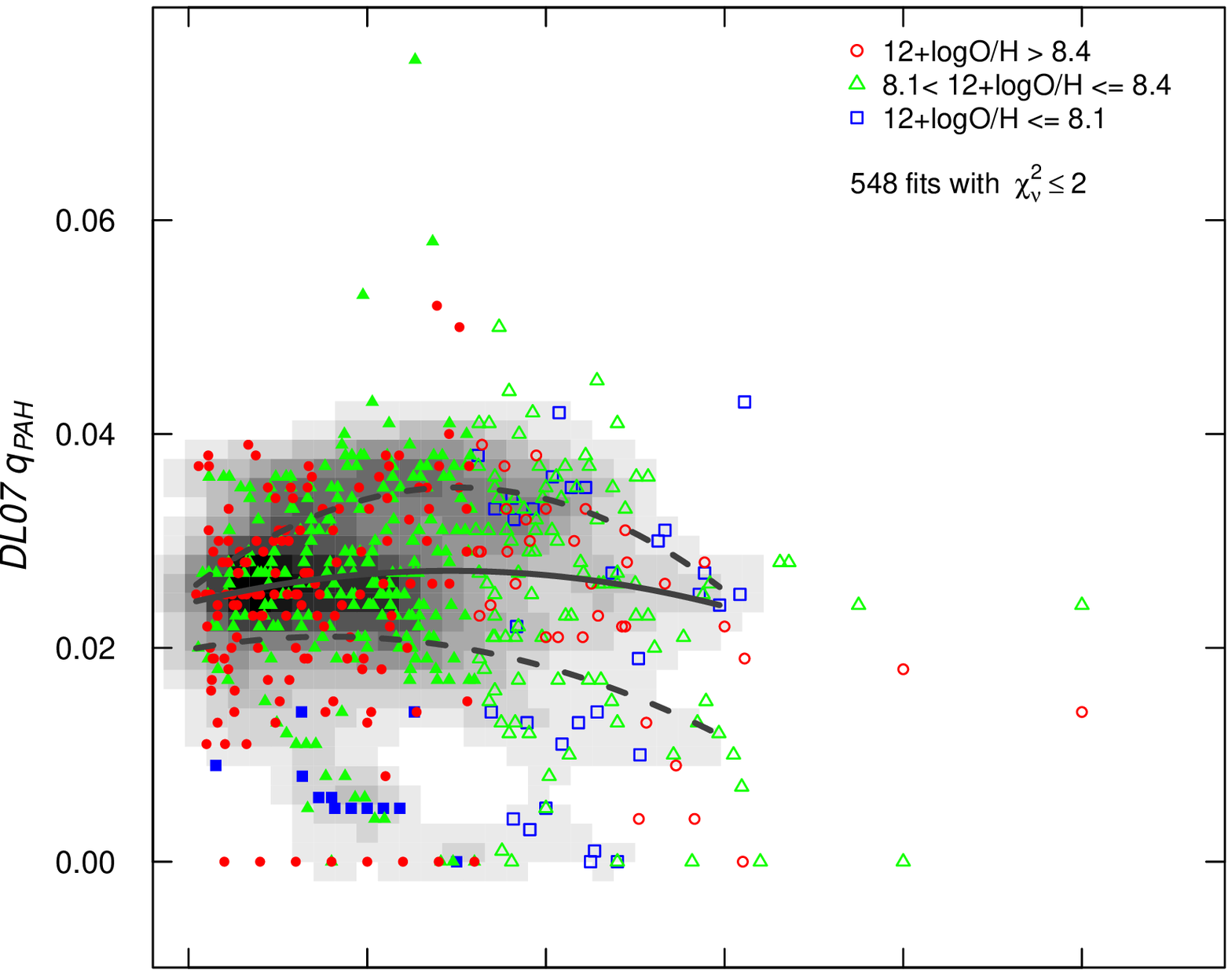}
}
}
%\vspace{-0.1cm}
\vspace{-1.9cm}
\centerline{
\hbox{
\includegraphics[angle=0,width=0.40\linewidth,bb=0 0 503 503]{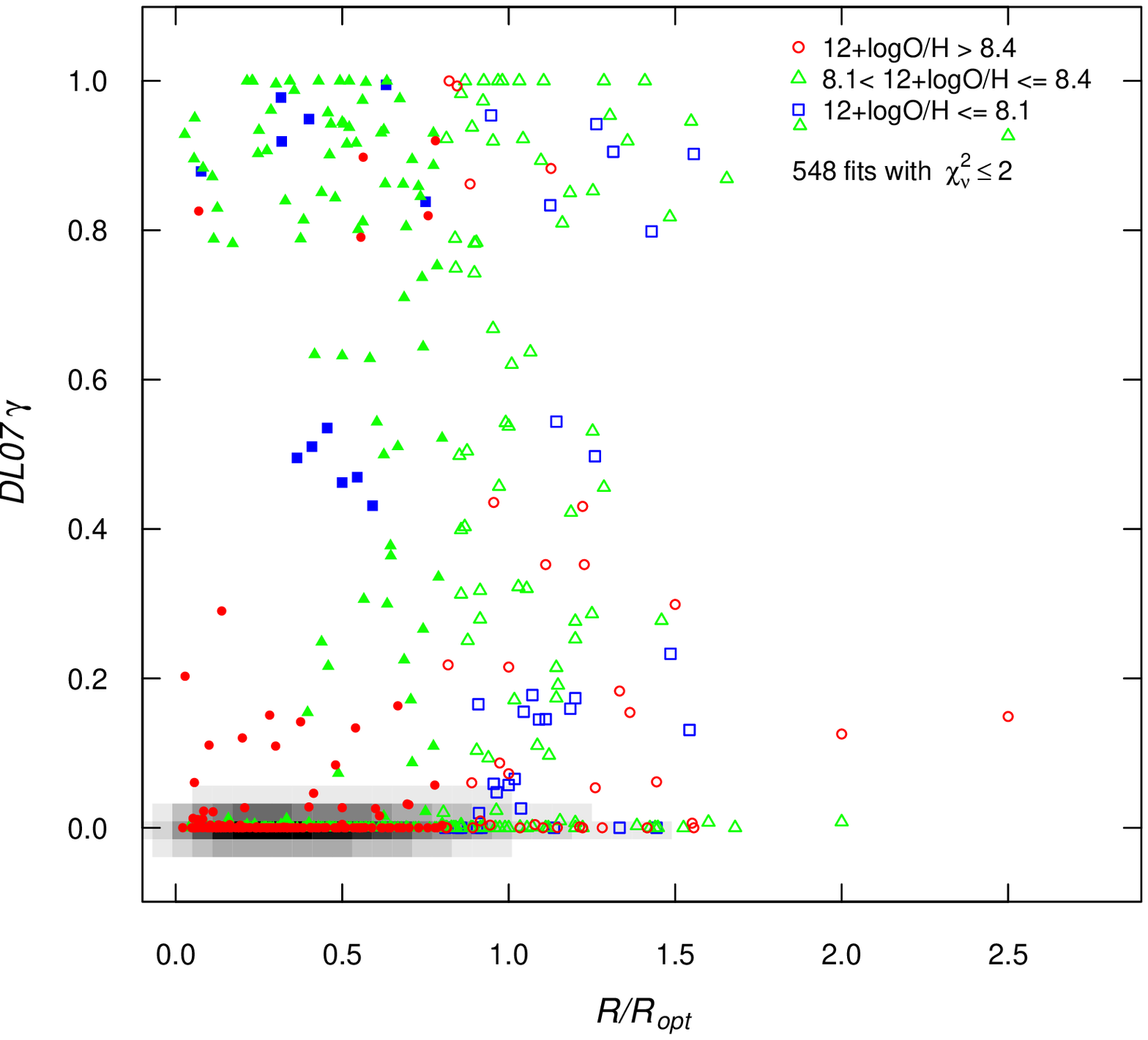}
\hspace{0.1cm}
\includegraphics[angle=0,width=0.40\linewidth,bb=0 0 503 503]{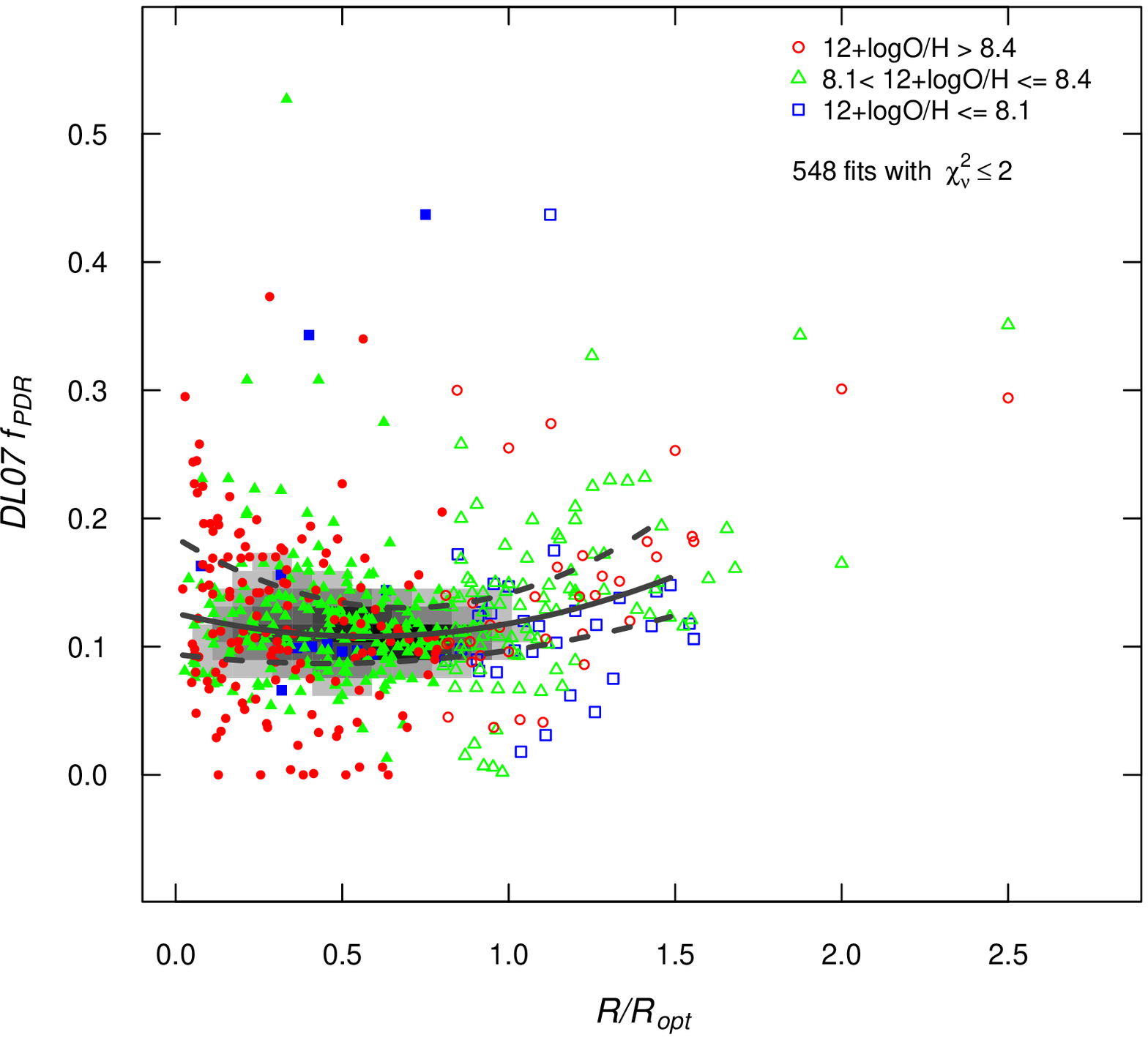}
}
}
\vspace{-\baselineskip}
\caption{Best-fit DL07 parameters plotted against normalized galactocentric distance.
The upper left panel shows \umin; the upper right \qpah;
the lower left $\gamma$; and the lower right PDR fraction \fpdr.
The underlying greyscales in each panel show the two-dimensional density distributions of the data.
%Assuming the linear radial decrease of \td\ shown in Fig.~\ref{fig:radial_tempropt_tau},
%the relation for \umin\,=\,(\td/17.9\,K)$^{(4+\beta)}$ is given
%as dashed curves with 
%$\beta$\,=\,1, 2, and 3 (from the lower to the upper curve).
%The relation for \umin\,=$U_0$\,(\td/21.1\,K)$^{(4+\beta)}$ is given
%as (the lower) dashed curves with 3 values of $\beta$ as before; in this case however,
%the difference in $\beta$ produces only minimal changes in
%the curves.
The upper left and right panels show the mean (solid curve) and quartiles (dashed curves) of
\umin\ (upper left), \qpah\ (upper right) and \fpdr\ (lower right). 
The (yellow) long-dashed curve in the upper left panel shows the
relation for \umin\,=$U_0$\,(\td/$T_0$)$^{(4+\beta)}$ 
for $U_0$\,=\,0.93, $T_0$\,=\,22\,K, and $\beta$\,=\,1.8 as described in the text.
As in previous figures, (red) circles correspond to
\logoh$>$8.4, (green) triangles to 8.1$<$\logoh$\leq$8.4,
and (blue) squares to \logoh$\leq$8.1;
filled symbols indicate $R$/\ropt$<$0.8.
\label{fig:radial_dl07}
}
\end{figure*}

Dust heating is expected to decrease with radius because of the
decline of stellar surface density.
The bulk of the dust in terms of mass is heated by \umin\ so we would expect an 
analogous radial decline of \umin.
Such a trend is shown in
Figure \ref{fig:radial_dl07} where we have plotted 
\umin\ as a function of normalized galactocentric distance,
as well as the other best-fit DL07 parameters,
\qpah, $\gamma$, and \fpdr.
The upper left panel shows that
in the KINGFISH galaxies, on average, \umin\
decreases roughly as a power law with $R/$\ropt.
%Such behavior would be
%consistent with the relation between \umin\ and \td, 
We can use the relation between \umin\ and \td: 
\begin{equation}
U_{\rm min}(R)\,=\,U_0\,\left(\frac{T_{\rm dust}(R)}{T_0}\right)^{(4+\beta)}
\label{eqn:umin}
\end{equation}
and combine it with the linear decrease of \td\ with $R$/\ropt\ 
(see Sect. \ref{sec:radial_opticaldepth} and Fig. \ref{fig:radial_tempropt_tau})
to roughly predict the radial decline of \umin. 
Fitting \tddl\ determined from the DL07 MBB fits to a linear trend
with radius, we find \tddl\,=\,$-1.74\ R/R_{\rm opt} + 22.8$;
inserting this expression into Eq. \ref{eqn:umin}, fixing $\beta$ to the
DL07 value of $\sim$1.8, and setting \tddl\ to the DL07 value for
$R$\,=\,0 of $\sim$22\,K gives $U_0$\,=\,0.93.
Eq. \ref{eqn:umin} with these values is shown as the (yellow) long-dashed
curve in the upper left panel of 
%as shown by the curves in the upper left panel of 
Fig. \ref{fig:radial_dl07}.
%The upper set of curves in the figure corresponds to
%three different $\beta$ values (1, 2, 3) with $T_0$\,=\,17.9\,K,
%the mean temperature of cool dust in the Milky Way determined
%by Planck with $\beta$\,=\,1.84 \citep[e.g.,][]{planck11c};
%the lower (grey) curves have $T_0$\,=\,21.1\,K, the mean temperature
%in the center of KINGFISH galaxies (see Fig. \ref{fig:radial_tempropt_tau}).
%The two sets of curves encompass the radial trends of \umin,
%and seem consistent with observed behavior.
The median and quartiles of the data are shown by solid and
short-dashed curves respectively.
The radial trend of \umin\ predicted by Eq. \ref{eqn:umin} follows
roughly the data, 
but does not capture the variation shown by the percentile curves,
probably because the linear decrease of \td\ 
imposed above is only a crude approximation of
the radial variation of dust temperature.
We have checked that the behavior at $R\ga$\ropt\ is not governed
by only a few galaxies; there are in fact 92 radial data points 
in 32 galaxies beyond this radius with DL07 \redchi$\leq$2.

%\afterpage{\clearpage}

%---------------------------------------------------------------
\subsection{PAH and PDR fractions}
\label{sec:radial_dl07_pah}

Figure \ref{fig:radial_dl07} shows that there is a large
spread for both the
PAH fraction \qpah\ and the fraction of dust exposed to
ISRF $>$ \umin\ \fpdr; neither parameter decreases systematically with galaxy radius.
However, \qpah, shown in the upper right panel, 
tends to be slightly lower beyond \ropt\ than for $R$/\ropt\,$\leq$\,1,
while \fpdr, shown in the lower right, tends to be higher
(the medians and quartile deviations are also shown in Fig. \ref{fig:radial_dl07}).
%There is no apparent trend in \fpdr\ with metallicity.
High \fpdr\ could be related to the high-\td\ low-\taud\ regime illustrated in
Fig. \ref{fig:radial_tempropt_tau} in which tenuous dust in the outer
regions tends to be warmer, but possibly less rich in PAHs. 
It is possible that this radial increase in the PDR fraction is related to 
far-ultraviolet extended disks \citep[e.g.,][]{gildepaz05,bigiel10,holwerda12b},
and the ongoing star formation associated with \hi\ there
\citep[e.g.,][]{bigiel10,holwerda12b,cortese12}.

%---------------------------------------------------------------
\section{Discussion}
\label{sec:discussion}

Much previous work has been focused on characterizing dust grain
populations by fitting FIR SEDs.
In particular, the MBBV approach has been used to establish that
emissivity indices are not constant either globally in galaxies
or in individual regions.
However, as mentioned in Sect. \ref{sec:intro},
the interpretation of such variations is still open to debate.
Our KINGFISH data allow a new perspective 
on how the properties of dust can be quantified in nearby galaxies.
The radial profiles probe a wide range of ISRF levels;
\umin\ can be as high as 20 in the centers of NGC\,1377\footnote{This galaxy
is optically thick for $\lambda\la 30$\,\micron\ so the DL07 models
may not be strictly appropriate because of the inherent assumption
that the dust emission is optically thin.} and NGC\,2146,
but falls to \umin\,=\,0.1 in the centers of dwarf galaxies 
(e.g., DDO\,154, DDO\,165, Holmberg\,II, IC\,2574, M\,81 Dwarf B)
and in the outskirts of virtually all spirals and early-type galaxies.
Thus, our sample can quantify trends with $\beta$
and \td\ over a factor of $\sim$200 in ISRF intensities:
from 10 times lower than the Solar Neighborhood value to almost 20 times higher.
This section is devoted to an analysis of the relation of
fitted emissivity indices $\beta$, temperature \td, and the parameters
of the DL07 models, together with the insights they give
about the physical characteristics of dust emission.

%---------------------------------------------------------------
\subsection{The temperature-emissivity degeneracy}
\label{sec:degen}

MBBV fits of FIR dust SEDs are notorious for the degeneracy
between dust temperature \td\ and apparent emissivity index $\beta$ 
\citep[e.g.,][see also Sect. \ref{sec:mbb}]{galametz12}.
As mentioned before,
such trends have been generally interpreted as spurious, 
a signature of temperature mixing along the LOS combined with uncertainties
in flux measurements, rather than intrinsic variations in 
dust properties 
\citep[e.g.,][]{shetty09a,shetty09b,juvela12,juvela13}.
Indeed, \citet{kirkpatrick14} showed that by limiting analysis to
data with high S/N ($\geq$10) it is possible to significantly reduce the
correlation between \td\ and $\beta$.

\begin{figure*}[!ht]
\centerline{
\hbox{
\includegraphics[angle=0,height=0.4\linewidth,bb=18 280 592 718]{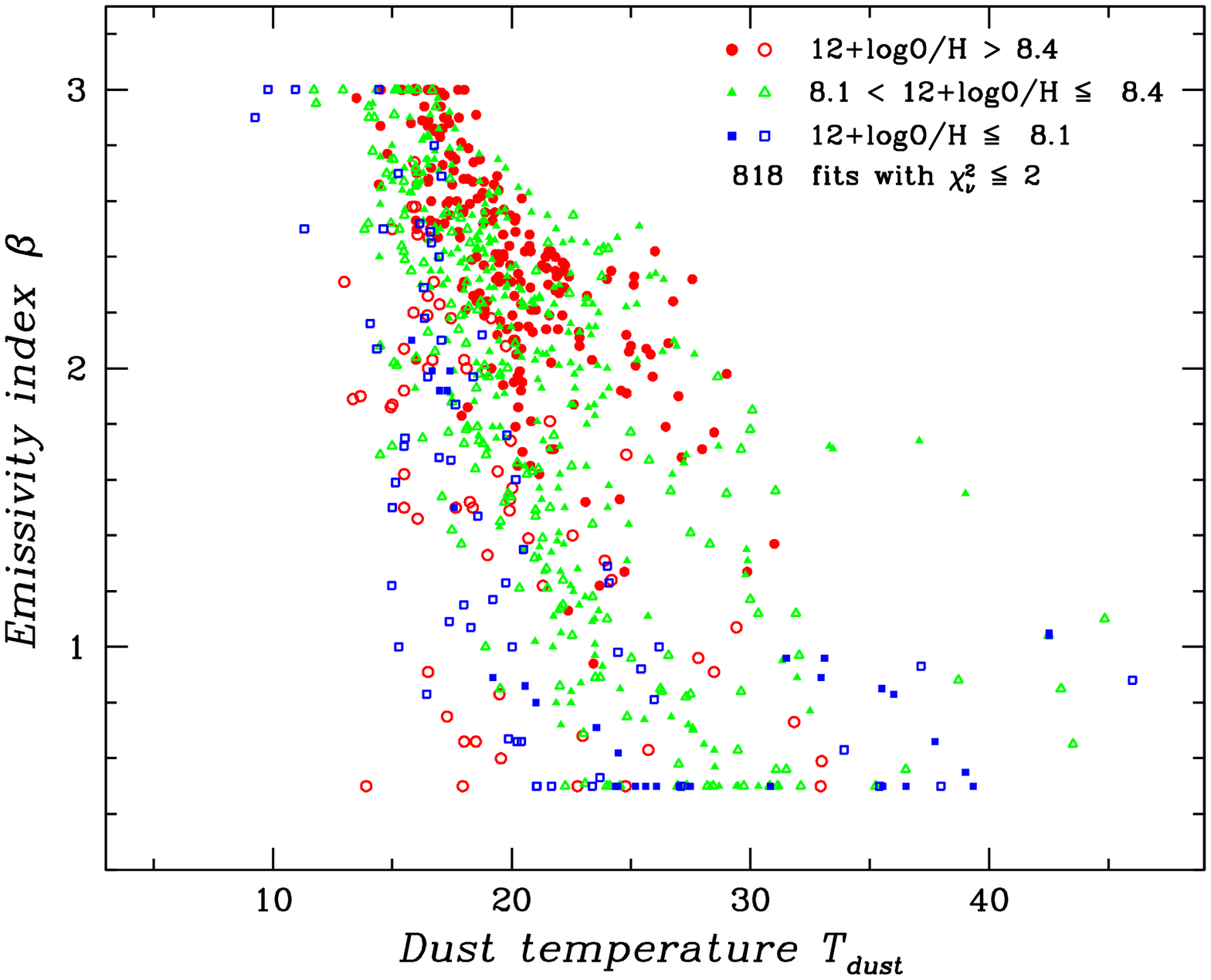}
%\hspace{-1.45cm}
\hspace{-0.5cm}
\includegraphics[angle=0,height=0.4\linewidth,bb=18 280 592 718]{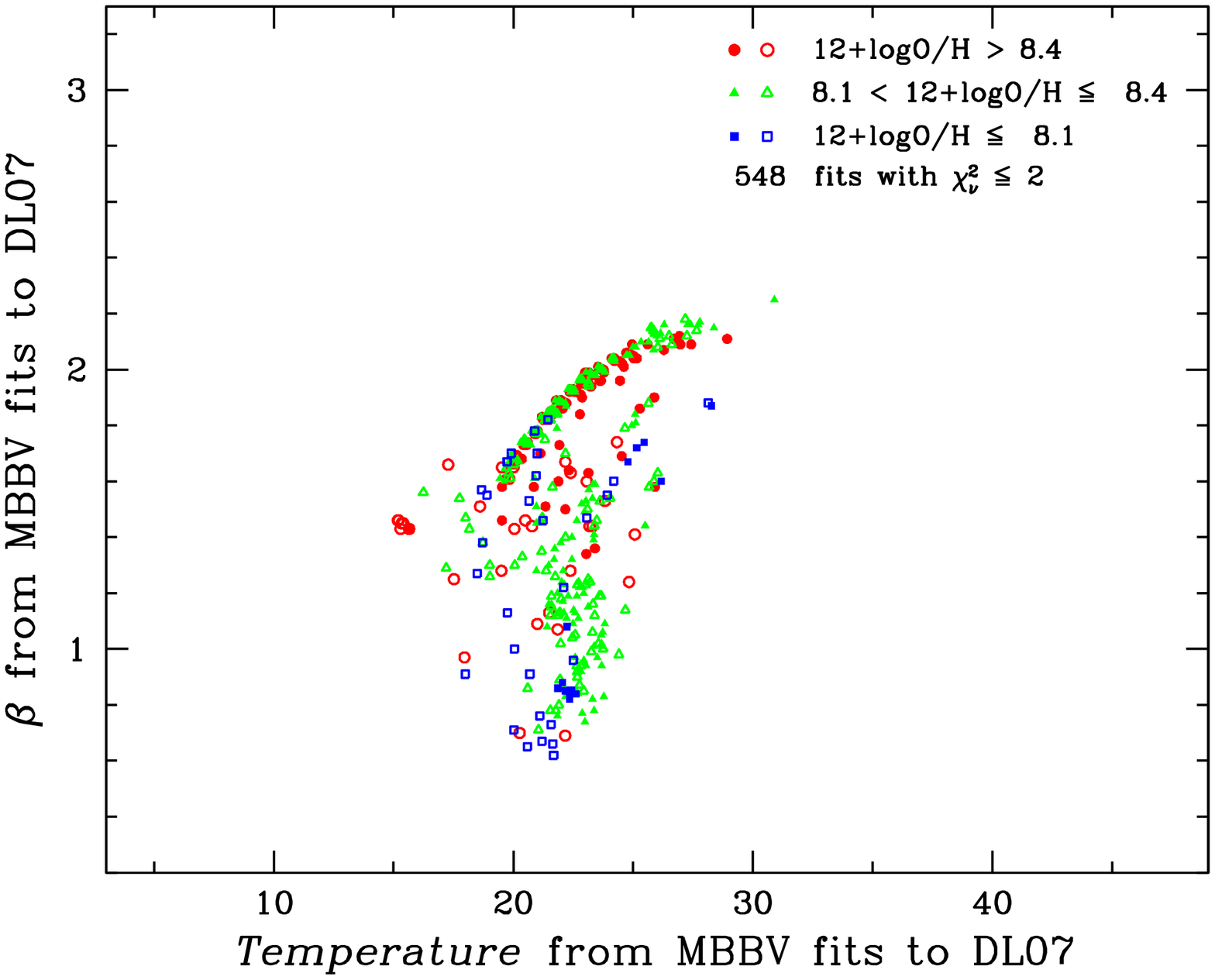}
}
}
\caption{Fitted MBBV emissivity index $\beta$ plotted against fitted dust temperature \td.
The left panel shows the MBBV fits to the data, and the right the MBBV fits to the best-fit
DL07 models.
Data are distinguished by O/H with (red) circles showing
\logoh$>$8.4, (green) triangles 8.0$<$\logoh$\leq$8.4, and (blue) squares \logoh$\leq$8.1.
Filled symbols correspond to positions with normalized (to optical radius \ropt) radii 
within $R$/\ropt$\leq$0.8, and open symbols to larger radii.
\label{fig:betat}
}
\end{figure*}

\begin{figure*}[!t]
\centerline{
\hbox{
\includegraphics[angle=0,width=0.5\linewidth,bb=18 280 592 718]{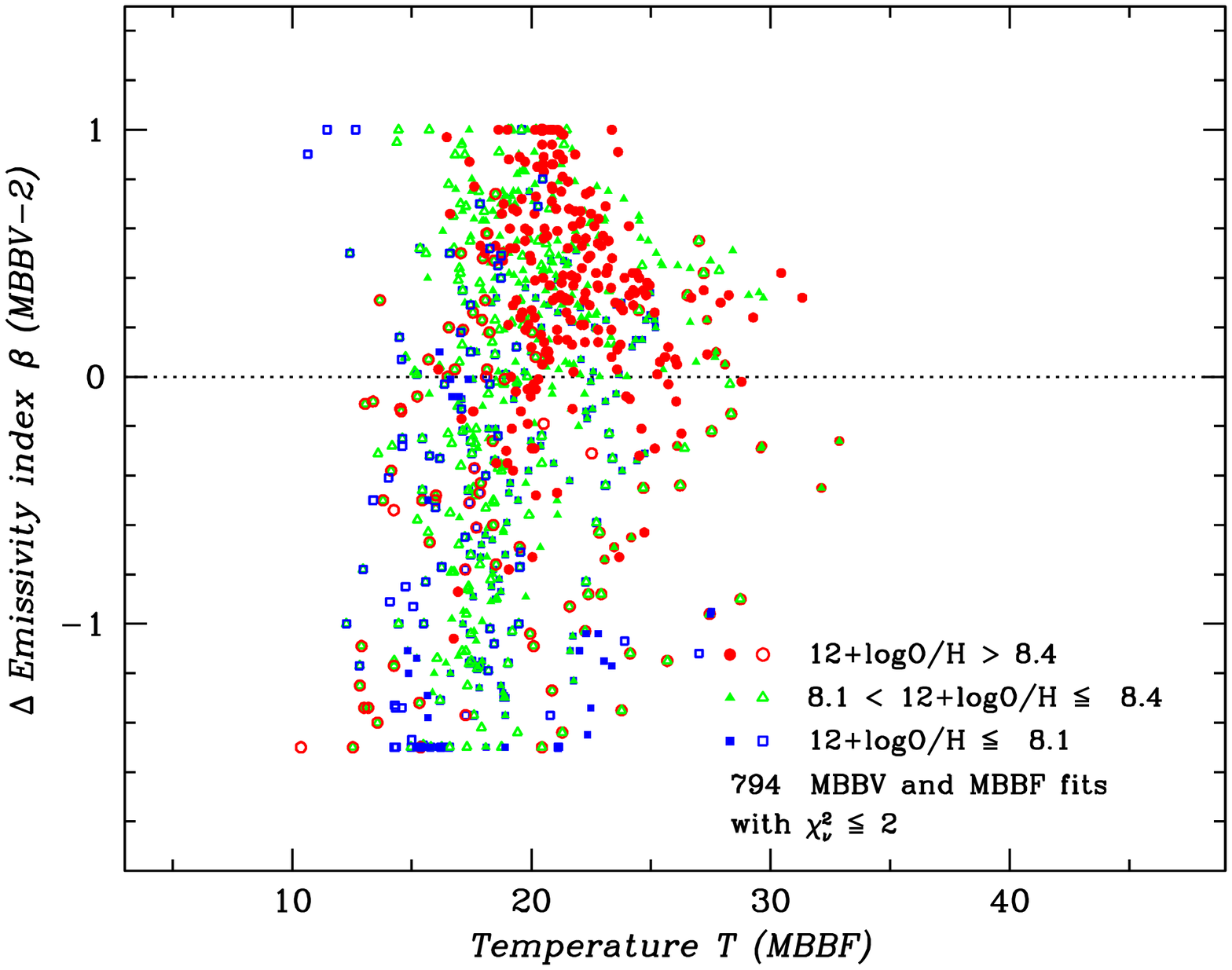}
\hspace{-1.4cm}
\includegraphics[angle=0,width=0.5\linewidth,bb=18 280 592 718]{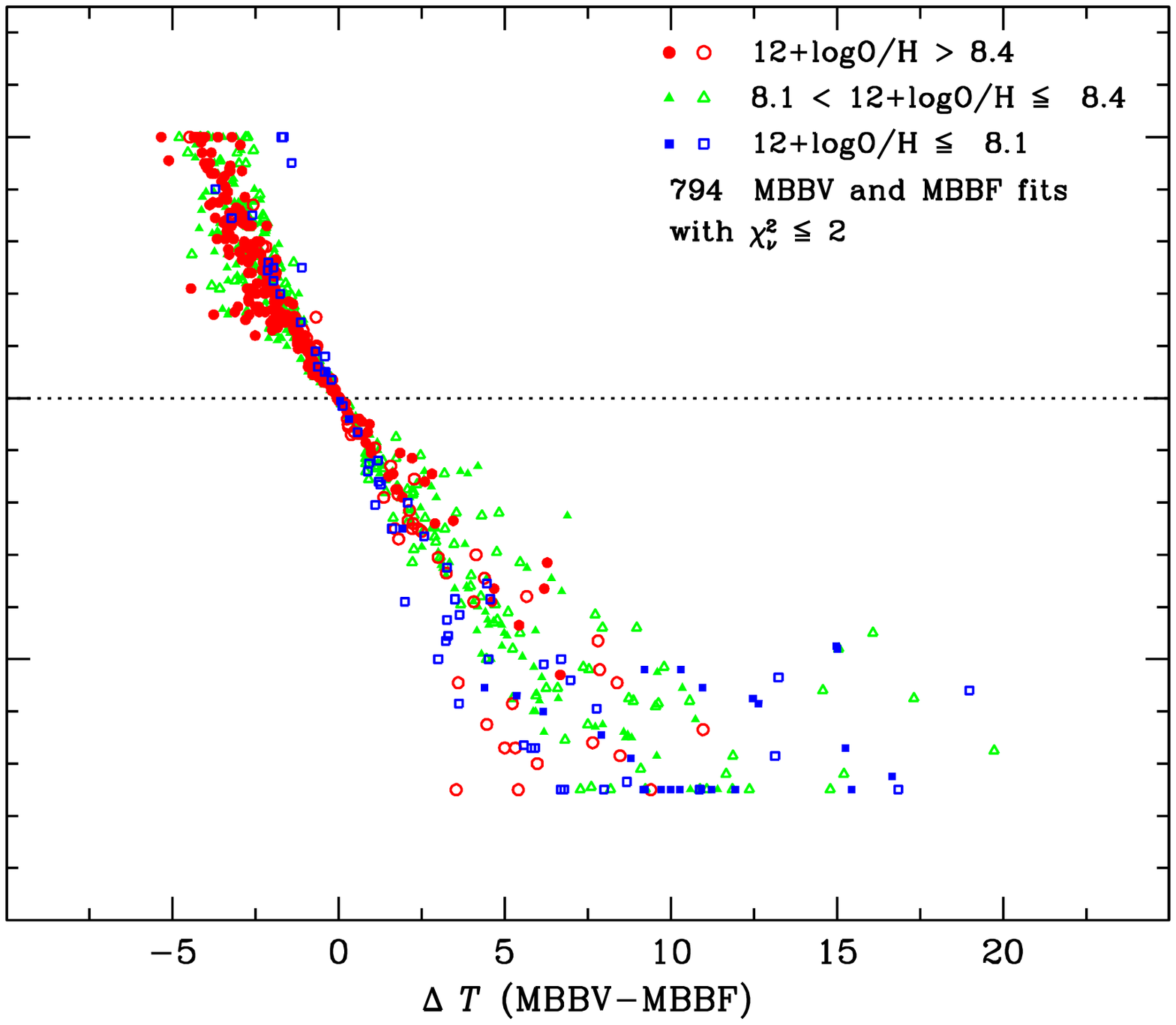}
}}
\vspace{0.5cm}
\centerline{
\hbox{
\includegraphics[angle=0,width=0.5\linewidth,bb=18 280 592 718]{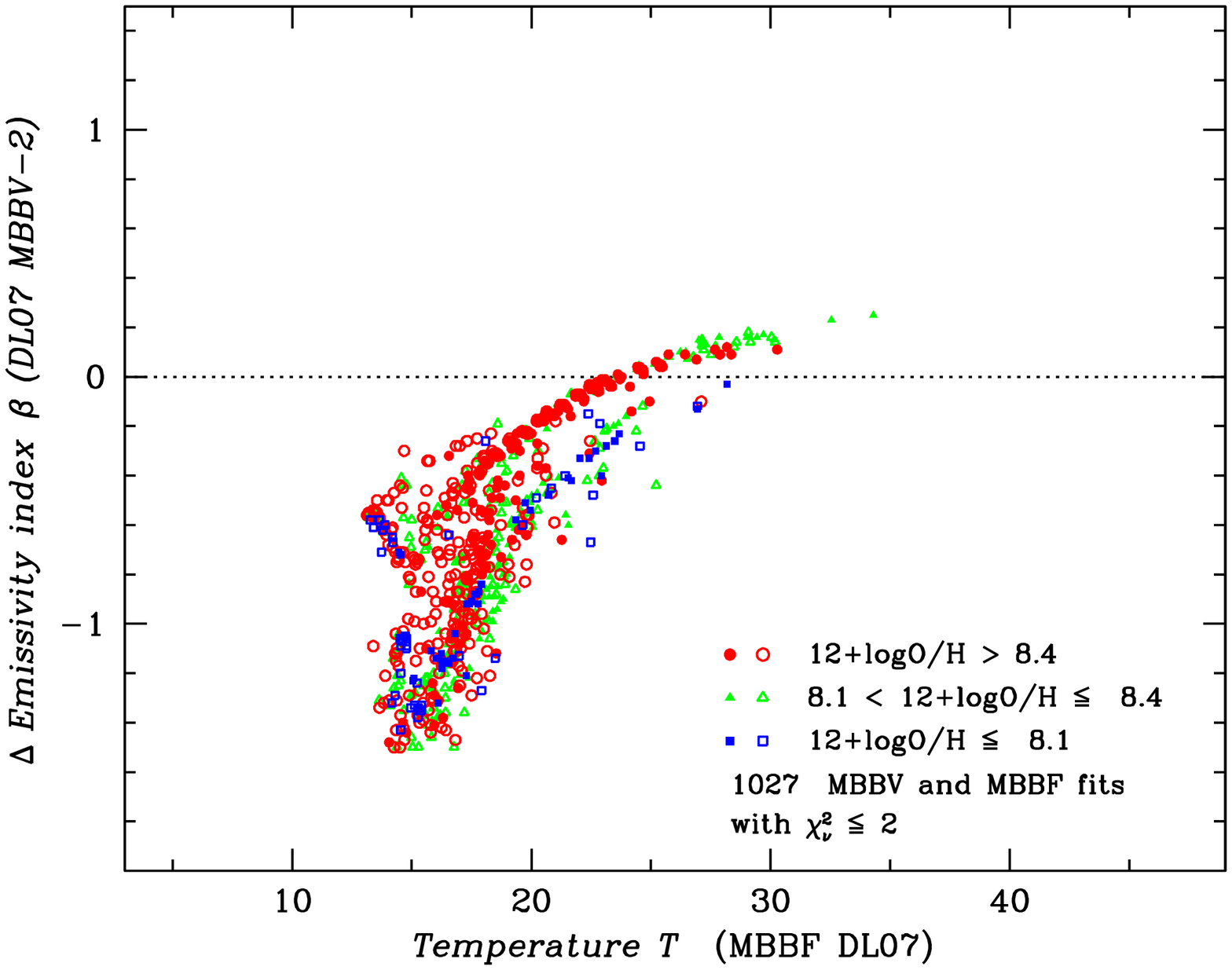}
\hspace{-1.4cm}
\includegraphics[angle=0,width=0.5\linewidth,bb=18 280 592 718]{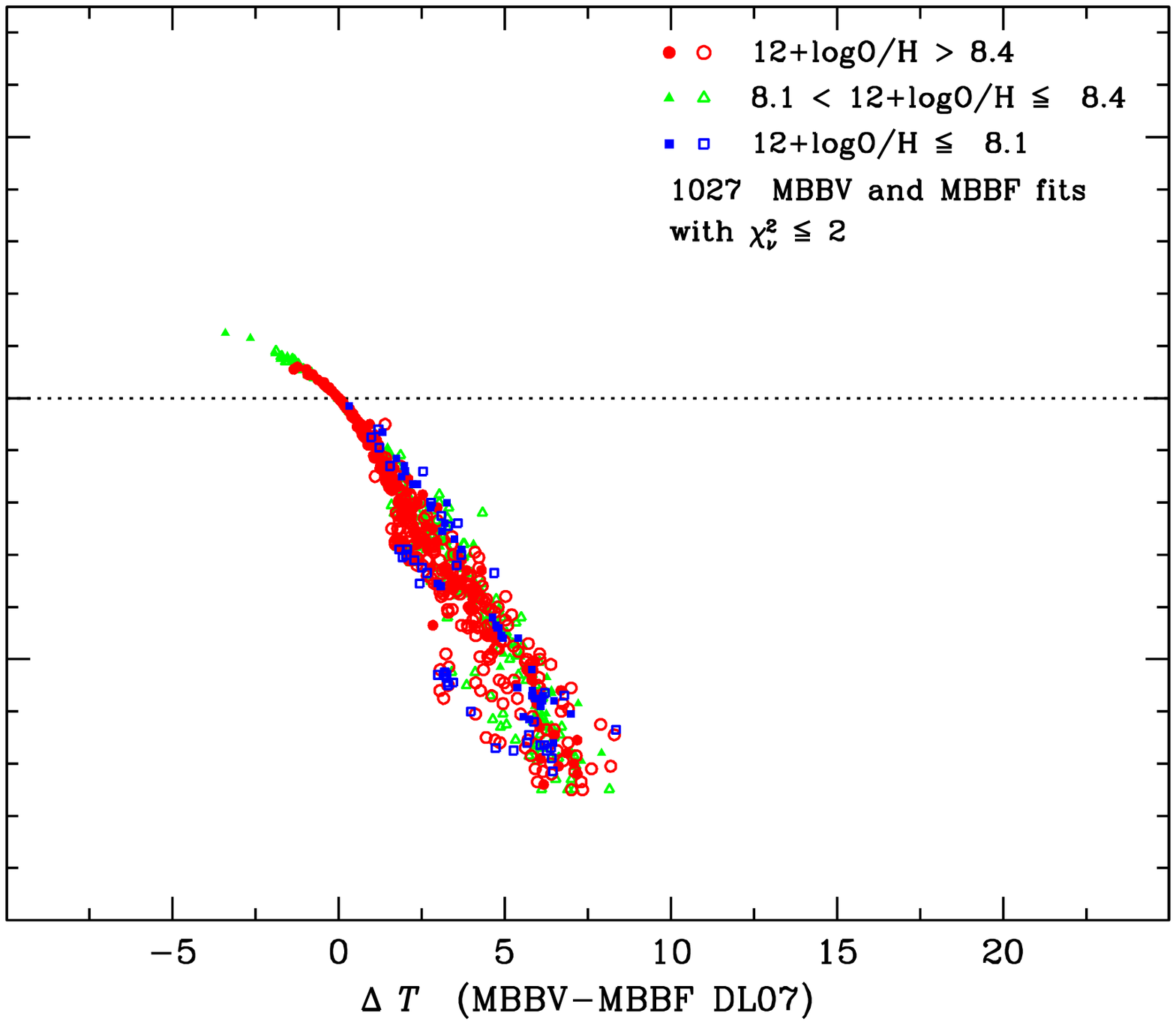}
}}
\caption{Top left panel: Difference of $\beta$ from the MBBV fits to the data and MBBF with $\beta\,=\,2$
vs. temperature from MBBF fits. Top right panel: the same but vs. the difference in MBBV \td\ - MBBF \td.
Bottom panels: The same as for the top panels but for the MBBV and MBBF fits to the DL07 best-fit models.
Points are coded by their oxygen abundance, with filled symbols corresponding
to $R$/\ropt$\leq$0.8 and open ones to larger radii.
Only those MBBV and MBBF fits with \redchi$\leq$2 are shown.
\label{fig:bbtt}
}
\end{figure*}

The relation between these two parameters in our MBBV fits to the KINGFISH profiles
and in the MBBV fits of the analogous best-fit DL07 models is shown in
Fig. \ref{fig:betat}.
Despite our grid technique to mitigate the degeneracy between $\beta$ and \td,
there is still some vestige of it remaining in our MBBV fits.
The left panel of Fig. \ref{fig:betat} shows that higher values of $\beta$
tend to be associated with lower \td.
However, for $\beta$\,=\,0.5, the lowest value allowed by our fitting algorithm,
\td\ can be as low as $\sim$14\,K and as high as $\ga$40\,K.
Thus, our fits partly reduce the known degeneracy between these parameters
because they do not rely on traditional curve fitting.
In Fig. \ref{fig:betat}, data points are coded by metallicity and the most metal-rich
inner regions (filled circles) of KINGFISH galaxies generally have $\beta\ga$2.

The right panel of Fig. \ref{fig:betat} shows the MBBV fits to the best-fit DL07
models.
The MBBV DL07 \betadl\ never exceed $\beta\sim$2, as expected
because of the intrinsic nature of the DL07 grain populations.
%Here the $\beta-$\td\ degeneracy is evident, if at all, only for \betadl$\la$1.5, and in any
%case with significant scatter.
Because the ``canonical'' $\beta$-\td\ degeneracy is virtually absent in the MBBV DL07 fits, 
we conclude, in accordance with previous work, 
that data uncertainties play an important role its generation.
%However, this same absence could point to a more minor role 
%of temperature mixing along the LOS. 
The MBBV fits in the left panel of Fig. \ref{fig:betat} show, if anything,
an increase in $\beta$ with {\it decreasing} \td;
the MBBV fits to the DL07 models in the right panel show, if anything,
an increase of $\beta$ with {\it increasing} \td.
The DL07 models, by definition, reproduce a range of dust 
temperatures through the distribution of ISRF intensities,
so if this were the primary cause of the degeneracy in the data,
we would expect to see the same trends in the $\beta$-\td\ relation
for the data and the best-fit DL07 models and we do not.
However, the MBBV fits of the DL07 models are restricted to a narrower range 
of temperatures than the MBBV fits to the data;
$\sim$56\% of the MBBV fits to the data with \redchi$\leq$2  
have \td$\leq$20\,K but only $\sim$25\% of the MBBV fits to the DL07 models
have similarly low temperatures.
Nevertheless, the data suffer from measurement noise and perhaps
combinations of \td\ and $\beta$ that are not present in the DL07
models.
Hence, we cannot draw definitive conclusions from these trends.

Figure \ref{fig:bbtt} 
compares the emissivities ($\Delta\,\beta$ relative to $\beta\,=\,2$) from the MBBV fits
with temperature \td.
The left panels show \td\ derived from MBBF fits, and the right panels
the difference $\Delta$\,\td\ of the temperature from variable-$\beta$ fits
and fixed-$\beta$ fits.
The \betadl\ of the DL07 best-fit models never exceeds $\sim$2, while
the data sometimes need $\beta\sim3$; hence the upper panels with the data show a 
larger excursion in $\Delta\,\beta$ than the lower ones (with the DL07 best-fit models).
%Although the upper right panel of Fig. \ref{fig:bbtt}
%shows essentially the same behavior as Fig. \ref{fig:betat} 
%(with a constant offset in $\beta$),
The upper left panel of Fig. \ref{fig:bbtt} gives a trend of $\Delta\,\beta$
with \td\ derived from fixed-$\beta$ MBBF fits
that is similar to the trends of the DL07 models in the lower panels.
The upper right panel show that MBBF temperatures exceed those in MBBV fits for
$\beta\geq$2, while they fall below them for lower values of $\beta$.

Such behavior of warmer \td\ from MBBF fits with increasing $\Delta\,\beta$
would be expected if temperature mixing were at work in the data.
First, single-temperature MBBF fits are as reasonable a representation
of the observed SEDs as the MBBV ones;
$\sim89-90$\% of the observed SEDs with good S/N are well approximated by both kinds
of fits (827/920 vs. 818/920, see Sect. \ref{sec:mbbdata}).
This is perhaps a surprising result given the additional free parameter in the MBBV fits.
Second, single-temperature MBBV fits tend to compensate for broader SEDs by flattening
$\beta$, which because of noise and the mathematical form of the MBB function 
causes warmer temperatures (that peak at shorter wavelengths) to be
associated with flatter $\beta$, thus creating the ``canonical'' degeneracy
of \td\ and $\beta$. 
However, if the data were truly encompassing a range of temperatures at long
wavelengths, fixed-$\beta$ MBBF fitting would result in a trend similar to that observed
in the upper right panel of Fig. \ref{fig:bbtt}.
This is because our wavelength coverage combined with an apparently shallower 
(than $\beta\,=\,2$, below dotted line) observed slope makes the MBBF fit try to compensate 
by lowering \td\ (moving the peak towards longer wavelengths); this would
make the curve around the peak emission broader than it would be at higher 
temperatures farther from the peak toward the blue.
On the other hand, the MBBF fit would compensate an apparently steeper slope  
by pushing the peak toward shorter wavelengths, thus raising the fitted \td.
%The range in best-fit temperatures would also be more confined than for MBBV 
%fits, as seen in the upper left panel of Fig. \ref{fig:bbtt}.
We investigate this point further in the next section.
 
%---------------------------------------------------------------
\subsection{Temperature mixing along the line-of-sight}
\label{sec:mixing}

\begin{figure*}[!ht]
\centerline{
\hbox{
\includegraphics[angle=0,width=0.5\linewidth,bb=18 280 592 718]{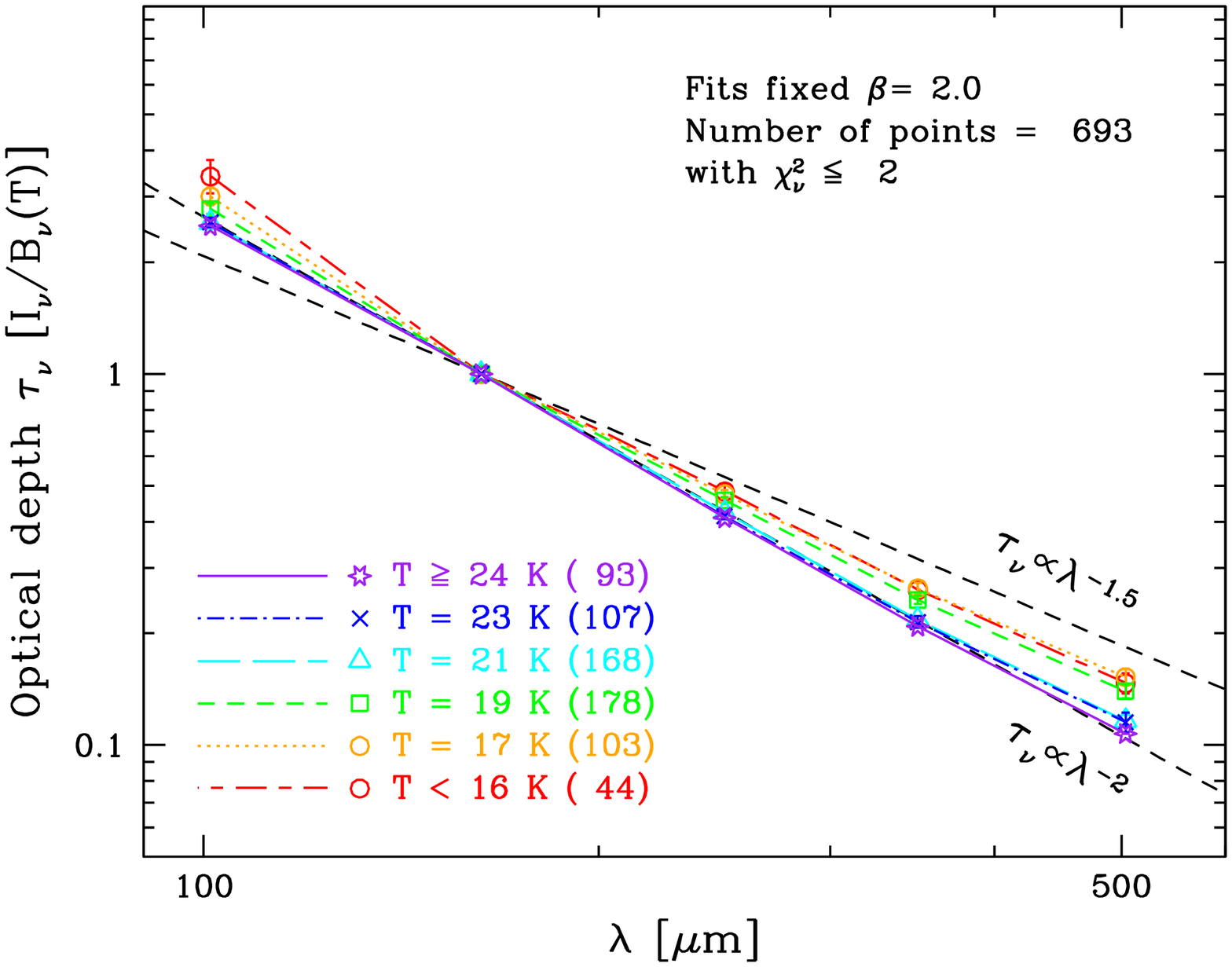}
\hspace{-1.4cm}
\includegraphics[angle=0,width=0.5\linewidth,bb=18 280 592 718]{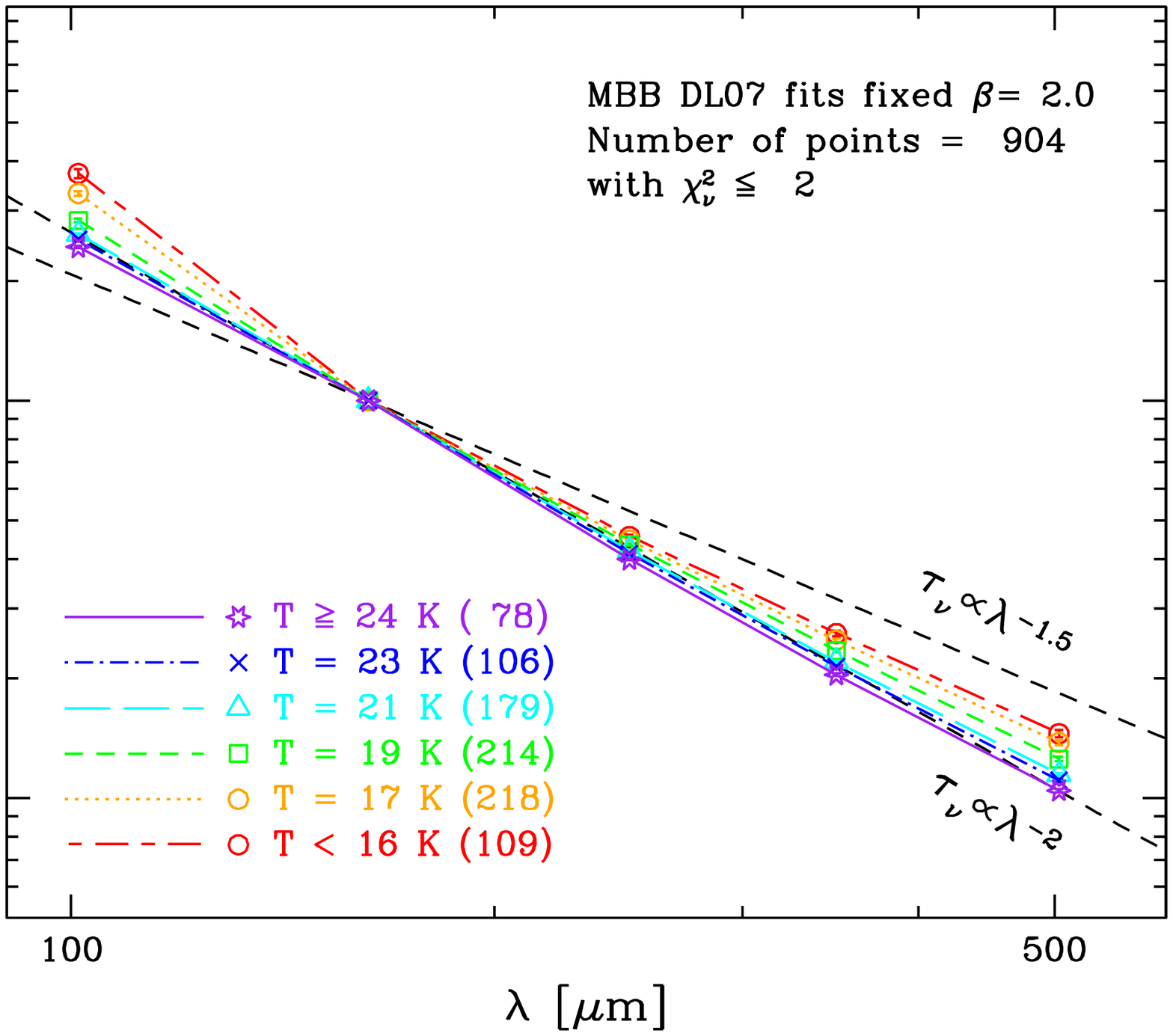}
}
}
\caption{Optical depth \taud\ normalized by $\tau_{160 \mu {\rm m}}$ as a function of wavelength $\lambda$.
The left panel shows the data, and the right panel the DL07 best-fit models.
Temperatures for the \taud\ calculation were obtained from MBBF fits
with $\beta$\,=\,2, and are shown binned to 2\,K, with the numbers of radial
points are given in parentheses.
Only those MBBF fits with \redchi\,$\leq$\,2 are considered in the calculations.
The dashed lines correspond to two values of $\beta$\,=\,1.5, 2.
\label{fig:mixing}
}
\end{figure*}

%As discussed in Sect. \ref{sec:mbb} and the previous section,
%the degeneracy between \td\ and $\beta$ could result from noise
%in the data and temperature mixing along the LOS
%\citep{shetty09a,shetty09b,juvela12,juvela13}. 
%On the other hand, 
%there is evidence that physical differences in grain population
%produce differences in FIR-submm spectral slope, $\beta$.
%In the Galaxy, dust grain properties
%differ in molecular regions relative to those dominated
%by atomic gas; this could give rise to different $\beta$ values
%because of different grain mixtures that react differently to variations 
%in the ISRF \citep[e.g.,][]{finkbeiner99,planck11c,planck11d}.
%It has also been established that
%some galaxies such as the Magellanic Clouds and M\,33
%have systematically low values of 
%$\beta$ \citep[e.g.,][]{galliano11,planck11a,tabata14}
%which have been attributed to variations in metallicity.
%Laboratory experiments show that different grain chemical 
%compositions have different $\beta$, confirming a flattening of $\beta$
%at long wavelengths and at high temperatures.
%\citep[e.g.,][]{mennella95,agladze96,mennella98,coupeaud11}. 
%Thus, although the form of the degeneracy of MBB-derived $\beta$\
%and \td\ is almost certainly primarily due to noise in the data
%and systematics in the fitting techniques,
%laboratory experiments and observations both suggest
%that there may be a break in the FIR-submm dust spectrum, and
%that there are physical reasons to expect different values
%of $\beta$ and a variation with \td.

We first want to establish whether observed trends of emissivity
index $\beta$ can be attributed to temperature variations 
along the LOS.
Following \citet{paradis12}, we have calculated the
average variation with wavelength of
dust emissivities in terms of the optical depth \taud.
We calculate \taud\ as the surface brightness $I_\nu$ ($=\,F_\nu/\Omega$)
at each wavelength divided by the blackbody at the best-fit fixed-$\beta$ MBBF \td.
%Because we want to test the reality of $\beta$ flatter than 2,
Like \citet{paradis12},
for the derivation of \taud\ we have derived the temperature \td\
using MBBF fits with $\beta \equiv$\,2 (see Sect. \ref{sec:mbb}).
If temperature mixing is in truth causing the trends between
$\beta$ and \td, we would expect the DL07 models to show the same
behavior as the data because in these models there is a multitude
of cool-dust temperatures for every LOS.
Accordingly, we have calculated the spectral trend of \taud\ also
for the best-fit DL07 models, again using MBBF fits with \betadl $\equiv$\,2;
in addition to the usual \redchi\ requirement, we also specify
that both PACS fluxes (100\,\micron\ and 160\,\micron) have S/N$\geq$3
(see Sect. \ref{sec:mbbdata}).
The results are shown in Fig. \ref{fig:mixing};
the left panel shows the data and the right panel the best-fit DL07 models.
For the figure, only the radial points with \redchi$\leq$2 have been binned
in temperature, and normalized to \taud\ at 160\,\micron;
there are fewer fits with \redchi$\leq$2 (693 vs. 818)
because of the additional requirement of PACS points with S/N$\geq$3.
%\footnote{These numbers
%include the additional requirement of PACS points with S/N$\geq$3.}) relative to the MBBV
%fits because of the increase incapacity of fixed-$\beta$ MBBF fits to accommodate
%the trends in the data.

\begin{figure*}[!ht]
\centerline{
\hbox{
\includegraphics[angle=0,width=0.5\linewidth,bb=18 280 592 718]{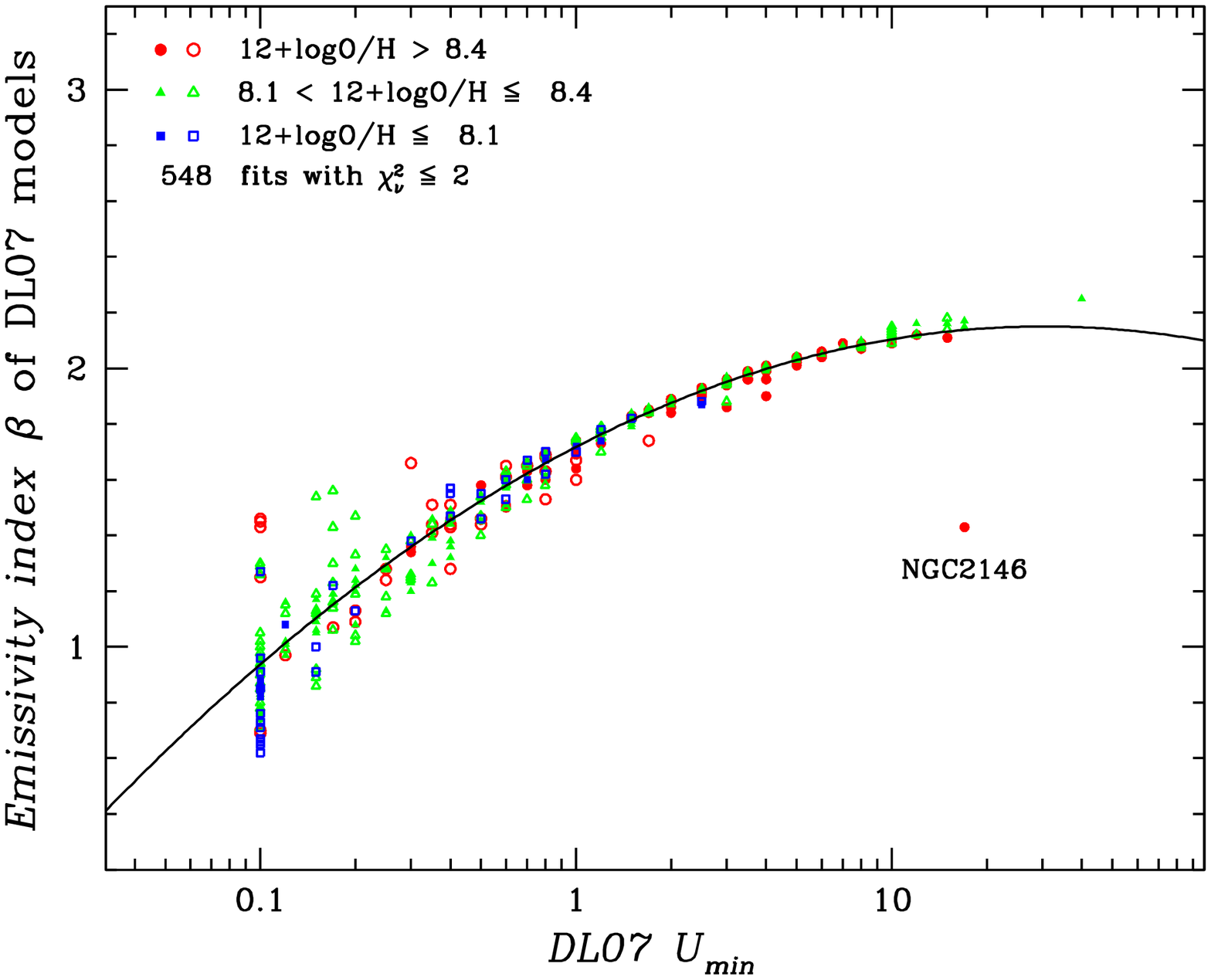}
\hspace{-1.4cm}
\includegraphics[angle=0,width=0.5\linewidth,bb=18 280 592 718]{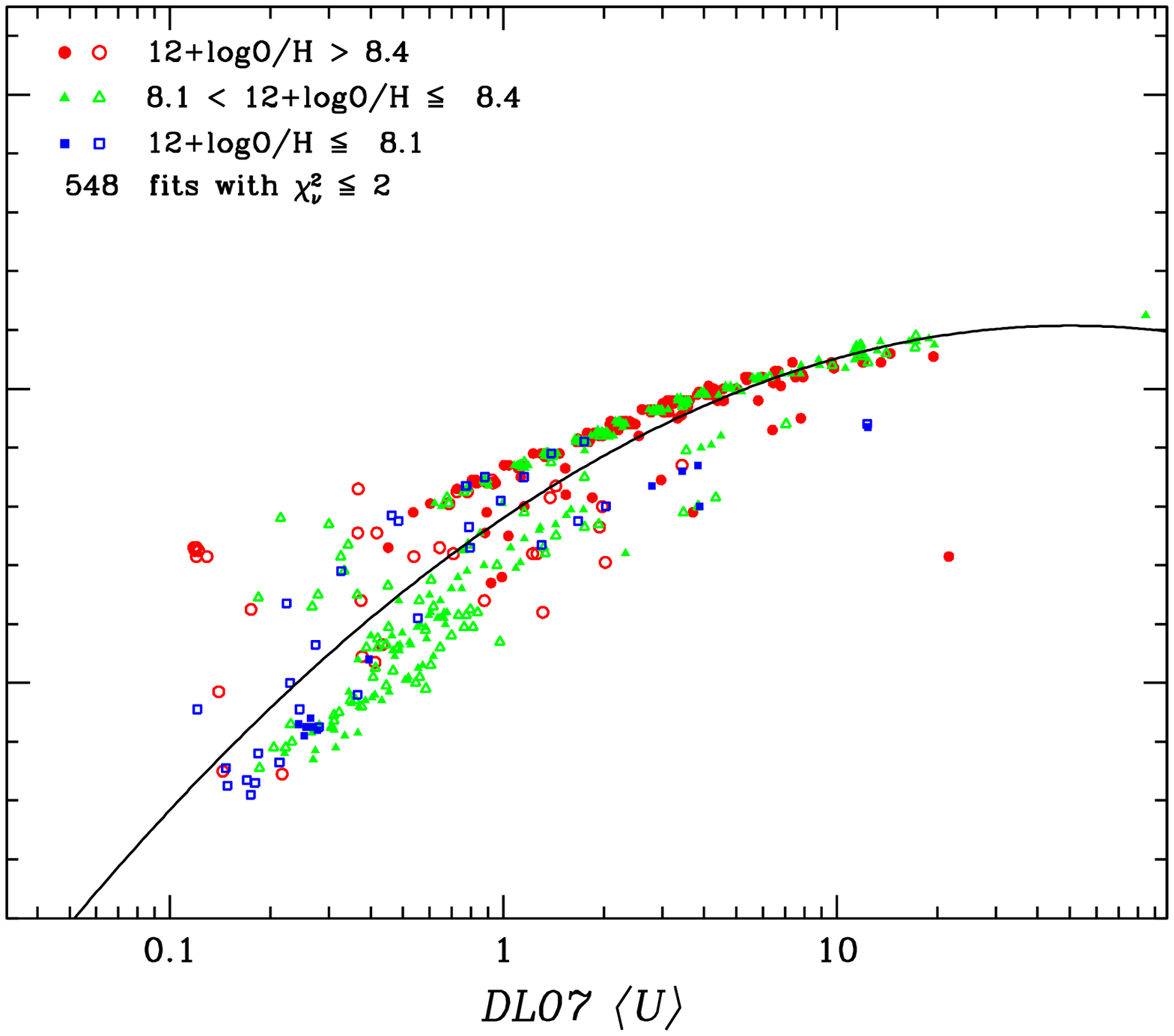}
}
}
\caption{Best-fit DL07 emissivity index \betadl\ plotted against \umin\ (in the left panel)
and \ubar\ (in the right).
Only those MBBV DL07 fits with \redchi\,$\leq$\,2 are considered in the plot.
As before, points are coded by oxygen abundance.
The curve in the left panel is the second-order polynomial best-fit of DL07 MBBV $\beta$ to 
$\log_{10}$(\umin): 
$\beta_{\rm DL07}\,=\,1.72 + 0.58\,\log(U_{\rm min}) - 0.20\,\log^2(U_{\rm min})$.
The right panel shows the analogous curve for $\log_{10}$(\ubar):
$\beta_{\rm DL07}\,=\,1.56 + 0.77\,\log(\langle U \rangle) - 0.23\,\log^2(\langle U \rangle)$.
\label{fig:betaumin}
}
\end{figure*}

\begin{figure*}[!ht]
\centerline{
\hbox{
\includegraphics[angle=0,width=0.5\linewidth,bb=18 280 592 718]{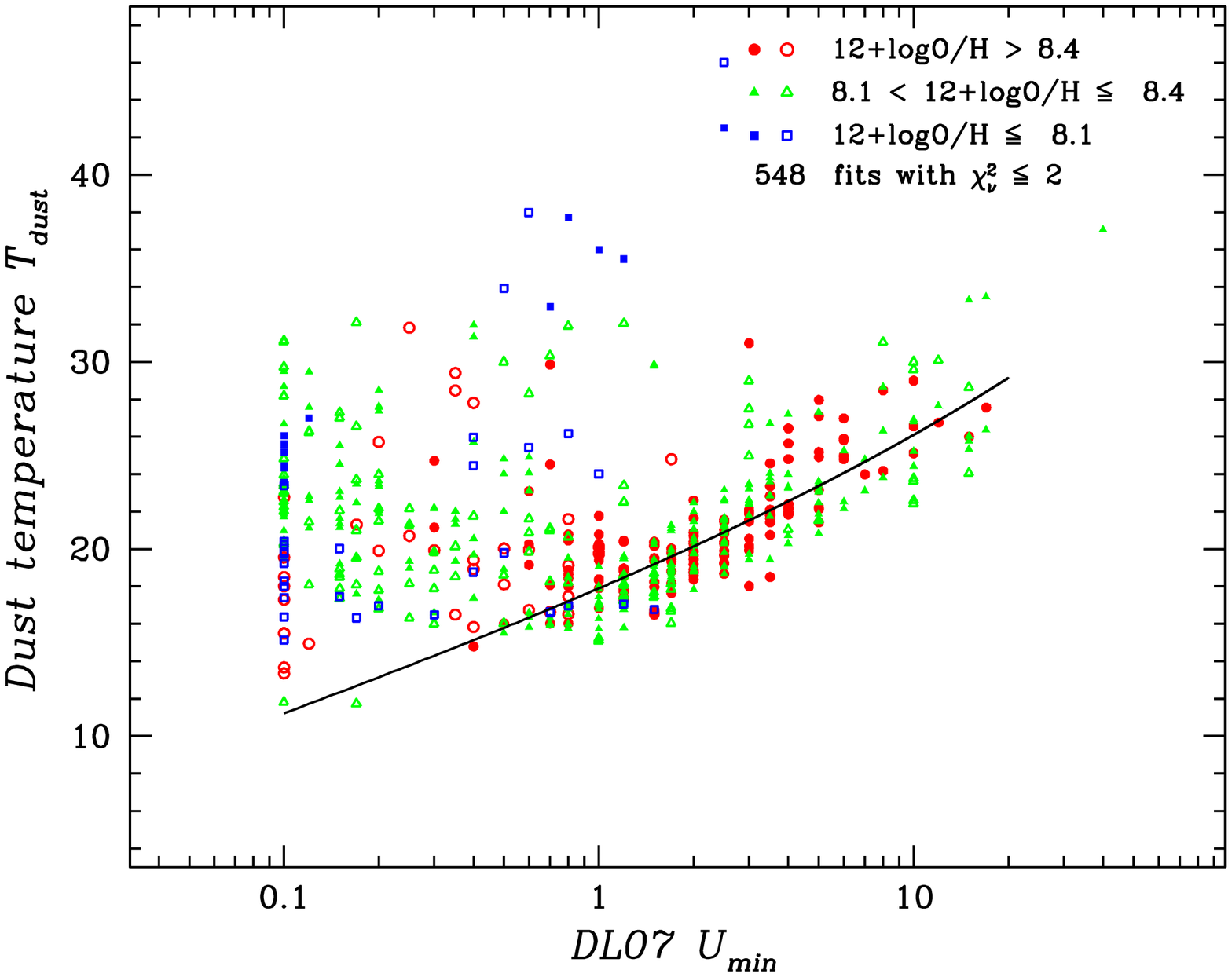}
\hspace{0.2cm}
\includegraphics[angle=0,width=0.5\linewidth,bb=18 280 592 718]{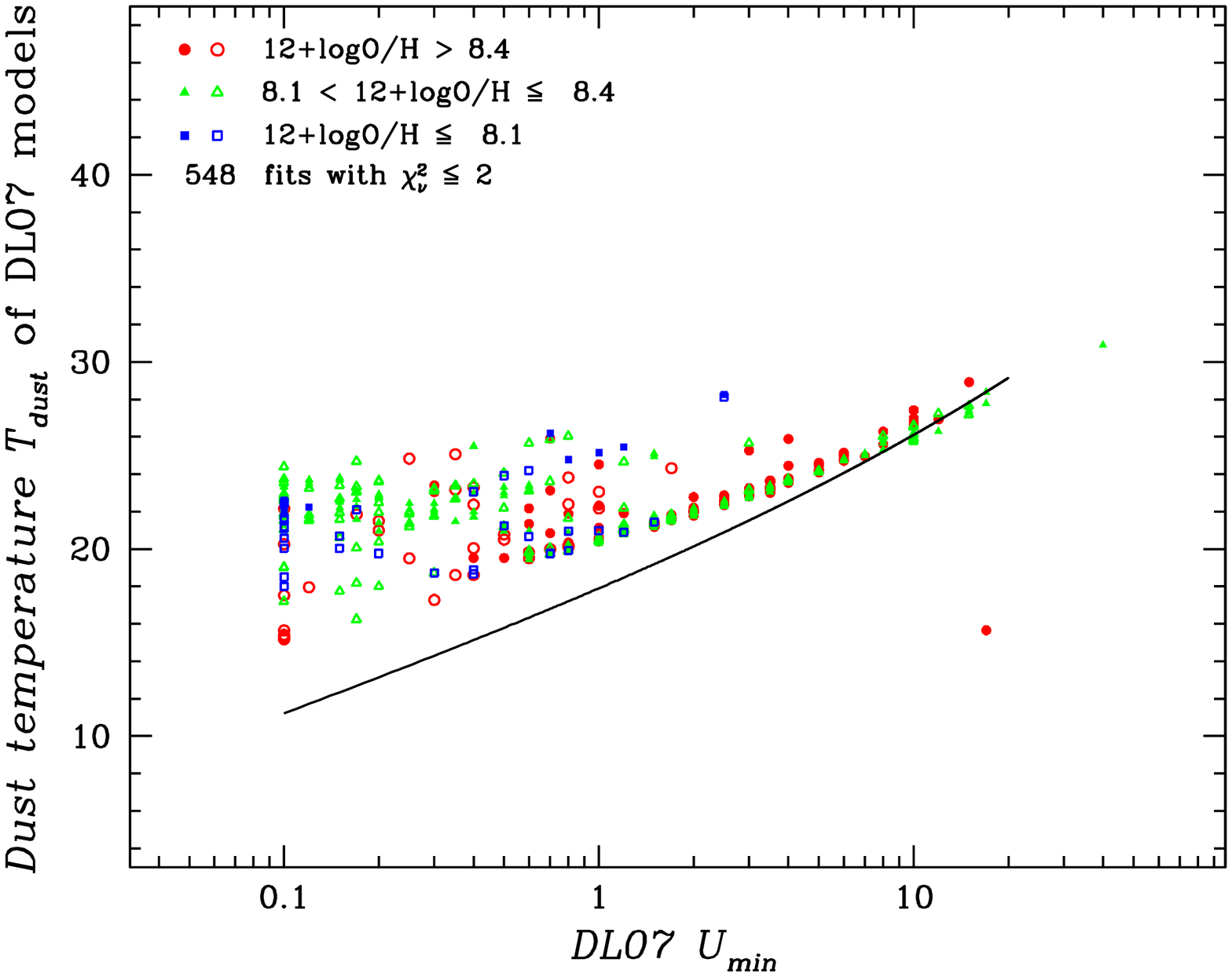}
}
}
\caption{Best-fit MBBV data dust temperature \td\ (in the left panel)
and best-fit MBBV DL07 \td\ (in the right) plotted against \umin.
Only those MBBV DL07 fits with \redchi\,$\leq$\,2 are considered in the plot.
As before, points are coded by oxygen abundance.
The curves in both panels are $T_{\rm dust}\propto U_{\rm min}^{[1/(4+\beta)]}$
assuming the relation between \betadl\ and \umin\ (see text and Fig. \ref{fig:betaumin}).
\label{fig:tempumin}
}
\end{figure*}

Figure \ref{fig:mixing}
shows an interesting feature in both the data and 
the DL07 models;
the slope of the FIR spectrum is flatter toward longer wavelengths
similarly to the Galaxy data analyzed by \citet{paradis12}.
However, instead of the flatter slope occurring at {\it high} temperatures
(\td$\ga$30\,K), in the KINGFISH data 
flatter $\beta$ is associated with {\it low} temperatures (\td$\la$20\,K).
This is the opposite of the trend expected from the $\beta$-\td\
degeneracy curve which has lower $\beta$ associated with {\it high} \td\
(see Fig. \ref{fig:betat}), but
similar to the trend of flatter $\beta$ and lower \td\ seen in the upper
left panel of Fig. \ref{fig:bbtt}.
The difference between Fig. \ref{fig:betat} and Fig. \ref{fig:mixing} is
that the parameters in the former come from MBBV fits and in the latter from MBBF fits with
$\beta\,=\,2$ as illustrated in Fig. \ref{fig:bbtt}.
%this apparent contradiction does not emerge in the single fits 
%plotted in Fig. \ref{fig:betat} but
%rather when the residuals from MBBF fits are binned in temperature
%as in Fig. \ref{fig:mixing}

We find that the best-fit DL07 models show the same
behavior as the data, namely flatter slopes at longer wavelengths corresponding to
low \td\ (see Figs. \ref{fig:betat}, \ref{fig:bbtt}, and \ref{fig:mixing}). 
This would imply that temperature mixing is causing the trend in the
data because of the similarity of the behavior of \td\ and $\beta$.
Thus our results contrast with the conclusions of \citet{paradis12}
who found that the \citet{dale02} models did not show
the same behavior as their data for Galaxy and thus that temperature mixing along the LOS
was not at work.
We propose that the KINGFISH profiles probe a wider range of physical
conditions than was possible with previous data.
In addition, the differences may be caused by the
different assumptions for $dM/dU$ (see Eq. \ref{eqn:dl07} for the DL07 approach) of
the \citet{dale02} models which are missing the bulk heating with \umin. 

Nevertheless, the spatial scales over which dust properties are integrated
must play a role in the degree of temperature mixing; the observations of \citet{paradis12} 
are along the Galactic plane and KINGFISH data are averaged over $\sim$1.9\,kpc
at the median sample distance (see Sect. \ref{sec:profiles}).
Temperature mixing is expected to be less important when the lines-of-sight
are averaged over spatial scales commensurate with less complex dust heating.
In the next section we compare \umin, \td, and $\beta$ 
to try to better understand these trends between $\beta$ and \td.

%---------------------------------------------------------------
\subsection{Radiation field, apparent emissivity index, and dust temperature}
\label{sec:umin}

We have seen in Sect. \ref{sec:radial} that the radial trends of DL07 \umin\
and \tddl\ are related in a power-law fashion, as expected for 
dust in thermal equilibrium with the ambient radiation field, 
\td$\propto U_{\rm min}^{[1/(4+\beta)]}$ (see curves in Fig. \ref{fig:radial_dl07}).
Here we explore whether the emissivity index of the best-fit DL07 models \betadl\ is related to
\umin\ in an analogous way.
Such a connection would be expected given the relation between \tddl\ and \umin,
and would help explain how the DL07 models can produce an apparent emissivity index $\beta\la 1$.
Figure \ref{fig:betaumin} plots \betadl\ against \umin.
The two parameters are closely related,  
as shown by the best fit second-order polynomial in the left panel of Fig. \ref{fig:betaumin}:
\betadl\,=\,$1.72 + 0.58\,\log(U_{\rm min}) - 0.20\,\log^2(U_{\rm min})$.
The right panel of Fig. \ref{fig:betaumin} shows the variation of \betadl\ with \ubar;
%the \umin\ curve is unable to accommodate the data, and the best fit is instead:
the best-fit curve is
\betadl\,=\,$1.56 + 0.77\,\log(\langle U \rangle) - 0.23\,\log^2(\langle U \rangle)$.
The mean of the residuals for \betadl\ from the \umin\ curve is 0.1;
thus for dust that behaves like the dust in KINGFISH galaxies, it is possible
to estimate $\beta$ from \umin, and vice versa.
The single outlier, the central point of NGC\,2146, does not follow this
trend, but this galaxy contains a dusty outflow along the minor axis
\citep{heckman00,kreckel14}
so these (optically thin) models may not be applicable. 

Despite the mean emissivity index $\beta\sim$2 of the DL07 models,
through low values of \umin\
they are very good at imitating flat FIR-submm SEDs with apparent $\beta\la 1$. 
Most of the variation in \betadl\ is for \umin$\la$1, a regime which  
was not well sampled by previous data;
the wide range in \umin\ covered by the KINGFISH profiles lets this result emerge.
At low \umin, there is a larger fraction of cool dust that emits at longer wavelengths; 
this tends to shift the peak wavelength,
broaden the SED, and flatten its apparent FIR-submm slope to $\beta\la 1$.
At high \umin, most of the dust is warmer, emitting radiation 
toward shorter wavelengths ($\la$200\,\micron) thus causing the slope of
the FIR-submm SED to assume its ``native'' value of $\beta\sim$2.
Both the shifting of the peak wavelength and the broadening of the SED contribute
to the apparently flatter slopes;
when the SED peaks at longer wavelengths, we are no longer in the Rayleigh-Jeans
regime where slopes should reflect true grain emissivities.
%\citep[e.g.,][]{planck11c}.
The association between $\beta$ and \ubar\ is looser than for \umin,
presumably because most of the ISM emitting dust in galaxies tends to be heated
by an ISRF around \umin\ rather than  \ubar. 
%it is important to remember that dust emission is always luminosity-weighted,
%so what we are really considering are luminosity-weighted averages of dust SEDs.

Because low \umin\ would be expected to correspond to low \td\ [or \tddl], 
we can understand
the trend of flatter SEDs with cooler temperatures shown in Fig. \ref{fig:mixing}.
This is also illustrated in Fig. \ref{fig:tempumin} where dust temperature
is plotted against \umin: \td\ from MBBV fits to the data are given in the left panel, 
and \tddl\ from MBBV fits to the DL07 models in the right.
The curves in Fig. \ref{fig:tempumin} correspond to 
\td$\propto U_{\rm min}^{[1/(4+\beta)]}$, assuming the polynomial curve relating \betadl\ and \umin\
shown in the left panel of Fig. \ref{fig:betaumin}.
At low values of \umin, there is a large range of \td\ as shown particularly in
the left panel where \td\ from data fits is plotted.
However, for \umin$\ga$1, the bulk of the data is close to the curve;
\td\ is expected to be lower for lower \umin\ (and equivalently for lower $\beta$
as shown in Fig. \ref{fig:betaumin}).
The DL07 dust temperatures \tddl\ (right panel of Fig. \ref{fig:tempumin}) all exceed
the curve (except for large values of \umin$\sim$10), and 
the trend of low \umin\ and low \tddl\ (and \betadl) is less pronounced.
Such behavior is consistent with the 
%data-DL07 temperature trends shown in Fig. \ref{fig:bbtt}.
narrower range of the DL07 MBBV temperatures relative to the data shown in
Fig. \ref{fig:bbtt} and discussed in Sect. \ref{sec:degen}.

%---------------------------------------------------------------
\subsection{Potential causes of temperature mixing}
\label{sec:causes}

Like much previous work \citep[e.g.,][]{galametz12,tabata14,kirkpatrick14},
we have found radial variations of $\beta$ and \td.
In the KINGFISH profiles, we also find radial variations of \umin, consistent with
what would be expected from considering the effects of dust reprocessing on $\beta$ and \td;
taking all galaxies together, $\beta$ and \td\ vary with a power-law dependence on \umin.
We have shown that in the KINGFISH profiles taken individually, $\beta$ and \td\ are weakly
related in the usual degeneracy with low values of $\beta$ (flatter slopes) associated with
high values of \td.
Nevertheless, when the MBBF fits to the profiles are binned in temperature,
flatter $\beta$ corresponds to lower \td\ (see Fig. \ref{fig:mixing}).
Moreover, despite their average dust emissivity index $\beta\sim$2 the DL07 models are able
to reproduce quite well the SEDs with apparently flatter slopes $\beta\la 1$;
this is because decreasing ISRF intensities with \umin$\la$1 produce dust SEDs
with increasingly flat apparent emissivity indices achieving $\beta\la 1$ for \umin$\approx$0.1.

\begin{figure}[!ht]
\centerline{
\hbox{
\includegraphics[angle=0,width=\linewidth,bb=0 50 566 600]{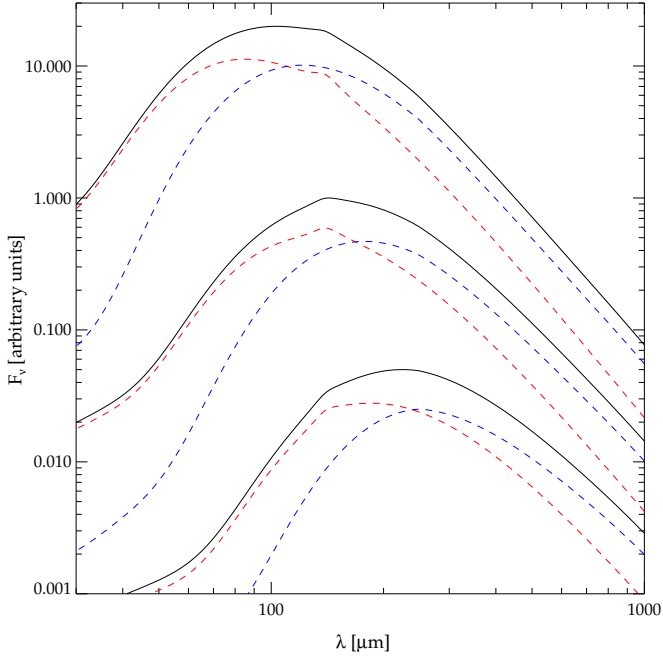}
}
}
\caption{SEDs of DL07 models with ISRF$\equiv$\umin.
Three values of \umin\ are shown: from lower to upper curves \umin\,=\,0.1, 1.0, 10.0.
Total emission is given by the solid (black) curve.
The blue dashed curves correspond to the contribution from silicates and the red to 
carbonaceous grains (including graphite, ionized and neutral PAHs).
\label{fig:simone}
}
\end{figure}

Clearly apparently flat $\beta$ SEDs can be achieved with low \umin\ even with 
the DL07 dust models that have much higher intrinsic emissivities.
We have argued that the reason for this is temperature mixing, and here
we examine two separate phenomena which could be driving the mixing:
the ISRF and grain composition.
First, a spread of temperatures is caused by the distribution of the intensities of the ISRF 
heating the dust. 
Low \umin\ means that a larger fraction of the dust tends to be cooler, with SEDs that peak at 
longer wavelengths \citep[$\lambda\ga$200\,\micron, see also][]{ciesla14}.
This is illustrated in Fig. \ref{fig:simone} where we show separately the SEDs of 
the DL07 grain populations (silicates and carbonaceous grains). 
The dust in Fig. \ref{fig:simone} is heated by a single radiation field, \umin, with
increasing \umin\ intensities associated with increasingly luminous SEDs (from \umin\,=\,0.1 to 1 to 10).
Lower \umin\ results in longer peak wavelengths.
Thus, in addition to adding more cool dust to the SED, 
low \umin\ also implies that our $\lambda\leq$500\,\micron\
data are not sampling well the Rayleigh-Jeans portion
of the spectrum where the slope of the SED converges to the 
limiting value of $\beta+2$.
%intrinsic value of $\beta$ \citep[e.g.,][]{planck11c}.
This results in slopes that are apparently flatter with smaller values of $\beta$.
Adding a more intense ISRF to \umin, such as with a distribution of $U$ (e.g., Eq. \ref{eqn:dl07}),
would broaden the SED even more and move the peak to even shorter wavelengths. 
Second, the different chemical compositions of the grains themselves react differently
to different \umin, resulting in broader SEDs.
As shown in Fig. \ref{fig:simone},
the peak wavelength changes with grain type, with silicate grains peaking toward
longer wavelengths.
The spread between the two peaks is larger for low \umin\ (lowest curve).
Both the ISRF distribution and the different grain properties contribute to the shape of the 
dust SED, and conspire to cause the apparently flatter slopes $\beta$ associated with lower \umin.

As a final check, because of previous suggestions that flatter $\beta$
may be associated with low metallicity \citep[e.g.,][]{galliano11,tabata14},
we have looked for correlations of $\beta$ with metallicity in the KINGFISH sample.
There is little evidence for such a correlation;
$\beta$ at \logoh\,$\la$\,8 ranges from 0.5 to 3.0, the same as its range
at \logoh\,$\ga$\,8.5.

%This means that temperature mixing which produces a larger fraction of cool dust
%can produce flat values of $\beta$ in FIR-submm SEDs,
%but this is not what is causing the $\beta$-\td\ degeneracy;
%low $\beta$ values are associated with cooler dust while in the canonical
%degeneracy low $\beta$ values are associated with warmer dust.
%Noise plays an important role in the $\beta$-\td\ degeneracy, but
%temperature mixing caused by low \umin\ does something else, namely flatten the
%long-wavelength slope because of increasing fractions of very cool dust. 

Although small values of $\beta$ can be attributed to low \umin\ and temperature
mixing in the form of more cool dust at low \td, large values of $\beta\ga$2
cannot be easily explained by such a phenomenon. 
Such high apparent values of $\beta$ may be due to real 
steepening of the dust SED in the FIR relative to the submm, with true
$\beta\sim$2.5, but flattened by temperature mixing along the LOS to slightly 
smaller $\beta$ values in the luminosity-weighted fits.
If the emitting dust is indeed characterized by $\beta >$\,2.5 in the
100$-$500\micron\ range, the interpretation is not clear.  Laboratory
studies of carbonaceous materials 
\citep{mennella95,mennella98}
and various amorphous silicates 
\citep{agladze96,mennella98,coupeaud11}
generally find $\beta <$\,2.2
at \td $\la$30\,K 
\citep[except for sample E of ][which had $\beta\,=\,$2.5 at \td\,=\,10\,K]{coupeaud11}.
%Large values of $\beta$ are also consistent with detailed
%studies of long-wavelength emission of Galactic cold clumps
%\citep[e.g.,][]{planck11b,planck11c}. 
More work on dust emission with submm ($\lambda\ga$800\,\micron) constraints is needed 
to better explore apparently high $\beta>$2 in nearby galaxies \citep[e.g.,][]{galametz14}.

%---------------------------------------------------------------
\subsection{Model assessment and far-infrared deviations}
\label{sec:models}

\begin{figure*}[!t]
\centerline{
\hbox{
\includegraphics[angle=0,width=0.38\linewidth,bb=18 280 592 718]{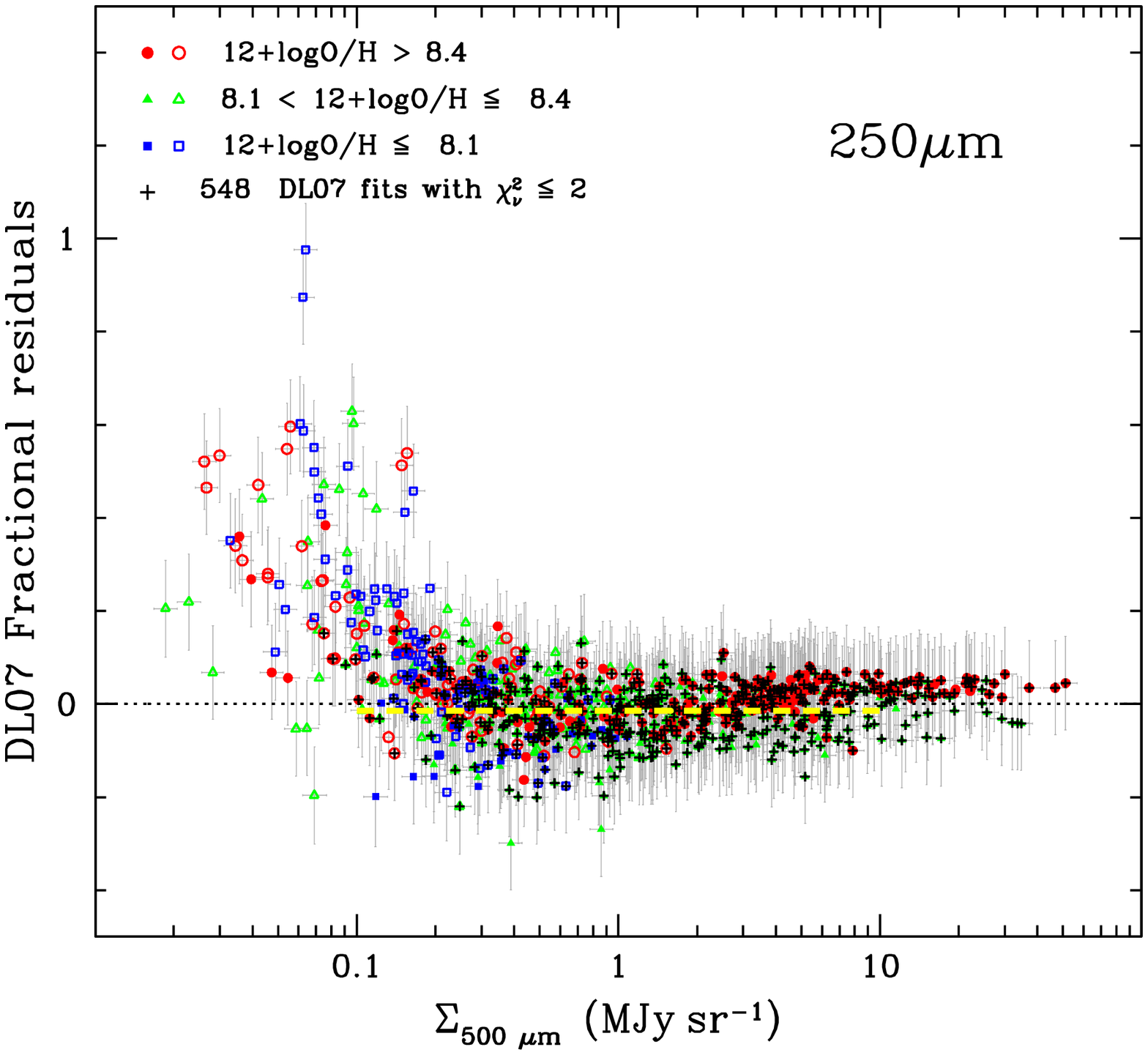}
\hspace{-1.35cm}
\includegraphics[angle=0,width=0.38\linewidth,bb=18 280 592 718]{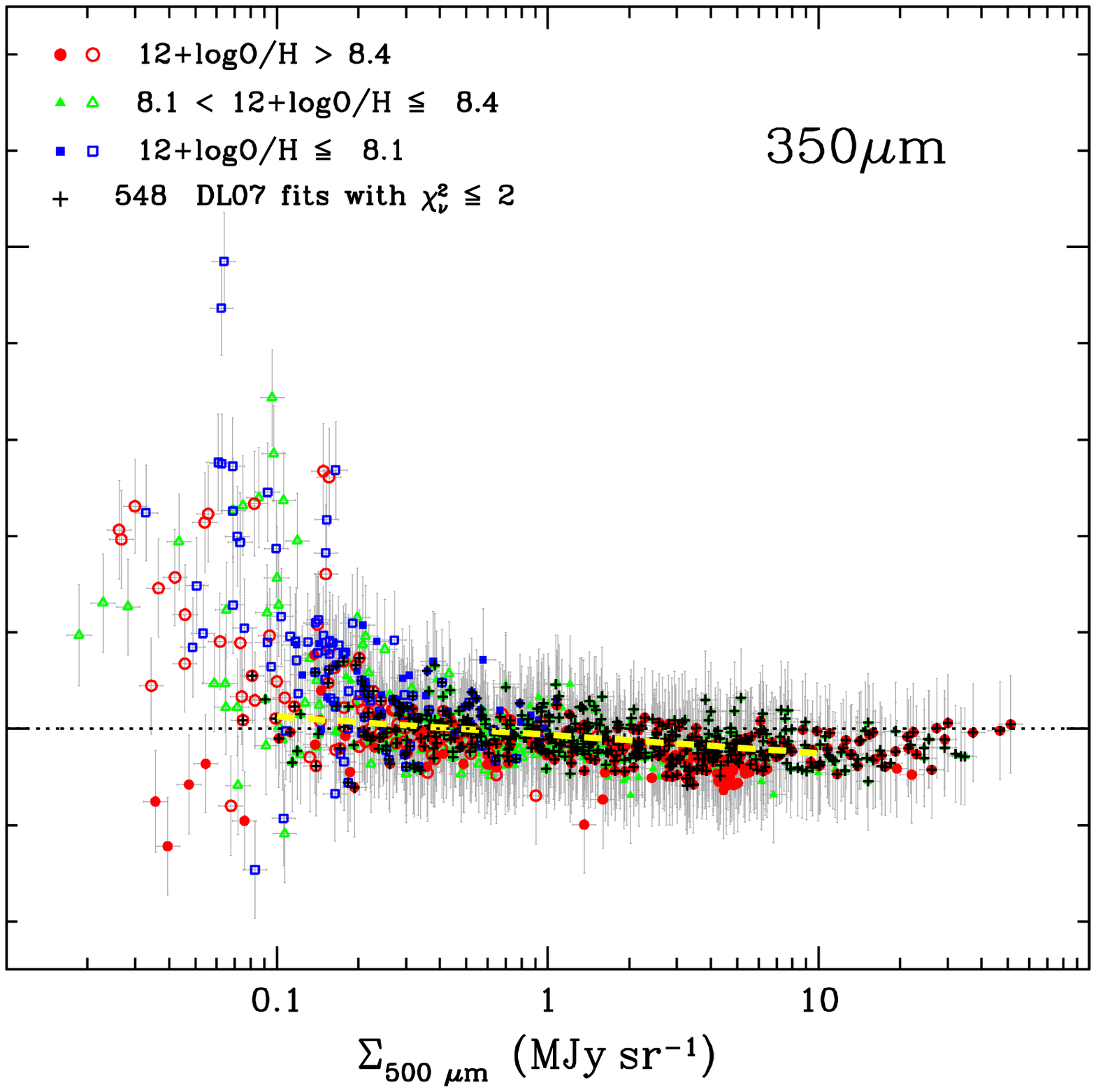}
\hspace{-1.35cm}
\includegraphics[angle=0,width=0.38\linewidth,bb=18 280 592 718]{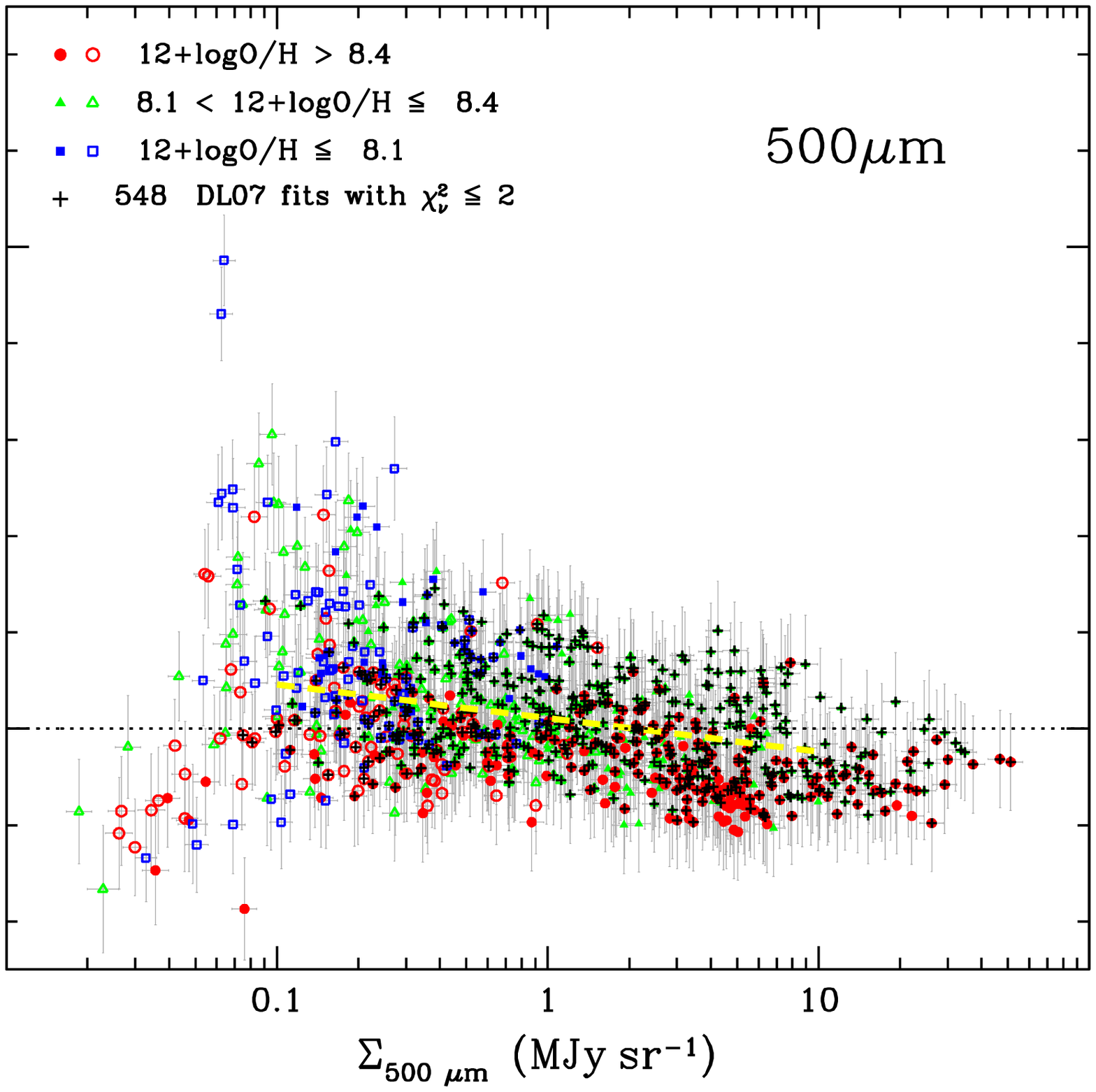}
}
}
\caption{SPIRE fractional residuals of the DL07 best fits plotted against the
500\,\micron\ surface brightness $\Sigma_{500}$. As explained in the text,
fractional residuals are defined
as ($F_\nu$-$F_{\rm DL07}$)/$F_\nu$; 250\,\micron\ residuals are shown in the left panel,
350\,\micron\ in the middle, and 500\,\micron\ in the right.
As in previous figures, data points are distinguished by O/H with (red) circles showing
\logoh$>$8.4, (green) triangles 8.0$<$\logoh$\leq$8.4, and (blue) squares \logoh$\leq$8.1.
Filled symbols correspond to positions with normalized (to optical radius \ropt) radii 
within $R$/\ropt$\leq$0.8, and open symbols to larger radii.
Unlike previous figures, here we show all data with S/N$\geq$3 independently of their
\redchi; the DL07 fits with \redchi$\leq$2 are indicated by $+$.
The (yellow) dashed lines give the linear regressions described in the text.
\label{fig:residuals_surf}
}
\end{figure*}
\begin{figure*}[!t]
\centerline{
\hbox{
\includegraphics[angle=0,width=0.38\linewidth,bb=18 280 592 718]{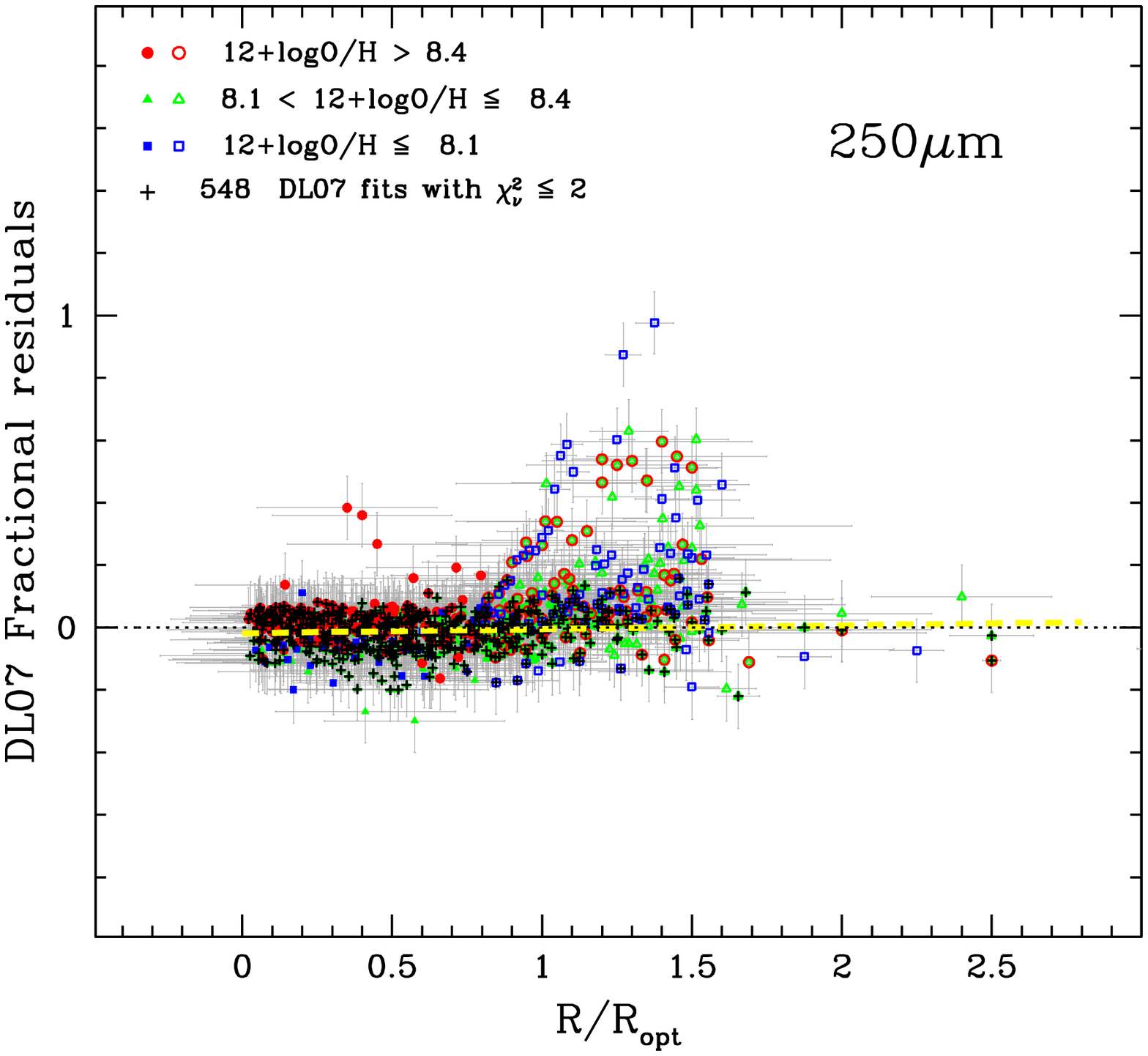}
\hspace{-1.35cm}
\includegraphics[angle=0,width=0.38\linewidth,bb=18 280 592 718]{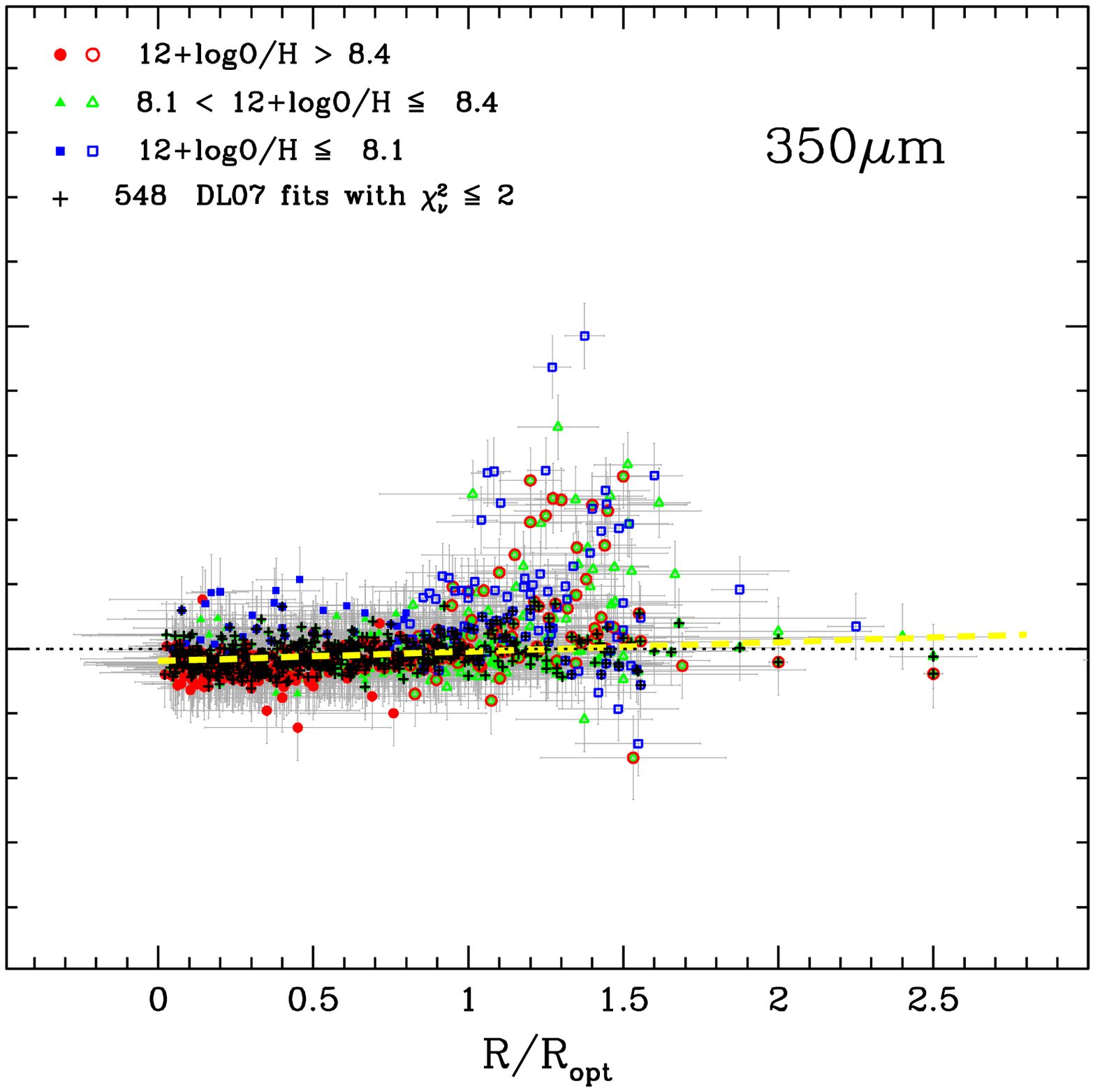}
\hspace{-1.35cm}
\includegraphics[angle=0,width=0.38\linewidth,bb=18 280 592 718]{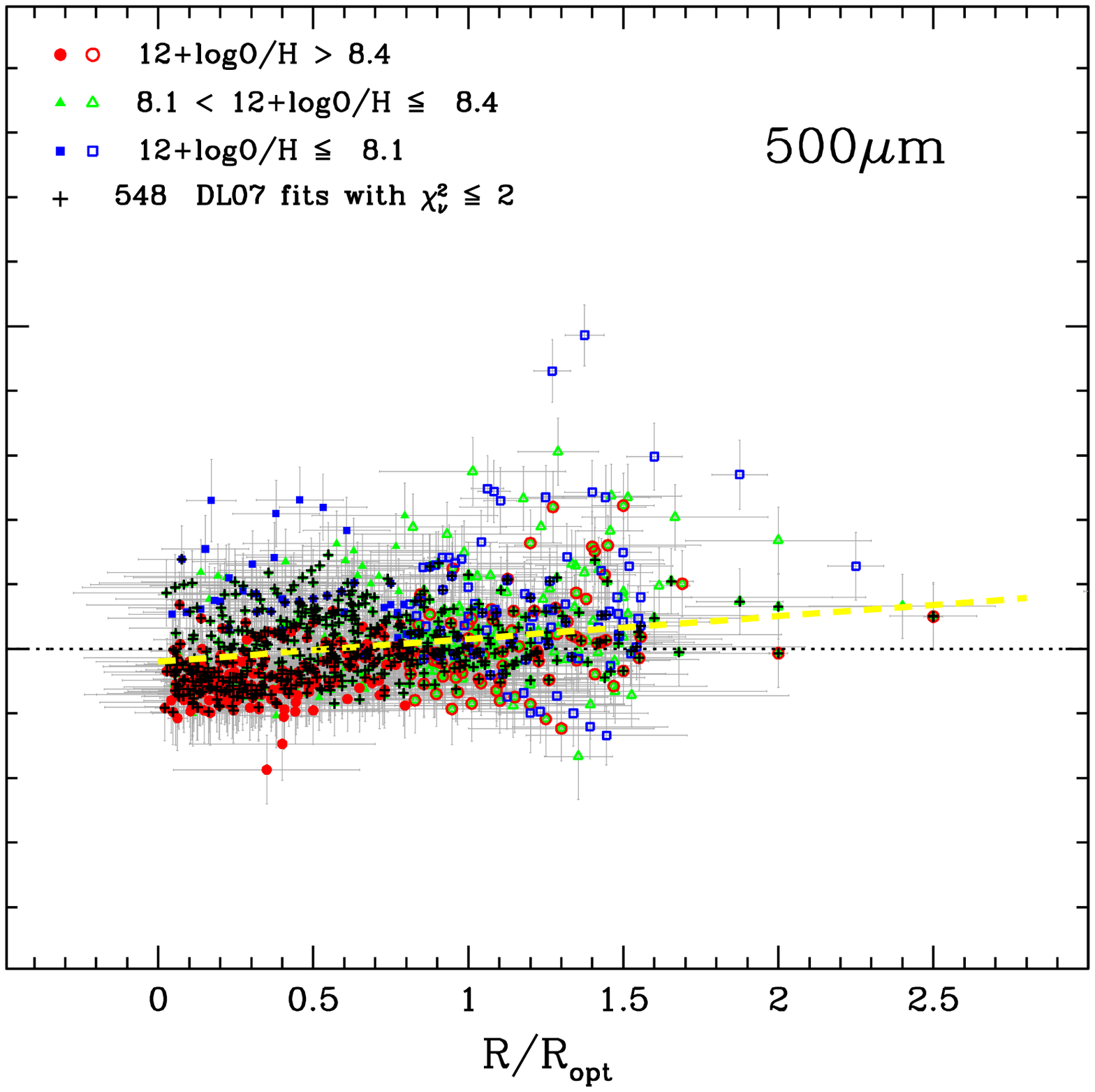}
}
}
\caption{SPIRE fractional residuals of the DL07 best fits plotted against the
normalized radius, $R$/\ropt. As explained in the text,
fractional residuals are defined
as ($F_\nu$-$F_{\rm DL07}$)/$F_\nu$; 250\,\micron\ residuals are shown in the left panel,
350\,\micron\ in the middle, and 500\,\micron\ in the right.
As in Fig. \ref{fig:residuals_surf}, data points are coded by O/H with (red) circles showing
\logoh$>$8.4, (green) triangles 8.0$<$\logoh$\leq$8.4, and (blue) squares \logoh$\leq$8.1.
Filled symbols correspond to positions with normalized (to optical radius \ropt) radii 
within $R$/\ropt$\leq$0.8, and open symbols to larger radii.
As in Fig. \ref{fig:residuals_surf}, we show all data with S/N$\geq$3 independently of their
\redchi; the DL07 fits with \redchi$\leq$2 are indicated by $+$.
The (yellow) dashed lines give the linear regressions described in the text.
\label{fig:residuals_ropt}
}
\end{figure*}

\begin{figure*}[!t]
\centerline{
\hbox{
\includegraphics[angle=0,width=0.38\linewidth,bb=18 280 592 718]{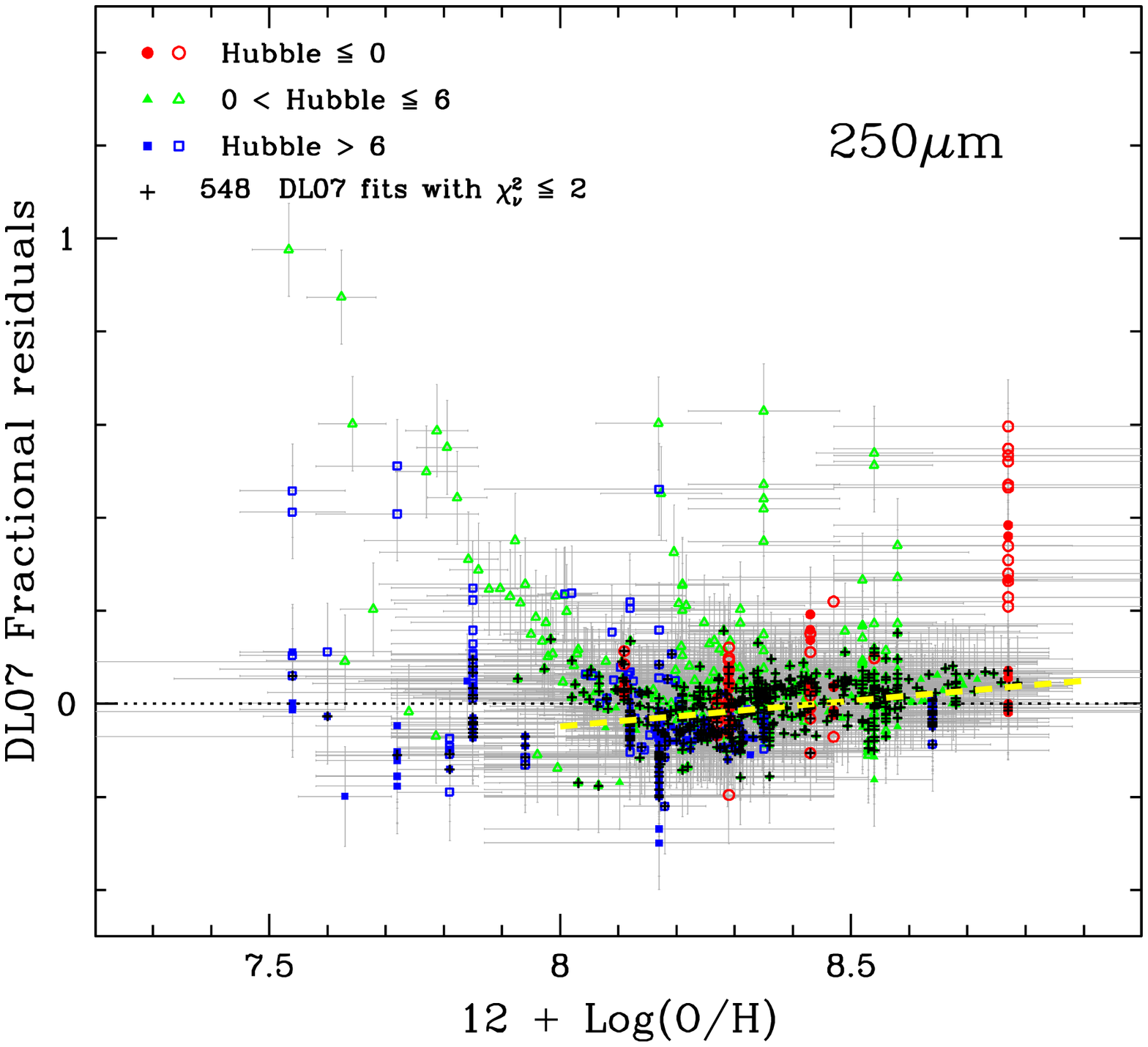}
\hspace{-1.35cm}
\includegraphics[angle=0,width=0.38\linewidth,bb=18 280 592 718]{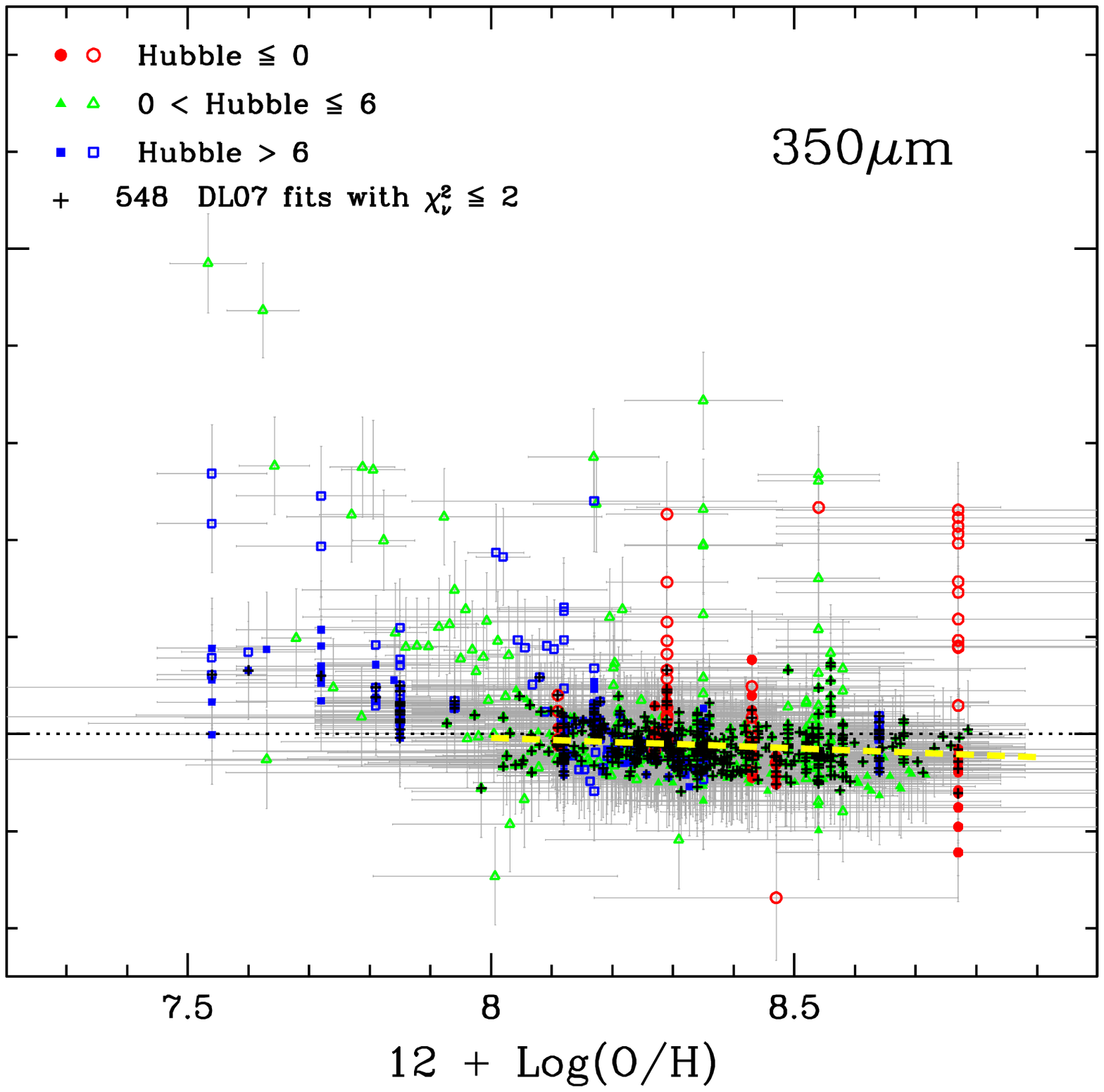}
\hspace{-1.35cm}
\includegraphics[angle=0,width=0.38\linewidth,bb=18 280 592 718]{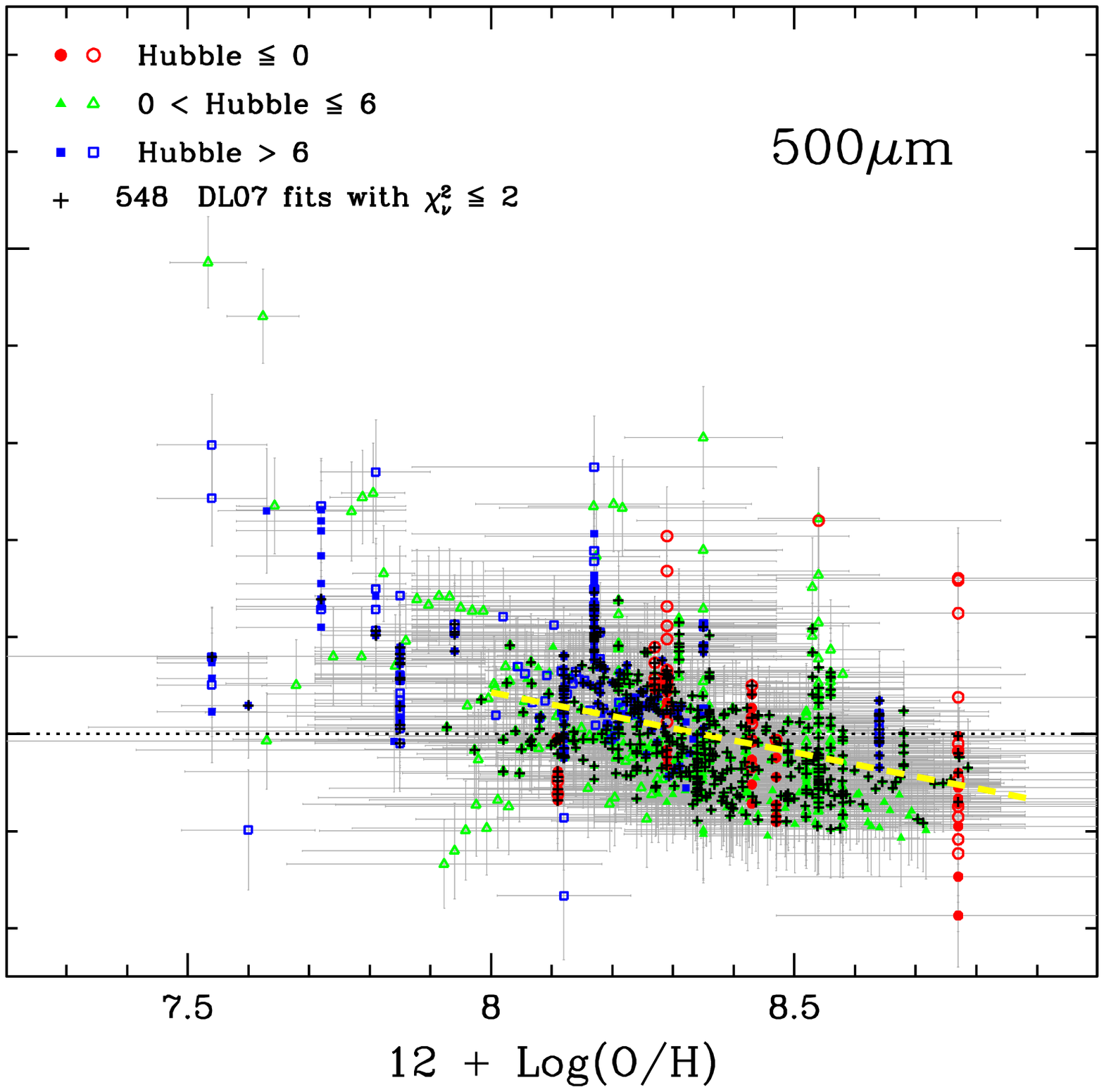}
}
}
\caption{SPIRE fractional residuals of the DL07 best fits plotted against the
oxygen abundance, \logoh. As explained in the text,
fractional residuals are defined
as ($F_\nu$-$F_{\rm DL07}$)/$F_\nu$; 250\,\micron\ residuals are shown in the left panel,
350\,\micron\ in the middle, and 500\,\micron\ in the right.
Data points are coded by Hubble type with (red) circles corresponding to
early types (T$\leq$0), (green) triangles to spirals (0$<$T$\leq$6), and
(blue) squares to late types (T$>$6).
Filled symbols correspond to positions with normalized (to optical radius \ropt) radii 
within $R$/\ropt$\leq$0.8, and open symbols to larger radii.
As in Fig. \ref{fig:residuals_surf} all points with S/N$\geq$3 are plotted,
with the DL07 fits with \redchi$\leq$2 shown by $+$.
The (yellow) dashed lines give the linear regressions described in the text.
\label{fig:residuals_oh}
}
\end{figure*}

As briefly discussed in Sect. \ref{sec:colors}, the DL07 models roughly
reproduce the SPIRE colors to within the uncertainties, but there may be
systematic variations.
Because of the importance of SPIRE wavelengths for understanding cold dust emission,
and possibly constraining physical properties,
in this Section we explore the degree to which the SPIRE fluxes are well fit
by the DL07 models.
For each radial point, we have defined the SPIRE residual as ($F_\nu$-$F_{\rm DL07}$)/$F_\nu$.
Figure \ref{fig:residuals_surf} shows these residuals at 250\,\micron, 350\,\micron,
and 500\,\micron\ plotted versus 500\,\micron\ surface brightness $\Sigma_{500}$.
As in previous figures, the points are coded by O/H with filled symbols corresponding
to locations with $R$/\ropt$\leq$0.8 and open ones to larger radii;
moreover, only points with S/N $\geq$ 3 are plotted.
%None of the fits of the radial SEDs included the 500\,\micron\
%point implying that, by construction, the 250\,\micron\ and 350\,\micron\
%data should be closer to the models than the data at 500\,\micron. 
Unlike previous figures, in Fig. \ref{fig:residuals_surf} (and Fig. \ref{fig:residuals_oh})
we have shown all the data with S/N$\geq$3; DL07 fits with \redchi$\leq$2 are 
highlighted with $+$ signs.

Even for the DL07 fits with \redchi$\leq$2,
there are systematic variations of the residuals with $\Sigma_{500}$ as shown
in Fig. \ref{fig:residuals_surf}. 
Over the range in $\Sigma_{500}$ shown in Fig. \ref{fig:residuals_surf} 
(0.1$\leq\Sigma_{500}\leq$10\,\mjysr),
the SPIRE residuals are well correlated with $\Sigma_{500}$ at 350\,\micron\ and 500\,\micron.
However, at 250\,\micron, except for very low $\Sigma_{500}$ ($\la$0.1\,\mjysr),
the DL07 models well approximate the data;
the best-fit slope for the residuals is 0.0 with an intercept of $-0.014$.
At 350\,\micron\ and 500\,\micron, where the slopes are non-zero,
the significance of the trend with $\Sigma_{500}$ is $>$99.999\% for DL07 fits with \redchi$\leq$2.
However, the excursions are minor:
at 350\,\micron, the mean residual is $\sim$3\% positive at $\Sigma_{500}$\,=\,0.1\,\mjysr;
at 500\,\micron, the mean residual is $\sim$9\% at $\Sigma_{500}$\,=\,0.1\,\mjysr\ 
and $\sim$6\% at  $\Sigma_{500}$\,=\,0.3\,\mjysr.
Nevertheless,
Fig. \ref{fig:residuals_surf} shows that at low 500\,\micron\ surface brightness,
the DL07 fits tend to have \redchi$>$2 at low $\Sigma_{500}$;
there is increased scatter which may be related to metallicity (see below).

Similar trends are seen in Fig. \ref{fig:residuals_ropt} where we have plotted
residuals against normalized optical radius, $R$/\ropt.
The DL07 models follow the data well at 250\,\micron\ with
a regression slope of 0.0.
However, as before, at 350\,\micron\ and 500\,\micron, the trends are highly significant
($>$99.999\% for DL07 fits with \redchi$\leq$2) but with minimal excursions:
at \ropt\,=\,1.5, $\la$1\% and $\sim$7\% at 350\,\micron\ and 500\,\micron,
respectively.
The trends with $R$/\ropt\ are probably reflecting those with surface brightness 
$\Sigma_{500}$ because of the tendency for low surface brightnesses to occur in 
the outer radii of galaxies.

Figure \ref{fig:residuals_oh} gives the same DL07 residuals as in Figs. 
\ref{fig:residuals_surf} and \ref{fig:residuals_ropt}
but plotted against oxygen abundance, \logoh.
As before, there are systematic trends revealed by highly significant ($>$99.999\%)
correlations between SPIRE residuals and O/H.
For \logoh\,=\,8, the mean residual is $\sim$5\% negative at 250\,\micron,
and $\sim$9\% positive at 500\,\micron.
Interestingly, large 350\,\micron\ residuals 
are spread over a large range in O/H, unlike 
those with respect to $\Sigma_{500}$ which are large only for $\Sigma_{500}\la$0.1\,\mjysr.
Moreover, there is no linear trend of the 350\,\micron\ residuals relative to O/H, unlike
for those with respect to $\Sigma_{500}$.
This could be an indication that the DL07 best fits are trying to split the
differences for O/H among the SPIRE bands;
thus there naturally would be a deficit at 250\,\micron, neutrality at 350\,\micron,
and a positive excess at 500\,\micron.

If the DL07 dust models were adjusted by these minute amounts at \logoh\,=\,8
(5\% smaller at 250\,\micron,
1\% smaller at 350\,\micron, $\sim$9\% larger at 500\,\micron), 
the SPIRE colors shown in 
Fig. \ref{fig:spirecolors} would be shifted $\sim$4\% down and $\sim$10\% to the left,
toward the apparent $\beta\sim$1 curve, which would be roughly appropriate for low metallicity.
Correcting for $\Sigma_{500}$, the shifts would be of similar amplitude, again consistent
with $\beta\sim$1.
If instead, we consider the metal-rich and high $\Sigma_{500}$ adjustments, the trends
are of similar amplitude but opposite sign, thus moving the colors $\sim$5\% up and $\sim$3\% to the right,
toward the apparent $\beta\sim$2 curve.
Adjusting the DL07 models by the small corrections suggested by the mean DL07 SPIRE residuals would 
bring the models to better agreement with the data.

\subsection{500\,\micron\ excess}
\label{sec:excess}

Because the DL07 models are constrained at SPIRE wavelengths only indirectly through grain
properties, the agreement between the models and the data is quite good.
Nevertheless, at low metallicities and low surface brightnesses,
the residuals at 500\,\micron\ are slightly positive ($\la$9\%) at a high significance level.
If we consider the DL07 fits with \redchi$>$2, the excesses at all SPIRE wavelengths are even
larger, although it is difficult to define systematic trends.
Other work has also found evidence for a 500\,\micron\ excess in
low-metallicity systems such as the LMC and dwarf galaxies
\citep[e.g.,][]{gordon10,galliano11,galametz11}. 
\citet{ciesla14} find a similar trend in the \hers\ Reference Survey 
with the DL07 models underestimating the 500\,\micron\ flux for low-mass systems
(which would be related to low metallicity).
We conclude that there is evidence in the KINGFISH profiles for a very weak submm excess,
$\la$10\%, at low surface brightnesses and at low metallicity \citep[c.f.,][]{kirkpatrick13}.
Because of metallicity gradients, and the resulting interdependence of metallicity and surface 
brightness in spiral disks, a partial correlation analysis would be necessary to establish 
whether the excess results primarily from low metallicity or from low surface brightness.
Longer wavelength data are needed \citep[e.g.,][]{galametz14}
to establish the existence of a systematic submm excess and the degree to which it 
depends on the grain properties of the models.

%---------------------------------------------------------------
\section{Summary and conclusions}
\label{sec:conclusions}

We have analyzed the entire collection of radial surface brightness profiles for 61 KINGFISH
galaxies both in terms of radial trends and SED properties.
By fitting the radial profiles with exponentials, we find that
the 250\,\micron\ scalelength is on average comparable to that of the
stars, as measured by the 3.6\,\micron\ scalelength.
In the KINGFISH galaxies, except for isolated cases, the dust tends to be
distributed in the same way as the stars.

We have also fitted the SEDs of each annular region with single-temperature MBB and DL07 models.
To better understand the relation between physical parameters of dust emission
and the apparent \td\ and emissivity index $\beta$,
the best-fit DL07 models themselves have also been fit with single-temperature MBB models. 
The analysis of the radial trends of these parameters shows that
dust temperature \td, dust optical depth \taud, and \umin\ all tend to 
decrease with radius.
The PDR fraction, \fpdr, shows a slight increase at large radii, perhaps
indicating the presence of extended UV disks in some galaxies that could be responsible
for PDR-like emission.

The analysis of the MBBV fits to the DL07 models shows that the models are well able
to reproduce flat spectral slopes with $\beta\la 1$.
Our methodology for the MBB fitting to some extent mitigates the usual correlation or degeneracy
between \td\ and $\beta$, and through an analysis of \taud\ and temperature binning,
we find that shallow slopes ($\beta\la 1$) in the data are
associated with cool \td\ in a similar way as the DL07 models.
Our results also show that the minimum ISRF intensity, \umin, responsible for heating
the bulk of the dust in most galaxies is closely related to the apparent emissivity
index $\beta$, with lower \umin\ associated with flatter $\beta$.
Hence, we conclude that temperature mixing is a major cause of trends of $\beta$
commonly seen in MBB fitting of IR SEDs of galaxies.
Temperature mixing may arise from the distribution of ISRF intensities responsible
for heating the dust, or from the different properties of the grain populations, or
both.
It is therefore difficult to ascribe variations in $\beta$ to real physical
properties of dust grains.

Finally, 
we assess the ability of the DL07 models to fit the observed SPIRE fluxes,
and find generally very good agreement.
However, there is some evidence for a small 500\,\micron\ excess, $\sim$10\%, for
regions of low dust surface brightness and low metallicity.

The detailed study of dust emission and grain properties in galaxies is only
in its infancy.
More work is needed at high spatial resolution and long wavelengths to establish 
whether or not current dust models are able to
accommodate the observations in physical regimes that are more extreme than
those usually encountered in the disks of spiral galaxies. 

%%%%%%%%%%%%%%%%%%%%%%%% acknowledgments
\begin{acknowledgements}
We dedicate this paper to the memory of Charles~W. Engelbracht, 
whose excellence and commitment as a scientist were fundamental for this work, and
without whom the SPIRE images analyzed here would not have existed. 
We thank the anonymous referee for concise comments which
improved the clarity of the paper.
SB and LKH acknowledge support from PRIN-INAF 2012/13.
BTD was supported in part by NSF grant AST-1408723.
Use was made of the NASA/IPAC Extragalactic Database (NED).
\end{acknowledgements}
%%%%%%%%%%%%%%%%%%%%%%%%%%%%%%%%%%%%%

\appendix

\section{The radial surface brightness profiles, continued}
 \label{app:profiles}
 
\setcounter{figure}{0}

\begin{figure*}[!ht]
\centerline{
\hbox{
\includegraphics[width=0.45\linewidth,bb=18 308 588 716]{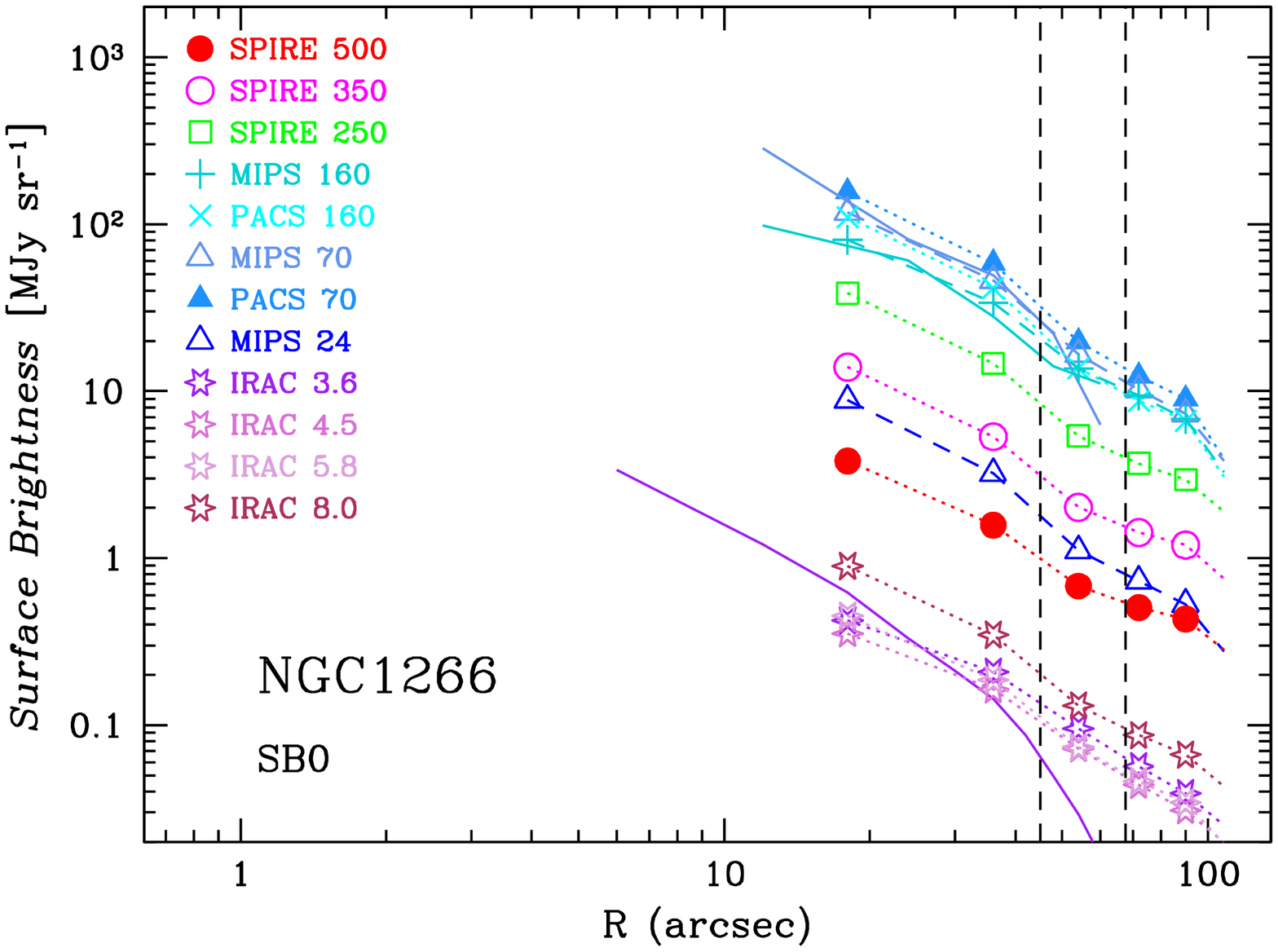}
%\hspace{0.1\linewidth}
\hspace{0.05\linewidth}
\includegraphics[width=0.45\linewidth,bb=18 167 592 718]{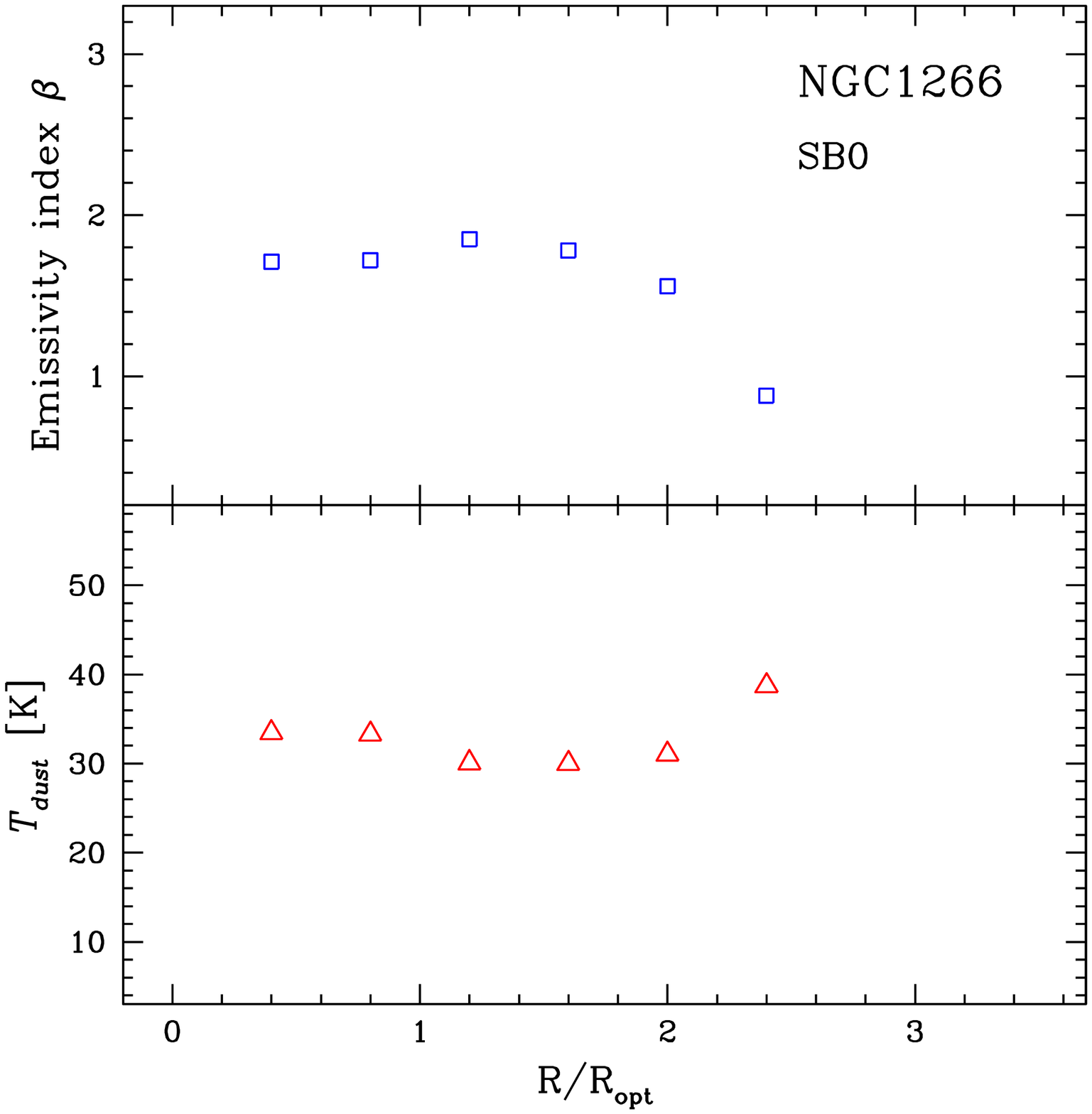}
}
}
\vspace{2.15\baselineskip}
\vspace{-1.15\baselineskip}
\centerline{
\hbox{
\includegraphics[width=0.22\linewidth,bb=18 308 588 716]{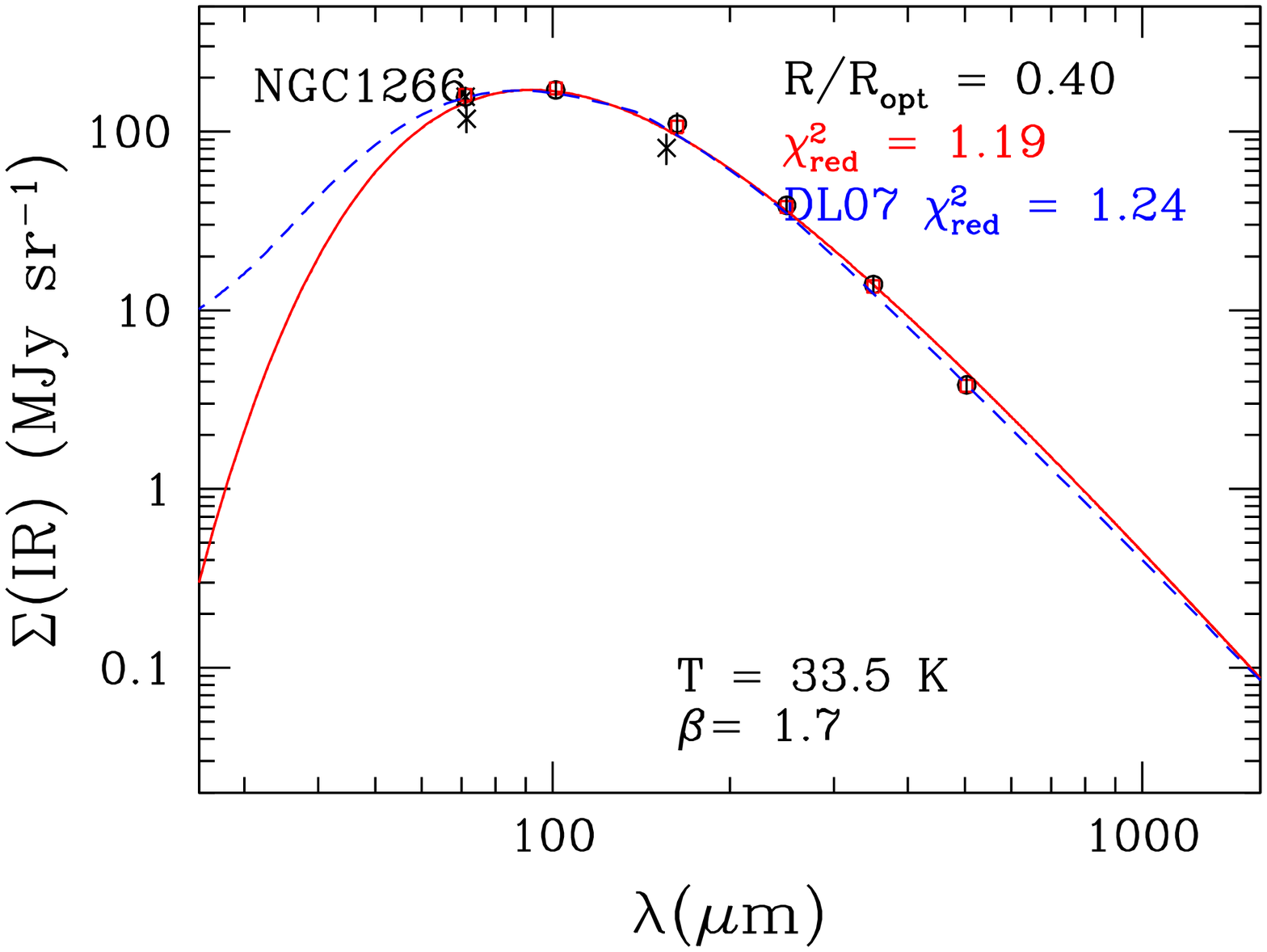}
\hspace{-0.045\linewidth}
\includegraphics[width=0.22\linewidth,bb=18 308 588 716]{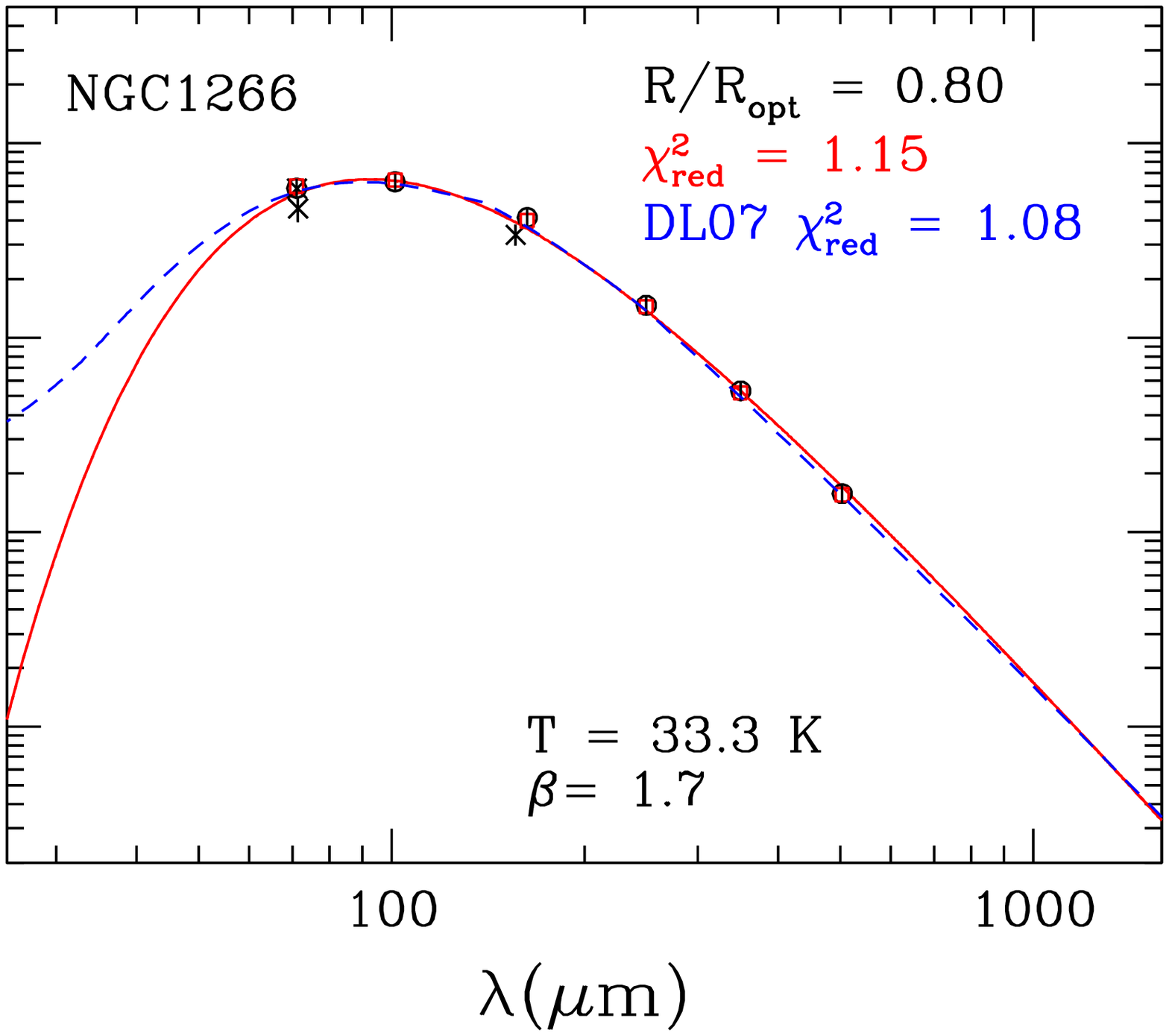}
\hspace{-0.045\linewidth}
\includegraphics[width=0.22\linewidth,bb=18 308 588 716]{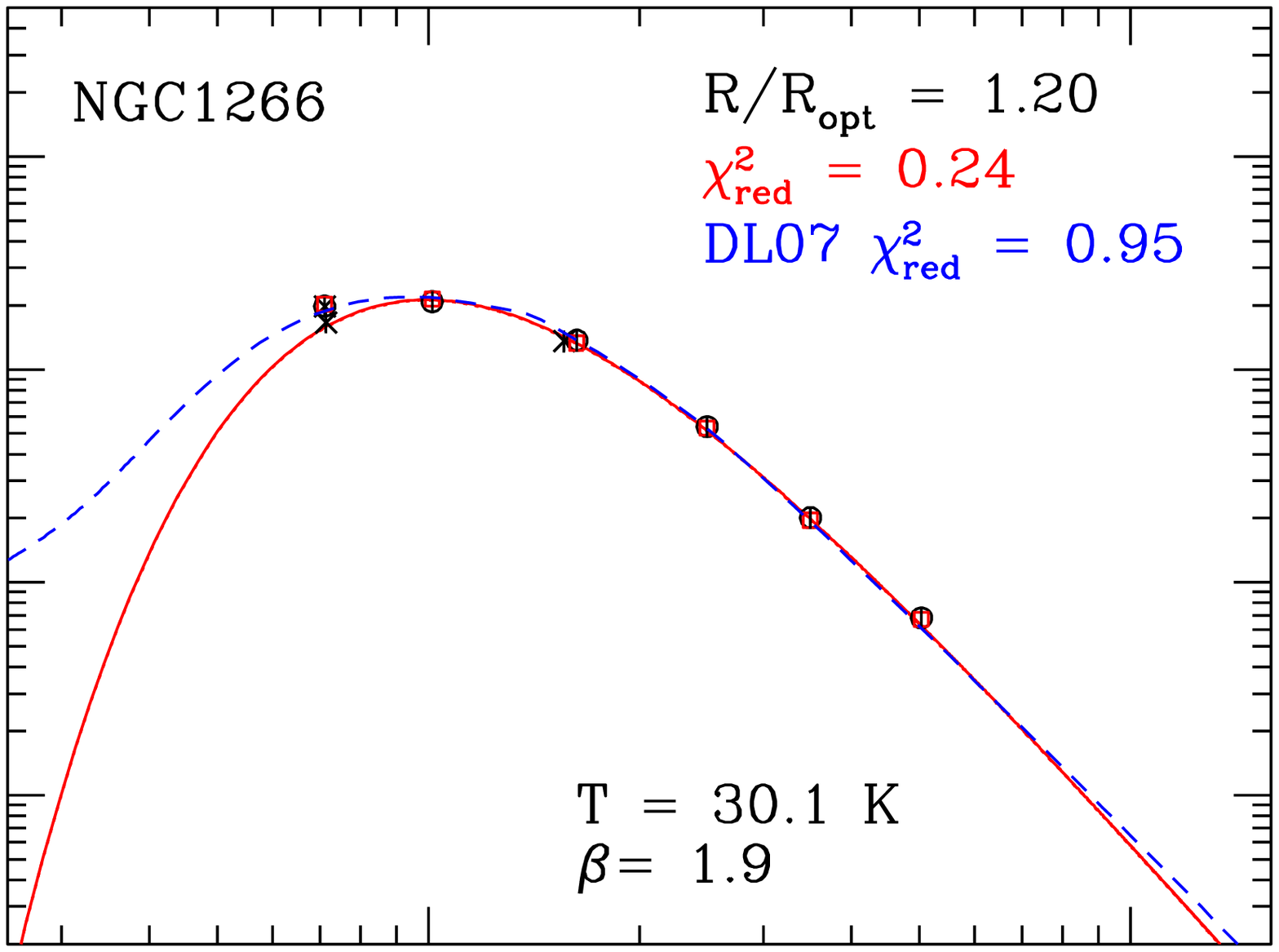}
\hspace{-0.045\linewidth}
\includegraphics[width=0.22\linewidth,bb=18 308 588 716]{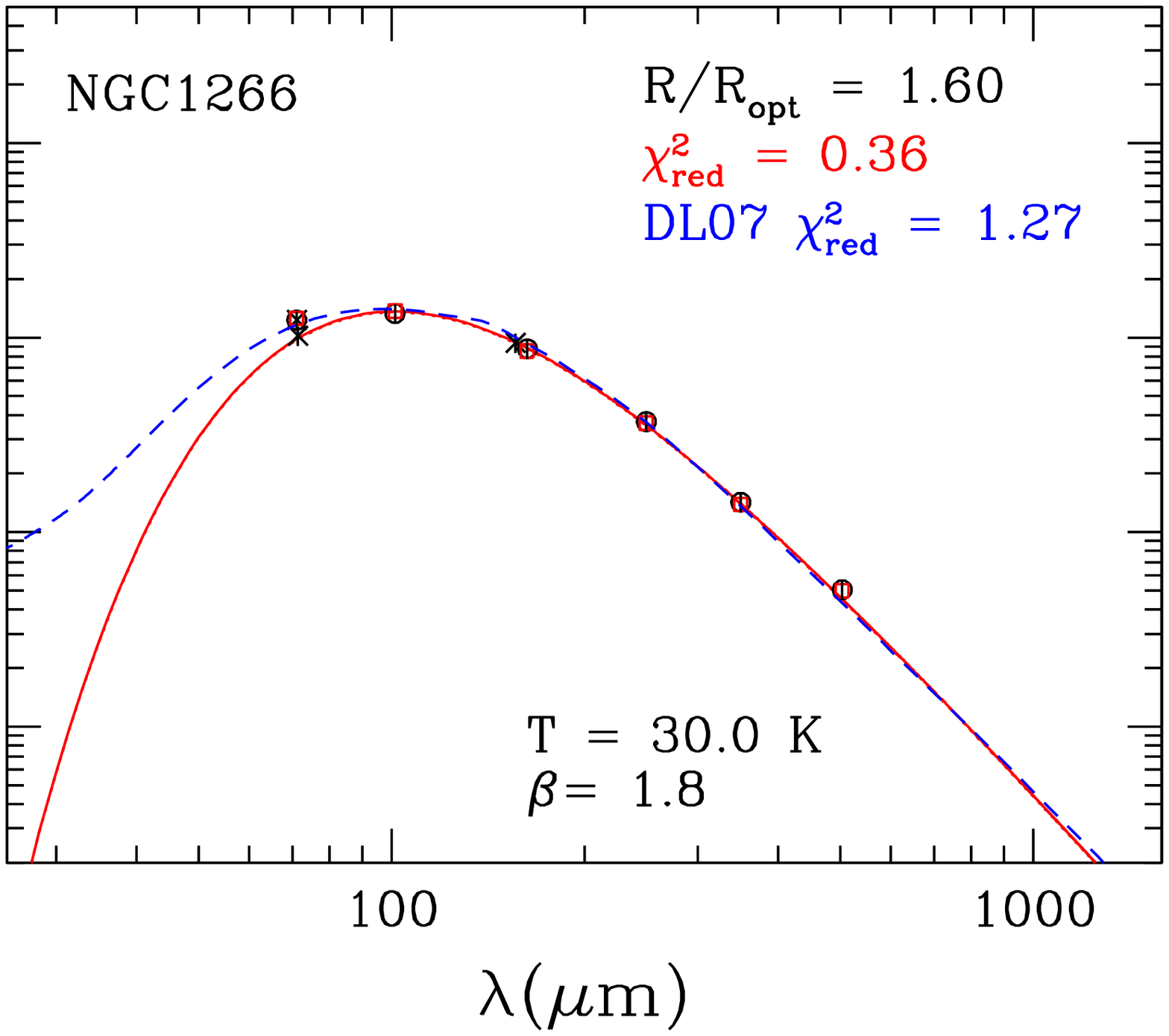}
\hspace{-0.045\linewidth}
\includegraphics[width=0.22\linewidth,bb=18 308 588 716]{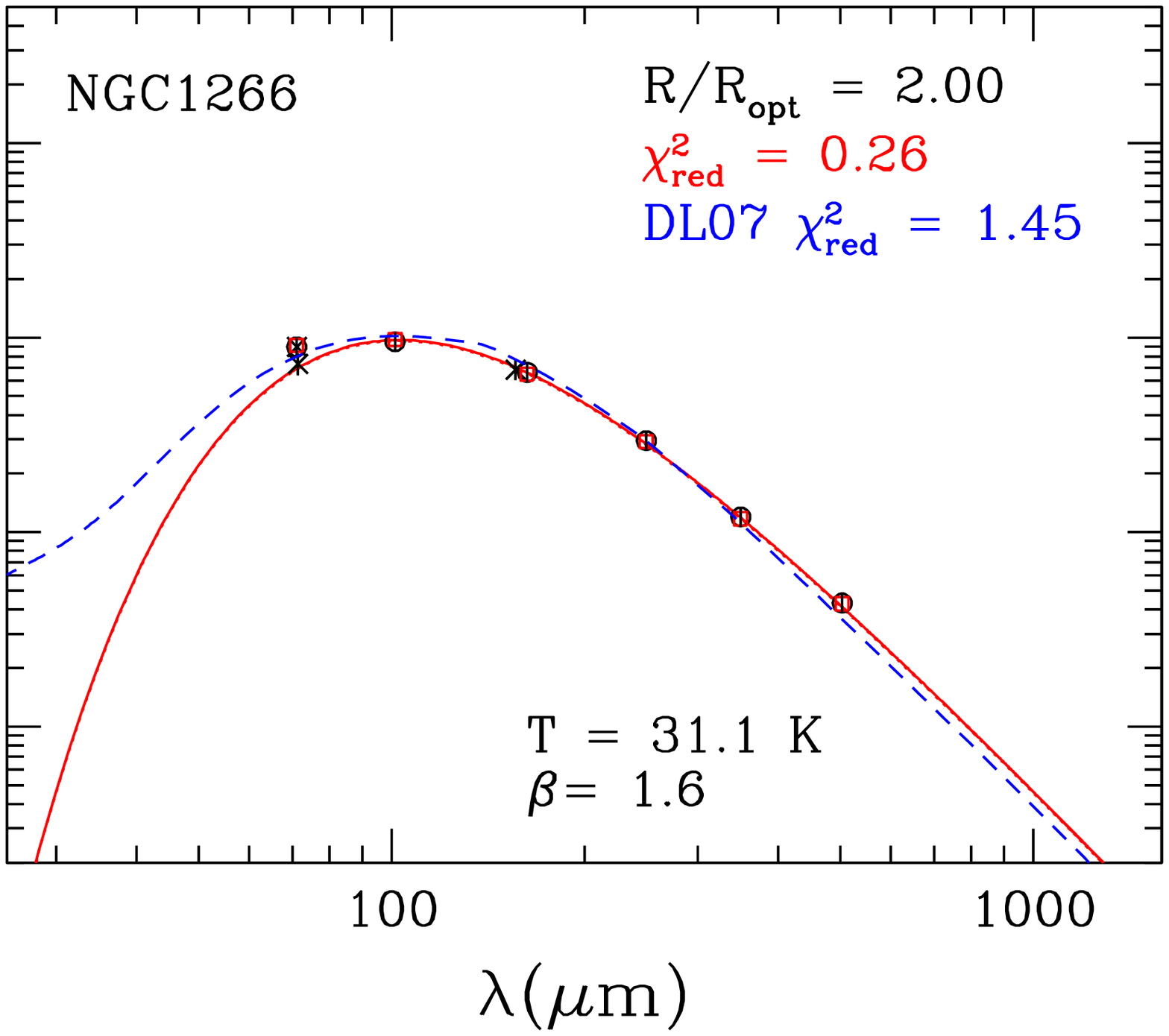}
}
}
\vspace{-1.15\baselineskip}
\centerline{
\hbox{
\includegraphics[width=0.22\linewidth,bb=18 308 588 716]{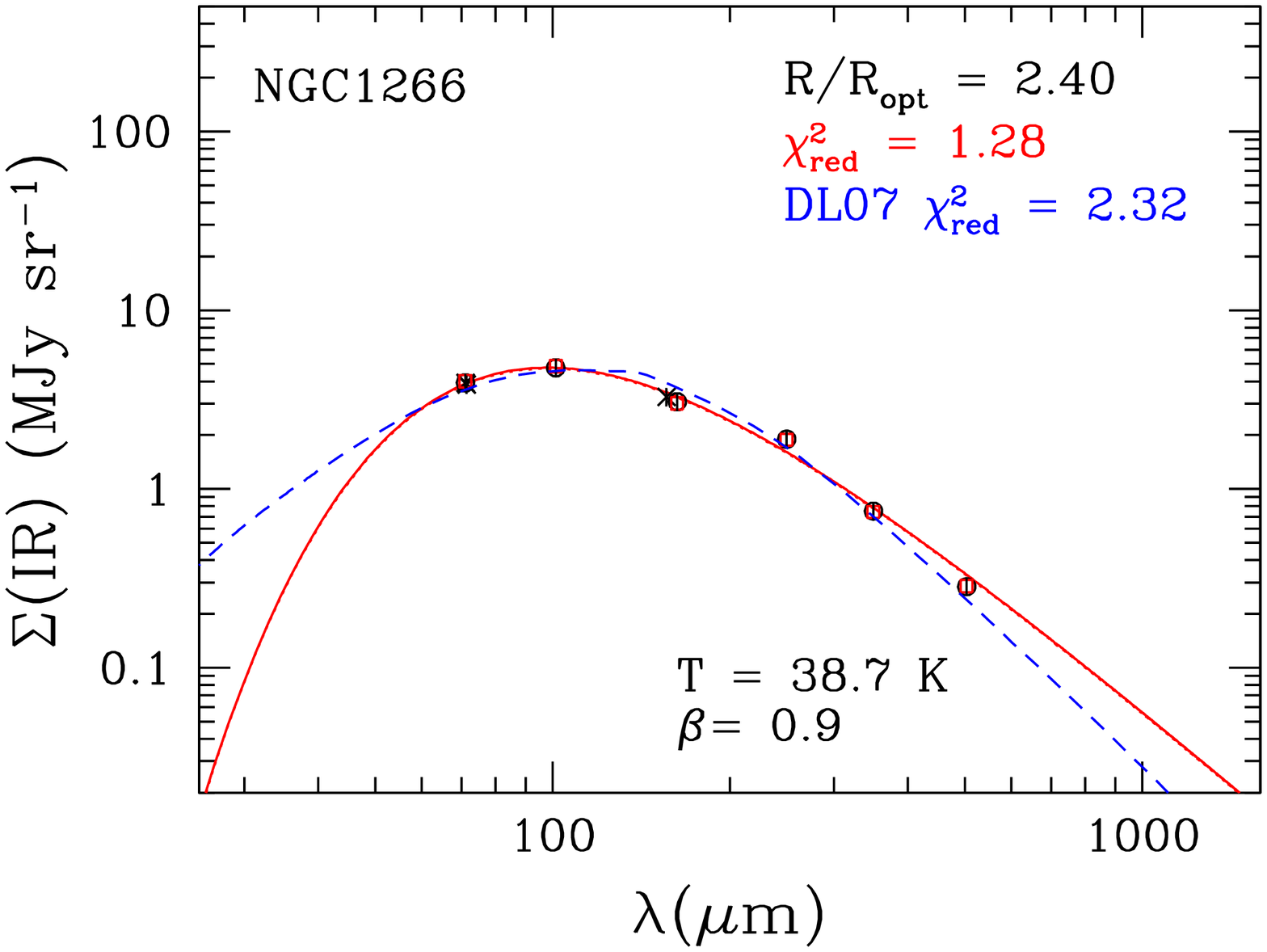}
}
}
\caption{{\bf (b)} as Fig. \ref{fig:radial}(a), but NGC\,1266 (SB0) radial profiles. 
}
\end{figure*}
\setcounter{figure}{0}
\begin{figure*}[!ht]
\centerline{
\hbox{
\includegraphics[width=0.45\linewidth,bb=18 308 588 716]{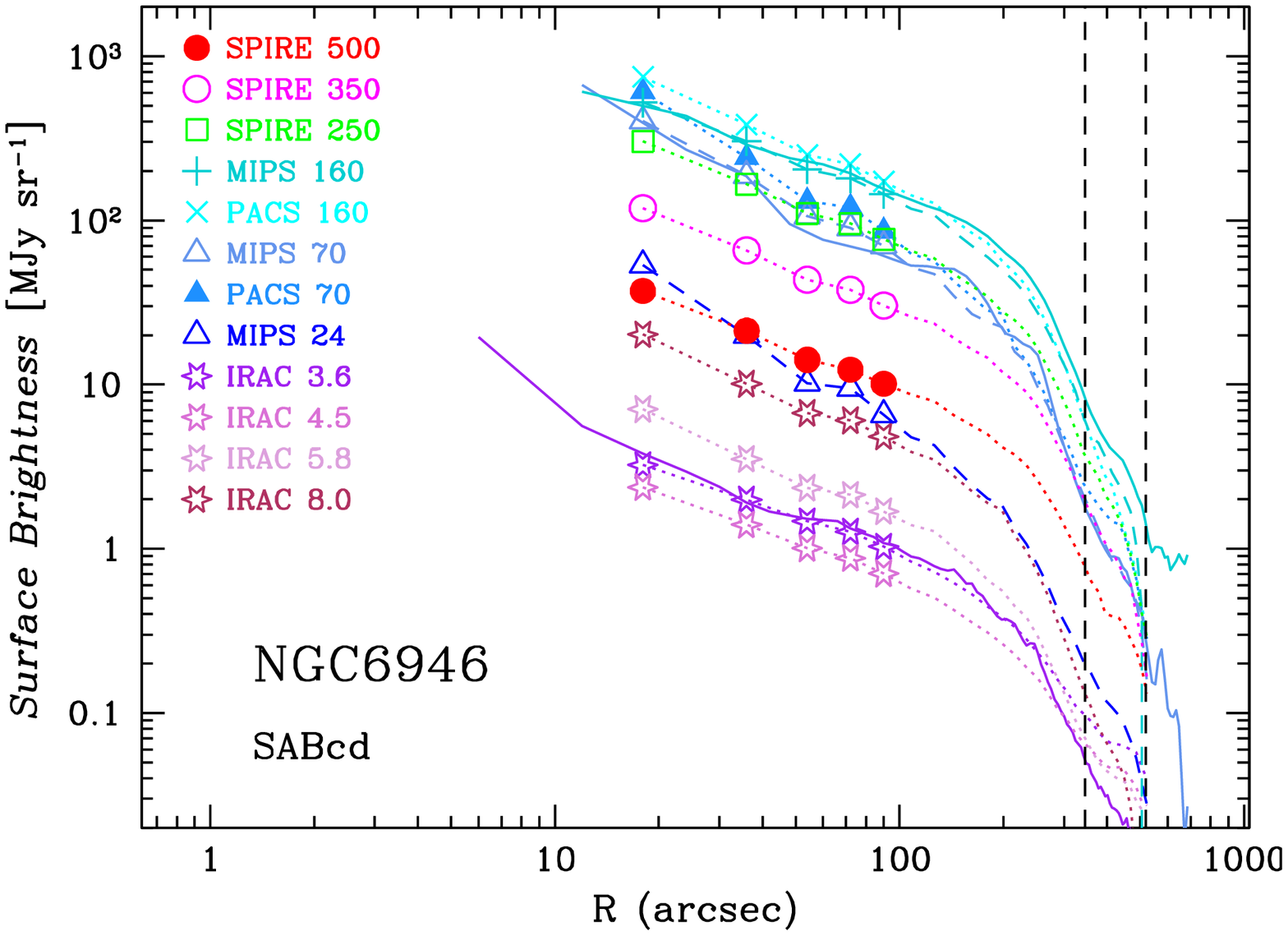}
%\hspace{0.1\linewidth}
\hspace{0.05\linewidth}
\includegraphics[width=0.45\linewidth,bb=18 167 592 718]{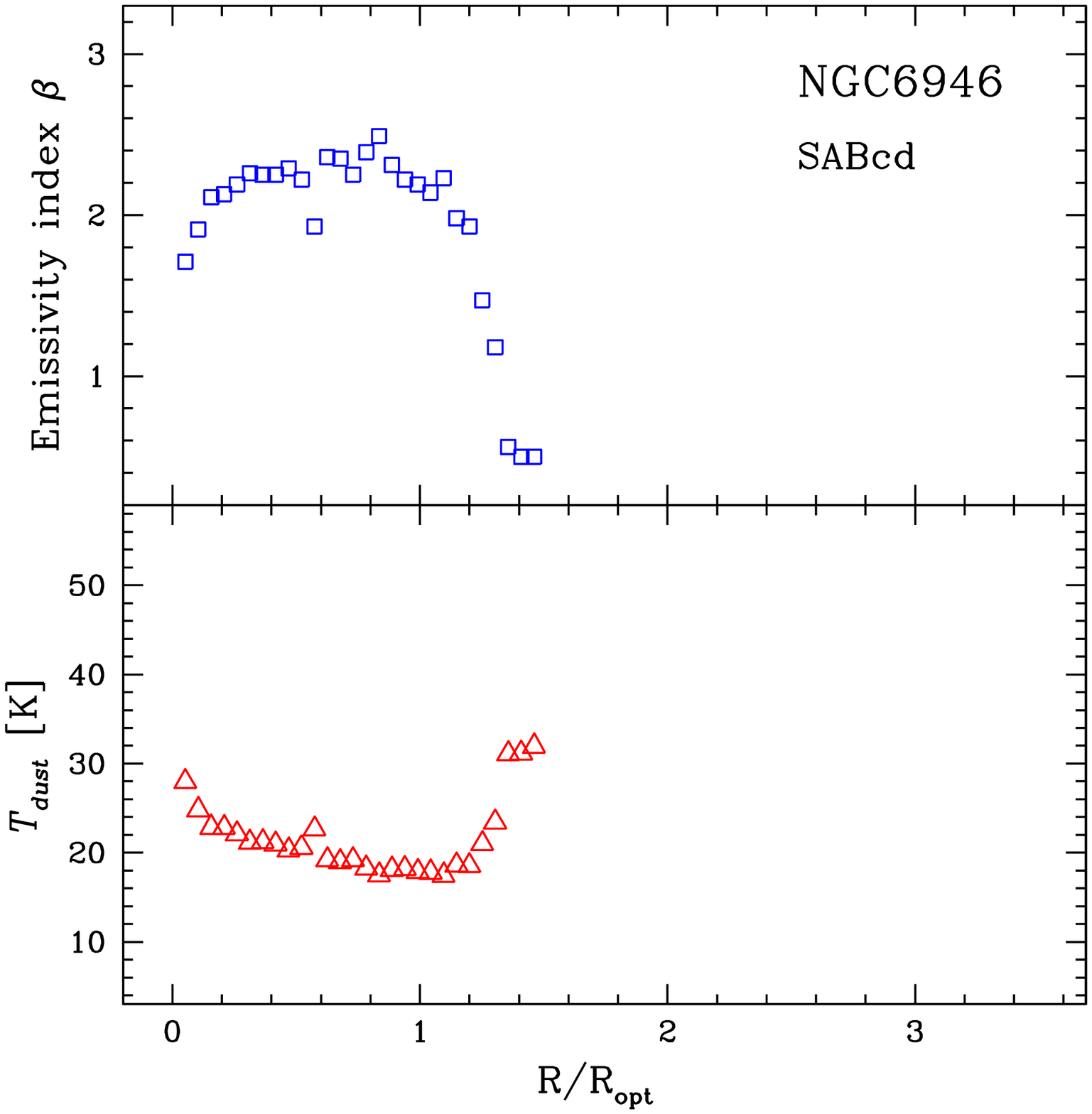}
}
}
\vspace{2.15\baselineskip}
\vspace{-1.15\baselineskip}
\centerline{
\hbox{
\includegraphics[width=0.22\linewidth,bb=18 308 588 716]{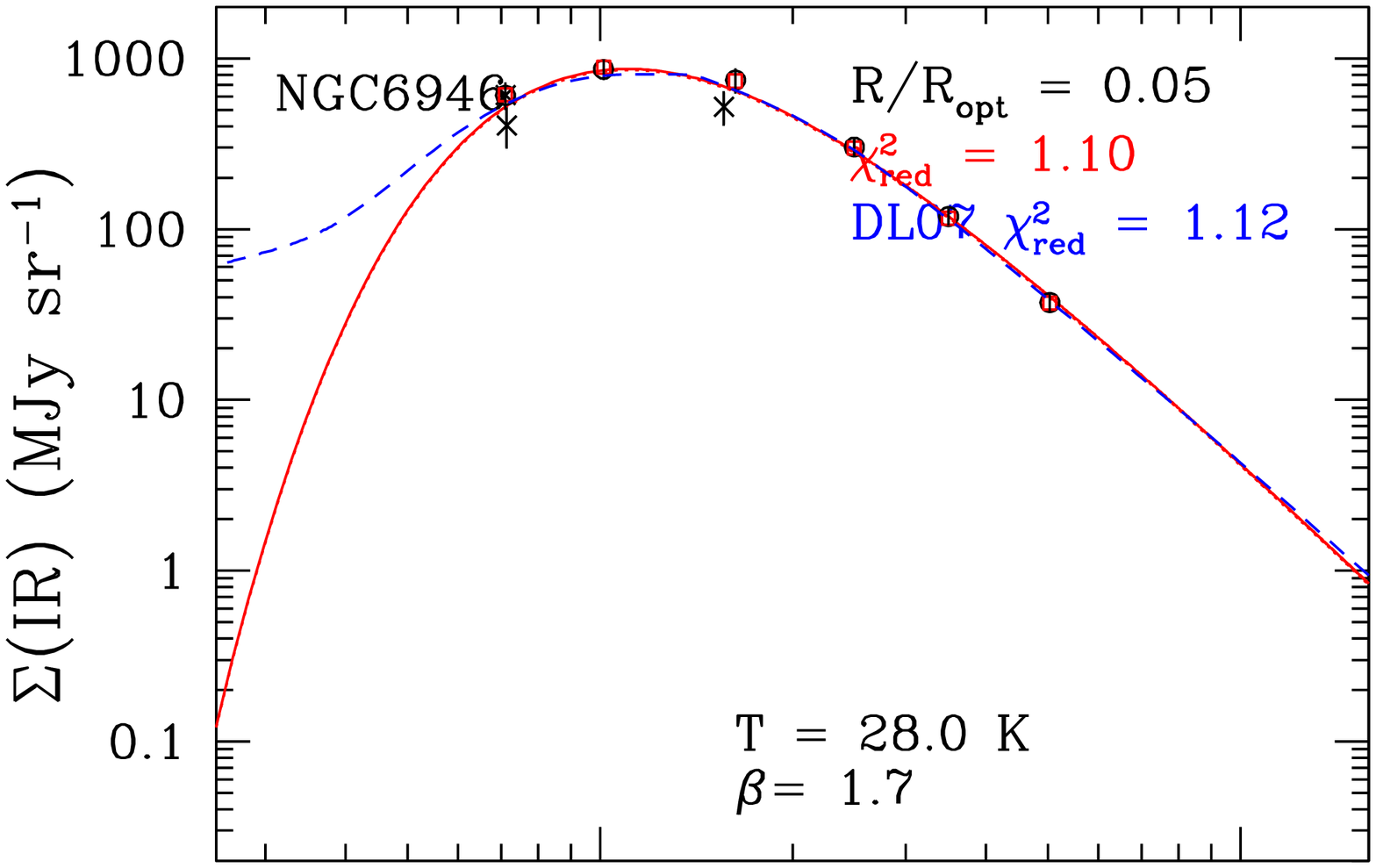}
\hspace{-0.045\linewidth}
\includegraphics[width=0.22\linewidth,bb=18 308 588 716]{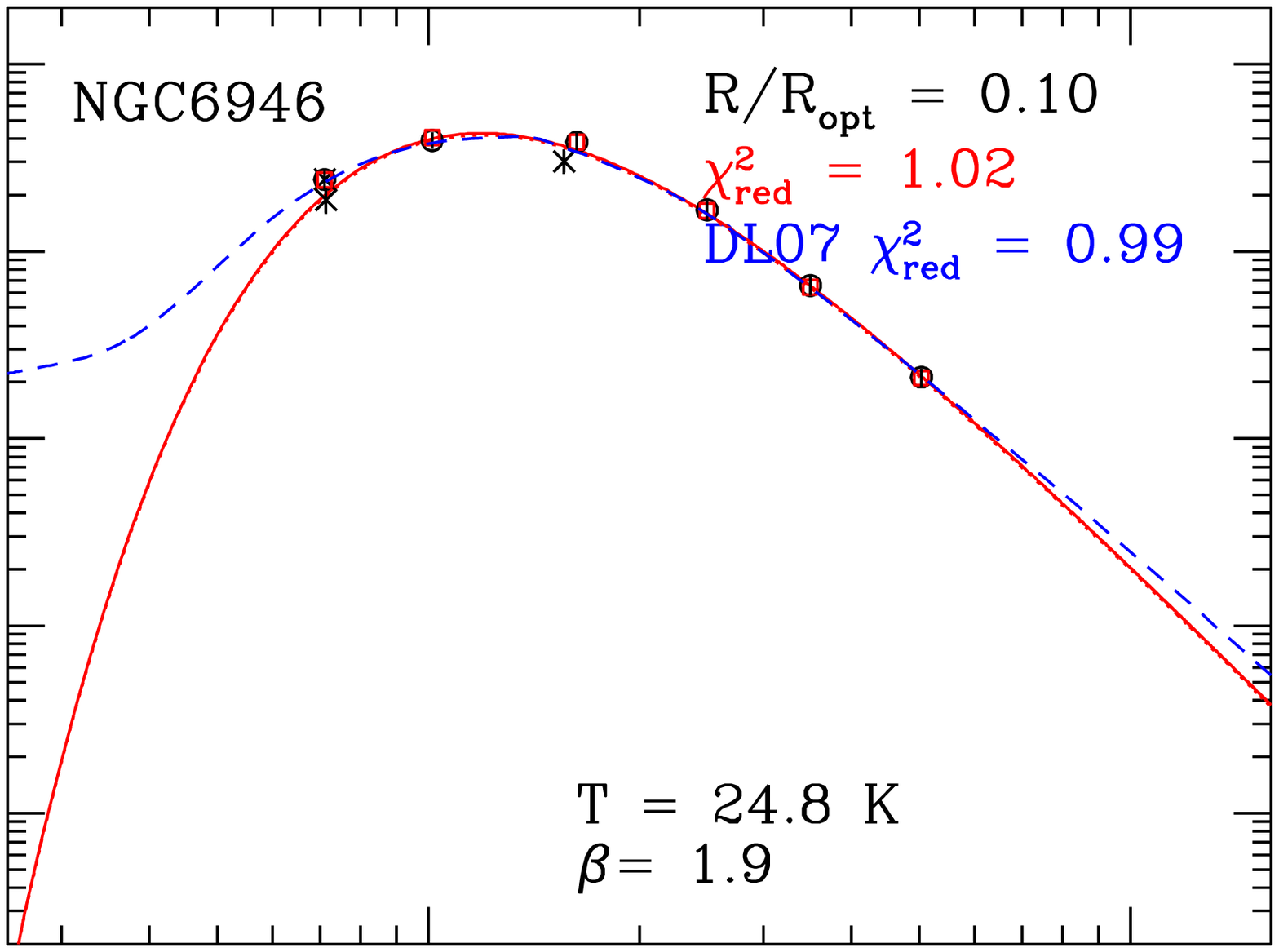}
\hspace{-0.045\linewidth}
\includegraphics[width=0.22\linewidth,bb=18 308 588 716]{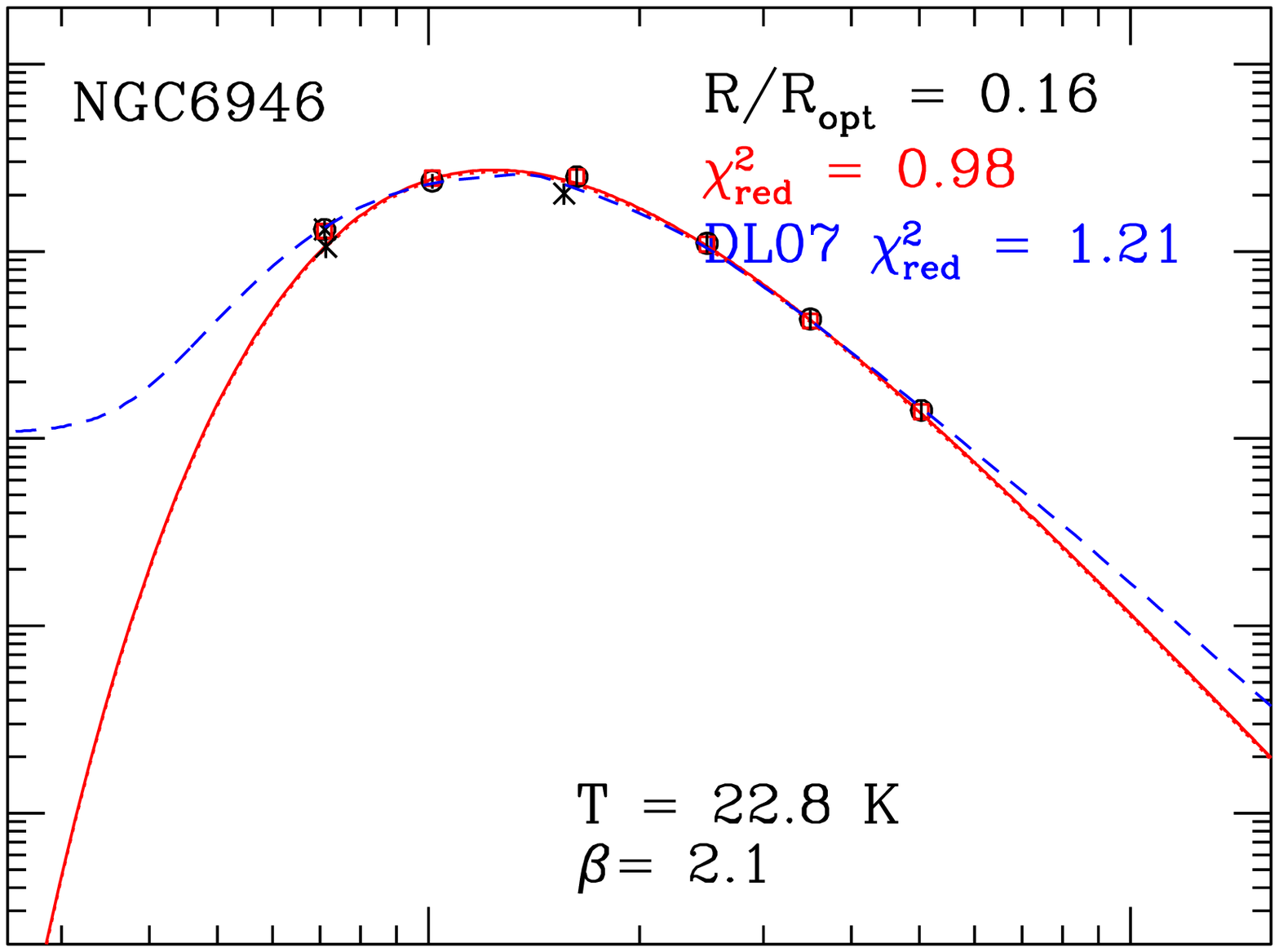}
\hspace{-0.045\linewidth}
\includegraphics[width=0.22\linewidth,bb=18 308 588 716]{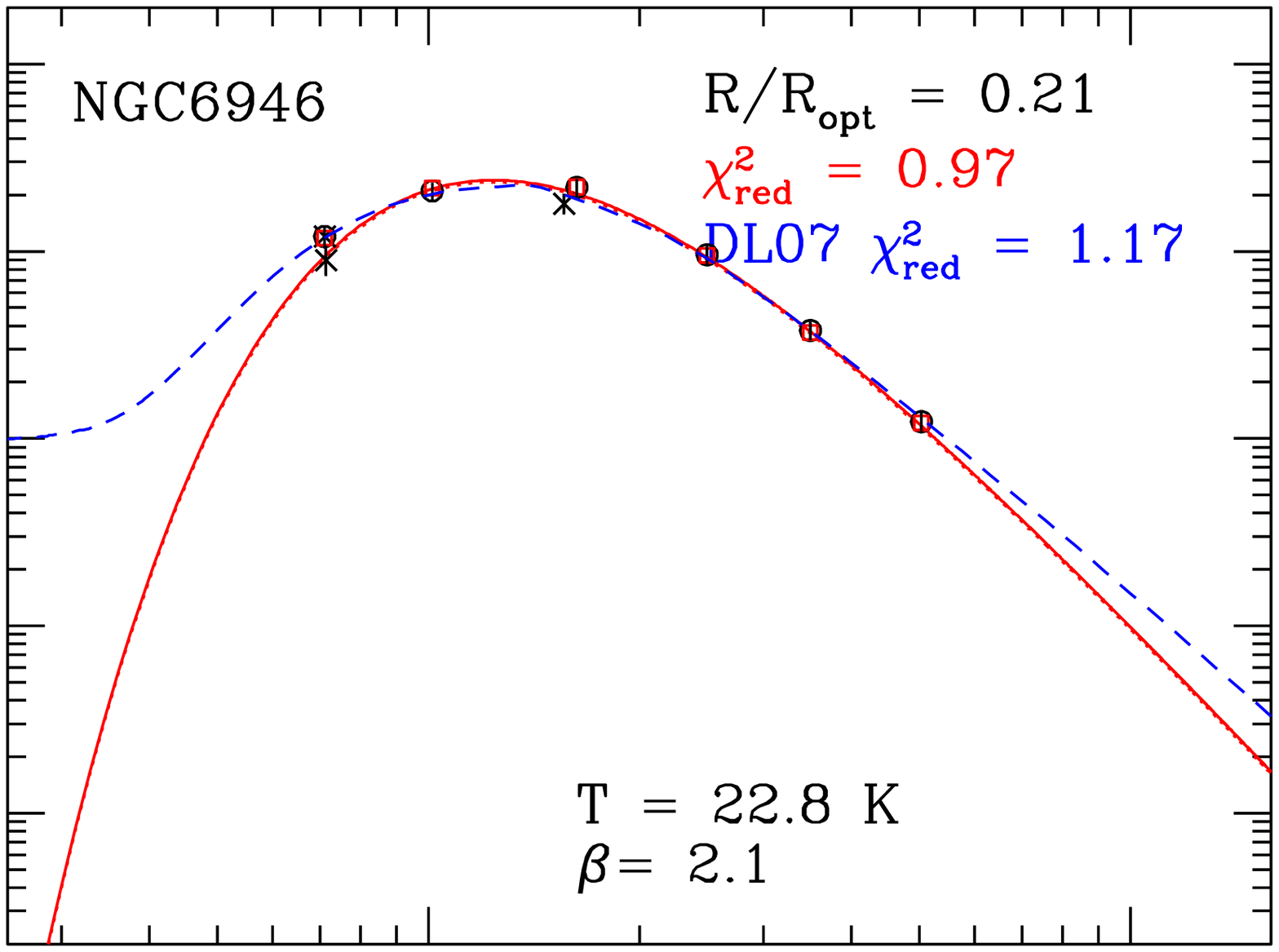}
\hspace{-0.045\linewidth}
\includegraphics[width=0.22\linewidth,bb=18 308 588 716]{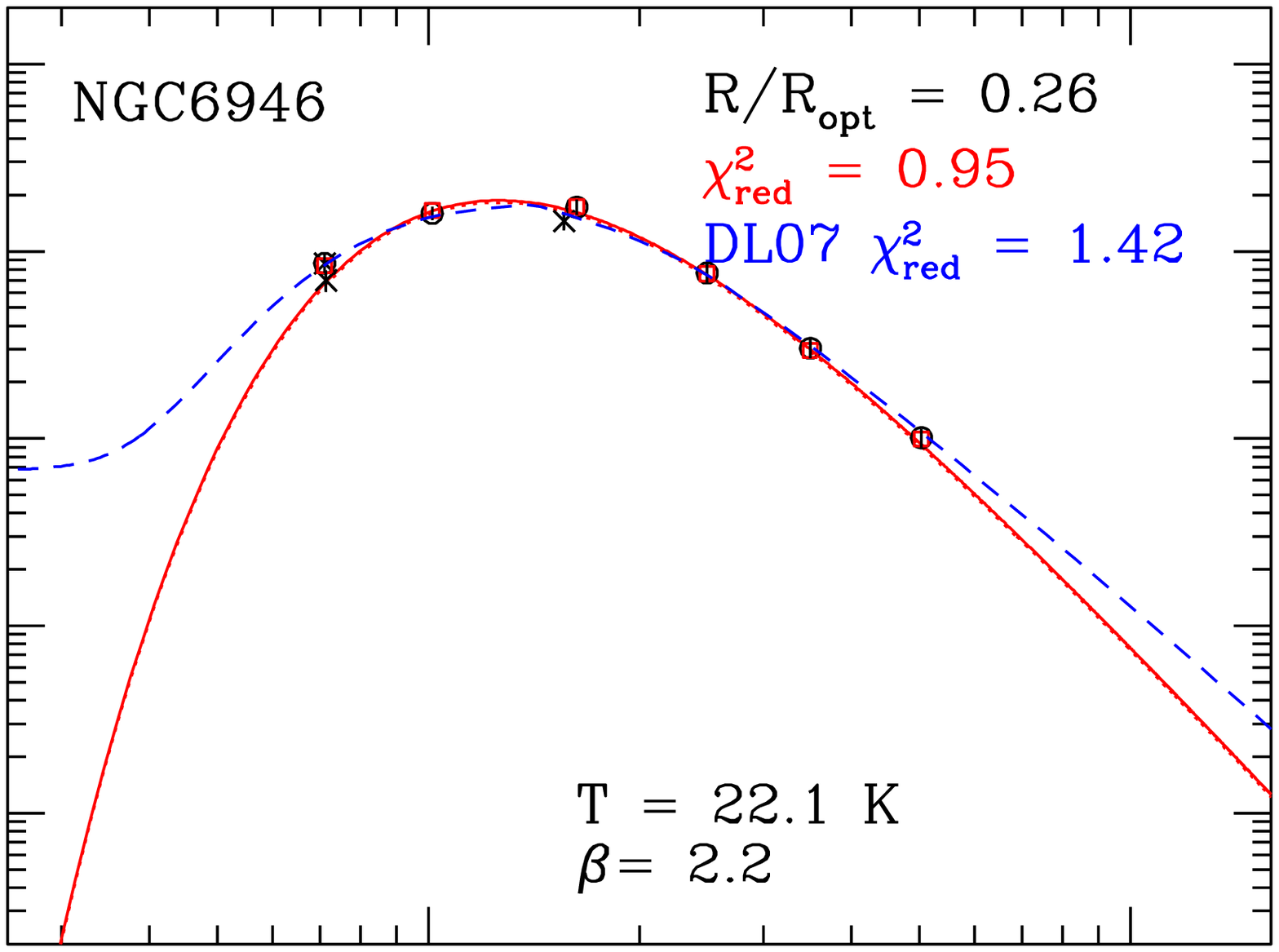}
}
}
\vspace{-1.15\baselineskip}
\centerline{
\hbox{
\includegraphics[width=0.22\linewidth,bb=18 308 588 716]{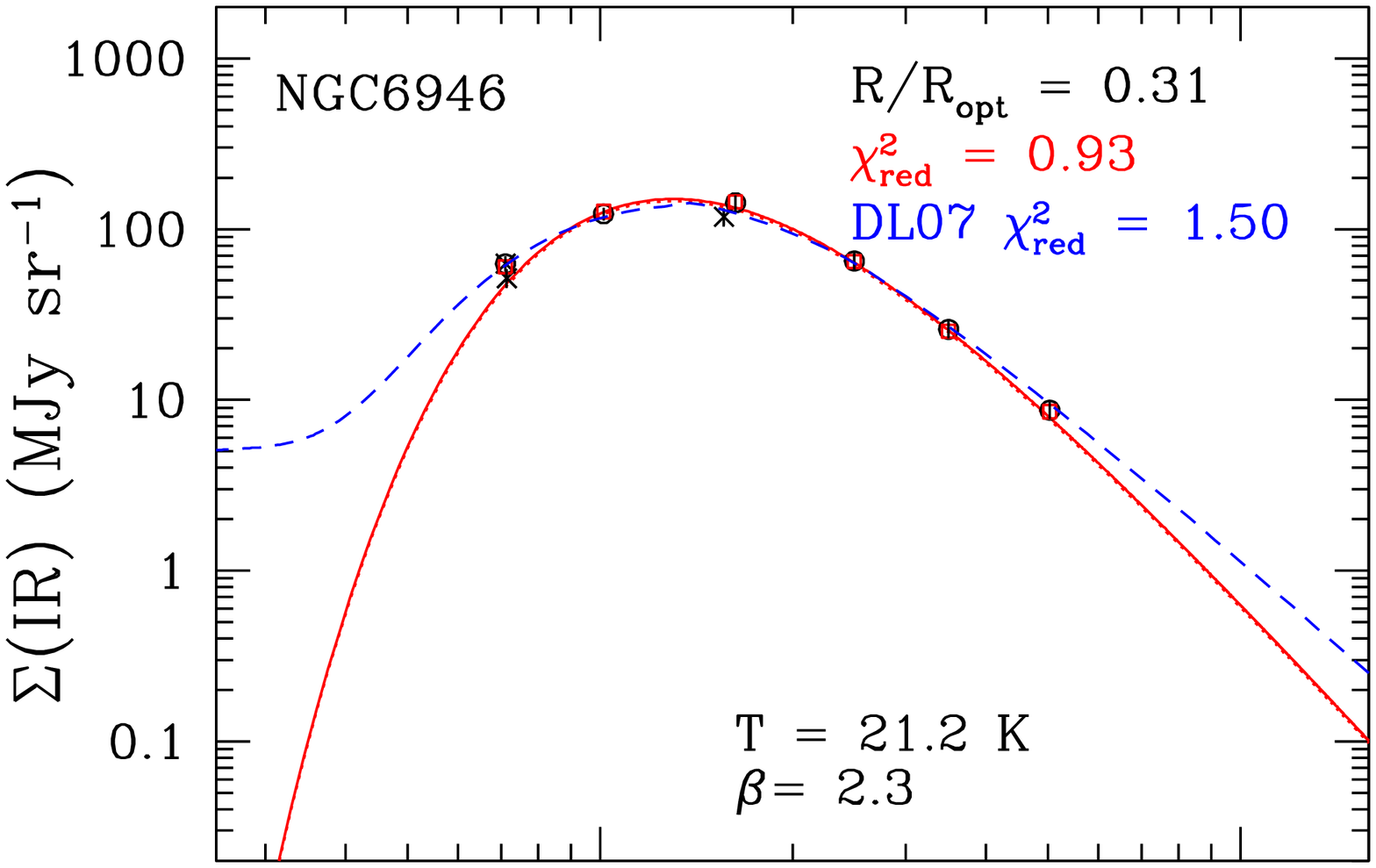}
\hspace{-0.045\linewidth}
\includegraphics[width=0.22\linewidth,bb=18 308 588 716]{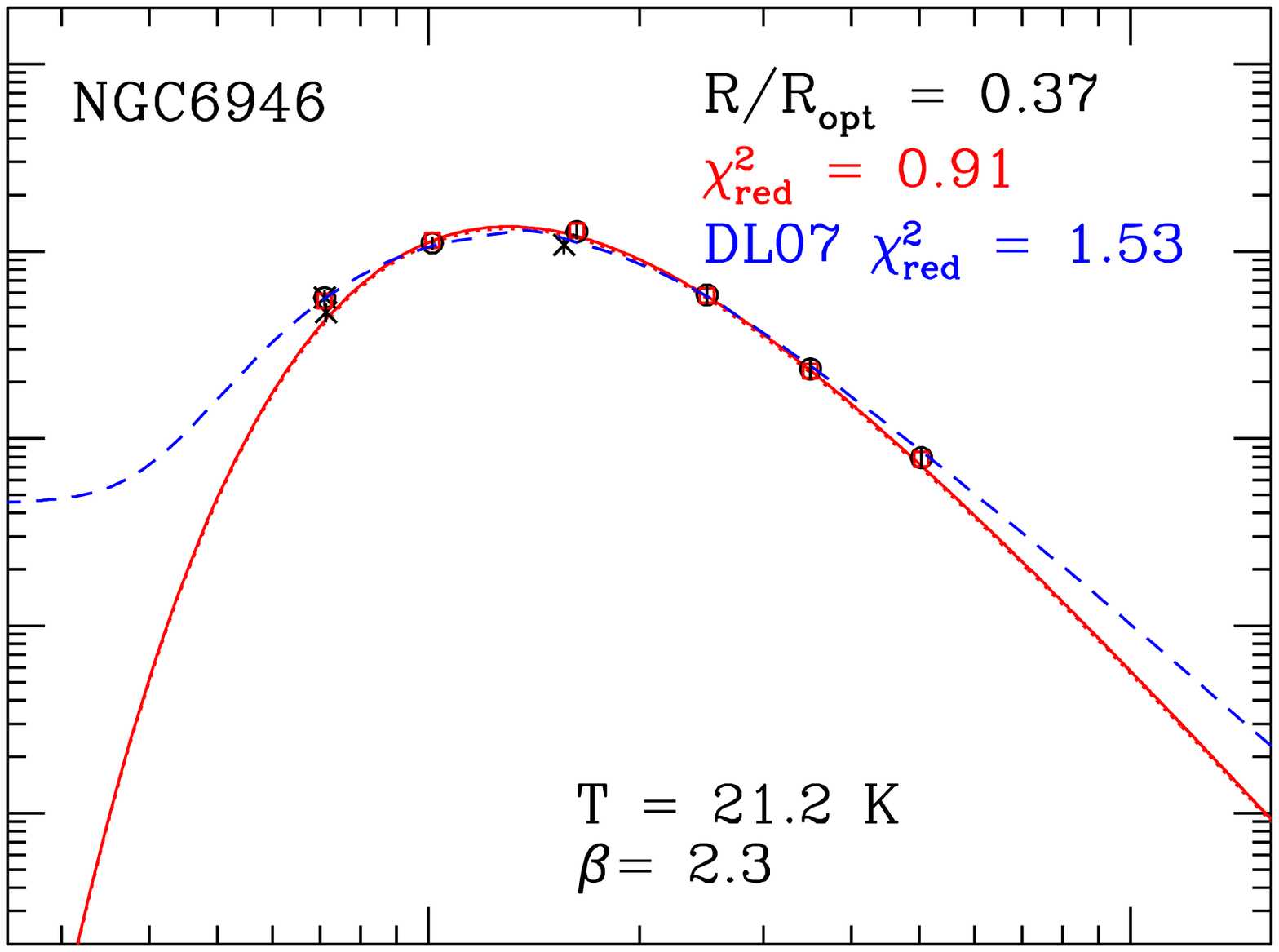}
\hspace{-0.045\linewidth}
\includegraphics[width=0.22\linewidth,bb=18 308 588 716]{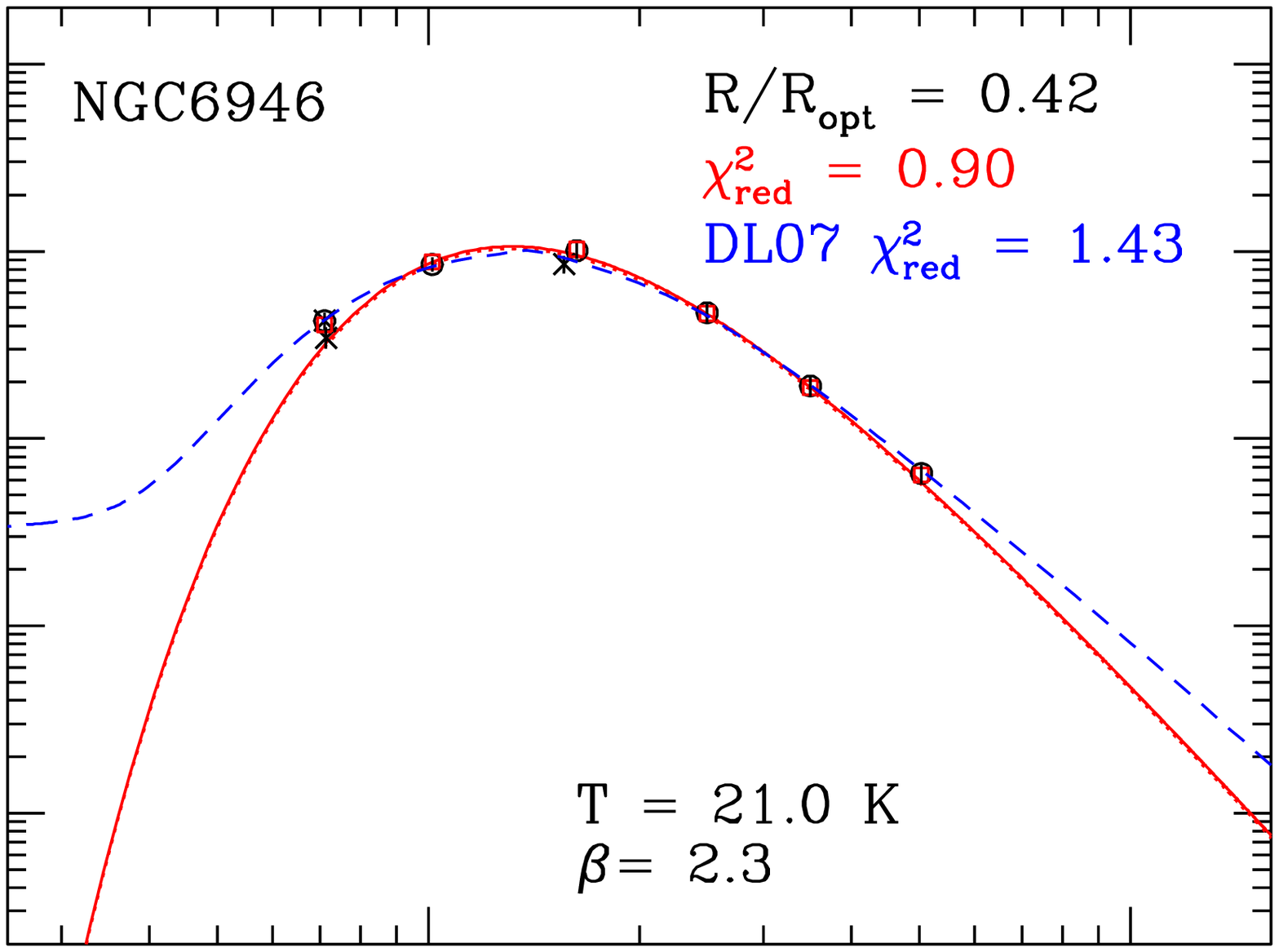}
\hspace{-0.045\linewidth}
\includegraphics[width=0.22\linewidth,bb=18 308 588 716]{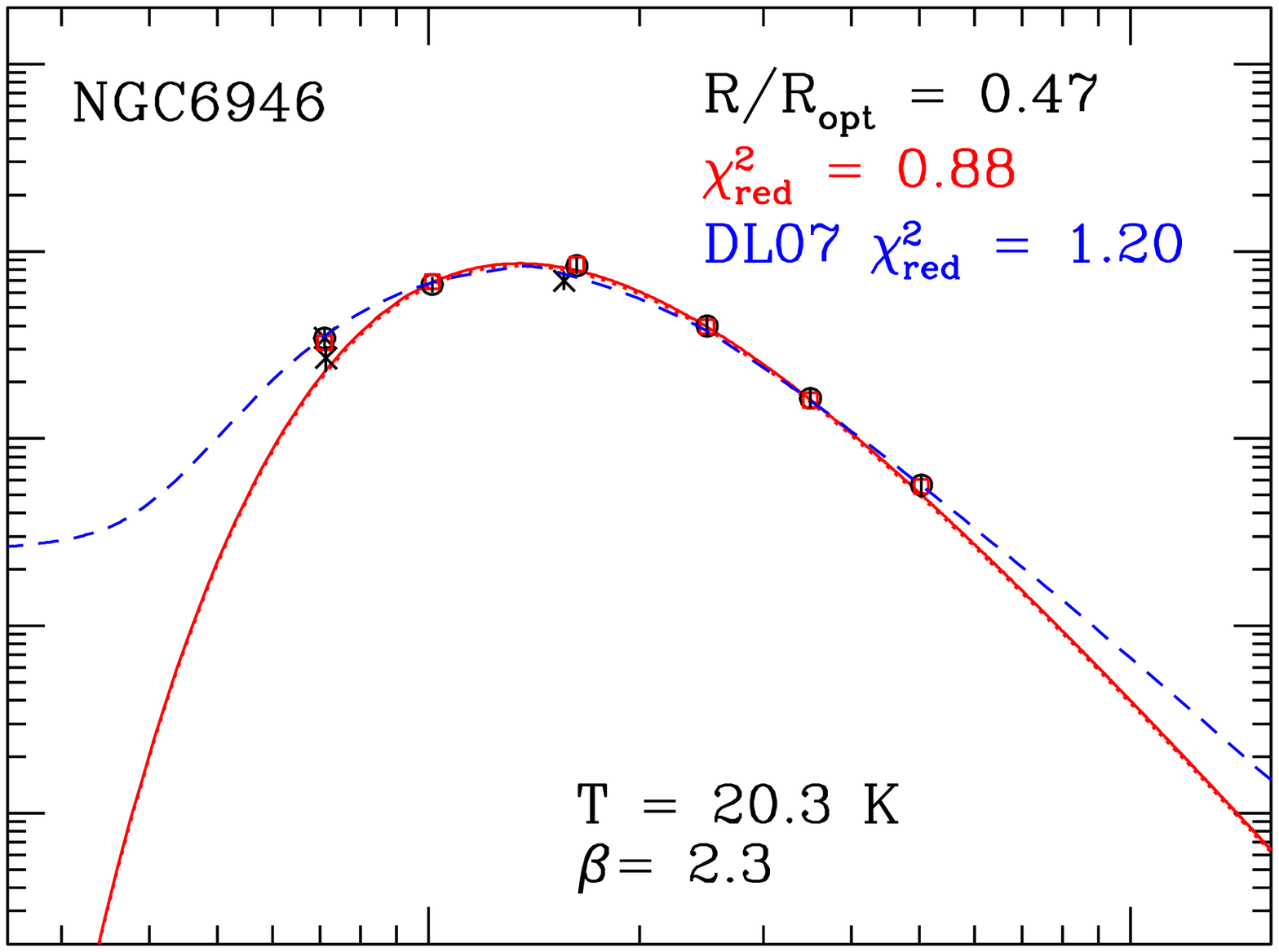}
\hspace{-0.045\linewidth}
\includegraphics[width=0.22\linewidth,bb=18 308 588 716]{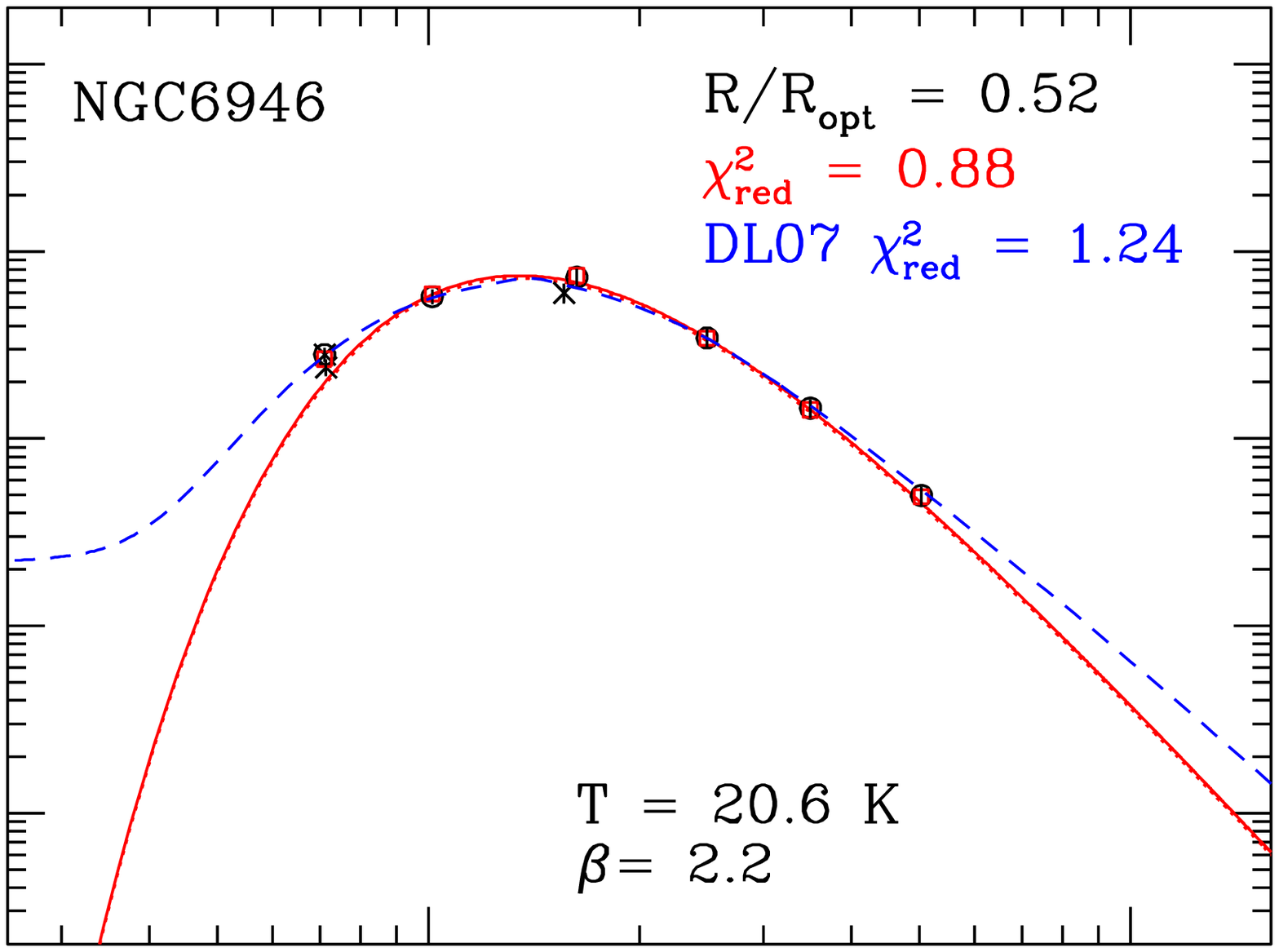}
}
}
\vspace{-1.15\baselineskip}
\centerline{
\hbox{
\includegraphics[width=0.22\linewidth,bb=18 308 588 716]{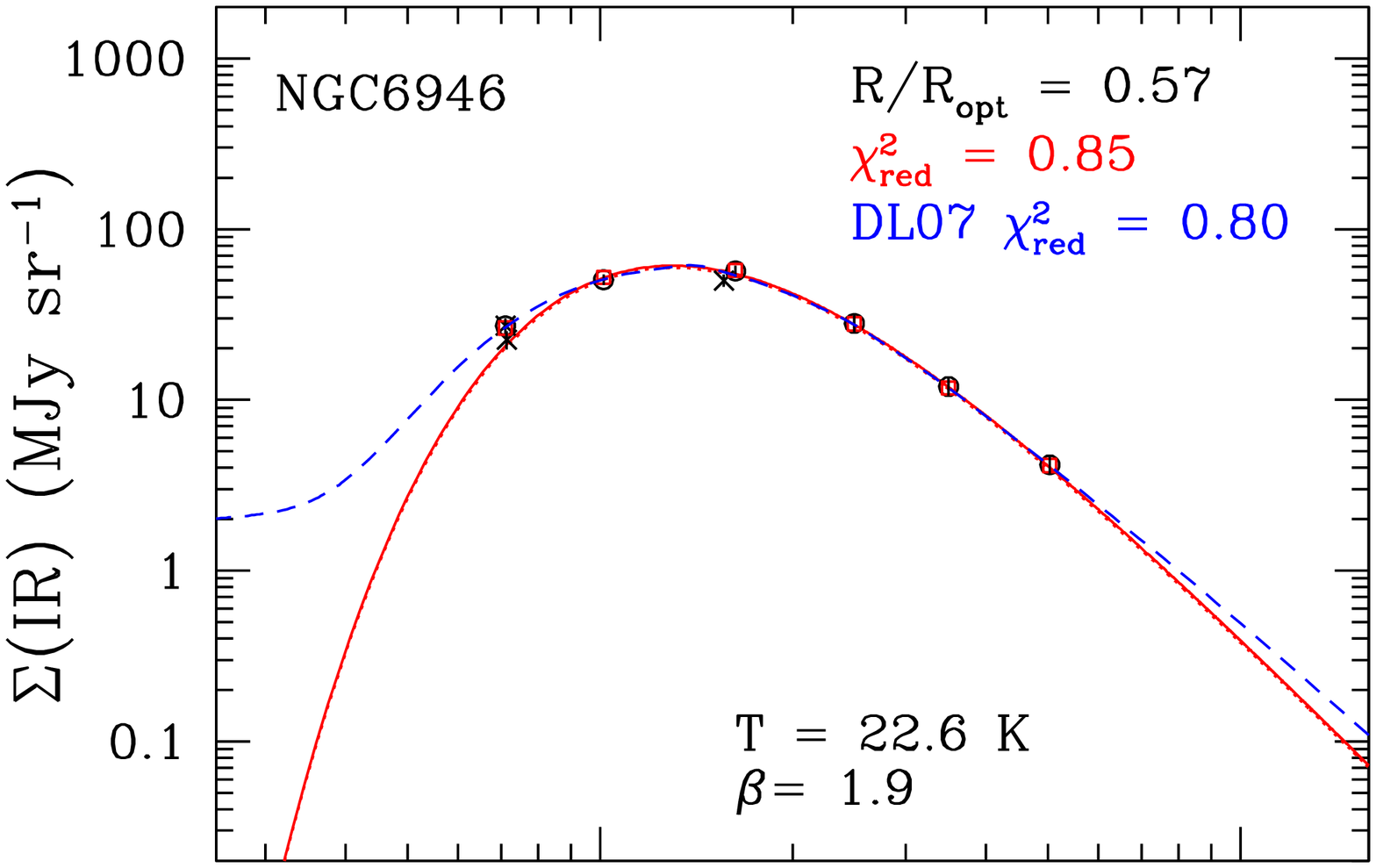}
\hspace{-0.045\linewidth}
\includegraphics[width=0.22\linewidth,bb=18 308 588 716]{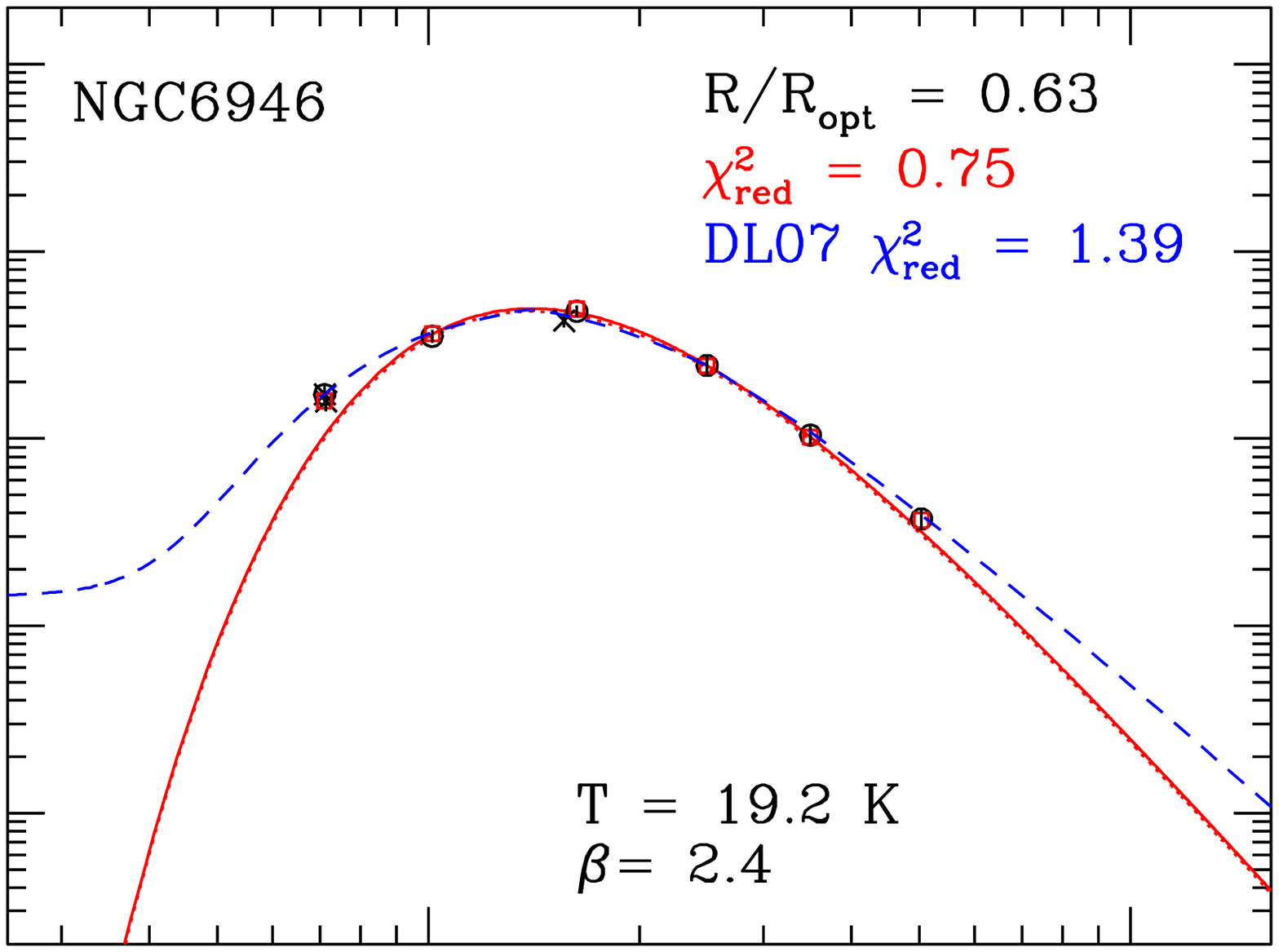}
\hspace{-0.045\linewidth}
\includegraphics[width=0.22\linewidth,bb=18 308 588 716]{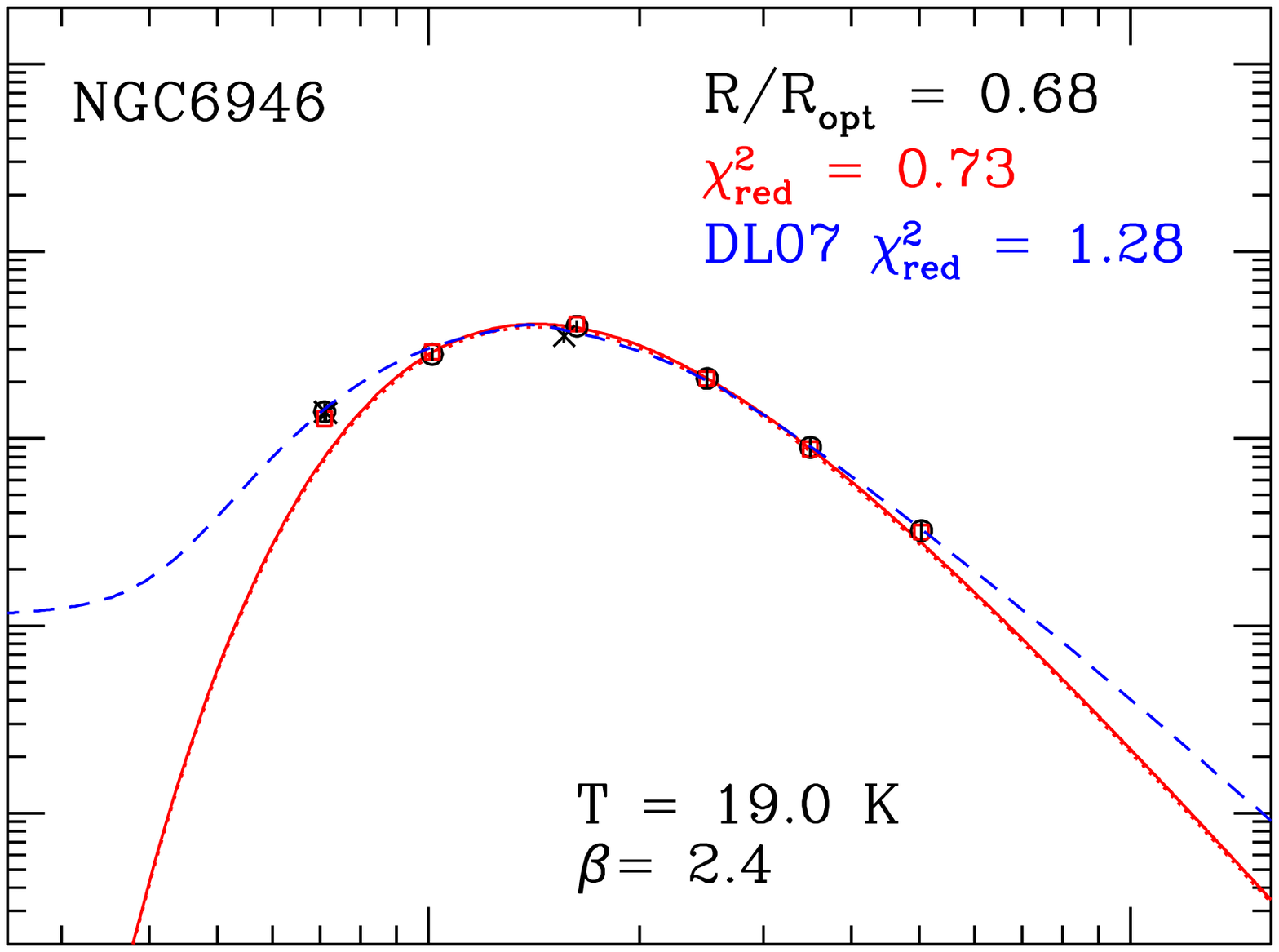}
\hspace{-0.045\linewidth}
\includegraphics[width=0.22\linewidth,bb=18 308 588 716]{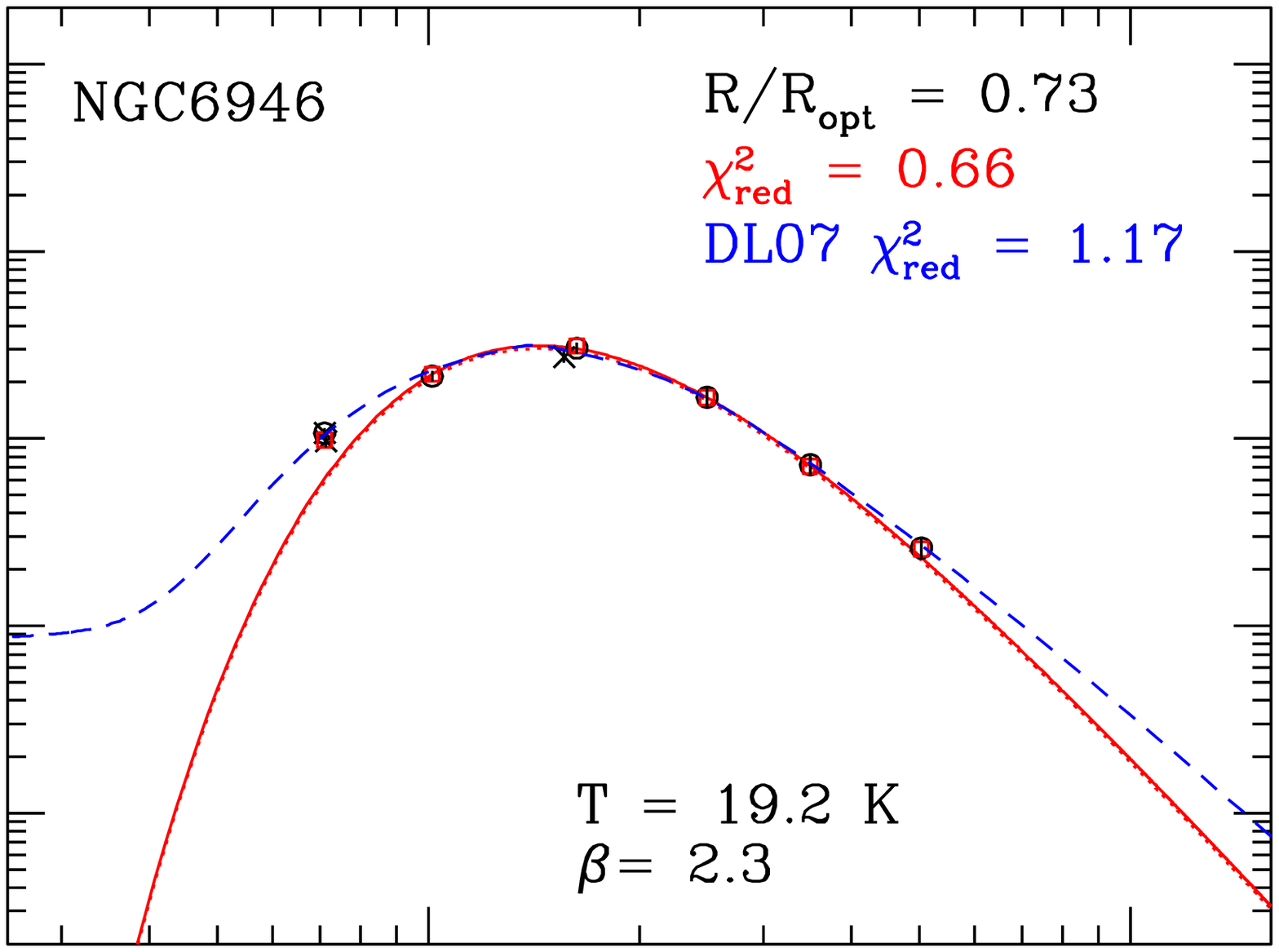}
\hspace{-0.045\linewidth}
\includegraphics[width=0.22\linewidth,bb=18 308 588 716]{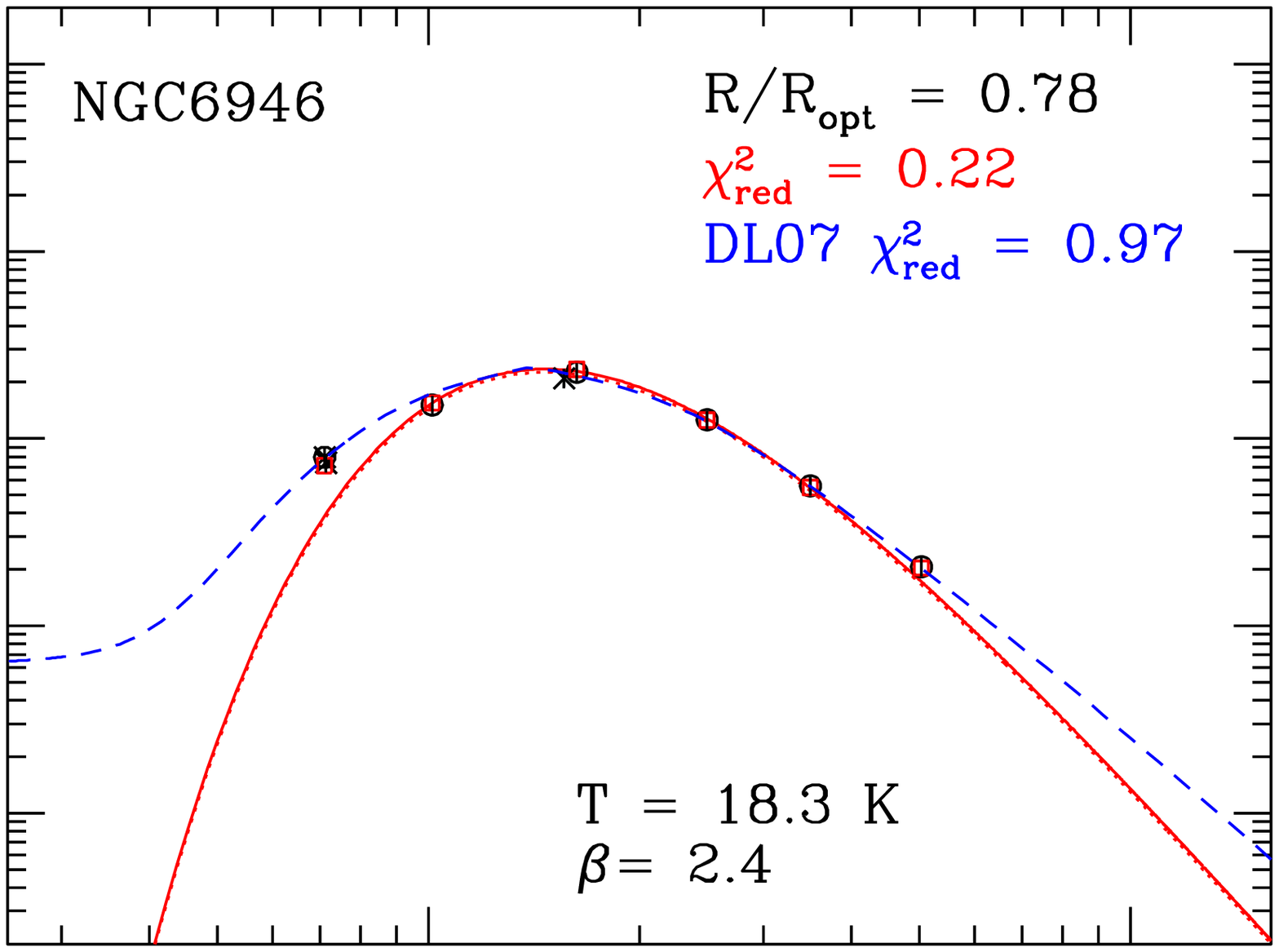}
}
}
\vspace{-1.15\baselineskip}
\centerline{
\hbox{
\includegraphics[width=0.22\linewidth,bb=18 308 588 716]{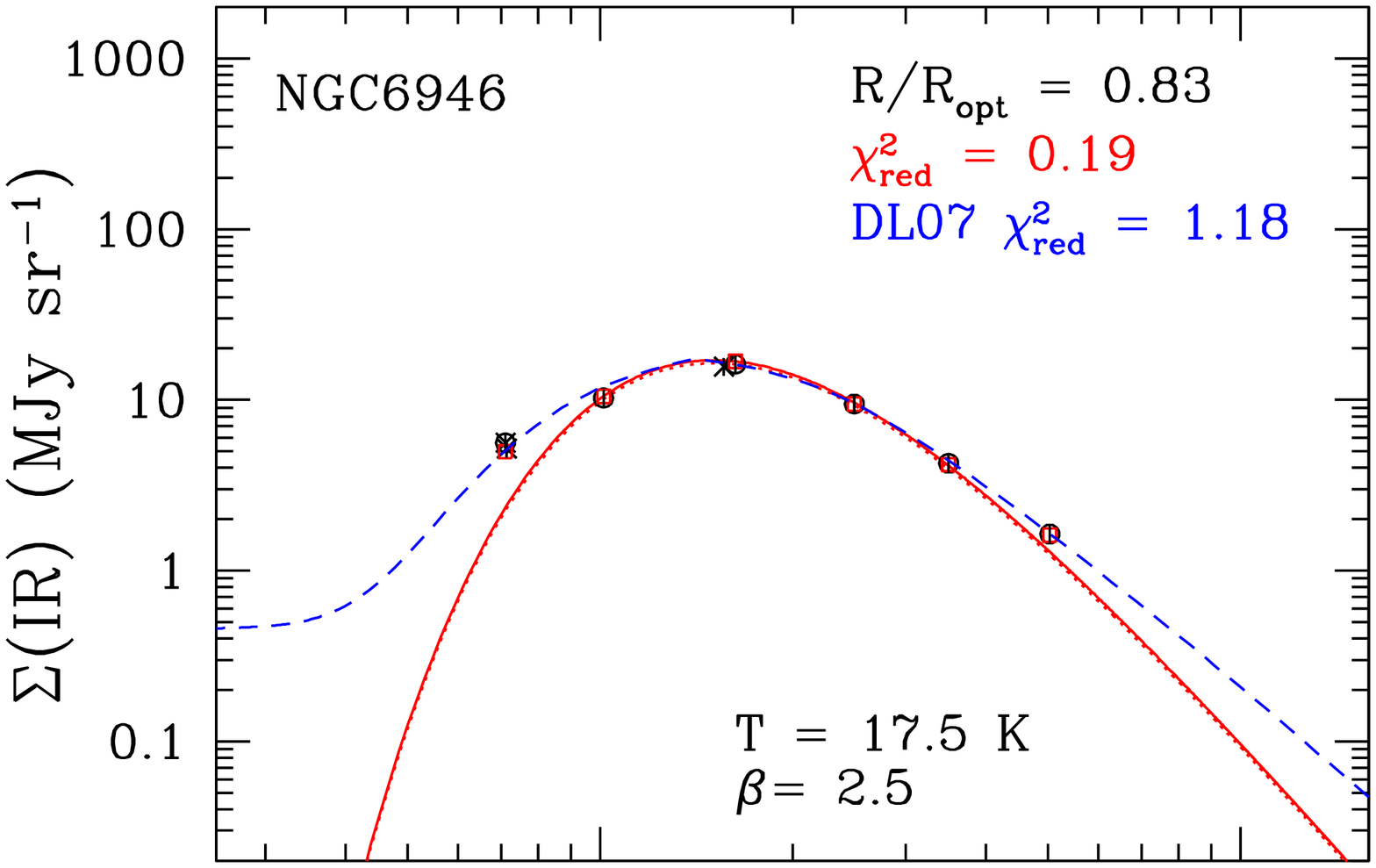}
\hspace{-0.045\linewidth}
\includegraphics[width=0.22\linewidth,bb=18 308 588 716]{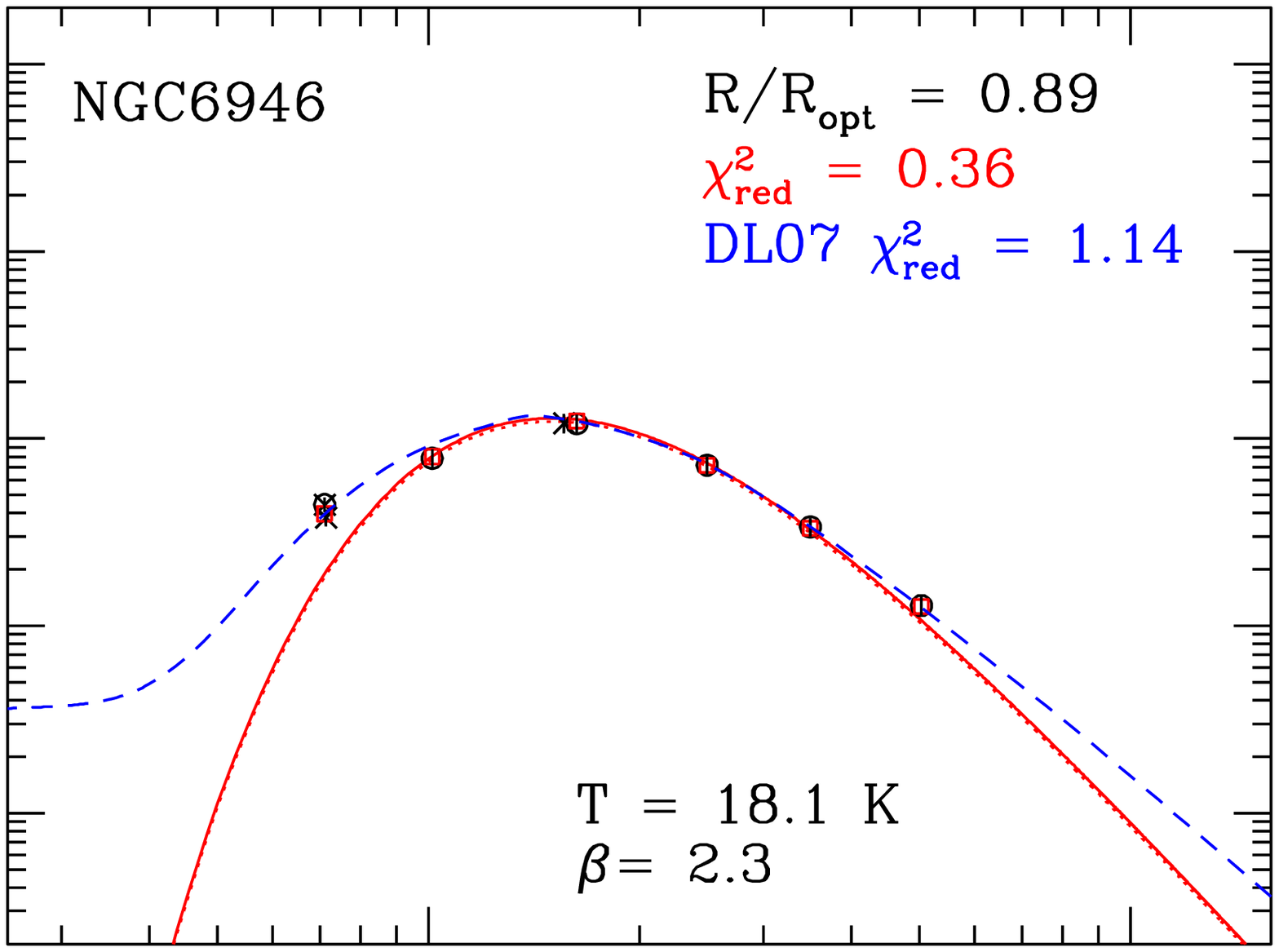}
\hspace{-0.045\linewidth}
\includegraphics[width=0.22\linewidth,bb=18 308 588 716]{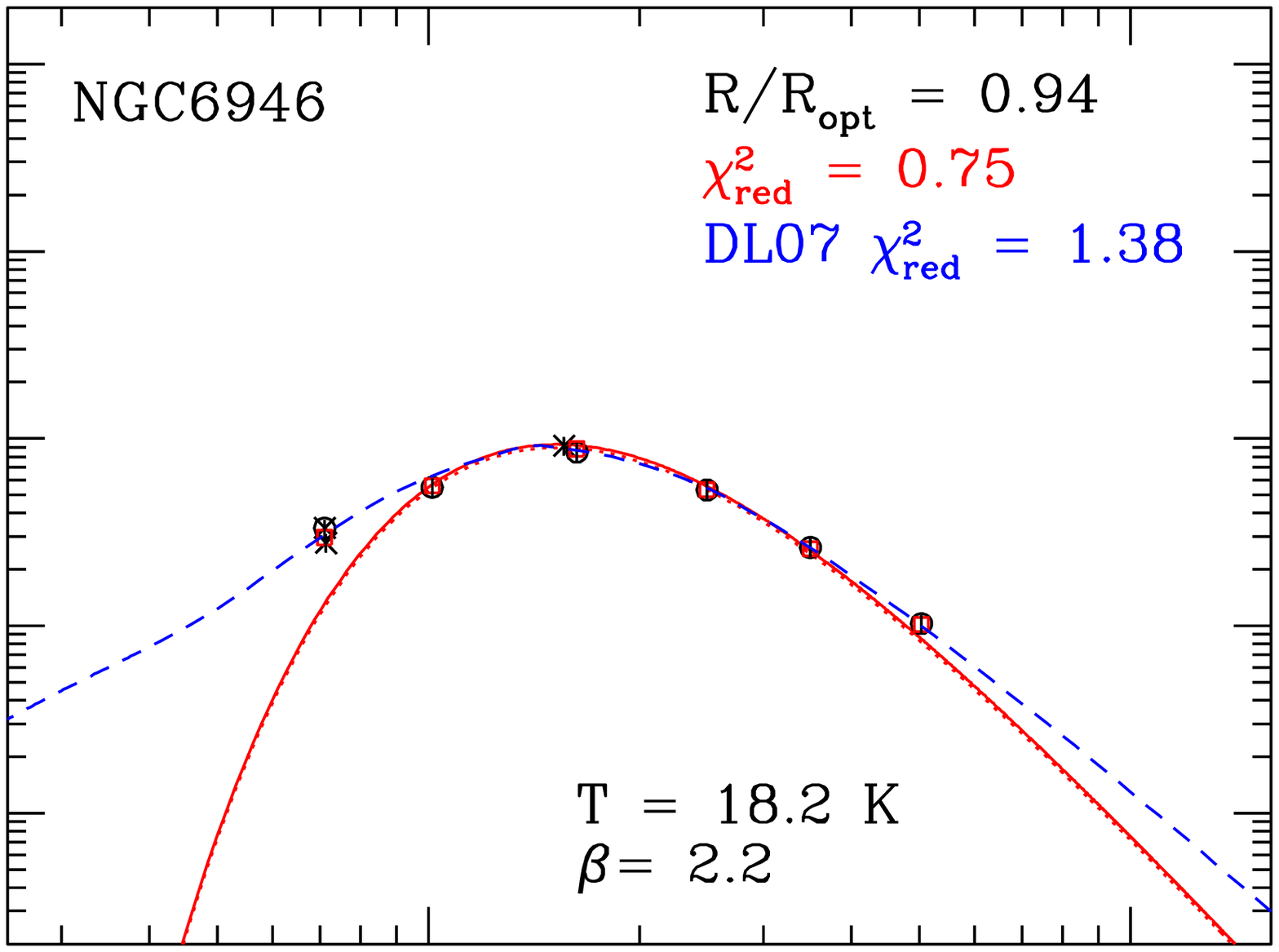}
\hspace{-0.045\linewidth}
\includegraphics[width=0.22\linewidth,bb=18 308 588 716]{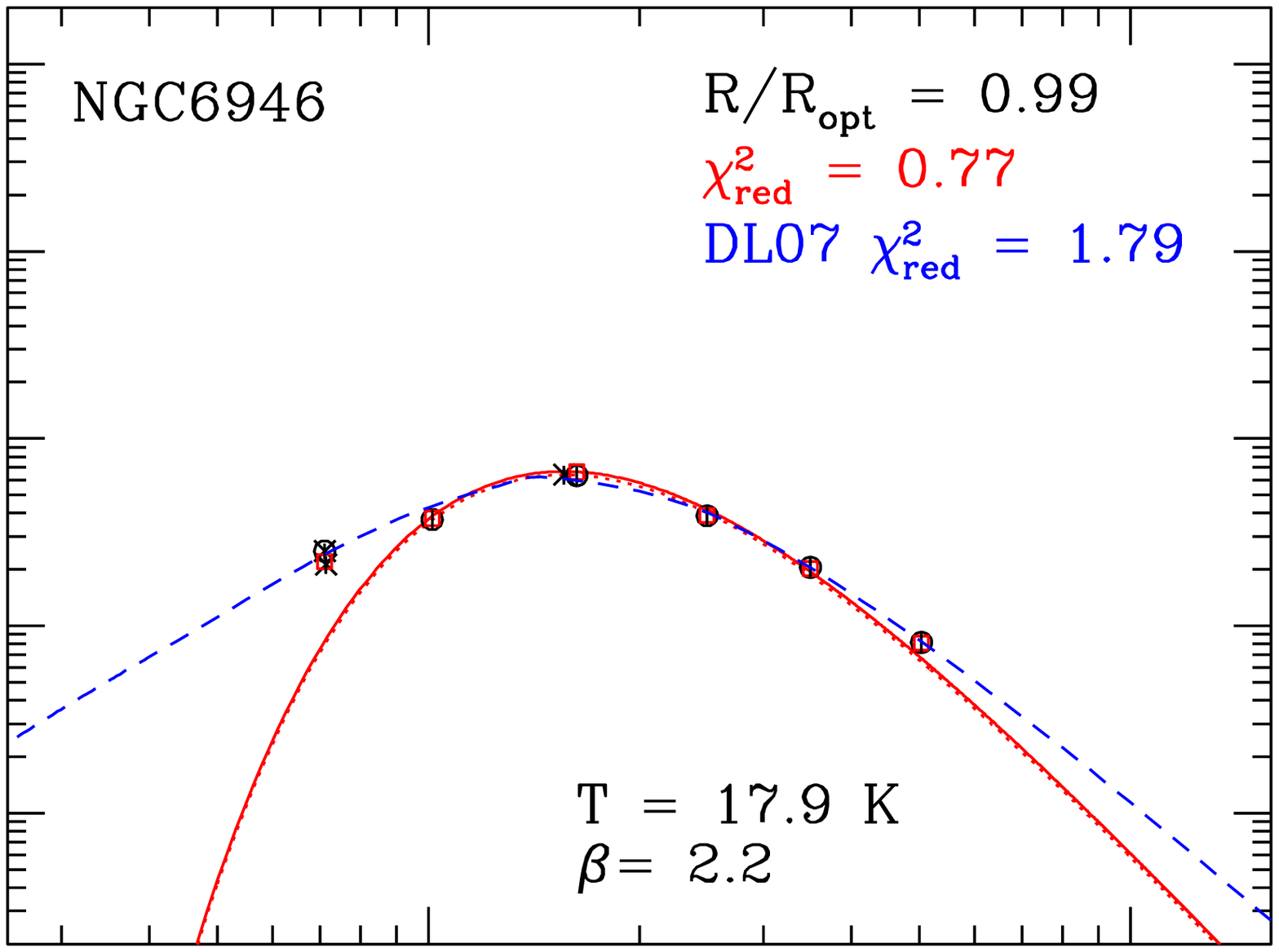}
\hspace{-0.045\linewidth}
\includegraphics[width=0.22\linewidth,bb=18 308 588 716]{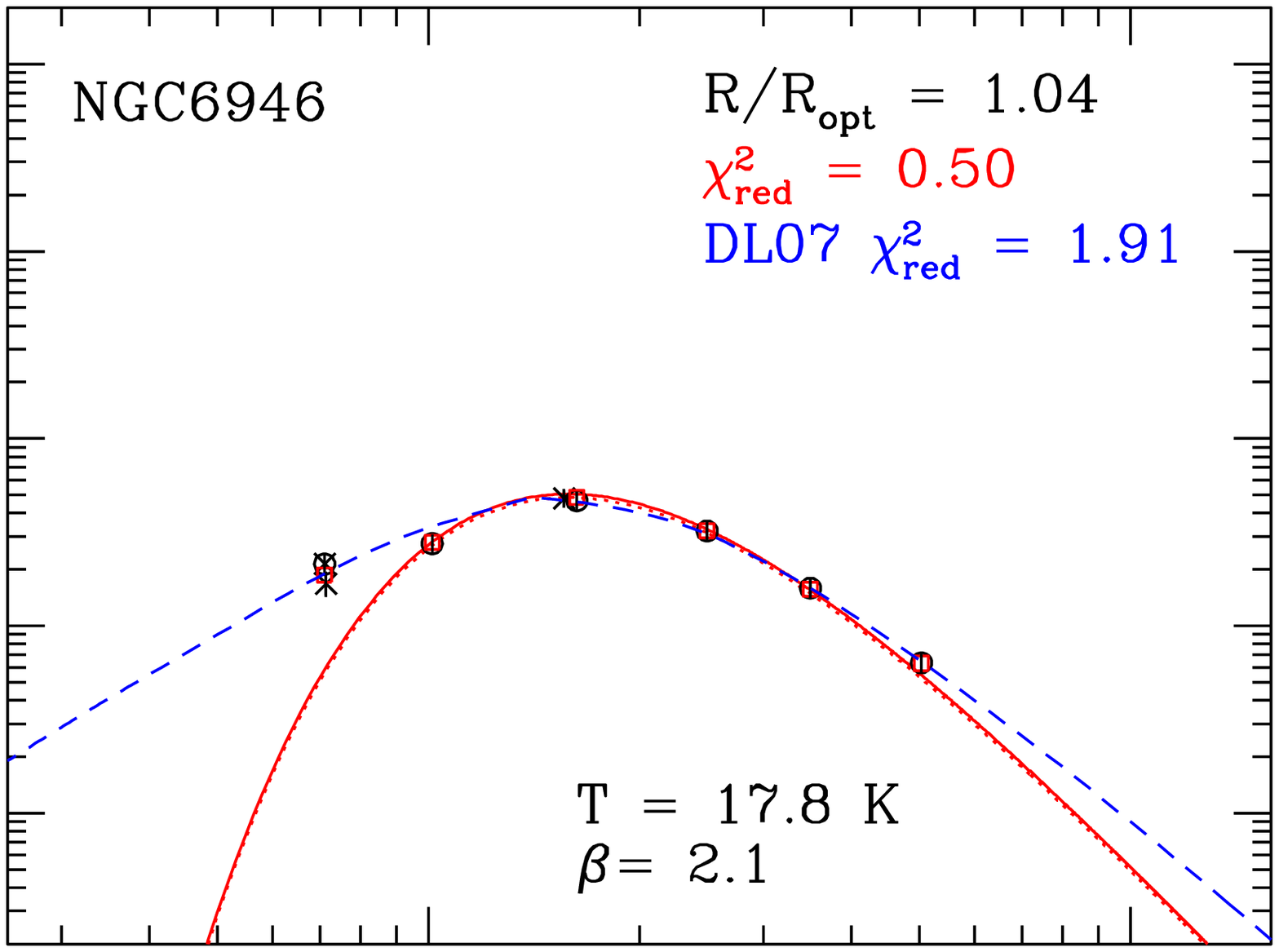}
}
}
\vspace{-1.15\baselineskip}
\centerline{
\hbox{
\includegraphics[width=0.22\linewidth,bb=18 308 588 716]{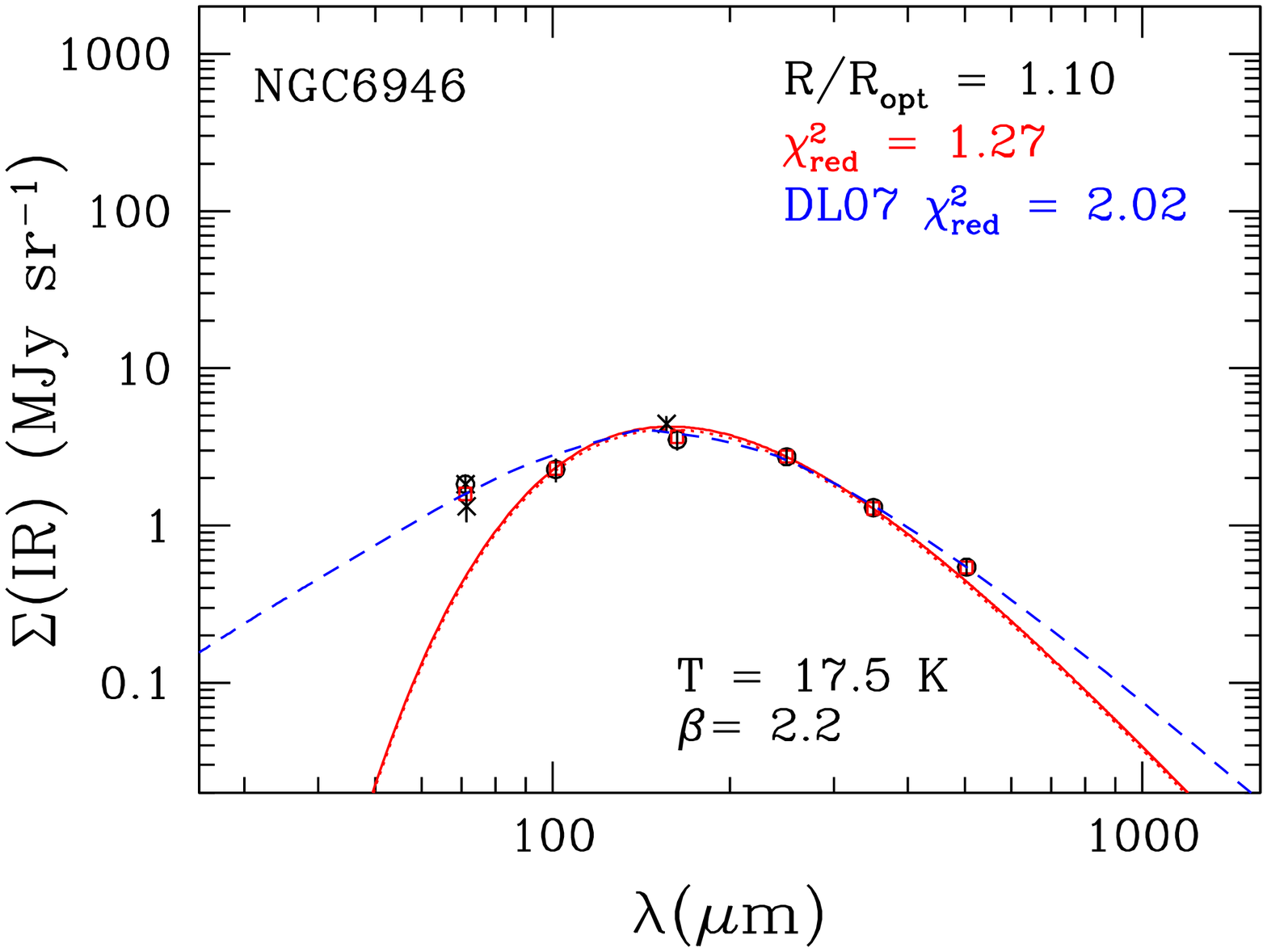}
\hspace{-0.045\linewidth}
\includegraphics[width=0.22\linewidth,bb=18 308 588 716]{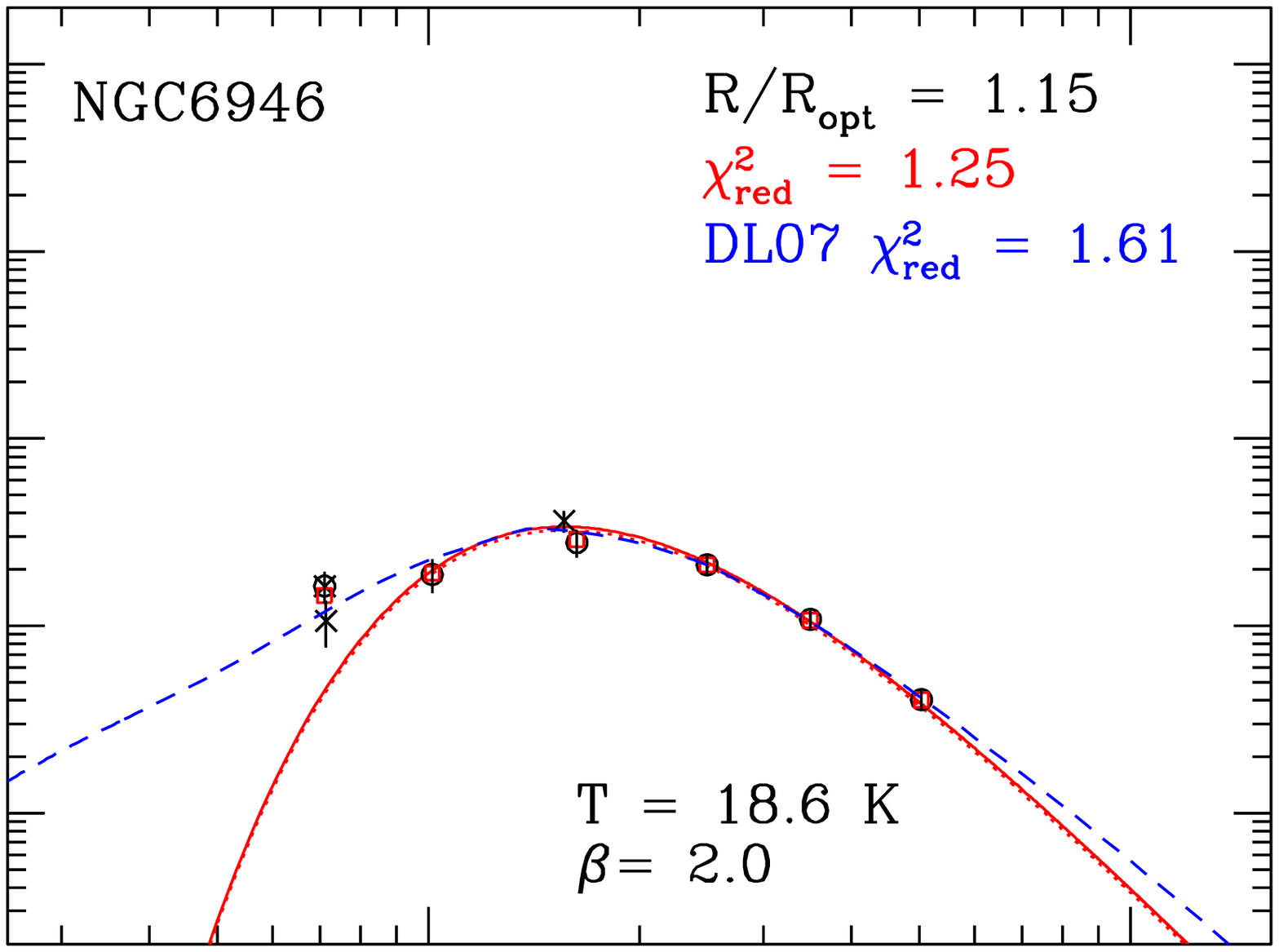}
\hspace{-0.045\linewidth}
\includegraphics[width=0.22\linewidth,bb=18 308 588 716]{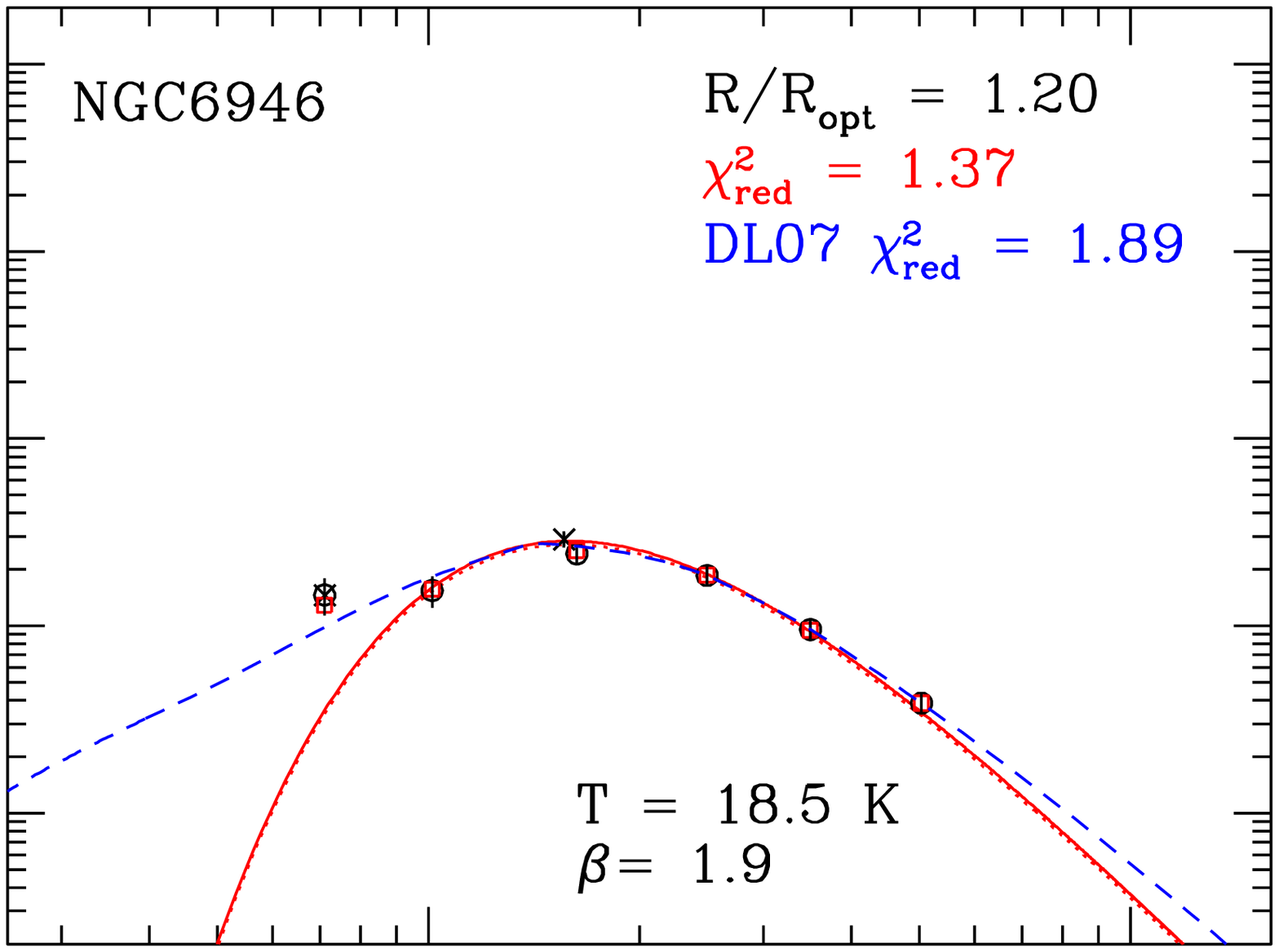}
\hspace{-0.045\linewidth}
\includegraphics[width=0.22\linewidth,bb=18 308 588 716]{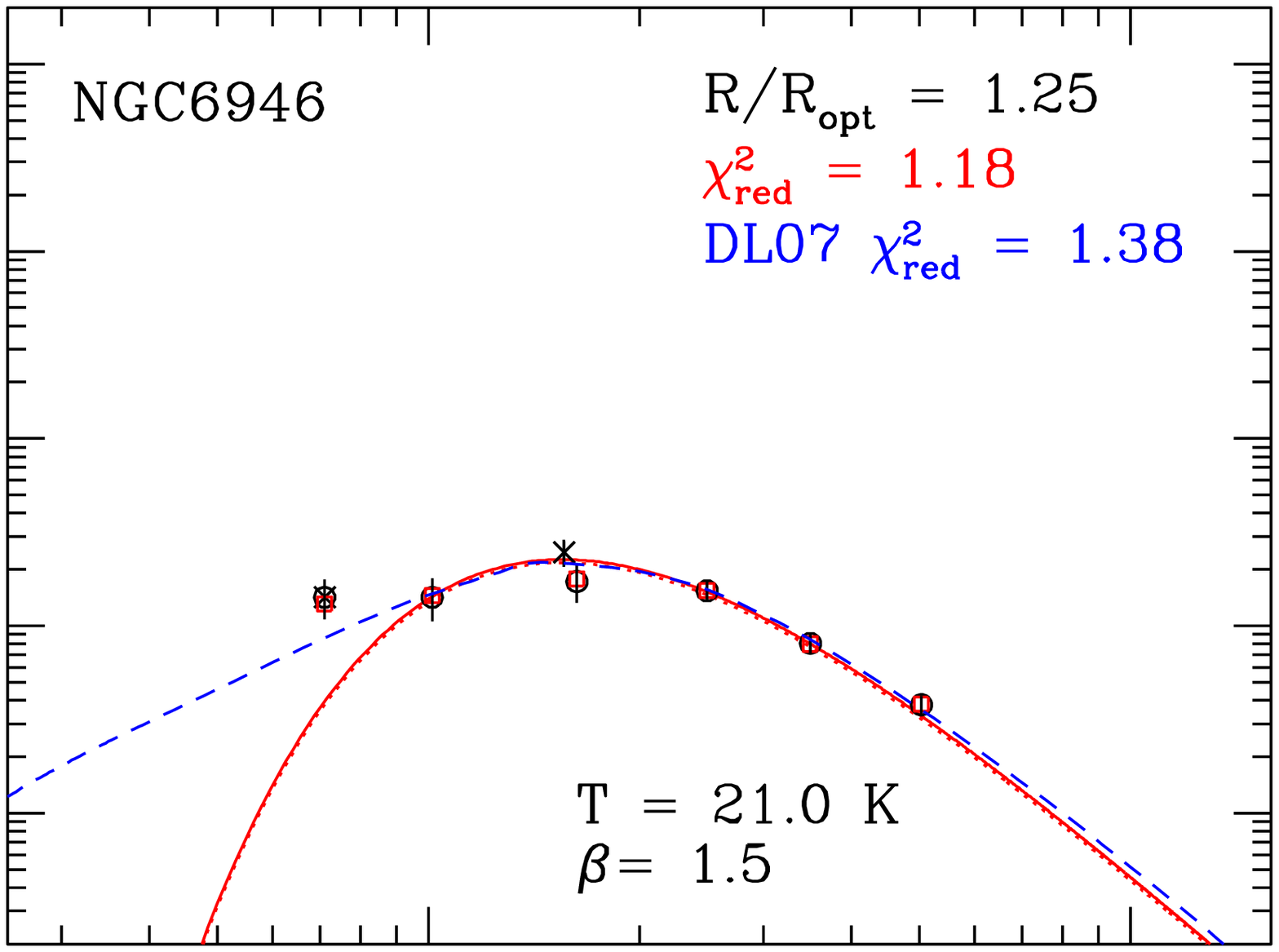}
\hspace{-0.045\linewidth}
\includegraphics[width=0.22\linewidth,bb=18 308 588 716]{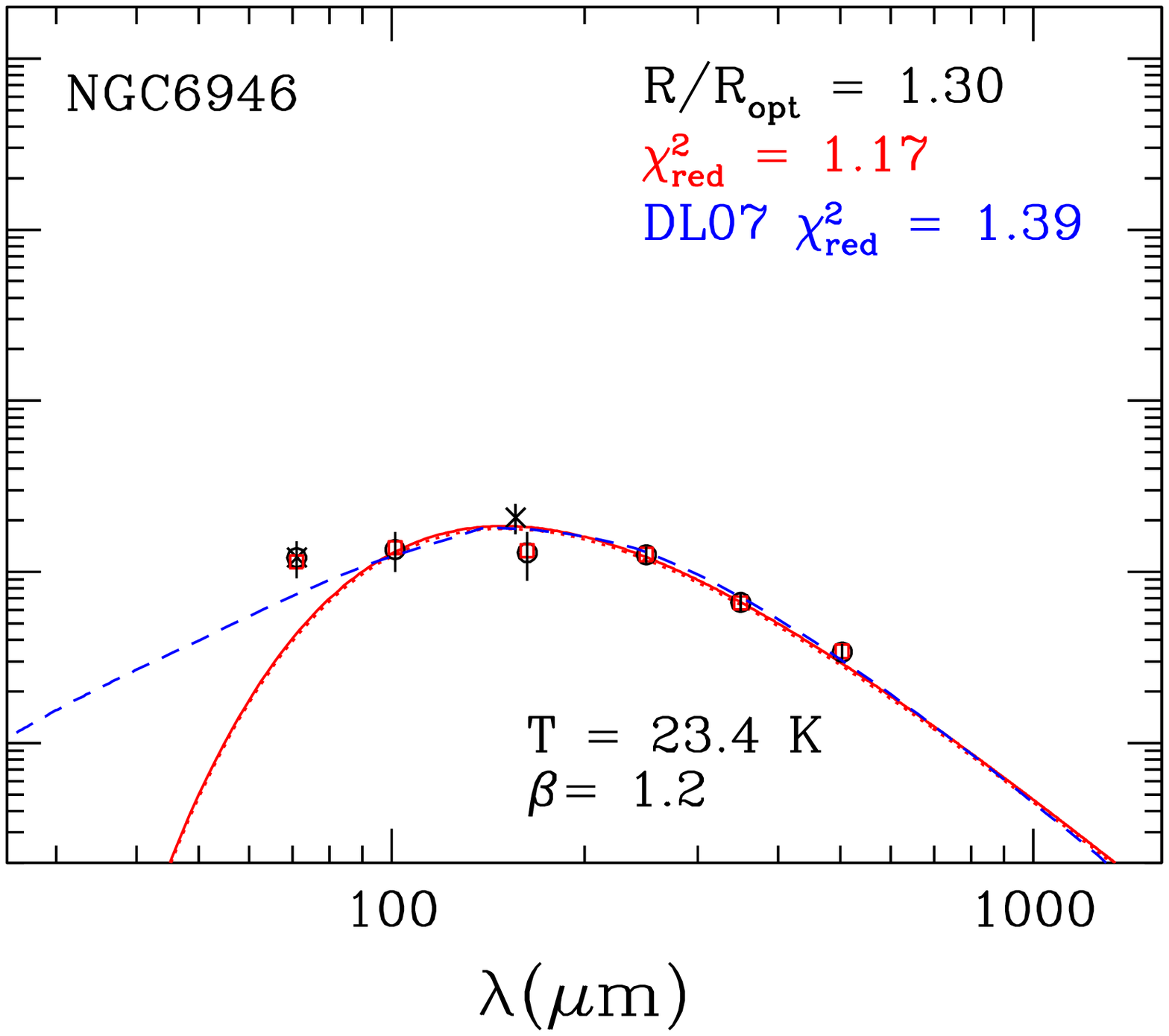}
}
}
\vspace{-1.15\baselineskip}
\centerline{
\hbox{
\includegraphics[width=0.22\linewidth,bb=18 308 588 716]{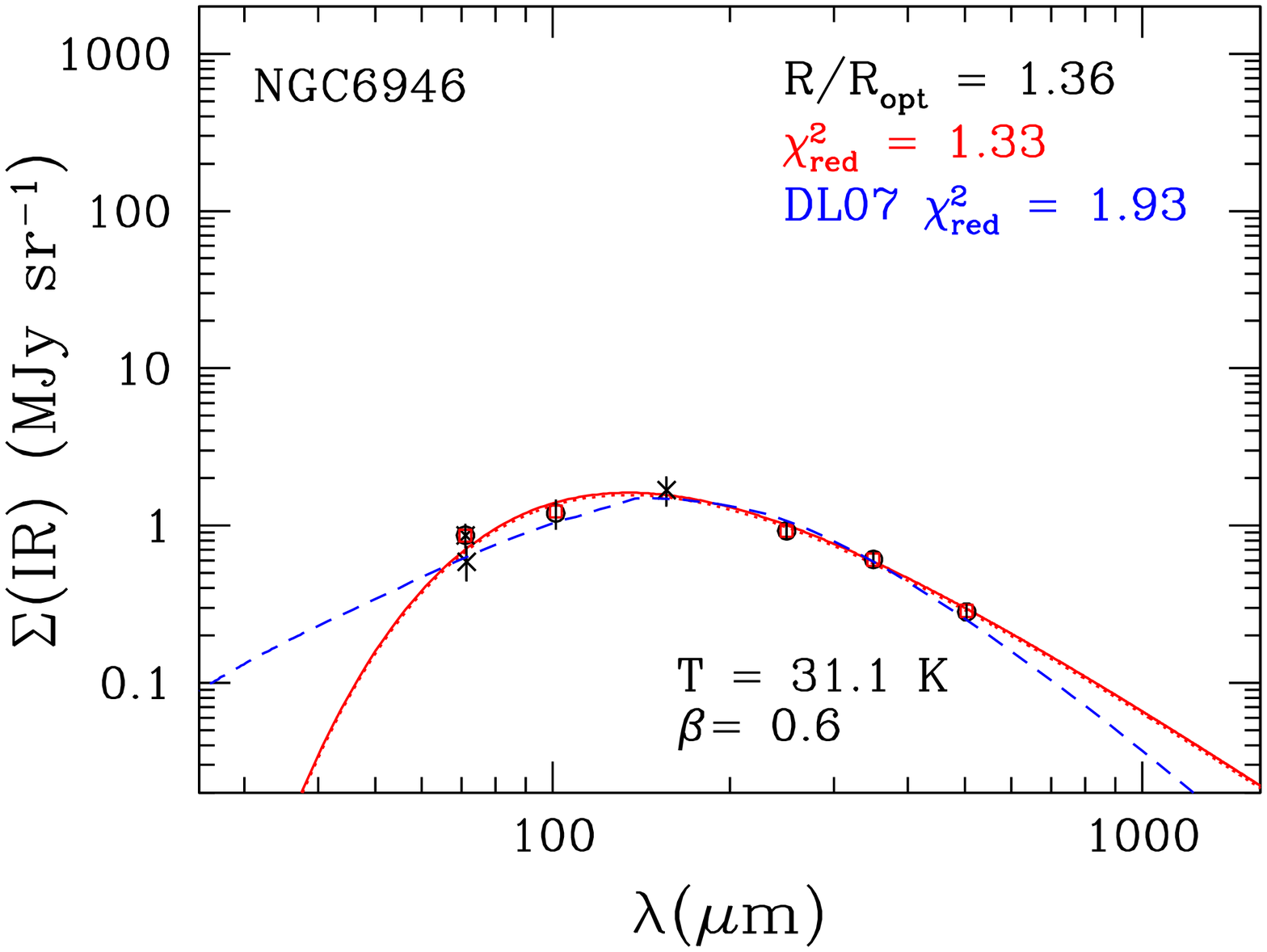}
\hspace{-0.045\linewidth}
\includegraphics[width=0.22\linewidth,bb=18 308 588 716]{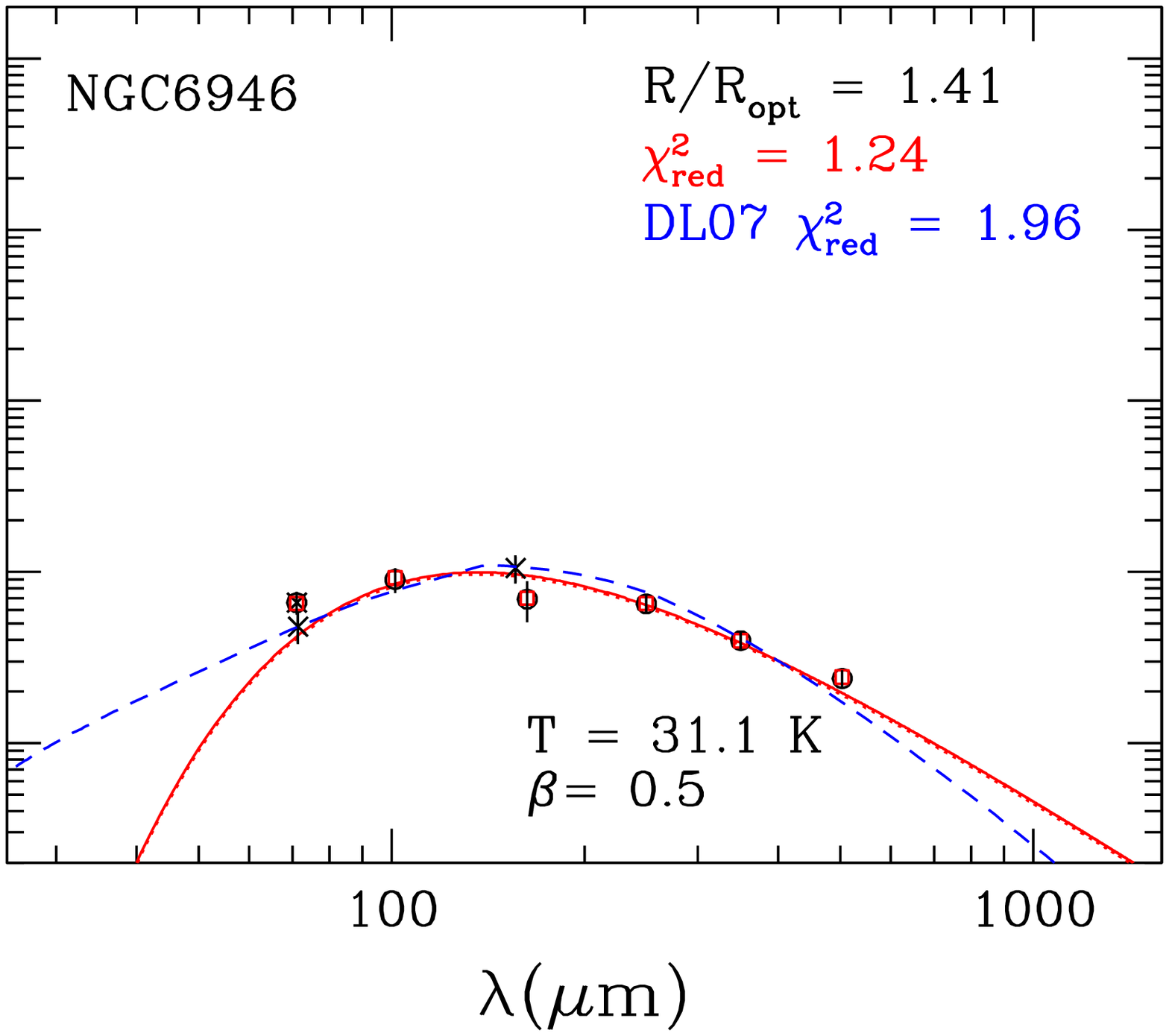}
\hspace{-0.045\linewidth}
\includegraphics[width=0.22\linewidth,bb=18 308 588 716]{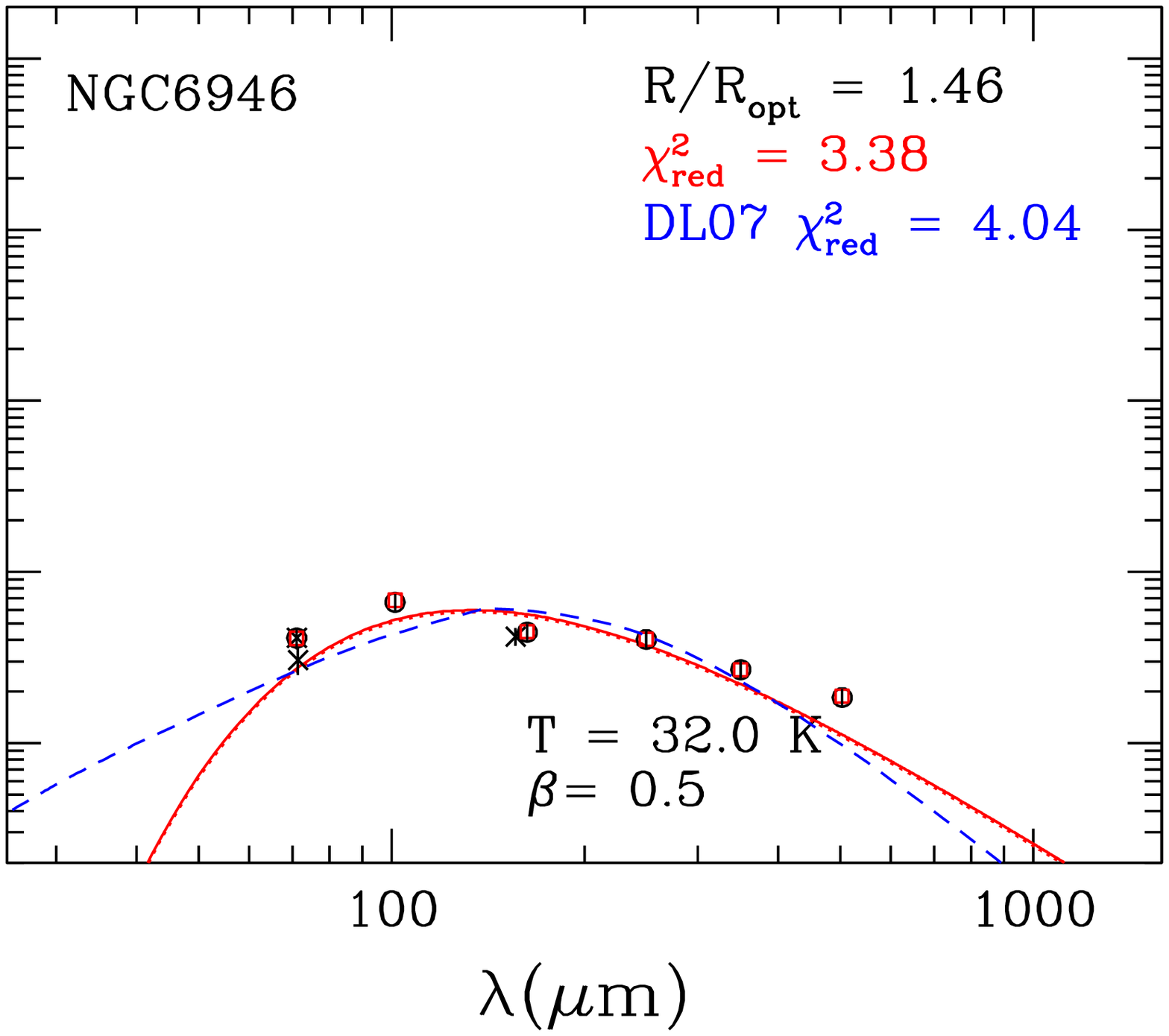}
}
}
%\vspace{-1.15\baselineskip}
%\centerline{
%\hbox{
%\includegraphics[width=0.22\linewidth,bb=18 308 588 716]{IC0342_radsed_31_m1.ps}
%\hspace{-0.045\linewidth}
%\includegraphics[width=0.22\linewidth,bb=18 308 588 716]{IC0342_radsed_32_m2.ps}
%\hspace{-0.045\linewidth}
%\includegraphics[width=0.22\linewidth,bb=18 308 588 716]{IC0342_radsed_33_m2.ps}
%\hspace{-0.045\linewidth}
%\includegraphics[width=0.22\linewidth,bb=18 308 588 716]{IC0342_radsed_34_m2.ps}
%\hspace{-0.045\linewidth}
%\includegraphics[width=0.22\linewidth,bb=18 308 588 716]{IC0342_radsed_35_m2.ps}
%}
%}
%\vspace{-1.15\baselineskip}
%\centerline{
%\hbox{
%\includegraphics[width=0.22\linewidth,bb=18 308 588 716]{IC0342_radsed_36_m1.ps}
%\hspace{-0.045\linewidth}
%\includegraphics[width=0.22\linewidth,bb=18 308 588 716]{NGC5457_radsed_37_m2.ps}
%\hspace{-0.045\linewidth}
%\includegraphics[width=0.22\linewidth,bb=18 308 588 716]{NGC5457_radsed_38_m2.ps}
%\hspace{-0.045\linewidth}
%\includegraphics[width=0.22\linewidth,bb=18 308 588 716]{NGC5457_radsed_39_m2.ps}
%\hspace{-0.045\linewidth}
%\includegraphics[width=0.22\linewidth,bb=18 308 588 716]{NGC5457_radsed_40_m2.ps}
%}
%}
%\caption{{\bf (c)} NGC\,5457 (SABcd) radial profiles. 
%\caption{{\bf (c)} IC\,342 (SABcd) radial profiles. 
\caption{{\bf (c)} as Fig. \ref{fig:radial}(a), but NGC\,6946 (SABcd) radial profiles. 
\label{fig:appradial}
}
\end{figure*}
\setcounter{figure}{0}
\begin{figure*}[!ht]
\centerline{
\hbox{
\includegraphics[width=0.45\linewidth,bb=18 308 588 716]{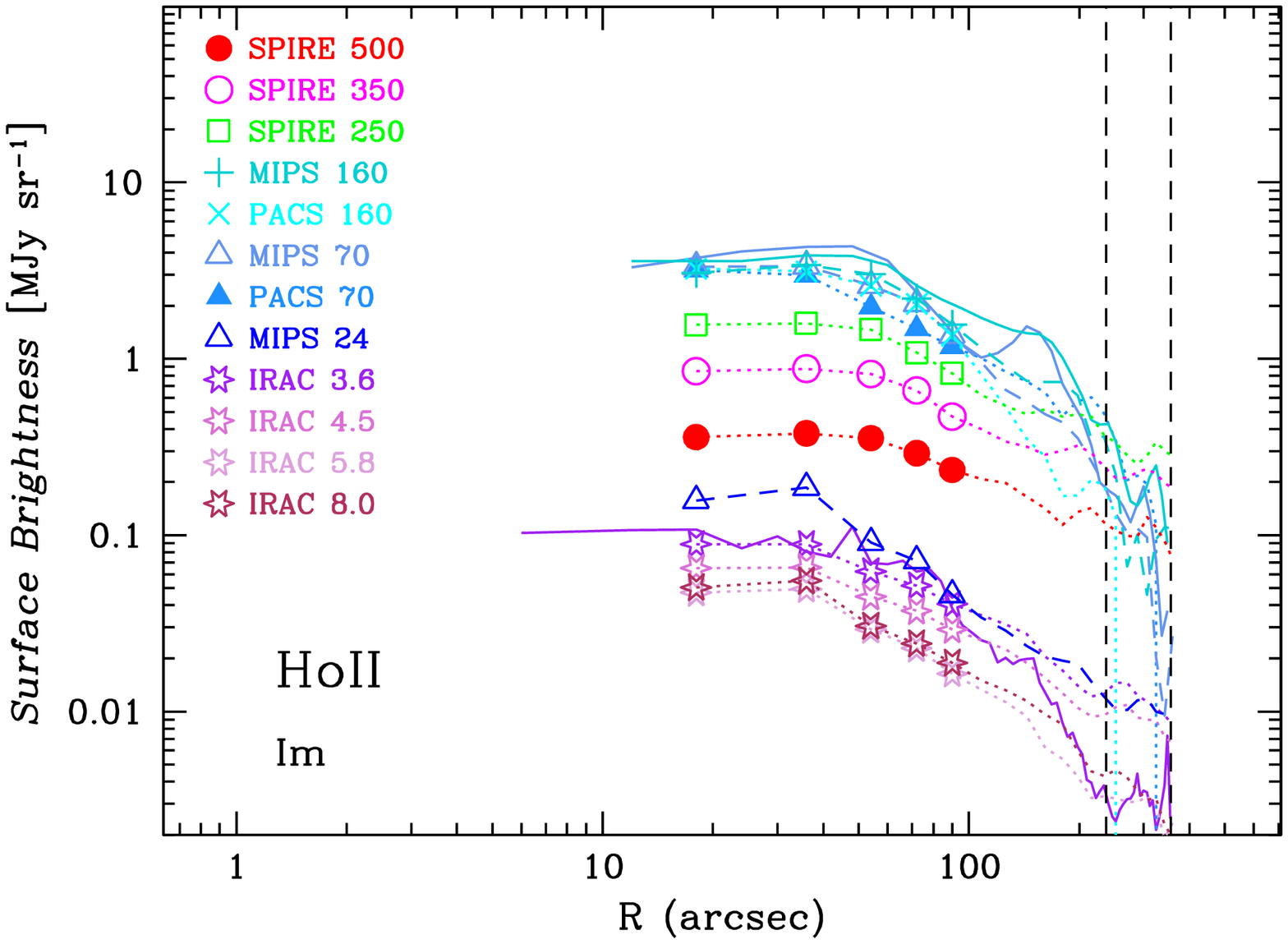}
%\hspace{0.1\linewidth}
\hspace{0.05\linewidth}
\includegraphics[width=0.45\linewidth,bb=18 167 592 718]{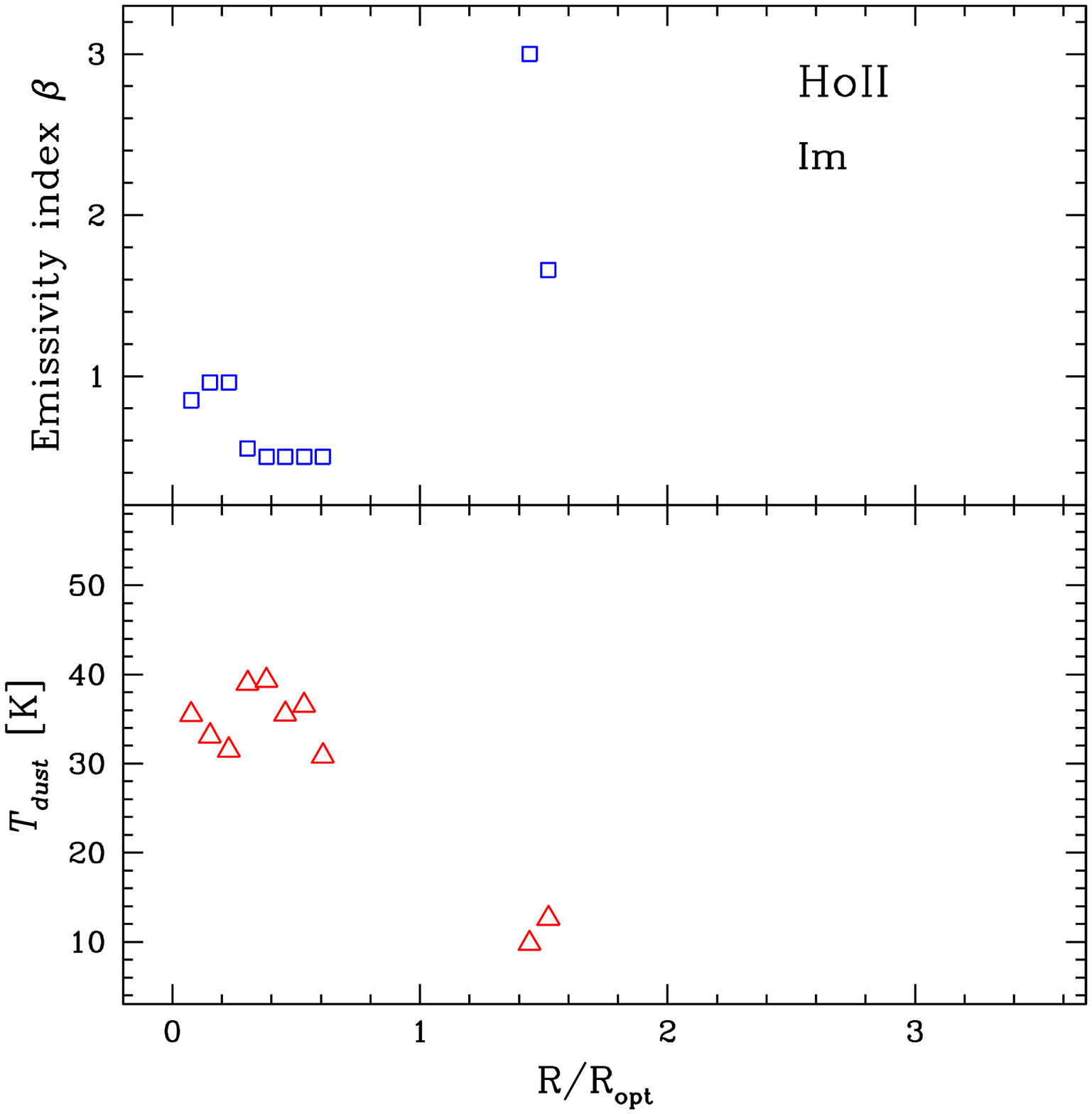}
}
}
\vspace{2.15\baselineskip}
\vspace{-1.15\baselineskip}
\centerline{
\hbox{
\includegraphics[width=0.22\linewidth,bb=18 308 588 716]{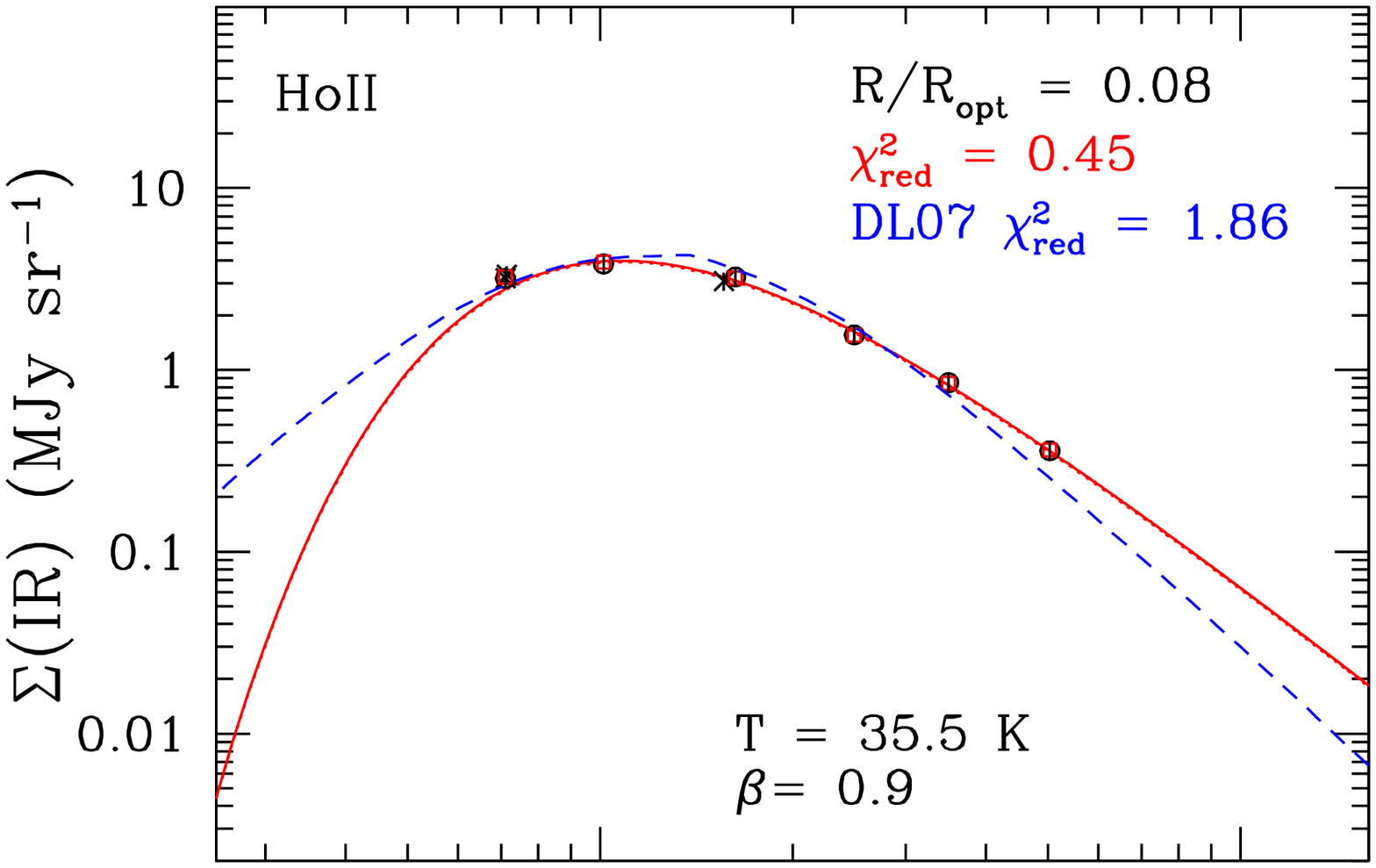}
\hspace{-0.045\linewidth}
\includegraphics[width=0.22\linewidth,bb=18 308 588 716]{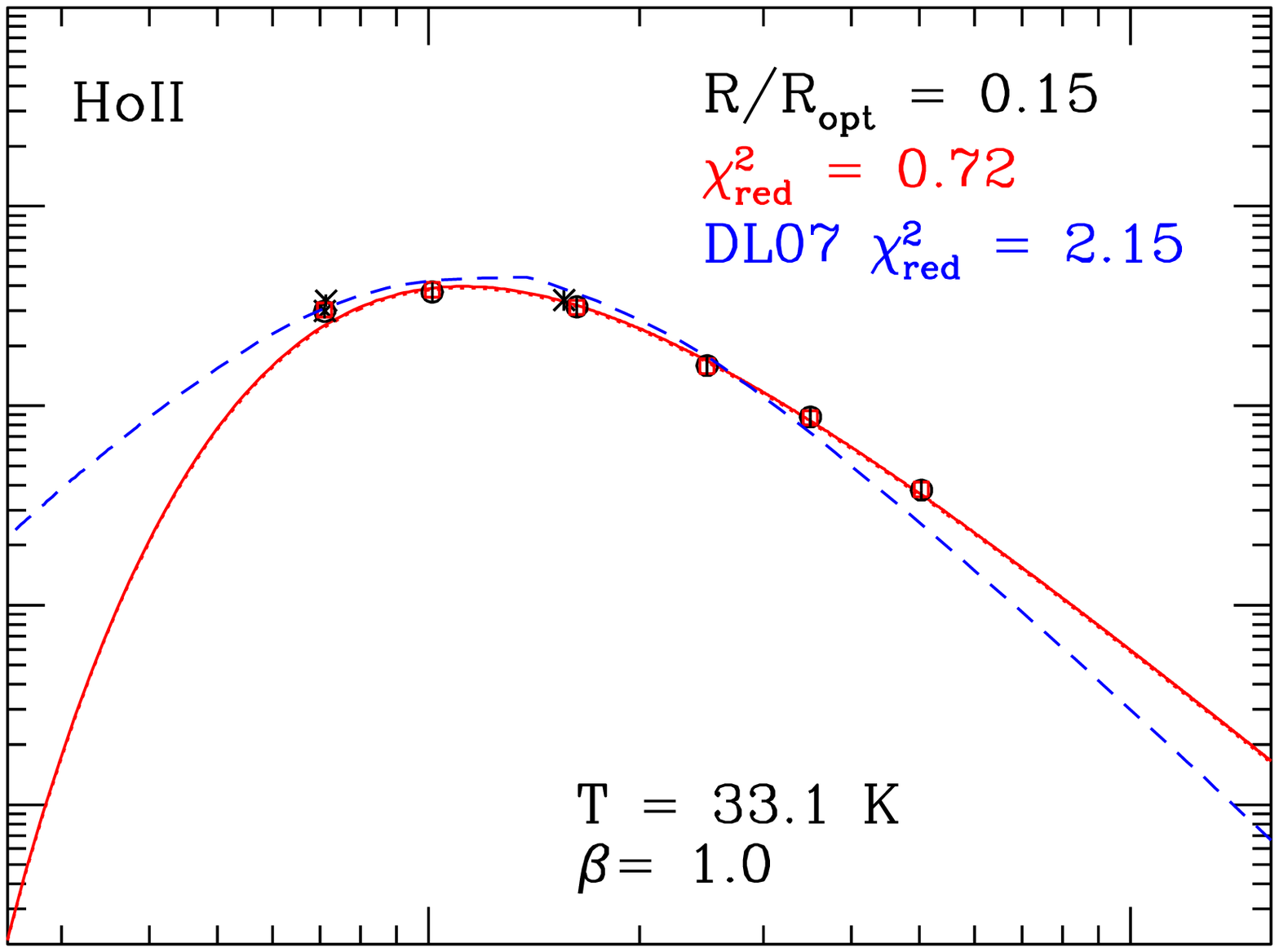}
\hspace{-0.045\linewidth}
\includegraphics[width=0.22\linewidth,bb=18 308 588 716]{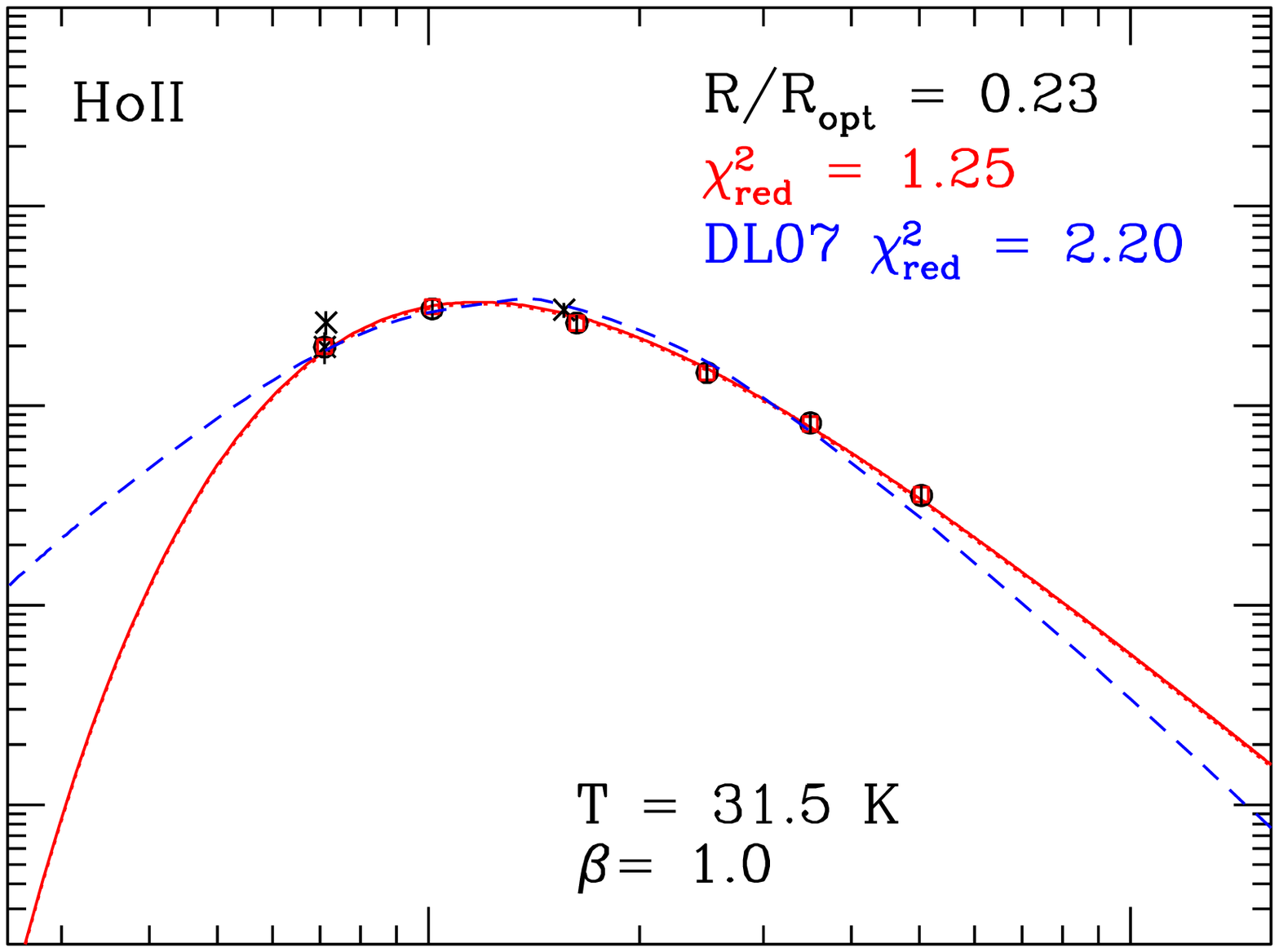}
\hspace{-0.045\linewidth}
\includegraphics[width=0.22\linewidth,bb=18 308 588 716]{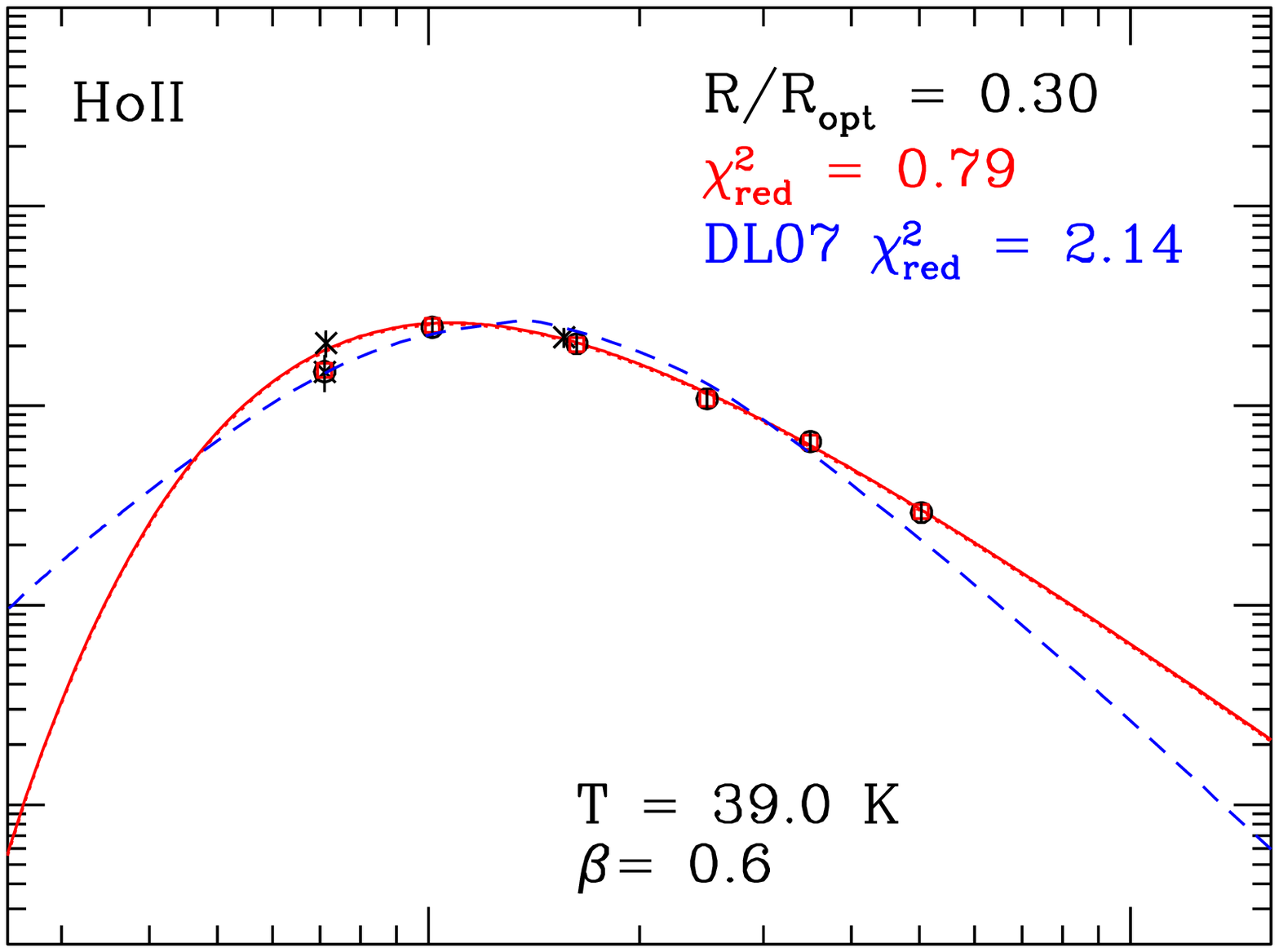}
\hspace{-0.045\linewidth}
\includegraphics[width=0.22\linewidth,bb=18 308 588 716]{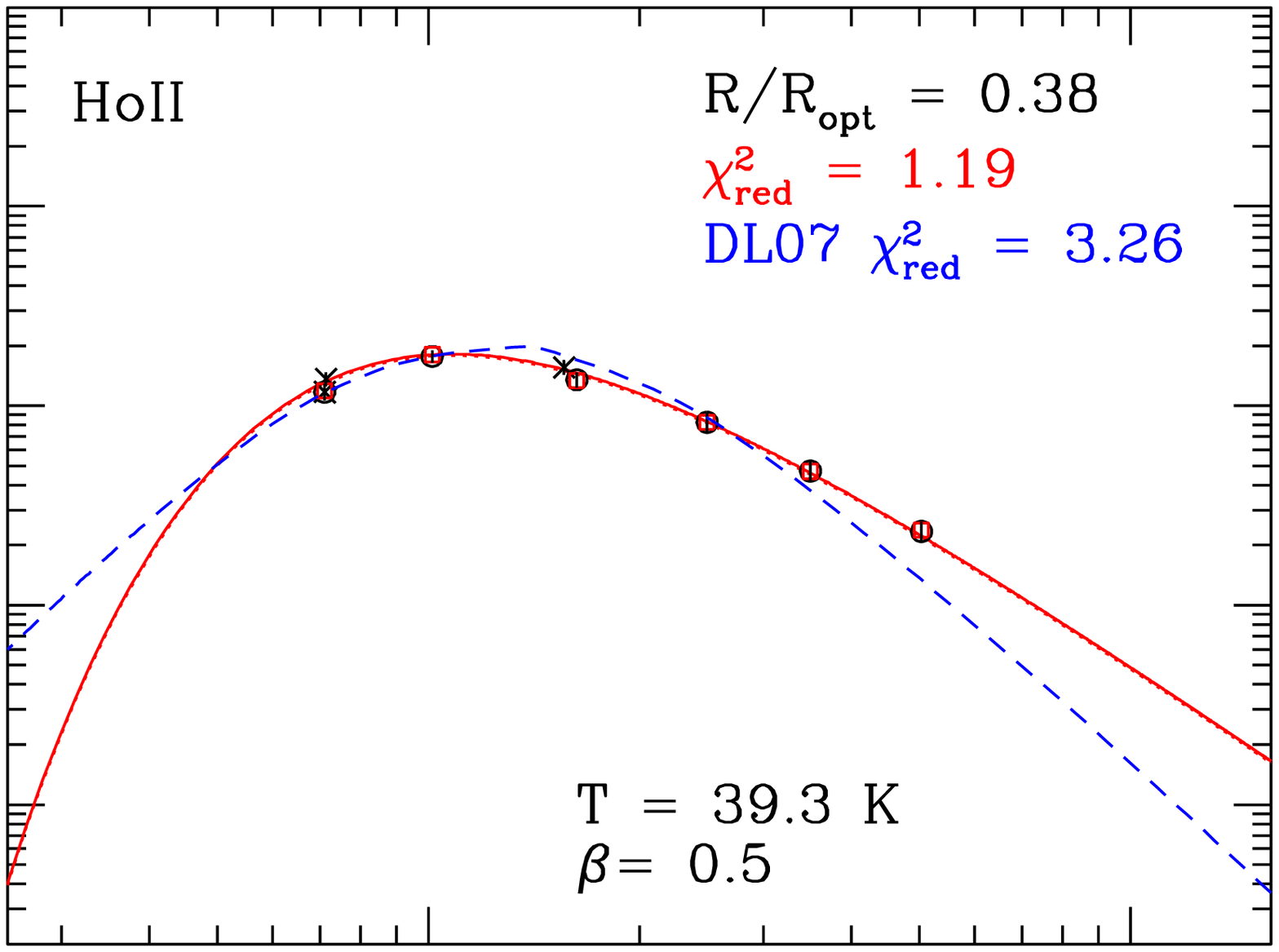}
}
}
\vspace{-1.15\baselineskip}
\centerline{
\hbox{
\includegraphics[width=0.22\linewidth,bb=18 308 588 716]{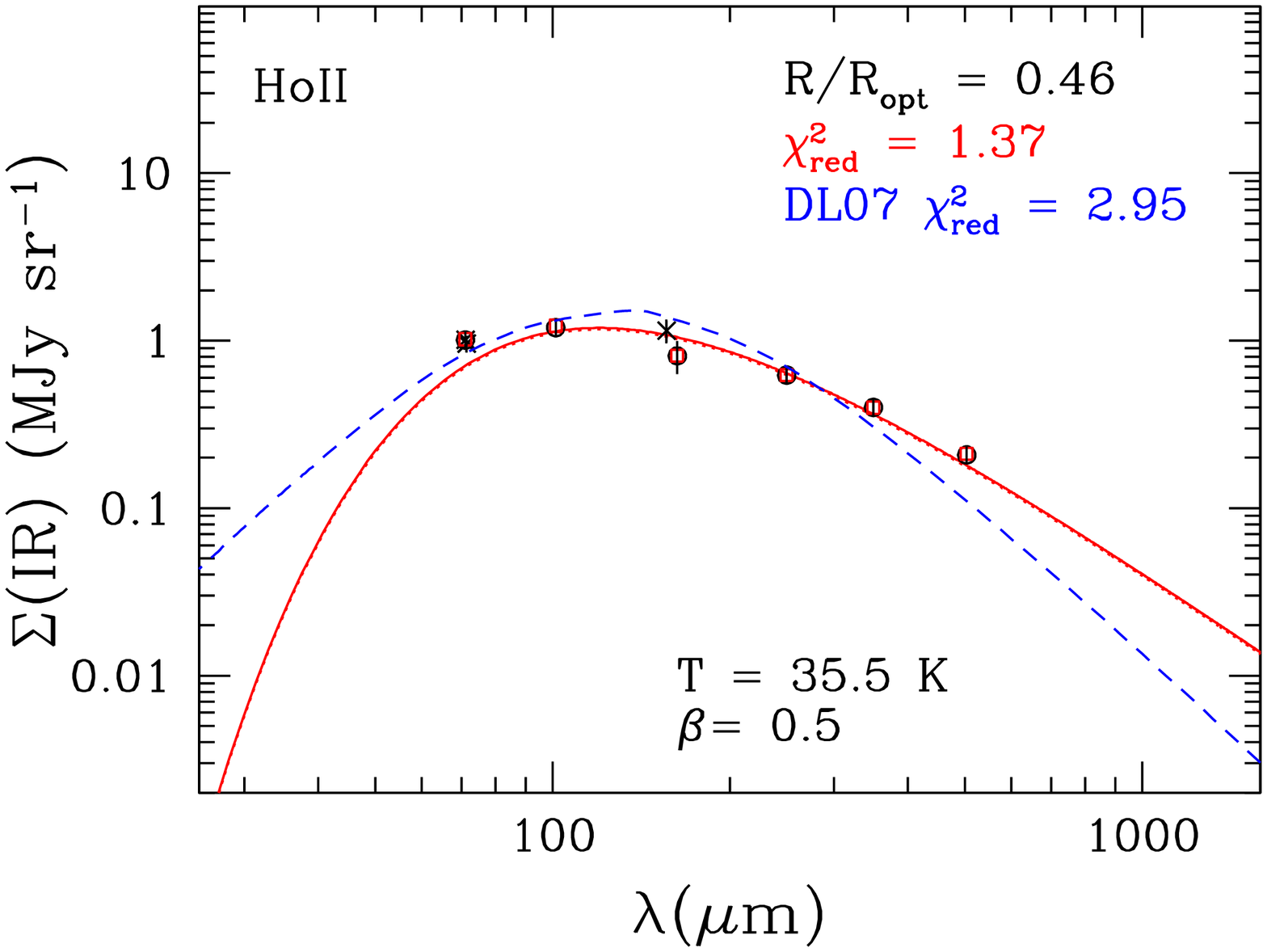}
\hspace{-0.045\linewidth}
\includegraphics[width=0.22\linewidth,bb=18 308 588 716]{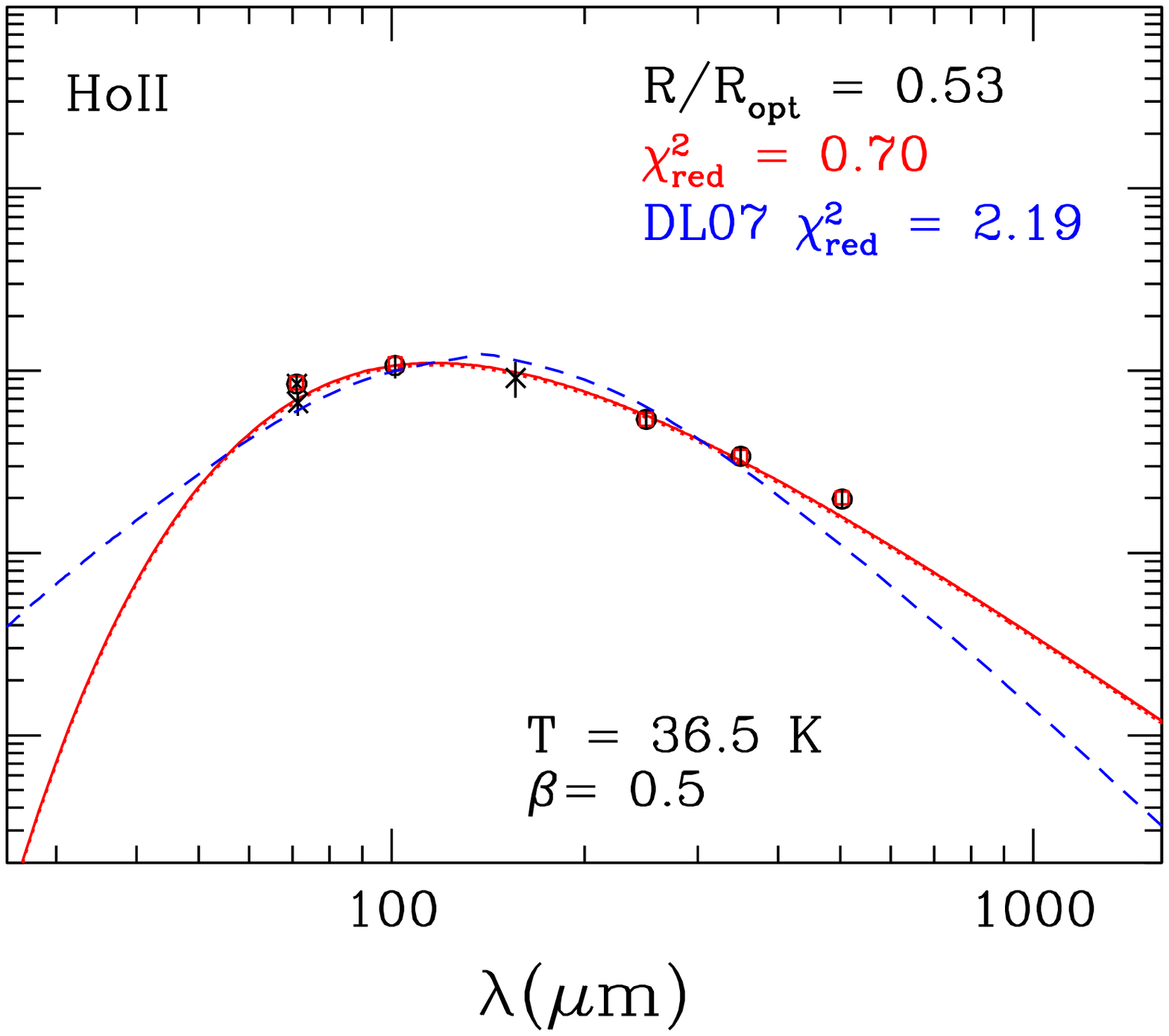}
\hspace{-0.045\linewidth}
\includegraphics[width=0.22\linewidth,bb=18 308 588 716]{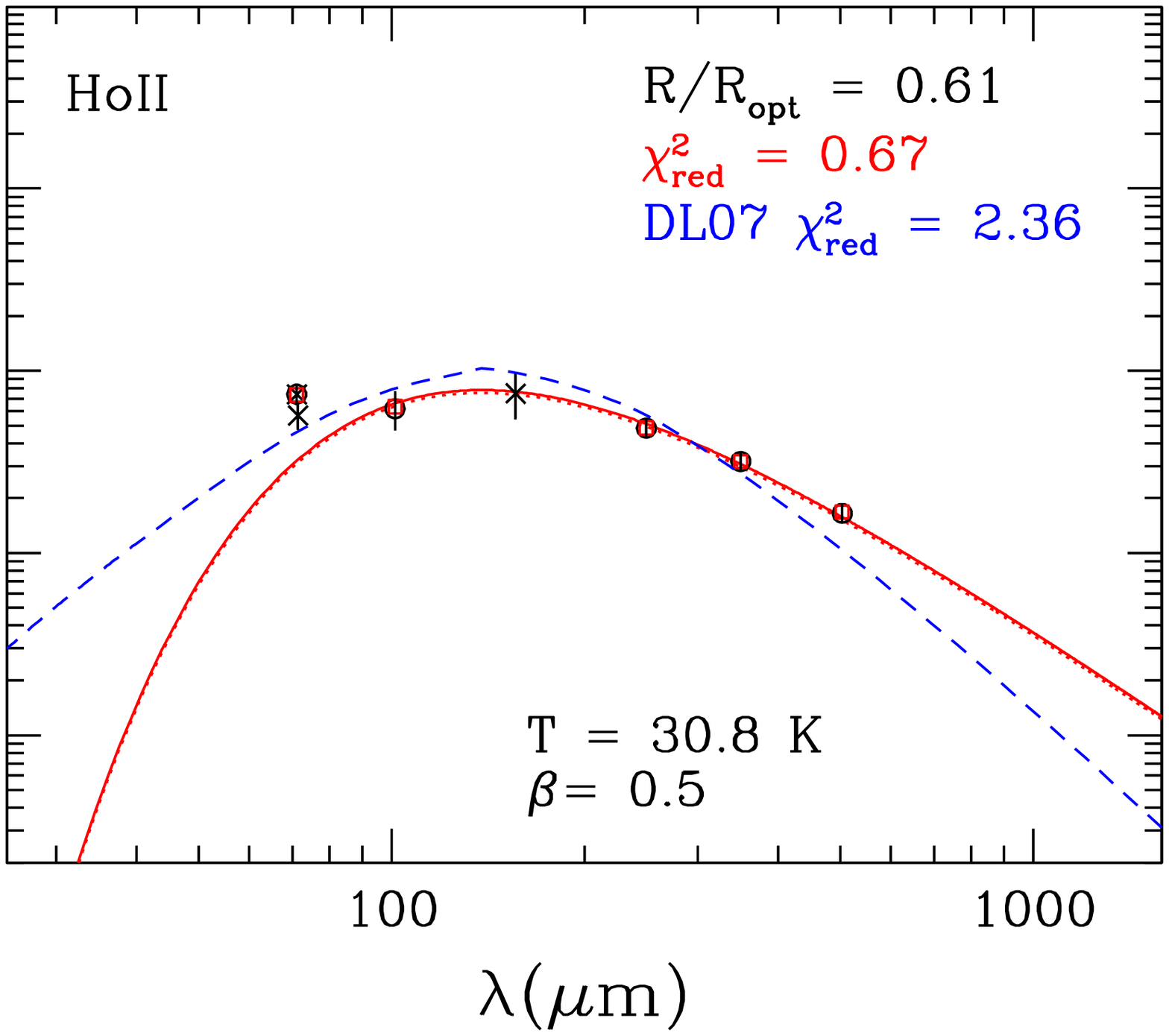}
\hspace{-0.045\linewidth}
\includegraphics[width=0.22\linewidth,bb=18 308 588 716]{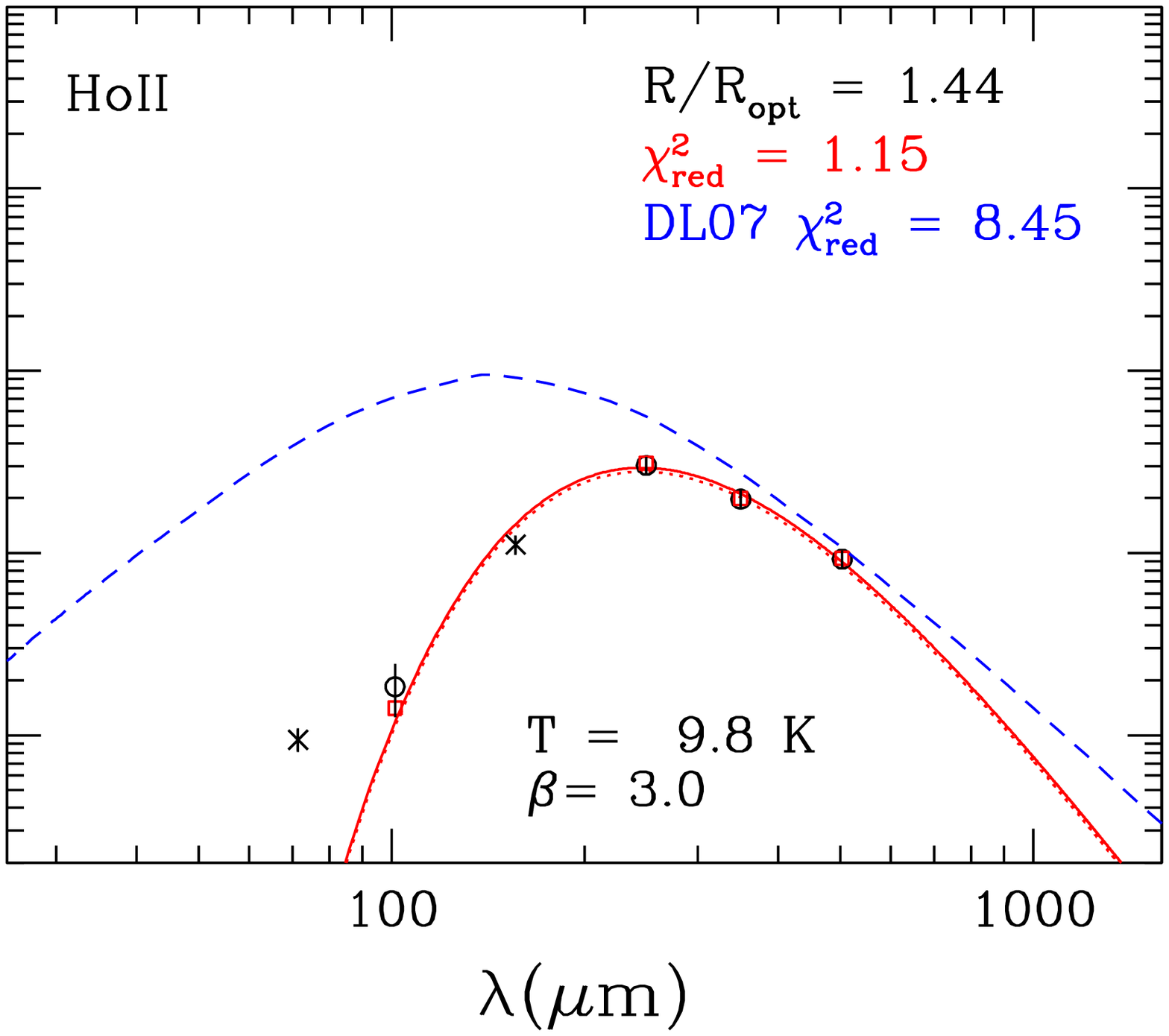}
}
}
\caption{{\bf (d)} as Fig. \ref{fig:radial}(a), but Holmberg\,II (Im) radial profiles. 
}
\end{figure*}

\begin{figure*}[!ht]
\centerline{
\hbox{
\includegraphics[width=0.45\linewidth,bb=18 308 588 716]{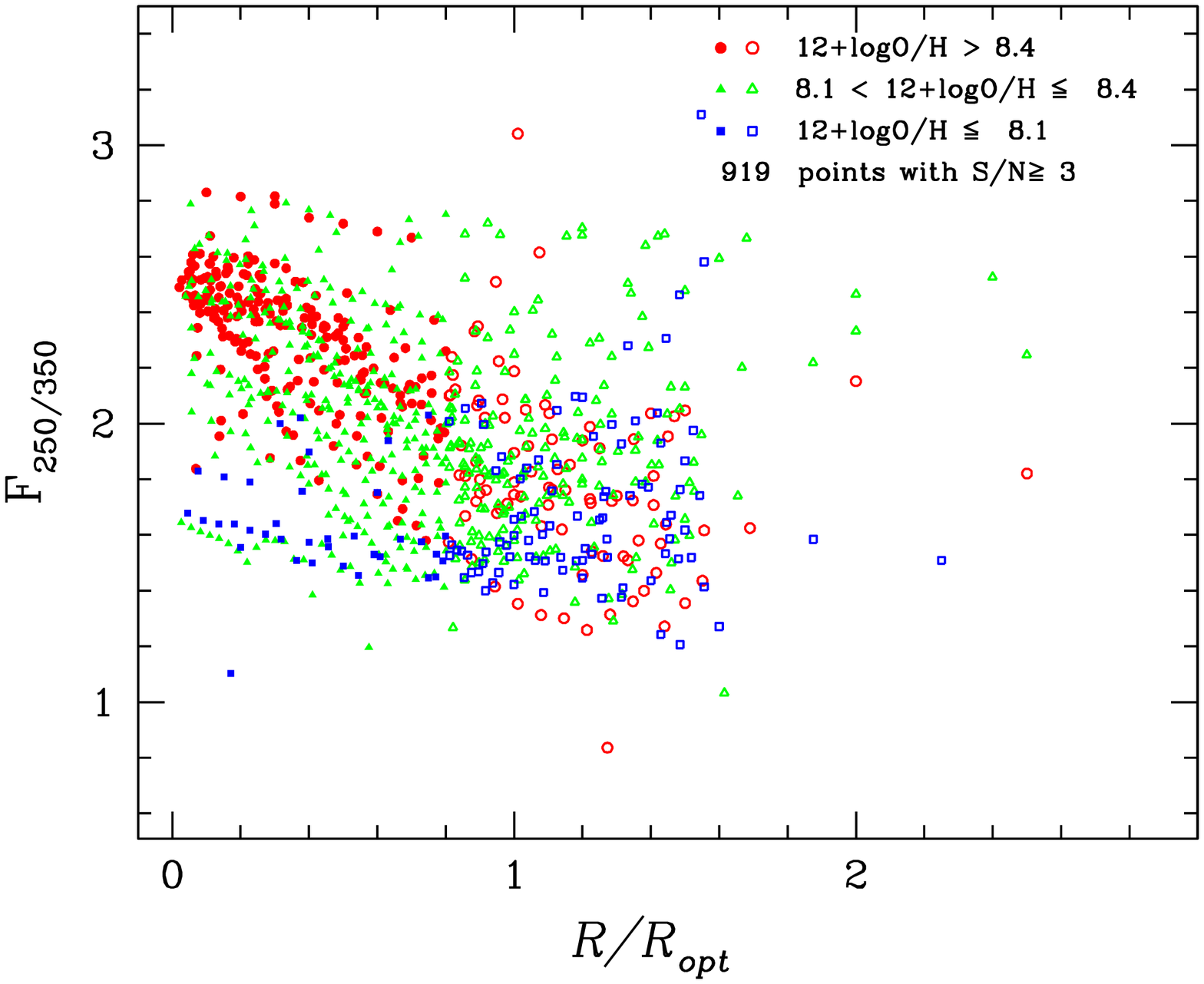}
%\hspace{0.1\linewidth}
\hspace{0.05\linewidth}
\includegraphics[width=0.45\linewidth,bb=18 308 588 716]{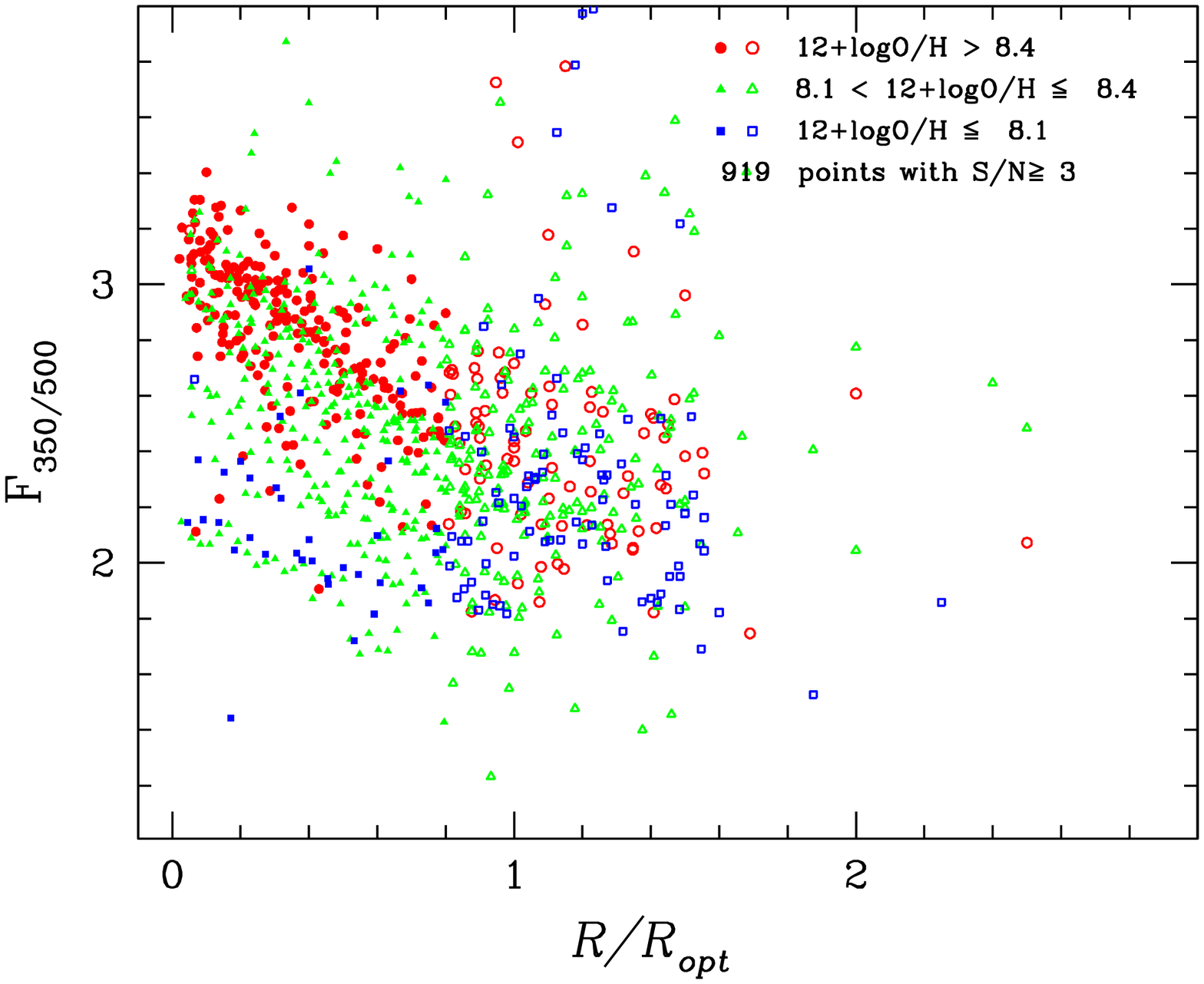}
}
}
\caption{SPIRE colors F$_{250}$/F$_{350}$ (left panel)
and F$_{350}$/F$_{500}$ (right) vs. galactocentric distance \ropt.
As in Fig. \ref{fig:radial_tauropt},
points are coded by their oxygen abundance, with filled symbols corresponding
to $R$/\ropt$\leq$0.8 and open ones to larger radii.
\label{fig:radial_colors}
}
\end{figure*}


\begin{thebibliography}{}

\bibitem[Agladze et al.(1996)]{agladze96} Agladze, N.~I., 
Sievers, A.~J., Jones, S.~A., Burlitch, J.~M., 
\& Beckwith, S.~V.~W.\ 1996, \apj, 462, 1026 

%\bibitem[Alatalo et al.(2014)]{alatolo14} Alatalo, K., Nyland, 
%K., Graves, G., et al.\ 2014, \apj, 780, 186 

\bibitem[Alatalo et al.(2011)]{alatolo11} Alatalo, K., Blitz, L., 
Young, L.~M., et al.\ 2011, \apj, 735, 88 

\bibitem[Alton et 
al.(1998)]{alton98} Alton, P.~B., Trewhella, M., Davies, J.~I., et al.\ 1998, \aap, 335, 807 

\bibitem[Amblard et al.(2014)]{amblard14} Amblard, A., Riguccini, 
L., Temi, P., et al.\ 2014, \apj, 783, 135 

\bibitem[Aniano et al.(2012)]{aniano12} Aniano, G., Draine, 
B.~T., Calzetti, D., et al.\ 2012, \apj, 756, 138 

\bibitem[Auld et al.(2013)]{auld13} Auld, R., Bianchi, S., 
Smith, M.~W.~L., et al.\ 2013, \mnras, 428, 1880 

\bibitem[Begum et al.(2006)]{begum06} Begum, A., Chengalur, 
J.~N., Karachentsev, I.~D., Kaisin, S.~S., 
\& Sharina, M.~E.\ 2006, \mnras, 365, 1220 

\bibitem[Bendo et al.(2012)]{bendo12} Bendo, G.~J., Boselli, 
A., Dariush, A., et al.\ 2012, \mnras, 419, 1833 

\bibitem[Bendo et 
al.(2010)]{bendo10} Bendo, G.~J., Wilson, C.~D., Pohlen, M., et al.\ 2010, \aap, 518, L65 

\bibitem[Bernard et 
al.(1999)]{bernard99} Bernard, J.~P., Abergel, A., Ristorcelli, I., et al.\ 1999, \aap, 347, 640 

\bibitem[Bianchi(2007)]{bianchi07} Bianchi, S.\ 2007, \aap, 471, 765 

\bibitem[Bianchi(2013)]{bianchi13} Bianchi, S.\ 2013, \aap, 552, A89 

\bibitem[Bianchi 
\& Xilouris(2011)]{bianchi11} Bianchi, S., \& Xilouris, E.~M.\ 2011, \aap, 531, L11 

\bibitem[Bigiel et al.(2008)]{bigiel08} Bigiel, F., Leroy, A., 
Walter, F., et al.\ 2008, \aj, 136, 2846 

\bibitem[Bigiel et al.(2010)]{bigiel10} Bigiel, F., Leroy, A., 
Seibert, M., et al.\ 2010, \apjl, 720, L31 % FUV disk in M83

%\bibitem[Bland-Hawthorn et al.(2005)]{bland05} Bland-Hawthorn, 
%J., Vlaji{\'c}, M., Freeman, K.~C., \& Draine, B.~T.\ 2005, \apj, 629, 239 

\bibitem[Boquien et al.(2011)]{boquien11} Boquien, M., Calzetti, 
D., Combes, F., et al.\ 2011, \aj, 142, 111 

\bibitem[Boselli et 
al.(2012)]{boselli12} Boselli, A., Ciesla, L., Cortese, L., et al.\ 2012, \aap, 540, A54 

\bibitem[Boselli et al.(2010)]{boselli10} Boselli, A., Eales, S., 
Cortese, L., et al.\ 2010, \pasp, 122, 261 % HRS

\bibitem[Bosma(1981)]{bosma81} Bosma, A.\ 1981, \aj, 86, 1791 

%\bibitem[Buat et al.(2010)]{buat10} Buat, V., Giovannoli, E., 
%Burgarella, D., et al.\ 2010, \mnras, 409, L1 

% \bibitem[Calzetti et al.(2000)]{calzetti00} Calzetti, D., Armus, 
% L., Bohlin, R.~C., et al.\ 2000, \apj, 533, 682 

% \bibitem[Cardelli et al.(1989)]{cardelli89} Cardelli, J.~A., 
% Clayton, G.~C., \& Mathis, J.~S.\ 1989, \apj, 345, 245 

%\bibitem[Chabrier(2003)]{chabrier03} Chabrier, G.\ 2003, \pasp, 
%115, 763 

%\bibitem[Chary \& Elbaz(2001)]{chary01} Chary, R., \& Elbaz, D.\ 2001, \apj, 556, 562 

\bibitem[Ciesla et 
al.(2014)]{ciesla14} Ciesla, L., Boquien, M., Boselli, A., et al.\ 2014, \aap, 565, A128 

\bibitem[Compi{\`e}gne et 
al.(2011)]{compiegne11} Compi{\`e}gne, M., Verstraete, L., Jones, A., et al.\ 2011, \aap, 525, A103 

%\bibitem[Corbelli et 
%al.(2012)]{corbelli12} Corbelli, E., Bianchi, S., Cortese, L., et al.\ 2012, \aap, 542, A32 

\bibitem[Cortese et 
al.(2012)]{cortese12} Cortese, L., Boissier, S., Boselli, A., et al.\ 2012, \aap, 544, A101 

\bibitem[Cortese et al.(2014)]{cortese14} Cortese, L., Fritz, J., 
Bianchi, S., et al.\ 2014, \mnras, in press (arXiv:1402.4524) 

\bibitem[Coupeaud et 
al.(2011)]{coupeaud11} Coupeaud, A., Demyk, K., Meny, C., et al.\ 2011, \aap, 535, A124 

\bibitem[Crosthwaite et al.(2000)]{crosthwaite00} Crosthwaite, L.~P., 
Turner, J.~L., \& Ho, P.~T.~P.\ 2000, \aj, 119, 1720 

\bibitem[Croxall et al.(2009)]{croxall09} Croxall, K.~V., van 
Zee, L., Lee, H., et al.\ 2009, \apj, 705, 723 

\bibitem[Daigle et al.(2006)]{daigle06} Daigle, O., Carignan, 
C., Amram, P., et al.\ 2006, \mnras, 367, 469 

\bibitem[Dale et al.(2012)]{dale12} Dale, D.~A., Aniano, G., 
Engelbracht, C.~W., et al.\ 2012, \apj, 745, 95 

\bibitem[Dale et al.(2009)]{dale09} Dale, D.~A., Cohen, S.~A., 
Johnson, L.~C., et al.\ 2009, \apj, 703, 517 % LVL

\bibitem[Dale et al.(1997)]{dale97} Dale, D.~A., Giovanelli, 
R., Haynes, M.~P., et al.\ 1997, \aj, 114, 455 

\bibitem[Dale 
\& Helou(2002)]{dale02} Dale, D.~A., \& Helou, G.\ 2002, \apj, 576, 159 

\bibitem[Dale et al.(2001)]{dale01} Dale, D.~A., Helou, G., 
Contursi, A., Silbermann, N.~A., \& Kolhatkar, S.\ 2001, \apj, 549, 215 

\bibitem[Davies et al.(2009)]{davies09} Davies, R.~I., 
Maciejewski, W., Hicks, E.~K.~S., et al.\ 2009, \apj, 702, 114 

\bibitem[de Blok et al.(2008)]{deblok08} de Blok, W.~J.~G., 
Walter, F., Brinks, E., et al.\ 2008, \aj, 136, 2648 

\bibitem[De Geyter et 
al.(2013)]{degeyter13} De Geyter, G., Baes, M., Fritz, J., \& Camps, P.\ 2013, \aap, 550, A74 

\bibitem[Draine 
\& Li(2001)]{draine01} Draine, B.~T., \& Li, A.\ 2001, \apj, 551, 807 

\bibitem[Draine 
\& Li(2007)]{dl07} Draine, B.~T., \& Li, A.\ 2007, \apj, 657, 810 % Post-Spitzer dust

\bibitem[Draine et al.(2014)]{draine14} Draine, B.~T., Aniano, 
G., Krause, O., et al.\ 2014, \apj, 780, 172 % M31

%\bibitem[Eales et al.(2012)]{eales12} Eales, S., Smith, 
%M.~W.~L., Auld, R., et al.\ 2012, \apj, 761, 168 

%\bibitem[Elmegreen 
%\& Hunter(2006)]{elmegreen06} Elmegreen, B.~G., \& Hunter, D.~A.\ 2006, \apj, 636, 712 

%\bibitem[Engelbracht et al.(2008)]{engelbracht08} Engelbracht, C.~W.,
%Rieke, G.~H., Gordon, K.~D., et al.\ 2008, \apj, 678, 804

\bibitem[Engelbracht et 
al.(2010)]{engelbracht10} Engelbracht, C.~W., Hunt, L.~K., Skibba, R.~A., et al.\ 2010, \aap, 518, L56 

%\bibitem[Erwin et al.(2005)]{erwin05} Erwin, P., Beckman, 
%J.~E., \& Pohlen, M.\ 2005, \apjl, 626, L81 

%\bibitem[Erwin et al.(2008)]{erwin08} Erwin, P., Pohlen, M., 
%\& Beckman, J.~E.\ 2008, \aj, 135, 20 

% \bibitem[Fathi et al.(2012)]{fathi12} Fathi, K., Gatchell, M., 
% Hatziminaoglou, E., \& Epinat, B.\ 2012, \mnras, 423, L112 

\bibitem[Fazio et al.(2004)]{fazio04} Fazio, G.~G., Hora, 
J.~L., Allen, L.~E., et al.\ 2004, \apjs, 154, 10 

\bibitem[Finkbeiner et al.(1999)]{finkbeiner99} Finkbeiner, D.~P., 
Davis, M., \& Schlegel, D.~J.\ 1999, \apj, 524, 867 

%\bibitem[Franceschini et al.(2008)]{franceschini08} 
%Franceschini, A., Rodighiero, G., \& Vaccari, M.\ 2008, \aap, 487, 837 

\bibitem[Freeman(1970)]{freeman70} Freeman, K.~C.\ 1970, \apj, 
160, 811 

\bibitem[Galametz et al.(2012)]{galametz12} Galametz, M., 
Kennicutt, R.~C., Albrecht, M., et al.\ 2012, \mnras, 425, 763 
% Mapping the cold dust temperatures and masses of nearby KINGFISH galaxies with Herschel

\bibitem[Galametz et al.(2014)]{galametz14} Galametz, M., 
Albrecht, M., Kennicutt, R., et al.\ 2014, \mnras, in press (arXiv:1401.3693) 
% Dissecting the origin of the submillimeter emission in nearby galaxies with Herschel and LABOCA

\bibitem[Galametz et 
al.(2011)]{galametz11} Galametz, M., Madden, S.~C., Galliano, F., et al.\ 2011, \aap, 532, A56 

\bibitem[Galliano et 
al.(2011)]{galliano11} Galliano, F., Hony, S., Bernard, J.-P., et al.\ 2011, \aap, 536, A88 
% LMC beta=1.7

%\bibitem[Garn 
%\& Best(2010)]{garn10a} Garn, T., \& Best, P.~N.\ 2010, \mnras, 409, 421 

%\bibitem[Garn et al.(2010)]{garn10b} Garn, T., Sobral, D., 
%Best, P.~N., et al.\ 2010, \mnras, 402, 2017 

\bibitem[Gil de Paz et al.(2005)]{gildepaz05} Gil de Paz, A., 
Madore, B.~F., Boissier, S., et al.\ 2005, \apjl, 627, L29  %NGC 4625

\bibitem[Giovanelli et al.(1994)]{giovanelli94} Giovanelli, R., 
Haynes, M.~P., Salzer, J.~J., et al.\ 1994, \aj, 107, 2036 

\bibitem[Gordon et 
al.(2010)]{gordon10} Gordon, K.~D., Galliano, F., Hony, S., et al.\ 2010, \aap, 518, L89 
% LMC beta=1.5

\bibitem[Gordon et al.(2011)]{gordon11} Gordon, K.~D., Meixner, 
M., Meade, M.~R., et al.\ 2011, \aj, 142, 102 % SAGE

\bibitem[Gordon et al.(2014)]{gordon14} Gordon, K.~D., et al.\ 2014, in preparation

\bibitem[Griffin et 
al.(2010)]{griffin10} Griffin, M.~J., Abergel, A., Abreu, A., et al.\ 2010, \aap, 518, L3 

\bibitem[Griffin et al.(2013)]{griffin13} Griffin, M.~J., North, 
C.~E., Schulz, B., et al.\ 2013, \mnras, 434, 992 

\bibitem[Groves et al.(2012)]{groves12} Groves, B., Krause, O., 
Sandstrom, K., et al.\ 2012, \mnras, 426, 892 

%\bibitem[Gruppioni et 
%al.(2010)]{gruppioni10} Gruppioni, C., et al.\ 2010, \aap, 518, L27 

%\bibitem[Hao 
%\& Yuan(2013)]{hao13} Hao, J.-M., \& Yuan, Y.-F.\ 2013, \apj, 772, 42 

\bibitem[Haas et 
al.(1998)]{haas98} Haas, M., Lemke, D., Stickel, M., et al.\ 1998, \aap, 338, L33 

%\bibitem[Hauser \& Dwek(2001)]{hauser01} Hauser, M.~G., \& Dwek, E.\ 2001, \araa, 39, 249 

%\bibitem[Hayward et al.(2012)]{hayward12} Hayward, C.~C., 
%Jonsson, P., Kere{\v s}, D., et al.\ 2012, \mnras, 424, 951 

\bibitem[Heckman et al.(2000)]{heckman00} Heckman, T.~M., 
Lehnert, M.~D., Strickland, D.~K., \& Armus, L.\ 2000, \apjs, 129, 493 

\bibitem[Hildebrand(1983)]{hildebrand83} Hildebrand, R.~H.\ 1983, 
\qjras, 24, 267 

\bibitem[Holmberg(1958)]{holmberg58} Holmberg, E.\ 1958, 
Meddelanden fran Lunds Astronomiska Observatorium Serie II, 136, 1 

\bibitem[Holwerda et 
al.(2012a)]{holwerda12a} Holwerda, B.~W., Bianchi, S., B{\"o}ker, T., et al.\ 2012a, \aap, 541, L5 

\bibitem[Holwerda et 
al.(2005)]{holwerda05} Holwerda, B.~W., Gonz{\'a}lez, R.~A., van der Kruit, P.~C., \& Allen, R.~J.\ 2005, \aap, 444, 109 

\bibitem[Holwerda et al.(2012b)]{holwerda12b} Holwerda, B.~W., 
Pirzkal, N., \& Heiner, J.~S.\ 2012b, \mnras, 427, 3159 % FUV and HI extended disks

\bibitem[Hubble(1926)]{hubble26} Hubble, E.~P.\ 1926, \apj, 64, 
321 

\bibitem[Hunt et 
al.(2004)]{hunt04} Hunt, L.~K., Pierini, D., \& Giovanardi, C.\ 2004, \aap, 414, 905 

\bibitem[Juvela et 
al.(2013)]{juvela13} Juvela, M., Montillaud, J., Ysard, N., \& Lunttila, T.\ 2013, \aap, 556, A63 

\bibitem[Juvela 
\& Ysard(2012)]{juvela12} Juvela, M., \& Ysard, N.\ 2012, \aap, 541, A33 

\bibitem[Karim et al.(2011)]{karim11} Karim, A., Schinnerer, 
E., Mart{\'{\i}}nez-Sansigre, A., et al.\ 2011, \apj, 730, 61 

\bibitem[Kelly et al.(2012)]{kelly12} Kelly, B.~C., Shetty, R., 
Stutz, A.~M., et al.\ 2012, \apj, 752, 55 

%\bibitem[Kennicutt(1998)]{kennicutt98} Kennicutt, R.~C., Jr.\ 1998, \araa, 36, 189 

\bibitem[Kennicutt et al.(2003)]{kennicutt03} Kennicutt, R.~C., 
Jr., Armus, L., Bendo, G., et al.\ 2003, \pasp, 115, 928 

\bibitem[Kennicutt et al.(2011)]{kennicutt11} Kennicutt, R.~C., 
Calzetti, D., Aniano, G., et al.\ 2011, \pasp, 123, 1347  % KINGFISH

%\bibitem[Kennicutt et al.(2008)]{kennicutt08} Kennicutt, R.~C., Jr.,
%Lee, J.~C., Funes, S.~J., Jos{\'e} G., Sakai, S., \& Akiyama, S.\ 2008,
%\apjs, 178, 247 % 11 HUGS Halpha

\bibitem[Kirkpatrick et al.(2013)]{kirkpatrick13} Kirkpatrick, A., 
Calzetti, D., Galametz, M., et al.\ 2013, \apj, 778, 51  % 500 um excess

\bibitem[Kirkpatrick et al.(2014)]{kirkpatrick14} Kirkpatrick, A., 
Calzetti, D., Galametz, M., et al.\ 2014, \apj, submitted  % dust heating

%\bibitem[Kormendy(2013)]{kormendy13} Kormendy, J.\ 2013, 
%arXiv:1311.2609 

\bibitem[Kreckel et al.(2014)]{kreckel14} Kreckel, K., Armus, L., 
Groves, B., et al.\ 2014, \apj, 790, 26 

\bibitem[Laureijs et 
al.(2000)]{laureijs00} Laureijs, R.~J., Watson, D., Metcalfe, L., et al.\ 2000, \aap, 359, 900 

%\bibitem[Lee et al.(2006)]{lee06} Lee, H., Skillman, E.~D., Cannon,
%J.~M., et al.\ 2006, \apj, 647, 970

\bibitem[Leroy et al.(2008)]{leroy08} Leroy, A.~K., Walter, F., 
Brinks, E., et al.\ 2008, \aj, 136, 2782 

\bibitem[Li et al.(2013)]{li13} Li, Y., Bresolin, F., 
\& Kennicutt, R.~C., Jr.\ 2013, \apj, 766, 17 

\bibitem[Li 
\& Draine(2001)]{li01} Li, A., \& Draine, B.~T.\ 2001, \apj, 554, 778 

%\bibitem[Magdis et al.(2011)]{magdis11} Magdis, G.~E., Daddi, 
%E., Elbaz, D., et al.\ 2011, \apjl, 740, L15 

\bibitem[Mathis et 
al.(1983)]{mathis83} Mathis, J.~S., Mezger, P.~G., \& Panagia, N.\ 1983, \aap, 128, 212 

%\bibitem[McMullin et al.(2007)]{mcmullin07}
%McMullin, J. P., Waters, B., Schiebel, D., Young, W., Golap, K. 2007,
%Astronomical Data Analysis Software and Systems XVI, 376, 127

\bibitem[Mennella et al.(1998)]{mennella98} Mennella, V., Brucato, 
J.~R., Colangeli, L., et al.\ 1998, \apj, 496, 1058 

\bibitem[Mennella et 
al.(1995)]{mennella95} Mennella, V., Colangeli, L., \& Bussoletti, E.\ 1995, \aap, 295, 165 

\bibitem[Meny et 
al.(2007)]{meny07} Meny, C., Gromov, V., Boudet, N., et al.\ 2007, \aap, 468, 171 

%\bibitem[Minchev et 
%al.(2012)]{minchev12} Minchev, I., Famaey, B., Quillen, A.~C., et al.\ 2012, \aap, 548, A126 

\bibitem[Moriondo et 
al.(1998)]{moriondo98} Moriondo, G., Giovanardi, C., \& Hunt, L.~K.\ 1998, \aaps, 130, 81 

\bibitem[Moustakas 
\& Kennicutt(2006)]{moustakas06} Moustakas, J., \& Kennicutt, R.~C., Jr.\ 2006, \apj, 651, 155 

\bibitem[Moustakas et al.(2010)]{moustakas10} Moustakas, J., 
Kennicutt, R.~C., Jr., Tremonti, C.~A., et al.\ 2010, \apjs, 190, 233 

\bibitem[Mu{\~n}oz-Mateos et al.(2007)]{munoz07} 
Mu{\~n}oz-Mateos, J.~C., Gil de Paz, A., Boissier, S., et al.\ 2007, \apj, 
658, 1006 

\bibitem[Mu{\~n}oz-Mateos et al.(2009a)]{munoz09a} 
Mu{\~n}oz-Mateos, J.~C., Gil de Paz, A., Boissier, S., et al.\ 2009a, \apj, 
701, 1965 % Paper II

\bibitem[Mu{\~n}oz-Mateos et al.(2009b)]{munoz09b} 
Mu{\~n}oz-Mateos, J.~C., Gil de Paz, A., Zamorano, J., et al.\ 2009b, \apj, 
703, 1569  % Paper I

\bibitem[Mu{\~n}oz-Mateos et al.(2013)]{munoz13} 
Mu{\~n}oz-Mateos, J.~C., Sheth, K., Gil de Paz, A., et al.\ 2013, \apj, 
771, 59 

\bibitem[Murphy et al.(2008)]{murphy08} Murphy, E.~J., Helou, 
G., Kenney, J.~D.~P., Armus, L., \& Braun, R.\ 2008, \apj, 678, 828 

\bibitem[Nguyen et 
al.(2010)]{nguyen10} Nguyen, H.~T., Schulz, B., Levenson, L., et al.\ 2010, \aap, 518, L5 

\bibitem[Noeske et al.(2007)]{noeske07} Noeske, K.~G., Weiner, B.~J.,
Faber, S.~M., et al.\ 2007, \apjl, 660, L43

\bibitem[Oh et al.(2008)]{oh08} Oh, S.-H., de Blok, 
W.~J.~G., Walter, F., Brinks, E., 
\& Kennicutt, R.~C., Jr.\ 2008, \aj, 136, 2761 

\bibitem[Ott(2010)]{hipe}
Ott, S. 2010, ASP Conference Series, 434, 139 

%\bibitem[Pappalardo et 
%al.(2012)]{pappalardo12} Pappalardo, C., Bianchi, S., Corbelli, E., et al.\ 2012, \aap, 545, A75 

\bibitem[Paradis et 
al.(2009)]{paradis09} Paradis, D., Bernard, J.-P., \& M{\'e}ny, C.\ 2009, \aap, 506, 745 

\bibitem[Paradis et 
al.(2012)]{paradis12} Paradis, D., Paladini, R., Noriega-Crespo, A., et al.\ 2012, \aap, 537, A113 

\bibitem[Pellegrini et al.(2013)]{pellegrini13} Pellegrini, E.~W., 
Smith (PI, J.~D., Wolfire, M.~G., et al.\ 2013, \apjl, 779, L19 

\bibitem[Pilbratt et 
al.(2010)]{pilbratt10} Pilbratt, G.~L., Riedinger, J.~R., Passvogel, T., et al.\ 2010, \aap, 518, L1 

\bibitem[Pilyugin 
\& Thuan(2005)]{pt05} Pilyugin, L.~S., \& Thuan, T.~X.\ 2005, \apj, 631, 231 

\bibitem[Pilyugin et al.(2007)]{pilyugin07} Pilyugin, L.~S., 
Thuan, T.~X., \& V{\'{\i}}lchez, J.~M.\ 2007, \mnras, 376, 353 % IC 342

\bibitem[Planck Collaboration et 
al.(2011a)]{planck11a} Planck Collaboration, Ade, P.~A.~R., Aghanim, N., et al.\ 2011a, \aap, 536, A17 

\bibitem[Planck Collaboration et 
al.(2011b)]{planck11b} Planck Collaboration, Ade, P.~A.~R., Aghanim, N., et al.\ 2011b, \aap, 536, A22 

\bibitem[Planck Collaboration et 
al.(2011c)]{planck11c} Planck Collaboration, Ade, P.~A.~R., Aghanim, N., et al.\ 2011c, \aap, 536, A23 

\bibitem[Planck Collaboration et 
al.(2011d)]{planck11d} Planck Collaboration, Ade, P.~A.~R., Aghanim, N., et al.\ 2011d, \aap, 536, A24 

\bibitem[Planck Collaboration et 
al.(2014)]{planck14} Planck Collaboration, Ade, P.~A.~R., Aghanim, N., et al.\ 2014, \aap, 564, A45 

\bibitem[Poglitsch et 
al.(2010)]{poglitsch10} Poglitsch, A., Waelkens, C., Geis, N., et al.\ 2010, \aap, 518, L2 

%\bibitem[Pohlen et 
%al.(2010)]{pohlen10} Pohlen, M., Cortese, L., Smith, M.~W.~L., et al.\ 2010, \aap, 518, L72 

%\bibitem[Pohlen 
%\& Trujillo(2006)]{pohlen06} Pohlen, M., \& Trujillo, I.\ 2006, \aap, 454, 759 

\bibitem[Pollack et al.(1994)]{pollack94} Pollack, J.~B., 
Hollenbach, D., Beckwith, S., et al.\ 1994, \apj, 421, 615 

\bibitem[Rieke et al.(2004)]{rieke04} Rieke, G.~H., Young, 
E.~T., Engelbracht, C.~W., et al.\ 2004, \apjs, 154, 25 

%\bibitem[Ro{\v s}kar et al.(2008)]{roskar08} Ro{\v s}kar, R., 
%Debattista, V.~P., Stinson, G.~S., et al.\ 2008, \apjl, 675, L65 

\bibitem[Roussel(2013)]{roussel13} Roussel, H.\ 2013, \pasp, 125, 
1126 

\bibitem[Roussel et al.(2006)]{roussel06} Roussel, H., Helou, G., 
Smith, J.~D., et al.\ 2006, \apj, 646, 841 

%\bibitem[Rowlands et al.(2012)]{rowlands12} Rowlands, K., Dunne, 
%L., Maddox, S., et al.\ 2012, \mnras, 419, 2545 

\bibitem[Salim et al.(2007)]{salim07} Salim, S., Rich, R.~M., Charlot,
S., et al.\ 2007, \apjs, 173, 267

%\bibitem[Salpeter(1955)]{salpeter55} Salpeter, E.~E.\ 1955, \apj, 
%121, 161 

%\bibitem[Santini et 
%al.(2010)]{santini10} Santini, P., Maiolino, R., Magnelli, B., et al.\ 2010, \aap, 518, L154 

\bibitem[Sauvage 
\& Thuan(1992)]{sauvage92} Sauvage, M., \& Thuan, T.~X.\ 1992, \apjl, 396, L69 

%\bibitem[Sellwood(2013)]{sellwood13} Sellwood, J.~A.\ 2013, 
%Reviews of Modern Physics, in press (arXiv:1310.0403) 

%\bibitem[Sellwood 
%\& Binney(2002)]{sellwood02} Sellwood, J.~A., \& Binney, J.~J.\ 2002, \mnras, 336, 785 

\bibitem[Shetty et al.(2009a)]{shetty09a} Shetty, R., Kauffmann, 
J., Schnee, S., \& Goodman, A.~A.\ 2009a, \apj, 696, 676 

\bibitem[Shetty et al.(2009b)]{shetty09b} Shetty, R., Kauffmann, 
J., Schnee, S., Goodman, A.~A., \& Ercolano, B.\ 2009b, \apj, 696, 2234 

\bibitem[Skibba et al.(2011)]{skibba11} Skibba, R.~A., 
Engelbracht, C.~W., Dale, D., et al.\ 2011, \apj, 738, 89 

\bibitem[Smith et al.(2012)]{smith12} Smith, M.~W.~L., Eales, 
S.~A., Gomez, H.~L., et al.\ 2012, \apj, 756, 40  % M31

\bibitem[Stepnik et 
al.(2003)]{stepnik03} Stepnik, B., Abergel, A., Bernard, J.-P., et al.\ 2003, \aap, 398, 551 

\bibitem[Storchi-Bergmann et al.(1994)]{storchi94} 
Storchi-Bergmann, T., Calzetti, D., \& Kinney, A.~L.\ 1994, \apj, 429, 572 

\bibitem[Tabatabaei et 
al.(2014)]{tabata14} Tabatabaei, F.~S., Braine, J., Xilouris, E.~M., et al.\ 2014, \aap, 561, A95 

\bibitem[Tamburro et al.(2008)]{tamburro08} Tamburro, D., Rix, 
H.-W., Walter, F., et al.\ 2008, \aj, 136, 2872 

\bibitem[Vader et al.(1993)]{vader93} Vader, J.~P., Frogel, 
J.~A., Terndrup, D.~M., \& Heisler, C.~A.\ 1993, \aj, 106, 1743 

%\bibitem[van den Bergh(1988)]{vandenbergh88} van den Bergh, S.\ 1988, 
%\pasp, 100, 344 

%\bibitem[van der Kruit 
%\& Freeman(2011)]{vdkruit11} van der Kruit, P.~C., \& Freeman, K.~C.\ 2011, \araa, 49, 301 

\bibitem[Verstappen et 
al.(2013)]{verstappen13} Verstappen, J., Fritz, J., Baes, M., et al.\ 2013, \aap, 556, A54 

\bibitem[Walter et al.(2008)]{walter08} Walter, F., Brinks, E., 
de Blok, W.~J.~G., et al.\ 2008, \aj, 136, 2563 

\bibitem[Weingartner 
\& Draine(2001)]{weingartner01} Weingartner, J.~C., \& Draine, B.~T.\ 2001, \apj, 548, 296 

\bibitem[Williams et al.(2009)]{williams09} Williams, B.~F., 
Dalcanton, J.~J., Dolphin, A.~E., Holtzman, J., 
\& Sarajedini, A.\ 2009, \apjl, 695, L15 

\bibitem[Xilouris et al.(1999)]{xilouris99} Xilouris, E.~M., 
Byun, Y.~I., Kylafis, N.~D., Paleologou, E.~V., \& Papamastorakis, J.\ 1999, \aap, 344, 868 

\bibitem[Ysard et 
al.(2012)]{ysard12} Ysard, N., Juvela, M., Demyk, K., et al.\ 2012, \aap, 542, A21 

\bibitem[Zibetti 
\& Groves(2011)]{zibetti11} Zibetti, S., \& Groves, B.\ 2011, \mnras, 417, 812 

\end{thebibliography}
\end{document}